\documentclass[11pt,oneside,a4paper,nofootinbib,longbibliography,superscriptaddress]{revtex4}

\usepackage[a4paper, top=2cm, bottom=2cm, left=2cm, right=2cm]{geometry}
\usepackage[linktocpage, colorlinks, citecolor=blue, linkcolor=blue, urlcolor=blue]{hyperref}
\usepackage{graphics}
\usepackage{subcaption} 
\usepackage[
 format = hang, 
 justification = centerlast, 
 textfont = it, 
 labelfont = bf,
 figurename = Figure,
 tablename = Table
 ]{caption}
\usepackage{xcolor}
\usepackage{enumerate}
\usepackage{multirow}
\usepackage{ragged2e}
\usepackage{float}
\usepackage{tabularx}
\usepackage[utf8]{inputenc}
\usepackage{MyCommands}
\usepackage{quoting,xparse}
\usepackage{setspace}
\usepackage{siunitx}
\usepackage{calc}
\usepackage{titlesec}
\usepackage[default]{gillius}

\titleformat{\section}{\raggedright\normalfont \Large \bfseries}{\thesection}{1em}{}

\renewcommand\thesection{\arabic{section}}

\numberwithin{equation}{section}

\renewcommand{\thefigure}{\textcolor{blue}{\arabic{figure}}}
\renewcommand{\figurename}{\textcolor{black}{\bfseries Figure}}

\renewcommand{\thetable}{\textcolor{blue}{\arabic{table}}}
\renewcommand{\tablename}{\textcolor{black}{\bfseries Table}}

\newcommand{\asection}[2]{
\setcounter{section}{#1}
\addtocounter{section}{-1}
\section{#2}
}

\begin{document}
\pagenumbering{gobble}
\title{Review on $f(\Q)$ Gravity}

\author{Lavinia Heisenberg}
\thanks{}
\email{heisenberg@thpyhs.uni-heidelberg.de}
\affiliation{Institut f\"{u}r Theoretische Physik, Philosophenweg 16, 69120 Heidelberg, Germany}

\begin{abstract}\noindent
     Recent years have witnessed a rise in interest in the geometrical trinity of General Relativity and its extensions. This interest has been fuelled by novel insights into the nature of gravity, the possibility to address computational and conceptual questions---such as the determination of black hole entropy or the definition of gravitational energy-momentum---from a new perspective. In particular, $f(\Q)$ gravity has also inspired numerous works on black holes, wormholes, and cosmology. In the latter case,  $f(\Q)$ models have the potential to elucidate phenomena in both early and late-time cosmology without necessitating the inclusion of dark energy, the inflaton field, or dark matter. Particularly noteworthy is the role of $f(\Q)$ theories in addressing cosmological tensions, presenting exciting possibilities for reshaping our understanding of gravity and its manifestations in cosmology. The emergence of intriguing new black hole solutions and the potential existence of wormhole solutions suggest the presence of novel physics within the realm of strong gravity. These phenomena have become increasingly measurable only in recent times, opening up exciting avenues for further exploration and discovery.
This review is tailored to students and researchers alike. It offers a self-contained and pedagogical introduction to metric-affine geometry--The mathematical foundation and indispensable tool upon which the geometrical trinity of General Relativity as well as its various extensions are built. 
 \end{abstract}

\clearpage
\maketitle
\pagenumbering{gobble}

\newpage
\tableofcontents
\newpage

\pagenumbering{roman} 
\section*{\Huge List of Symbols}\addcontentsline{toc}{section}{List of Symbols}

\bgroup 
\def\arraystretch{1.5} 

\begin{table}[H]
\raggedright
\begin{tabular}{p{0.33\textwidth}<{\raggedright}p{0.66\textwidth}<{\justifying}}
\raggedright
{\large\textbf{Manifolds and other Spaces}} &	\\ 
$\M$			& Spacetime manifold (connected, usually $4$-dimensional) \\
$\partial\M$    & Boundary of the manifold $\M$	\\
$T_p\M$, $T^*_p\M$ & Tangent and cotangent space at point $p\in\M$\\
$T\M$, $T^*\M$	& Tangent and cotangent bundle to $\M$ \\
$\Sigma$		& Co-dimension one hypersurface which is embedded in $\M$ (usually $3$-dimensional and spacelike) \\
$\mathbb{R}^n$	& Euclidean space of dimension $n$ \\
$\mathbb{S}^2$	& Topological $2$-sphere                                   
\end{tabular}
\end{table}

\begin{table}[H]
\raggedright
\begin{tabular}{p{0.33\textwidth}<{\raggedright}p{0.66\textwidth}<{\justifying}}
\raggedright
{\large\textbf{Maps}} &	\\ 
$\gamma$		& Curve on $\M$ defined as $\gamma:[0,1]\to\M$ \\
$\phi$			& Diffeomorphism, $\phi:\M\to\M$ \\
$\phi_{s}$	& $1$-parameter family of diffeomorphisms $\phi_{s}:\bbR\times\M\to\M$ with $\phi_{0} = \textsf{id}$ \\
$f$				& Real scalar field, $f:\M\to \bbR$
\end{tabular}
\end{table}

\begin{table}[H]
\raggedright
\begin{tabular}{p{0.33\textwidth}<{\raggedright}p{0.66\textwidth}<{\justifying}}
\raggedright
{\large\textbf{Connections}} &	\\ 
$\G{\alpha}{\mu\nu}$	& General affine connection (can have curvature, torsion, and non-metricity) \\
$\LC{\alpha}{\mu\nu}$ 	& Levi-Civita connection (unique torsionless and metric-compatible connection) \\
$\Lambda\ud{\mu}{\nu}$ 	& Matrix belonging to the real four-dimensional general linear group, $GL(4, \mathbb R)$ (used to parametrize flat connections) \\
$\zeta^\mu$				& Collection of four arbitrary functions (Not a vector! Used to parametrize flat and torsionless connections) 
\end{tabular}
\end{table}

\begin{table}[H]
\raggedright
\begin{tabular}{p{0.33\textwidth}<{\raggedright}p{0.66\textwidth}<{\justifying}}
\raggedright
{\large\textbf{Metrics}} &	\\ 
$\eta_{\mu\nu}$	& Minkowski metric, signature $(-,+,+,+)$ \\
$g_{\mu\nu}$	& Spacetime metric, signature $(-,+,+,+)$ \\
$g$, $|g|$		& Determinant of $g_{\mu\nu}$; Modulus of $g$ \\
$h_{ab}$		& Intrinsic metric of $\partial\M$ or $\Sigma$, signature $(+,+,+)$ if $\partial\M$ (or $\Sigma$) is spacelike, $(-,+,+)$ if $\partial\M$ (or $\Sigma$) is timelike \\
$h$, $|h|$		& Determinant of $h_{ab}$\,; Modulus of $h$
\end{tabular}
\end{table}

\begin{table}[H]
\raggedright
\begin{tabular}{p{0.33\textwidth}<{\raggedright}p{0.66\textwidth}<{\justifying}}
\raggedright
{\large\textbf{Derivative Operators}} &	\\ 
$\partial_\mu$	& Coordinate derivative, does not transform covariantly \\
$\mathcal{L}_v$	& Lie derivative along the vector field $v$ \\
$\nabla_\mu$	& Covariant derivative with respect to a general affine connection $\G{\alpha}{\mu\nu}$ \\
$\D_\mu$		& Covariant derivative with respect to the Levi-Civita connection $\LC{\alpha}{\mu\nu}$ \\
$\delta$		& Variational derivative                         
\end{tabular}
\end{table}

\begin{table}[H]
\raggedright
\begin{tabular}{p{0.33\textwidth}<{\raggedright}p{0.66\textwidth}<{\justifying}}
\raggedright
{\large\textbf{Tensor Fields}} &	\\ 
$R\ud{\alpha}{\mu\nu\rho}$ & Curvature tensor with respect to a general affine connection (sometimes also denoted by $R(\Gamma)\ud{\alpha}{\mu\nu\rho}$) \\
$\R\ud{\alpha}{\mu\nu\rho}$ & Curvature tensor with respect to the Levi-Civita connection (sometimes also denoted by $\R(g)\ud{\alpha}{\mu\nu\rho}$) \\
$\mathcal{H}\du{\alpha}{\mu\nu}$ & Hypermomentum of matter fields \\
$K\ud{\alpha}{\mu\nu}$	& Contortion tensor\\
$L\ud{\alpha}{\mu\nu}$	& Disformation tensor\\
$P\ud{\alpha}{\mu\nu}$, $\hat{P}\ud{\alpha}{\mu\nu}$ & Non-metricity conjugate\\
$Q_{\alpha\mu\nu}$		& Non-metricity tensor, $Q_{\alpha\mu\nu} \ce \nabla_\alpha g_{\mu\nu}$ \\
$S\du{\alpha}{\mu\nu}$, $\hat{S}\du{\alpha}{\mu\nu}$ & Torsion conjugate\\
$T\ud{\alpha}{\mu\nu}$	& Torsion tensor, $T\ud{\alpha}{\mu\nu} \ce 2\G{\alpha}{[\mu\nu]}$ \\
$G_{\mu\nu}$			& Einstein tensor, $G_{\mu\nu}\ce \R_{\mu\nu} - \frac12 \R\, g_{\mu\nu}$\\
$\mathcal{K}_{\mu\nu}$	& Extrinsic curvature tensor \\
$q_{\mu\nu}$, $\hat{q}_{\mu\nu}$			& Symmetric tensor defined by $\PD{\Q}{g^{\mu\nu}}$ while keeping $\nabla_\alpha g^{\mu\nu}$ fixed; Symmetric tensor defined with respect to $\hat{\Q}$. \\
$R_{\mu\nu}$			& Ricci tensor associated to the curvature tensor with respect to a general affine connection (sometimes also denoted by $R(\Gamma)_{\mu\nu}$), $R_{\mu\nu}\ce R\ud{\lambda}{\mu\lambda\nu}$ \\
$\R_{\mu\nu}$			& Ricci tensor associated to the curvature tensor with respect to the Levi-Civita connection (sometimes also denoted by $\R(g)_{\mu\nu}$), $\R_{\mu\nu}\ce \R\ud{\lambda}{\mu\lambda\nu}$ \\
$\T_{\mu\nu}$			& Energy-momentum tensor of matter fields
\end{tabular}
\end{table}

\begin{table}[H]
\raggedright
\begin{tabular}{p{0.33\textwidth}<{\raggedright}p{0.66\textwidth}<{\justifying}}
\raggedright
{\large\textbf{Vector Fields}} &	\\ 
$n^\mu$		& Unit normal vector to $\Sigma$ or $\partial\M$, normalized as $n^\mu n_\mu = -1$ if $\Sigma$ (or $\partial\M$) is spacelike, normalized as $n^\mu n_\mu = + 1$ if $\Sigma$ (or $\partial\M$) is timelike \\
$N^{a}$ 	& Shift vector field \\
$Q^\mu$, $\bar{Q}^\mu$	& The two independent traces of the non-metricity tensor, $Q^\mu \ce Q\udu{\mu}{\lambda}{\lambda}$ and $\bar{Q}^\mu \ce Q\du{\lambda}{\lambda\mu}$ \\
$T^\mu$		& Only trace of the torsion tensor, $T^\mu\ce T\ud{\lambda\mu}{\lambda}$
\end{tabular}
\end{table}

\begin{table}[H]
\raggedright
\begin{tabular}{p{0.33\textwidth}<{\raggedright}p{0.66\textwidth}<{\justifying}}
\raggedright
{\large\textbf{Scalar Fields}} &		\\ 
$\varepsilon$	& Norm of the normal vector $n^\mu$ to a co-dimension one hypersurface, $\varepsilon\ce n_\mu n^\mu$ and $\varepsilon = + 1$ if the hypersurface is timelike, $\varepsilon = -1$ if the hypersurface is spacelike\\
$N$ & Lapse function, assumed to be nowhere zero \\
$\mathcal{K}$	& Trace of the extrinsic curvature tensor, $\mathcal{K}\ce g^{\mu\nu}\mathcal{K}_{\mu\nu}$ \\                          
$\Q$, $\hat{\Q}$	& Non-metricity scalar of STEGR; Most general scalar which is quadratic in the non-metricity tensor (appears in STG	 and $\Q$ is a special case of~$\hat{\Q}$)\\
$R$, $\R$	& Ricci scalar of the affine connection, $R\ce g^{\mu\nu}R_{\mu\nu}$ (sometimes also denoted by $R(\Gamma)$); Ricci scalar of the Levi-Civita connection, $\R \ce g^{\mu\nu} \R_{\mu\nu}$ (sometimes also denoted by $\R(g)$)\\
$\mathbb{T}$, $\hat{\mathbb{T}}$		& Torsion scalar of TEGR; Most general scalar which is quadratic in the torsion tensor (appears in TG and $\mathbb{T}$ is a special case of $\hat{\mathbb{T}}$)\\
\end{tabular}
\end{table}

\begin{table}[H]
\raggedright
\begin{tabular}{p{0.33\textwidth}<{\raggedright}p{0.66\textwidth}<{\justifying}}
\raggedright
{\large\textbf{Other Symbolds}} &	\\ 
$\Lambda$	& Cosmological constant \\
$\ce$, $\ec$& Left side defined by right side; Right side defined by left side\\
$\equiv$	& Identity \\
$[\cdot,\cdot]$ 	& Commutator or Lie bracket \\
$\{\cdot,\cdot\}$ 	& Poisson bracket \\
$\tilde{T}$	& A tilde $\tilde{\phantom{.}}$ on top indicates that $T$ is a tensor density of weight $w=+1$
\end{tabular}
\end{table}

\egroup 

\newpage
\section*{\Huge Acronyms}\addcontentsline{toc}{section}{Acronyms}

\bgroup
\def\arraystretch{1.5}
\begin{table}[hbt!]
\raggedright
\begin{tabular}{p{0.33\textwidth}<{\raggedright}p{0.66\textwidth}<{\justifying}}
\raggedright
CG			&	\ul{C}oincident \ul{G}auge								\\
CGR			&	\ul{C}oincident \ul{G}eneral \ul{R}elativity				\\
EH			& 	\ul{E}instein-\ul{H}ilbert								\\
FCC			&	\ul{F}irst \ul{C}lass \ul{C}onstraint(s)					\\
GHY			&	\ul{G}ibbons-\ul{H}awking-\ul{Y}ork						\\
GR			&	\ul{G}eneral \ul{R}elativity								\\
GTEGR		& 	\ul{G}eneral \ul{T}eleparallel \ul{E}quivalent of \ul{G}eneral \ul{R}elativity\\
NGR			&	\ul{N}ewer \ul{G}eneral \ul{R}elativity					\\
SCC			&	\ul{S}econd \ul{C}lass \ul{C}onstraint(s)				\\
STEGR		&	\ul{S}ymmetric \ul{T}eleparallel \ul{E}quivalent of \ul{G}eneral \ul{R}elativity	\\	
STG			&	\ul{S}ymmetric \ul{T}eleparallel \ul{G}ravity	\\
TEGR		&	\ul{T}eleparallel \ul{E}quivalent of \ul{G}eneral \ul{R}elativity\\	
TG			&   \ul{T}eleparallel \ul{G}ravity  
\end{tabular}
\end{table}
\egroup

\newpage

\pagenumbering{gobble}
\pagenumbering{arabic}

\asection{1}{Introduction}\label{sec:Introduction}
In 1912, Einstein's study of static gravitational fields had led him to a bold hypothesis. A simple application of his equivalence principle, in conjunction with basic results of special relativity, suggested that the gravitational field is described by the metric tensor. He conjectured that this is also true beyond the static limit and thus embarked on a three year long journey, which culminated in November 1915 with the field equations of his General Relativity (GR). 

This feat was only possible after having learned what we nowadays call Riemannian geometry. Back then, this branch of mathematics was relatively new and many concepts we now take for granted were either not as clear-cut as they are now, or they were not even conceived yet. One such example is the concept of an affine connection, which was in part developed by mathematicians in response to the advent and success of GR. It is therefore not surprising that Einstein's original theory is based on the Riemann curvature tensor. This tensor is in fact fully determined by the metric and does not require the introduction of an independent affine connection.

In later years, Einstein would famously attempt the unification of GR and electromagnetism. By then, the concept of an affine connection had been introduced by mathematicians such as Weyl and Einstein made use of these new tools. Even tough his unification attempts were ultimately not successful, he developed the first theory where gravity is mediated by torsion, rather than by curvature~\cite{Einstein:1928}. This culminated in a whole class of so-called metric teleparallel theories of gravity~\cite{AldrovandiBook, Bahamonde:2021}.

Only decades later was it realized that teleparallel theories of gravity can also be formulated in flat, torsionless geometries, if one attributes gravitational phenomena to the so-called non-metricity tensor~\cite{Nester:1998}. Postulating that curvature vanishes, but allowing for torsion, or non-metricity, or both, leads to what we now call the geometric trinity of GR~\cite{Heisenberg:2018,BeltranJimenez:2019c, BeltranJimenez:2019d}: Three distinct but equivalent description of General Relativity. All these theories are rooted in the mathematical framework of metric-affine geometry~\cite{Hehl:1994, Blagojevi:2001}.

The geometric trinity, as well as its various extensions and modifications, have witnessed a rising interest and a flurry of research activities. Their popularity is due to two factors. First of all, having different but physically equivalent formulations of GR sheds new light on its foundations. It also allows to address old problems from a new perspective. For instance, issues regarding the definition of gravitational energy-momentum have gained new momentum due to developments in teleparallel theories of gravity~\cite{BeltranJimenez:2019b, Koivisto:2022, Gomes:2022, Gomes:2023}. So have questions regarding the computation of black hole entropy~\cite{BeltranJimenez:2018, BeltranJimenez:2019b, Heisenberg:2022b}. 

Secondly, the geometrical trinity has given rise to different extensions and modifications of gravity. There is a growing number of cosmological observations and tensions, which hint at physics beyond the standard $\Lambda$CDM model. While GR has passed every empirical test it has been subjected to, there remain phenomena which cannot be explained on the basis of GR alone. Most notably, the early- and late-time expansion of the universe requires the introduction of an inflaton field and dark energy, respectively. Furthermore, several observations strongly suggest the existence of dark matter. Rather than introducing new matter fields or exotic forms of energy, one can also attempt to explain these phenomena using modified theories of gravity. Indeed, a model known as $f(\Q)$ gravity has gained considerable popularity in the past couple of years and the bulk of the research efforts have been concentrated on cosmological applications~\cite{Lymperis:2022, Paul:2022, Narawade:2023, Narawade:2023b, Dimakis:2023}. This model has also been applied to large structure formation~\cite{Sokoliuk:2023}, the development of relativistic versions of Modified Newtonian Dynamics (MOND)~\cite{Milgrom:2019, DAmbrosio:2020b}, bouncing cosmologies~\cite{Bajardi:2020, Agrawal:2021, Gadbail:2023}, and even quantum cosmology~\cite{Dimakis:2021, Bajardi:2023}. A lot of effort has also gone into constraining or testing $f(\Q)$ models~\cite{Dialektopoulos:2019, Ayuso:2020, Barros:2020, Frusciante:2021, Aggarwal:2022, De:2022, Albuquerque:2022, Ferreira:2022, Koussour:2023, Najera:2023, Bouali:2023, Ferreira:2023, Subramaniam:2023b}. Extensions that involve incorporating boundary terms \cite{Capozziello:2023vne, De:2023xua, Paliathanasis:2023pqp} or non-minimally coupled scalar field \cite{Jarv:2018bgs, Harko:2018gxr} have also been explored.

Other very active area of research are black holes within $f(\Q)$ gravity~\cite{Banerjee:2021, Mustafa:2021, Parsaei:2022, Hassan:2022b, Hassan:2022, Hassan:2022c, Venkatesha:2023, Jan:2023, Godani:2023, Javed:2023, Mishra:2023, Chanda:2022}, modified stellar solutions~\cite{Wang:2021, Maurya:2022b, Maurya:2022, Errehymy:2022, Sokoliuk:2022, Calza:2022, Chanda:2022, Bhar:2023, Ditta:2023, Maurya:2023}, and wormholes~\cite{Banerjee:2021, Mustafa:2021, Parsaei:2022, Hassan:2022b, Hassan:2022, Hassan:2022c, Venkatesha:2023, Jan:2023, Godani:2023, Mustafa:2023, Mishra:2023}. Also in this regard, some thought has been given to how observational data could be used to constrain $f(\Q)$ gravity~\cite{DAgostino:2022}. The beyond-GR stellar solutions could play an important role in this regard.

However, $f(\Q)$ gravity, or teleparallel theories of gravity in general, have also stirred up new challenges. As a particular example we mention the Hamiltonian analysis of $f(\Q)$ gravity, which need to overcome certain technical challenges which may require new techniques~\cite{DAmbrosio:2020c, Hu:2022, DAmbrosio:2023}.

This review is dedicated to a pedagogical introduction into the subject of teleparallel theories of gravity and its extensions. The first two sections cover the necessary mathematical foundations, which are needed to formulate, understand, and work with teleparallel theories of gravity. Section~\ref{sec:Trinity} discusses the geometrical trinity of gravity in detail. In particular, we cover Einstein's original formulation of GR, the Teleparallel Equivalent of GR (TEGR), the Symmetric Teleparallel of Gravity (STEGR), Coincident GR (CGR), the General Teleparallel Equivalent of GR (GTEGR), theories of gravity which renounce the flatness condition, and finally we also discuss matter coupling. 

In section~\ref{sec:TrinityGeneralizations} we turn to modified theories of gravity, focusing mostly on general quadratic extensions of TEGR and STEGR. Non-linear extensions such as $f(\R)$, $f(\bbT)$, and $f(\bbG)$ are discussed only tangentially. An exception is made for $f(\Q)$ gravity, which is the main subject of section~\ref{sec:f(Q)}. A particular focus is laid on cosmology, black holes, and the Hamiltonian analysis as well as the open question regarding how many degrees of freedom the theory propagates. Finally, we conclude with a summary in section~\ref{sec:Summary}.

\newpage
\asection{2}{Fundamentals of Metric-Affine Geometries}\label{sec:Fundamentals}
This review is dedicated to the geometrical trinity of gravity and its extensions, with a particular focus on~$f(\Q)$ gravity. It is therefore indispensable to first talk about the geometric foundations which underpin these different descriptions of gravity. Our objective is to provide a didactical overview over the basic concepts of metric-affine geometry needed to formulate, understand, and work with the geometric trinity of gravity and its various extensions. We do not strive for mathematical rigour nor completeness and refer readers interested in mathematical aspects to the literature~\cite{HawkingEllisBook, WaldBook, BaezBook, NakaharaBook, CarrollBook}.

Our approach is to start from the most basic structure---a bare manifold with neither a metric nor a connection nor any other field defined on it---and to introduce step by step concepts and structures. The aim is to illustrate the meaning and physical relevance of each concept. This step-by-step approach also serves the purpose to highlight at which point it is necessary to introduce new structures---such as a metric or a connection---in order to deepen our description of the physical world.

\subsection{Manifolds, Diffeomorphisms, Curves, and Scalar Fields}
The world we inhabit seems to be four-dimensional and what we call ``spacetime'' is, at least in classical physics, well-described by a ``four-dimensional continuum'' in the sense that we need four numbers to label events. In pre-relativistic physics as well as in special relativistic physics, it is assumed that there is indeed a one-to-one correspondence between spacetime events and the topological space $\mathbb R^4$. Thus, a fixed spacetime topology is \textit{postulated}.

However, General Relativity (GR) teaches us that spacetime is dynamical and governed by its own field equations, in stark contrast to the absolute space and time of Newtonian physics or the rigid Minkowski spacetime of special relativity. Assuming any global properties of spacetime, such as its topology, would thus severely limit the possible solutions to Einstein's field equations and hide a wealth of interesting physical phenomena from us. No black hole or cosmological solutions could be found under such restrictive assumptions.

To overcome this obstacle, we introduce the concept of a \textbf{manifold}. As will be familiar to most readers, a real $n$-dimensional manifold\footnote{Technically, a real $n$-dimensional manifold $\M$ is a real $n$-dimensional topological space which is Hausdorff, paracompact, and locally homeomorphic to $n$-dimensional Euclidean space $\bbR^n$. However, most of these technical terms will not be relevant for us and we refer the mathematically inclined reader to standard books such as Hawking~\& Ellis~\cite{HawkingEllisBook} or Wald~\cite{WaldBook}. See also Carroll~\cite{CarrollBook} for a less technical introduction.} $\M$ can be thought of as a space which ``locally looks like Euclidean space~$\bbR^n$''. To be slightly more precise, $\M$ is a \textbf{topological space} which is \textit{locally homeomorphic} to the \textit{topological} space~$\bbR^n$. It is important to distinguish the topological space $\bbR^n$ from the vector space $\bbR^n$. In the former we can talk about points $p$ and their neighbourhoods, while in the latter we have a space with points in it which also satisfy certain axioms. Namely the axioms of how to do ``computations'' with these points such as add them together and multiply them by scalars. In other words, supplementing the topological space~$\bbR^n$ with vector space axioms turns points $p$ into vectors $\vec{p}$, loosely speaking.

In our current context, however, we are only interested in the topological aspects. Vectors will concern us in the next subsection. What our loose definition of a manifold means is therefore the following: The manifold $\M$ is a space inhabited by \textbf{points $\boldsymbol{p}$}. Since $\M$ is a topological space, the notion of neighbourhood is well-defined, which allows us to talk about what is happening ``locally'', i.e., in ``close proximity'' of the point $p$. By definition, there is a local homeomorphism, i.e., a map from one topological space to an other topological space, which maps $p$ and the points in its neighbourhood to a point and its neighbourhood in~$\bbR^n$. Since in~$\bbR^n$ we have a standard way of labelling each point unambiguously by $n$ numbers by laying out a coordinate grid, we have now a method to assign \textbf{coordinates} to the points in $\M$. 

Put simply: A bare manifold $\M$, i.e., a manifold \textit{without} any additional structure allows us first and foremost to assign coordinates to points $p$. These points have the physical interpretation of \textbf{spacetime events}. Of course, the assignment of coordinates is not unique. Even tough we have a standard way of labelling points with $n$ numbers in $\bbR^n$, two different persons might choose two different coordinate grids to do so. Let us denote a \textbf{coordinate system} by $\{x^\mu\}$. In order to relate one coordinate system to an other one, we introduce the concept of a \textbf{change of coordinates}. This is really a special case of the more general concept of a \textbf{diffeomorphism $\boldsymbol{\phi}$}, which is a smooth (i.e., infinitely differentiable) map $\phi:\N \to \M$ between the manifolds $\N$ and $\M$. A change of coordinates is then a diffeomorphism $\phi:\M \to \M$ between $\M$ and itself, which also has a smooth inverse and which maps $\{x^\mu\}$ onto $\{x'^\mu\} \ce \{\phi(x^\mu)\}$.

\begin{figure}[htb!]
	\centering
	\includegraphics[width=0.45\linewidth]{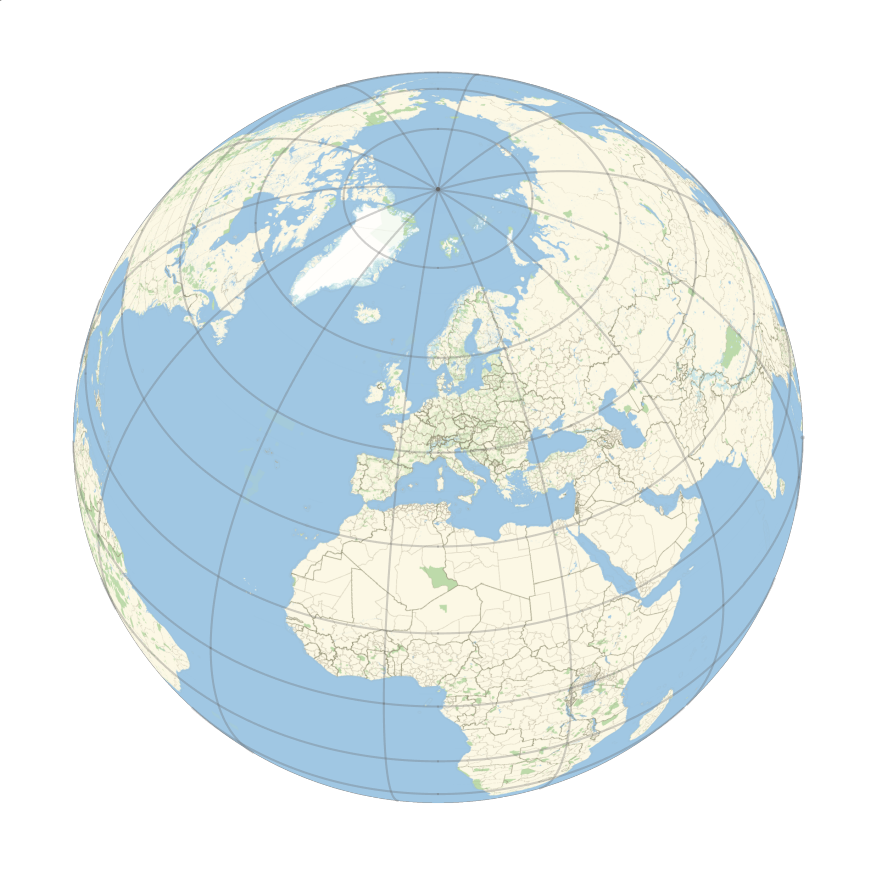}
	\caption{\protect The world modelled as the manifold $\mathbb S^2$. Longitude and latitude, which is just a specific choice for a coordinate system, allow us to label points (i.e., location) on the $2$-sphere. \hspace*{\fill}}
	\label{fig:1_World}
\end{figure}

Two points are worth emphasizing: In general, and in contrast with Newtonian or special relativistic physics, we need more than one coordinate system to cover a manifold $\M$. In fact, this is already the case for simple manifolds such as the example of $\bbS^2$ shown in Figure~\ref{fig:1_World}. In the technical jargon we say that we need an \textbf{atlas} in order to cover all of $\M$ with coordinates. However, this technical point will play no role for us and we will always simply talk about the coordinate system $\{x^\mu\}$. The second point is that coordinates have no intrinsic physical meaning and they only serve the purpose of labelling spacetime events. Ultimately, however, all physical observables have to be independent of the choice of coordinate system.

A common example of a manifold is the sphere $\mathbb S^2$. Figure~\ref{fig:1_World} shows a picture of our world modelled as a two-dimensional sphere. By introducing longitude and latitude we can label points, i.e., locations on the~$2$-sphere. 
However, longitude and latitude are just one particular example of a coordinate system, which is based on arbitrary choices\footnote{The prime meridian, which defines $0^\circ$ longitude, is defined as the one which passes through a certain point near the Royal Observatory in Greenwich, England. The equator is chosen to represent $0^\circ$ latitude.}. Other coordinate systems could be chosen without having any substantial effect, since coordinate systems are a mere matter of convenience and convention.

Given that coordinate transformations are generated by diffeomorphisms, which possess smooth inverses by definition, we can transform back and forth between coordinate systems without loosing information. Thus, all coordinate systems are on equal footing, reinforcing the notion that there are no preferred coordinate systems.

So far, we only have the bare manifold $\M$ at our disposal, without any additional structure or fields defined on it, and the concept of a diffeomorphism. There are two more concepts which are completely intrinsic (i.e., which do not require us to introduce any new structure) to $\M$ and which can be constructed using maps: Curves and scalar fields.

\begin{figure}[htb!]
	\centering
    \begin{subfigure}[b]{0.45\textwidth}
        \includegraphics[width=\textwidth]{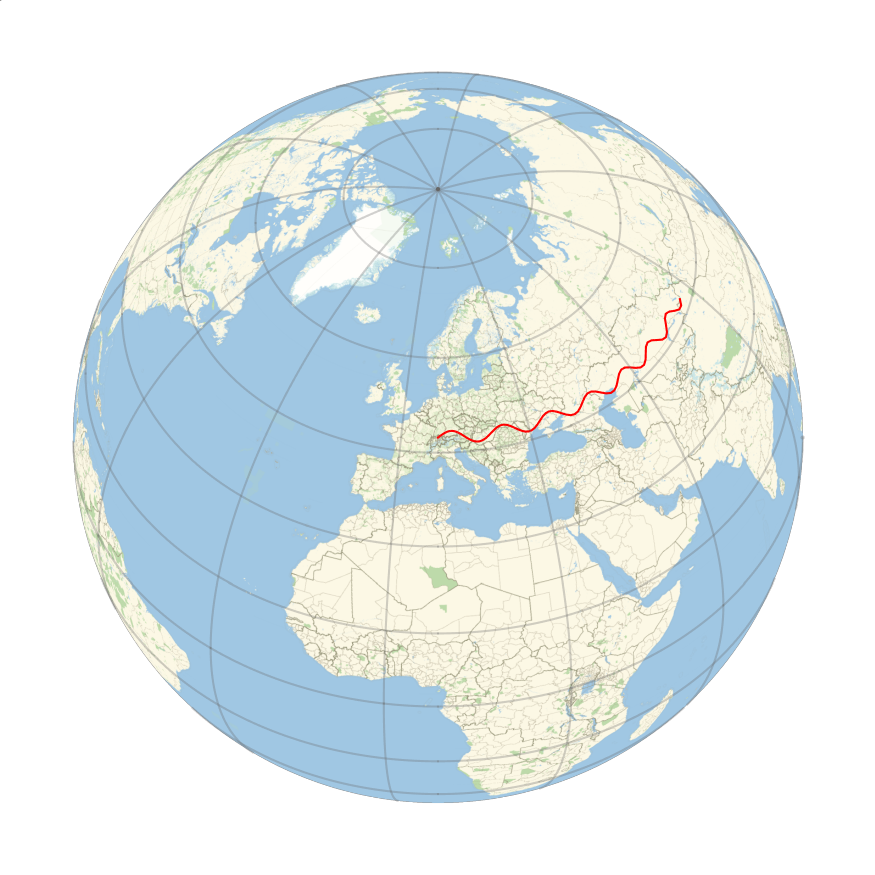}
        \caption{\protect A non-geodesic curve between two points in $\bbS^2$. \textcolor{white}{as a scalar field.} \hspace*{\fill}}
	    \label{fig:2_NonGeodesicPath}    
    \end{subfigure}
    \hfill
    \begin{subfigure}[b]{0.45\textwidth}
        \includegraphics[width=\textwidth]{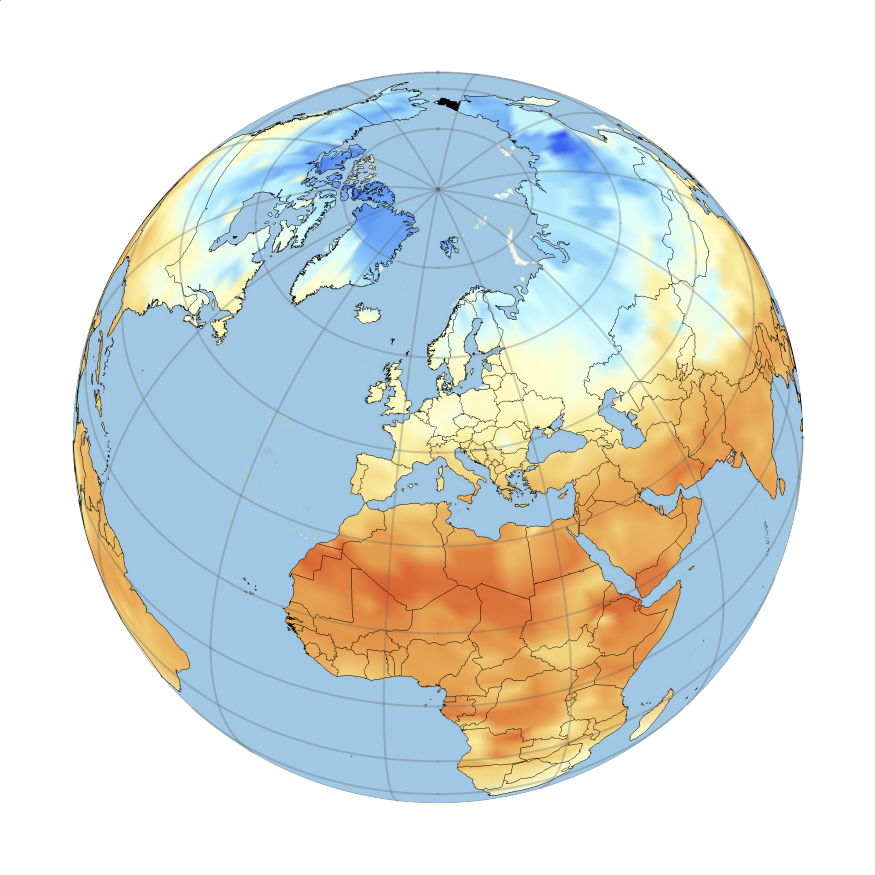}
        \caption{\protect The temperature field of the Earth as a scalar field. \hspace*{\fill}}
	    \label{fig:3_TemperatureField}          
    \end{subfigure}
    \caption{\protect Curves and scalar fields are two important examples of maps between manifolds (here between $\mathbb R$ and $\M$) which are of direct physical relevance. Curves are used to model observers and test particles, while scalar fields play an important role in inflationary cosmology, for instance. \hspace*{\fill}}
    \label{fig:MapsOnM}
\end{figure}

Curves provide us with a good model for \textbf{observers} and \textbf{test particles}. Mathematically, a \textbf{curve} is defined as a map $\gamma:I\to \M$ from an interval $I\subseteq \bbR$ into the manifold $\M$. We say that a curve is \textbf{parametrized by $\boldsymbol{s \in I}$}. What the map $\gamma$ ultimately does, is assign a point $\gamma(s)$ in $\M$ to every value of the parameter~$s$. Again, this concept is completely intrinsic to $\M$. Figure~\ref{fig:2_NonGeodesicPath} illustrates this concept and we emphasize that we cannot yet talk about ``the shortest path between two points'' (aka geodesics) since we have not yet introduced a metric. The concept of a metric is relegated to subsection~\ref{ssec:Metric} since, as we will see, many things can be done without having to resort to metrics.
We can translate the rather abstract notion of a curve as map from $I$ to $\M$ into the more familiar component language. All we need is the fact that (a) a coordinate system assigns to every point $p$ a set of $n$ numbers $x^\mu(p)$ and that (b) a curve assigns to every parameter value $s$ a point $\gamma(s)$ in $\M$. Thus, we can define the components of $\gamma$ with respect to the coordinate system $\{x^\mu\}$ as
\begin{align}\label{eq:DefComponentsCurve}
	\gamma^\mu(s) \ce x^\mu(\gamma(s))\,.
\end{align}
For all our purposes we can always assume that the curve $\gamma$ in question is differentiable. Therefore, we introduce for later convenient the shorthand notation
\begin{align}
	\dot{\gamma}^\mu(s) \ce \TD{\gamma^\mu(s)}{s}\,.
\end{align}
Now we turn to the second and last concept we can introduce on $\M$ using a map: A \textbf{scalar field $\boldsymbol{f}$} is a map $f:\M \to \bbR$. In simple words, the scalar field assigns to every point of $\M$ a real number. The temperature field of Earth shown in Figure~\ref{fig:3_TemperatureField} is an example of a scalar field. Again, the concept is intrinsic to $\M$ since we did not introduce any new structure.

In the next subsection we show how scalar fields and curves help us in defining vector fields, $1$-forms, and the spaces they live in. Namely the tangent and co-tangent space. Since these spaces are derived from $\M$ and other concepts intrinsic to $\M$, we ultimately find that tensor fields are concepts purely intrinsic to $\M$. We emphasize this point because in subsection~\ref{ssec:CovD} we will be forced for the first time to introduce a new structure which is not intrinsic or naturally present in $\M$. This refers to the concept of connection. Similarly, in subsection~\ref{ssec:Metric} we will be forced to recognize that also the metric is a concept which is not intrinsic or naturally present in $\M$. The affine structure described by the connection and the metric structure of a manifold described by the metric tensor are both concepts which have to be stipulated separately.

\subsection{Vector Fields, Tensor Fields, and Densities}\label{ssec:VFTFD}
\begin{figure}[htb!]
	\centering
	\includegraphics[width=0.45\linewidth]{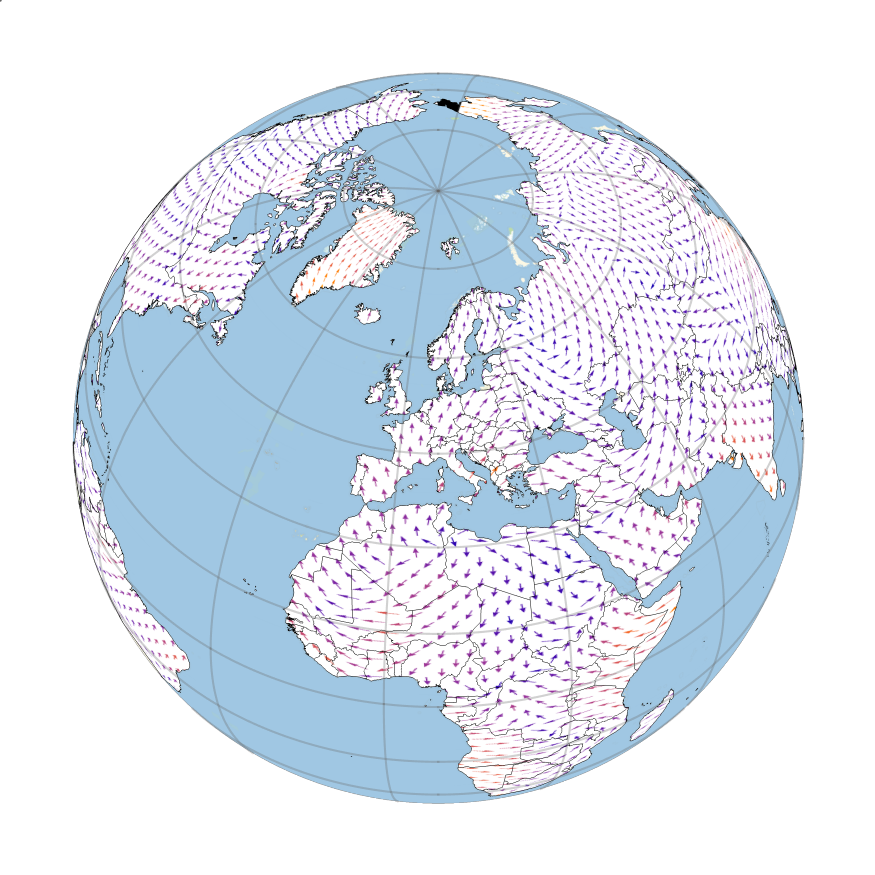}
	\caption{\protect The wind field on Earth is an example of a vector field. At each point on $\bbS^2$ we can assign a little arrow, with magnitude and direction, which is tangential to $\bbS^2$.  \hspace*{\fill}}
	\label{fig:4_WindField}
\end{figure}

Vector fields are omnipresent in physics and every physicist has an intuitive understanding as well as ample mental pictures of them. For instance, one picture that could come to mind is the one of a wind field on the surface of the Earth. Figure~\ref{fig:4_WindField} shows such a wind field, represented by an arrow at every point on~$\bbS^2$. 

How do we translate this intuitive mental picture of a vector field into mathematical language? How can we give meaning to these arrows in a way which is intrinsic to the manifold $\M$, i.e., in a way which does not refer to any structure that lies outside of $\M$?

The key is to realize that a vector allows us to define the directional derivative of scalar fields. This idea combines the intuitive notion that a vector has a direction with an object which is intrinsically defined on the manifold, namely the scalar field $f:\M\to \bbR$. In a given coordinate system, say $\{x^\mu\}$, we can write the directional derivative of $f$ (in this coordinate system) as
\begin{align}
	v^\mu \partial_\mu f\,,
\end{align}
where $v^\mu\in C^\infty(\M)$ are $n$ smooth functions of the coordinates. It is common to introduce the notations
\begin{align}\label{eq:DirectionalDerivative}
	v(f) &\ce v^\mu\partial_\mu f & \text{and} && v\ce v^\mu \partial_\mu\,.
\end{align}
The notion of directional derivative gives us the correct intuition to define vector fields. Let us temporarily forget the explicit coordinate-dependent expression~\eqref{eq:DirectionalDerivative}. Rather, we focus on the key properties of the directional derivative and distill a set of axioms from it in order to define what we mean by a vector field on the manifold $\M$: A \textbf{vector field $\boldsymbol{v}$} on $\M$ is a map which takes $f\in C^{\infty}(\M)$ as input, produces $v(f) \in C^{\infty}(\M)$ as output, and which satisfies
\begin{itemize}
	\item[] $\begin{array}{lllll} 
				\text{A1} & & v(c\, f_1 +  f_2) = c\, v(f_1) +  v(f_2) & & \text{(Linearity)} \\[5pt]
				\text{A2} & & v(f_1 f_2) = v(f_1)\, f_2 + f_1\, v(f_2)  & & \text{(Leibniz rule)} \\[5pt]
				\text{A3} & & (g\, v_1 + v_2)(f) = g\, v_1(f) + v_2(f) & & \text{(Vector addition and scalar multiplication)}
			 \end{array}$
\end{itemize}
for all scalar fields $f, f_1, f_2, g\in C^{\infty}(\M)$ and constants $c\in \bbR$. Observe that the definition is independent of any coordinate system! The vector field is simply a linear map between smooth functions. The Leibniz rule captures the notion of differentiation inherent to the directional derivative.

For concrete computations, it is nevertheless useful to have a coordinate-representation of a vector field. To that end, we define the \textbf{components of the vector field} $v$ with respect to a coordinate system $\{x^\mu\}$ as
\begin{align}
	v^\mu \ce v(x^\mu)\,,
\end{align}
where $x^\mu$ is the $\mu$-th coordinate. If we remember the coordinate expression of the directional derivative~\eqref{eq:DirectionalDerivative} again, we see that the above definition of vector component is consistent with~\eqref{eq:DirectionalDerivative}, which implies
\begin{align}
	v(x^\mu) = v^\alpha \partial_\alpha x^\mu = v^\alpha \delta\ud{\mu}{\alpha} = v^\mu\,.
\end{align}
In physics, we often call $v^\mu$ the vector field, rather than component of a vector field. However, it is important to remember that (i) vector fields are defined in a way which is independent of any coordinate system and (ii) the vector field $v$ can have different components with respect to different coordinate systems (more on this below).

\begin{figure}[htb!]
	\centering
	\includegraphics[width=0.75\linewidth]{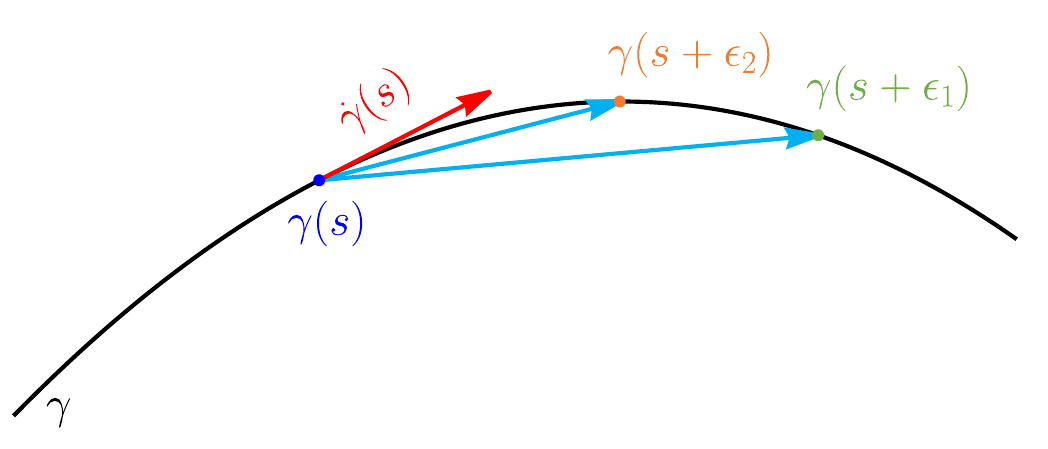}
	\caption{\protect On the process of generating tangent vectors: The smaller $\epsilon$ becomes, the better the separation between $\gamma(s)$ and $\gamma(s+\epsilon)$ is described by the vector $\epsilon\, \dot{\gamma}(s)$ tangential to the curve.  \hspace*{\fill}}
	\label{fig:5_TangentToCurve}
\end{figure}

The definition of vector field given above has the advantage of being coordinate-independent, but it is not clear how this idea relates to the intuitive conception that a vector field assigns an arrow to every point of~$\M$. To remedy that, we introduce the notion of \textbf{tangent vector}. Let $\gamma: [0,1]\to \M$ be a curve on $\M$ which is parametrized by $s$ and let $f\in C^\infty(\M)$ be a smooth function (scalar field). Then consider the derivative
\begin{align}\label{eq:TangentDerivative}
	\TD{}{s}f(\gamma(s)) = \TD{\gamma^\mu}{s}\PD{}{\gamma^\mu}f \equiv \dot{\gamma}^\mu \partial_\mu f\,,
\end{align}
where we used the product rule and the fact that given a coordinate system $x^\mu$, the curve $\gamma$ has components~$\gamma^\mu \ce x^\mu(\gamma)$. Looking at the right hand side of~\eqref{eq:TangentDerivative}, it is clear that the derivative we have just computed satisfies our abstract definition of vector field. Moreover, it is intuitively clear (see Figure~\ref{fig:5_TangentToCurve}) that $\TD{\gamma^\mu}{s}$ is a vector which is tangent to $\gamma$. Thus, the left hand side of~\eqref{eq:TangentDerivative} simply gives us the directional derivative of $f$ in the direction which is tangent to the curve $\gamma$. The advantage of this computation is that it gives us a clear relation with arrows and thus with the intuitive notion of vectors we know from $\bbR^n$. 

It is clear that~\eqref{eq:TangentDerivative} defines a vector field for every $f\in C^{\infty}(\M)$ and every curve $\gamma$. Moreover, one can show~\cite{BaezBook} that every vector field $v\in \rchi(\M)$ can be represented as in~\eqref{eq:TangentDerivative}.

Before proceeding, it is useful to introduce some terminology and define tangent vectors in slightly more abstract terms. Recall that a vector field $v$ is a map from $C^\infty(\M)$ to $C^\infty(\M)$. We define a \textbf{tangent vector at $\boldsymbol{p}$} to be a map $v_p$ from $C^\infty(\M)$ to $\bbR$. This is achieved by evaluating the vector field $v$ at the point $p\in \M$,
\begin{align}
	&v_p : C^\infty(\M)\to \bbR \notag\\
	&v_p(f) \ce \left.v(f)\right|_{p}\,.
\end{align}
In other words, a tangent vector $v_p$ is simply obtained by evaluating the smooth function $v(f)$ at the point~$p$, thus giving us a real number. The \textbf{set of all tangent vectors at $\boldsymbol{p}$} is called \textbf{tangent space at $\boldsymbol{p}$} and denoted by $T_p\M$. A two dimensional visulaisation of this concept is given in Figure~\ref{fig:5.1_TangentSpace} below.
\begin{figure}[htb!]
	\centering
	\includegraphics[width=0.75\linewidth]{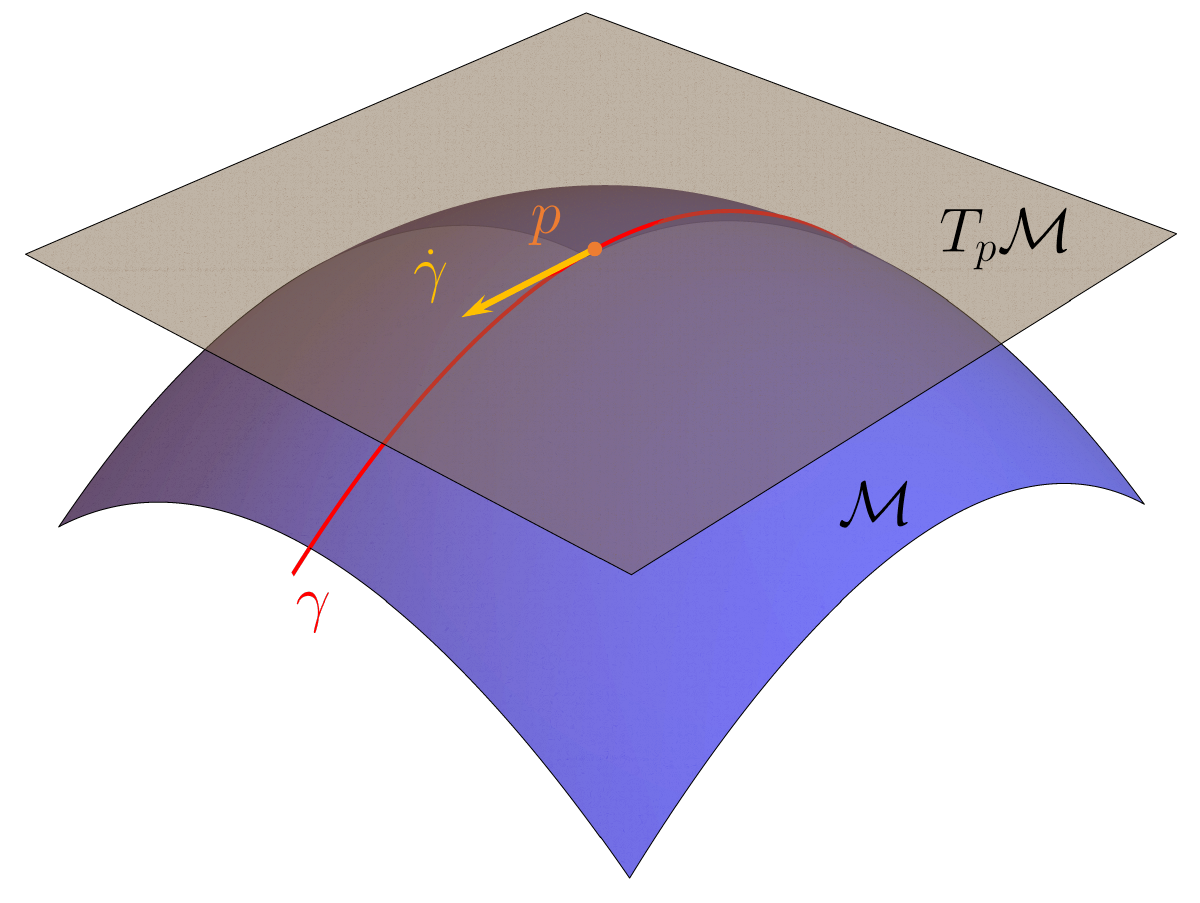}
	\caption{\protect The tangent space $T_p\M$ at $p$ is obtained by considering all linear independent vectors which are tangent to curves $\gamma$ passing through $p\in\M$. \hspace*{\fill}}
	\label{fig:5.1_TangentSpace}
\end{figure}

The space
\begin{align}
	T\M \ce \bigsqcup_{p\in\M} T_p \M \equiv \bigcup_{p\in\M} \{p\}\times T_p\M
\end{align}
is called the \textbf{tangent bundle} and can be thought of as the collection of all tangent spaces at every point of $\M$. We sometimes distinguish this from the \textbf{set of all smooth vector fields of $\M$} which we denote by $\mathcal{V}(\M)$. The distinction is not of great importance to us. What is important, however, is that the tangent space at $p$, i.e., $T_p\M$ is a \textbf{real, $n$-dimensional vector space}. This means that given two elements, say $v_p$ and $u_p$ of $T_p\M$, we can do everything we can do with regular vectors in $\bbR^n$. All elements of $T_p\M$ follow the rules of vector addition, multiplication by scalars, etc.

However, what we have \textit{not} yet defined, is a notion of scalar product. In Euclidean geometry, a scalar product takes two vectors as input and produces a real number as output. This real number comes attached with geometric meaning, since it provides us with a measure of angles between vectors and a measure for the magnitude of vectors. Generalizing this notion requires us to introduce a metric, which we will do in subsection~\ref{ssec:Metric}. Recall, however, that we know a second procedure from linear algebra which allows us to produce a number out of a vector. Namely, we can apply a linear functional, or, in other terms, pair a vector with a dual vector.

This leads us to the concept of \textbf{$\boldsymbol{1}$-form}, which are sometimes also referred to as \textbf{co-vectors}. Since $T_p\M$ is a real, $n$-dimensional vector space, it automatically possesses a real, $n$-dimensional \textbf{dual space $\boldsymbol{T^*\M}$}. This space is also called \textbf{co-tangent space at $\boldsymbol{p}$} and it consists of \textbf{linear functionals}. We recall that a linear functional $\omega$ is a map which takes a vector as input and produces a real number as output. More formally we can define it as the linear map
\begin{align}
	\omega:T_p\M &\to \bbR\notag\\
	v &\mapsto \langle\omega, v\rangle \in \bbR\,,
\end{align}
where $\omega$ is called a \textbf{$\boldsymbol{1}$-form} and the bracket $\langle\cdot,\cdot\rangle$ symbolizes the pairing of a $1$-form with a vector. Given a coordinate system $\{x^\mu\}$, we can define the components of $\omega$ as
\begin{align}
	\omega_\mu \ce \langle\omega,\partial_\mu\rangle\,,
\end{align}
i.e., we obtain the components by evaluating the linear functional on the basis elements of $T_p\M$. Since~$\omega$ is really a \textit{linear} map, we obtain for the pairing of $v$ with $\omega$ the following coordinate expression:
\begin{align}\label{eq:pairing}
	\langle \omega, v\rangle &= \langle \omega, v^\mu \partial_\mu \rangle \notag\\
	&= v^\mu\, \langle \omega, \partial_\mu\rangle \notag\\
	&= v^\mu \omega_\mu\,.
\end{align}
In the first line we simply expanded $v$ in its basis, then we used the linearity of $\langle \cdot,\cdot\rangle$, and finally the definition of $1$-form components we have just given. Notice that the contraction $\omega_\mu v^\mu$ does \textit{not} require a metric: The components of $v$ are naturally defined with an upper index, while the components of $\omega$ are naturally defined with  a lower index.

Given a coordinate system $\{x^\mu\}$, we can define the basis co-vectors of $T_p^*\M$ as $\dd x^\mu$ and write the $1$-form as
\begin{align}
	\omega \ce \omega_\mu \dd x^\mu\,.
\end{align}
These basis elements have to satisfy
\begin{align}
	\langle \dd x^\mu, \partial_\nu \rangle = \delta\ud{\mu}{\nu}
\end{align}
in order to be able to reproduce~\eqref{eq:pairing} and be consistent with the definitions we have given so far. Observe that we have defined vector fields as well as $1$-forms as \textit{linear maps}. This fact allows us to define more general tensors as \textbf{multilinear maps}. To do so, we define a \textbf{tensor of type $\boldsymbol{(p,q)}$} to be a multilinear map
\begin{align}
	S: \underbrace{T\M\otimes \dots \otimes T\M}_{p-\text{times}} \otimes \underbrace{T^*\M\otimes \dots \otimes T^*\M}_{q-\text{times}} \to \bbR
\end{align}
which takes $p$ vectors and $q$ co-vectors as input and produces a real number. This is sometimes written as
\begin{align}
	S(v_1, \dots, v_p, \omega_1, \dots, \omega_q)\,.
\end{align}
Since $S$ is a \textit{multilinear map}, i.e., since $S$ is linear in every one of its $p+q$ slots, it follows that in a coordinate system $\{x^\mu\}$ we can write
\begin{align}
	S(v_1, \dots, v_p, \omega_1, \dots, \omega_q) &= v_1^{\mu_1}\cdots v_p^{\mu_p} (\omega_1)_{\nu_1}\cdots (\omega_q)_{\nu_q}   S(\partial_{\mu_{1}}, \dots, \partial_{\mu_{p}}, \dd x^{\nu_{1}}, \dots , \dd x^{\nu_{q}})\notag\\
	&= v_1^{\mu_1}\cdots v_p^{\mu_p} (\omega_1)_{\nu_1}\cdots (\omega_q)_{\nu_q} S\du{\mu_{1}\cdots \mu_{p}}{\nu_{1}\cdots \nu_{q}}\,,
\end{align}
where in the last line we defined the components of $S$ as
\begin{align}
	S\du{\mu_{1}\cdots \mu_{p}}{\nu_{1}\cdots \nu_{q}} \ce S(\partial_{\mu_{1}}, \dots, \partial_{\mu_{p}}, \dd x^{\nu_{1}}, \dots , \dd x^{\nu_{q}})\,.
\end{align}
Due to their multilinearity, tensors have a very characteristic behaviour under changes of coordinates. We define a \textbf{change of coordinates} as a diffeomorphism which maps the coordinates $x^\mu$ to the new coordinates $x'^\mu(x)$. We will sometimes use the shorthand notation $x^\mu \mapsto x'^\mu(x)$. One can easily deduce that under such a change of coordinates partial derivatives transform as
\begin{align}
	\PD{}{x^\mu}  = \PD{x'^\lambda}{x^\mu}\PD{}{x'^{\lambda}} \ec J\ud{\lambda}{\mu} \PD{}{x'^\lambda}\,,
\end{align}
where in the last equation we have introduced the \textbf{Jacobian matrix $\boldsymbol{J\ud{\mu}{\nu}}$}, defined as
\begin{align}
	J\ud{\mu}{\nu} \ce \PD{x'^\mu}{x^\nu}\,.
\end{align}
Since $x'^\mu$ is generated from $x^\mu$ via a diffeomorphism, the Jacobian is never degenerate. This means it always possesses a well-defined inverse
\begin{align}
	(J^{-1})\ud{\mu}{\nu} \ce \PD{x^\mu}{x'^\nu}\,.
\end{align}
Now recall that we defined a vector field in a manner which is manifestly coordinate independent. Thus, we should have
\begin{align}
	v = v^\mu \partial_\mu \overset{!}{=} v'^\mu \partial'_\mu\,,
\end{align}
where $v'^\mu$ and $\partial'_\mu$ are the vector components and basis elements in the coordinate system $\{x'^\mu\}$. Since we know how partial derivatives transform under changes of coordinates, it follows that
\begin{align}
	v'^\nu \partial'_\nu = v^\mu J\ud{\nu}{\mu} \partial'_\nu = v^\mu \PD{x'^\nu}{x^\mu} \partial'_\nu\qquad \Longrightarrow \qquad v'^\nu = v^\mu \PD{x'^\nu}{x^\mu}\,. 
\end{align}
In other words, the components of the vector field in the new coordinate system are obtained by multiplying the old components by the Jacobian matrix,
\begin{align}\label{eq:TransformationVector}
	v'^\nu = J\ud{\nu}{\mu} v^\mu\,.
\end{align}
The transformation behaviour of $1$-forms follows now from simple considerations. Since we defined $1$-forms in a coordinate independent manner, and since they map vectors to real numbers, we have
\begin{align}
	\langle \omega, v\rangle = \omega_\mu v^\mu \overset{!}{=} \omega'_\mu v'^\mu\,.
\end{align}
Using~\eqref{eq:TransformationVector}, it then follows that
\begin{align}
	\omega_\mu (J^{-1})\ud{\mu}{\nu} v'^\nu = \omega'_\mu v'^\mu\qquad\Longrightarrow\qquad \omega'_\nu = (J^{-1})\ud{\mu}{\nu} \omega_\mu\,.
\end{align}
Knowing the transformation behaviour of vectors and $1$-forms immediately allows us to work out the transformation behaviour of tensors. All we have to do is exploit their multilinearity in order to find
\begin{align}\label{eq:TensorTransformation}
	S'^{\mu_{1}\dots \mu_{p}}{}_{\nu_{1}\cdots \nu_{q}} &= J\ud{\mu_{1}}{\alpha_{1}}\cdots J\ud{\mu_{p}}{\alpha_{p}} (J^{-1})\ud{\beta_{1}}{\nu_{1}} \cdots (J^{-1})\ud{\beta_{q}}{\nu_{q}}\, S\ud{\alpha_{1}\cdots \alpha_{p}}{\beta_{1}\cdots \beta_{q}}\,.
\end{align}
As a last concept, we introduce \textbf{tensor densities}. A tensor density is a tensor (this includes scalar fields, which are tensors of type $(0,0)$) which does \textit{not} transform according to~\eqref{eq:TensorTransformation}, because it picks up an even or odd power of the determinant of the Jacobian. Concretely, a \textbf{tensor density $\boldsymbol{\tilde{S}}$ of weight $\boldsymbol{w}$} transforms as
\begin{align}\label{eq:TensorDensity}
	\tilde{S}\ud{\alpha_{1}\cdots \alpha_{p}}{\beta_{1}\cdots \beta_{q}} &= \left(\det(J)\right)^{w}\,(J^{-1})\ud{\alpha_{1}}{\mu_{1}}\cdots (J^{-1})\ud{\alpha_{p}}{\mu_{p}} J\ud{\nu_{1}}{\beta_{1}} \cdots J\ud{\nu_{q}}{\beta_{q}}\,    \tilde{S}'^{\mu_{1}\dots \mu_{p}}{}_{\nu_{1}\cdots \nu_{q}}\,,
\end{align}
where $w$ is called the \textbf{density weight}. Notice our convention for defining the weight: The untransformed tensor density $\tilde{S}$ is on the left of this equation, while on the right we have the transformed tensors density $\tilde{S}'$ together with the Jacobian matrices and, importantly, the Jacobian determinant. Only in this form do we read off the density weight $w$. Notice that the weight can be positive, negative, or zero. A tensor density of weight zero is simply an ordinary tensor. Also, our convention is to denote tensor densities with a tilde on top, in order to highlight their special transformation behaviour. We only make an exception for Lagrangian densities $\mathcal{L}$, Hamiltonian densities $\mathcal{H}$, and the determinant of the metric, $g$.

Tensor densities play an important role when it comes to integration on manifolds. In order to guarantee that an integral constructed from tensorial objects is independent of the coordinate system we chose to represent these quantities in (and which we chose to perform the integration), the integrand has to transform as a scalar density of weight $w=+1$. We will later see that the square root of the metric determinant, $\sqrt{|g|}$, transforms as a tensor density of weight $w=+1$. Thus, integrals of the form
\begin{align}
	\int_\M \sqrt{|g|}\, f\, \dd^n x\,,
\end{align}
where $f$ is a scalar field, are coordinate-independent. Moreover, we will also encounter other tensor densities when we construct action functional for teleparallel theories of gravity in section~\ref{sec:Trinity}.

\subsection{The Flow of a Vector Field and the Lie Derivative}\label{ssec:LieDerivative}
In the previous subsection we mentioned that every vector field can be understood as the tangent vector to some curve. Indeed, if we are given a vector field $v$, we can find the corresponding curve by solving the first order differential equation
\begin{align}\label{eq:IntegralCurves}
	\dot{\gamma}(t) = v_{\gamma(t)}\,,
\end{align}
where $v_{\gamma(t)}$ is the vector $v$ evaluated at the point $\gamma(t)$. Since this is a first order ordinary differential equation, a solution always exists (at least locally). We call the curve $\gamma$ the \textbf{integral curve of $\boldsymbol{v}$}.

\begin{figure}[htb!]
	\centering
	\includegraphics[width=0.55\linewidth]{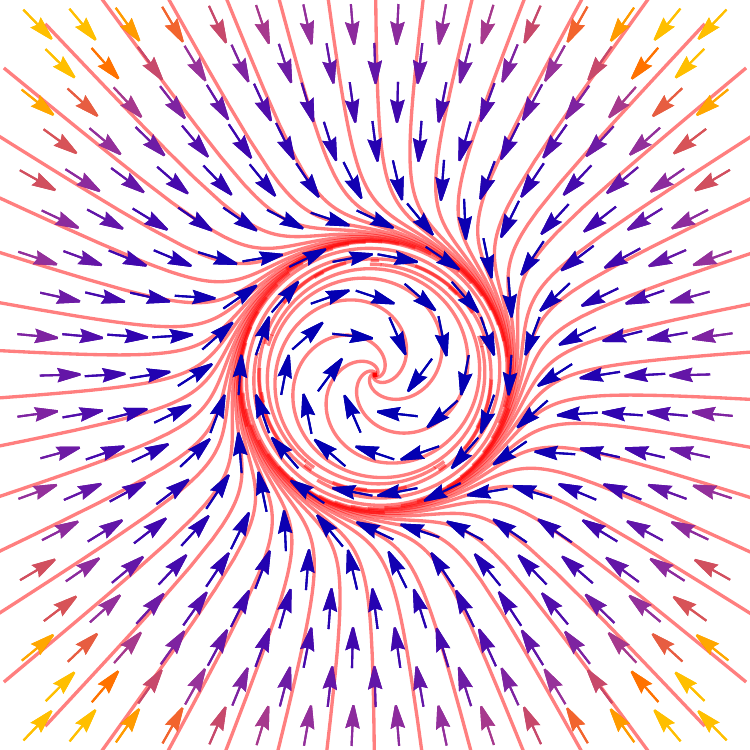}
	\caption{\protect The vector field $v = \left(y-x\left[1-x^2-y^2\right]^2\right)\partial_x + \left(-x-y\left[1-x^2-y^2\right]^2\right)\partial_y$ and the flow lines (integral curves) it generates. The flow lines are solutions of the coupled first order differential equations $\dot{x}(t) = y(t)-x(t) \left[1-x(t)^2-y(t)^2\right]^2$ and $\dot{y}(t) = -x(t)-y(t)\left[1-x(t)^2-y(t)^2\right]^2$. The vector field $v$ is tangential to the flow lines. \hspace*{\fill}}
	\label{fig:6_VectorFieldFlow}
\end{figure}

Globally, the integral curve may not exist because $\gamma$ can diverge in a finite amount of time. Nevertheless, locally we can visualize the vector field with its integral curves as in Figure~\ref{fig:6_VectorFieldFlow}. Qualitatively speaking, the integral curves describe the flow of some fluid, while $v$ assigns a velocity vector to each point in that fluid. 

\begin{figure}[htb!]
	\centering
	\includegraphics[width=0.9\linewidth]{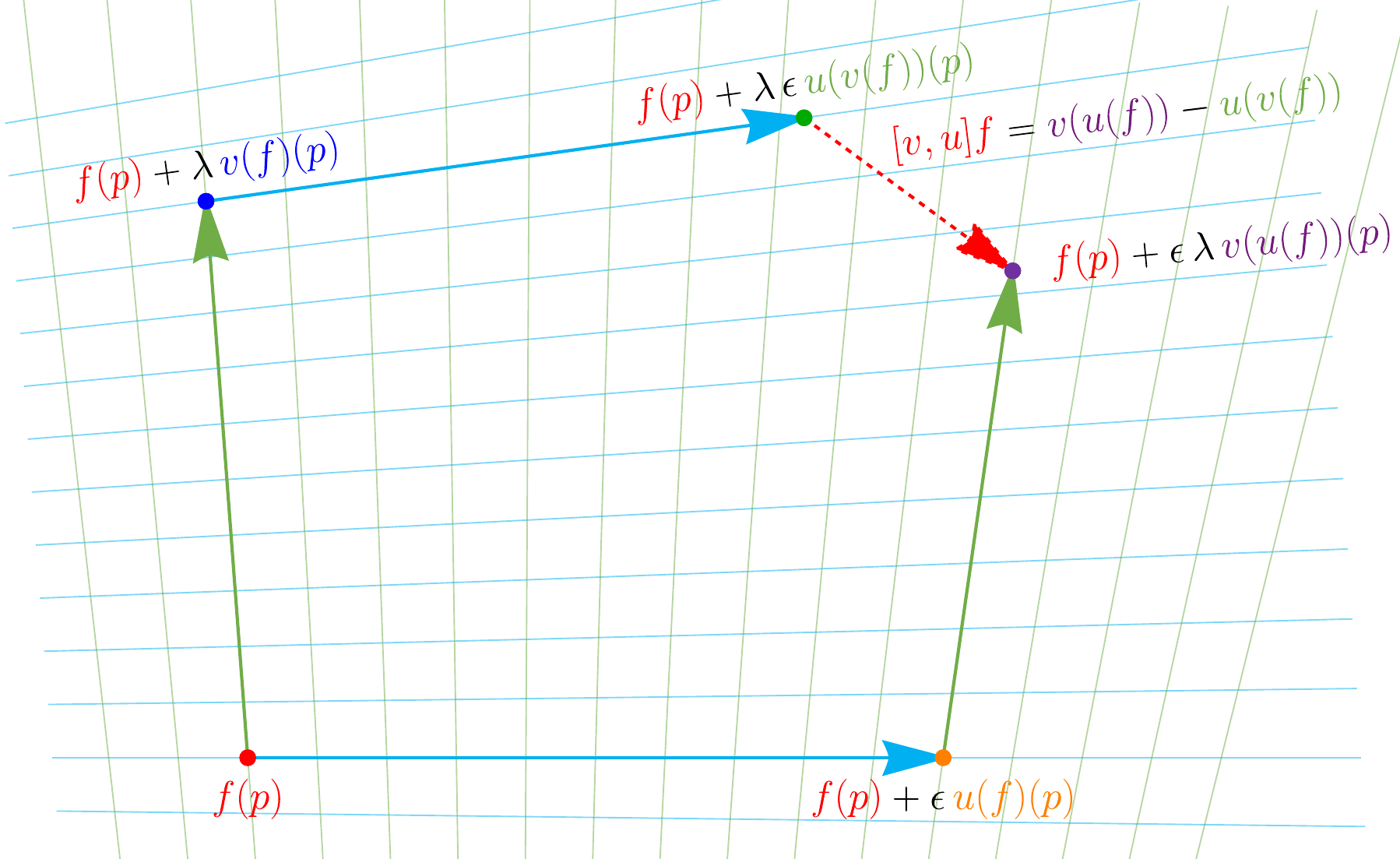}
	\caption{\protect Dragging $f$ a little bit along $v$ and then along $u$ results in a different outcome than dragging it along $u$ for a short while and then along $v$. The Lie bracket measures the failure for the so obtained parallelogram to close. \hspace*{\fill}}
	\label{fig:7_LieBracket}
\end{figure}

This idea of flow can also be expressed as a diffeomorphism which takes a point $p\in\M$ as input, and maps it to some other point in $\M$. This represents the movement or flow of $p$ on the manifold. To make this idea concrete, we introduce a $1$-parameter family of diffeomorphisms
\begin{align}
	\phi_t:\bbR\times\M &\to \M \notag\\
	(t, p) &\mapsto \phi_t(p)
\end{align}
with $\phi_0 = \text{id}$ and $\phi_s\circ\phi_t = \phi_{s+t}$ for all $s, t\in\bbR$. Importantly, if we work in the components language which is common in the physicist's literature, we have for a point $p$ with coordinates $x^\mu$
\begin{align}\label{eq:DiffeoComponents}
	\phi^\mu_t(x) &\ce  \phi_t(x^\mu(p)) &\text{and} && \phi^\mu_{t=0}(x) = x^\mu(p)\,.
\end{align}
Using this $1$-parameter family of diffeomorphisms, we can re-write equation~\eqref{eq:IntegralCurves} as
\begin{align}\label{eq:FlowEquation}
	\TD{}{t}\phi_t(p) = v_{\phi_t(p)}\,.
\end{align}
The family $\phi_t$ is called the \textbf{flow generated by $\boldsymbol{v}$}. Since we based the concept of flow on diffeomorphisms, it is easy to see that a flow not only affects the points on a manifold, but also tensors defined on it. The simplest example is the one of a scalar field $f$, which is carried along by the flow $\phi_t$ generated by the vector field $v$. The carried-along $f$, which we denote\footnote{Technically speaking, $\phi^*_t f$ is the \textbf{pull-back of $\boldsymbol{f}$ by $\boldsymbol{\phi_t}$}.} by $\phi^*_t f$, is defined as
\begin{align}
	(\phi_t^* f)(p) \ce f(\phi_t(p))\,.
\end{align}
In practice, we are often interested in the infinitesimal action of a flow on a vector field. Thus, we may expand $\phi^*_t f$ around $t=0$ up to first order. The first order derivative in this expansion is given by
\begin{align}
	\left.\TD{}{t}(\phi^*_t f)(p)\right|_{t=0} &= \left.\TD{}{t}f(\phi_t(p))\right|_{t=0} = \left.\PD{f}{\phi^\mu_t} \TD{\phi^\mu_t}{t}\right|_{t=0} \notag\\
	&= v^\mu_p \partial_\mu f = v_p(f)\,,
\end{align}
where we used equations~\eqref{eq:FlowEquation} and~\eqref{eq:DirectionalDerivative}, as well as $\left.\PD{f}{\phi^\mu_t}\right|_{t=0} = \PD{f}{x^\mu}$, which follows directly from~\eqref{eq:DiffeoComponents}. Thus, to first order, a scalar field which is carried along by a flow changes by the directional derivative generated by the vector field $v$. More concretely:
\begin{align}
	(\phi_t^* f)(p) = f(p) + t\, v_p(f) + \mathcal{O}(t^2)\,.
\end{align}
Similarly, we may ask how a vector field $u$ changes when it is carried along by the flow generated by $v$. We could give an abstract definition of the carried-along vector field $\phi^*_t u$ in terms of pull-backs. However, to keep the discussion lighter, we point out that in the components language, this essentially amounts to performing a change of coordinates:
\begin{align}
	(\phi^*_t u)(p) \quad \leadsto \quad \PD{\phi^\mu_{-t}(p)}{x^\nu} u^\nu(\phi_t(p))\;.
\end{align}
If we consider an infinitesimal change of coordinates, we can expand the above expression to first order in~$t$ around $t=0$. The first order term in this expansion is then given by
\begin{align}
	\left.\TD{}{t}\left(\PD{\phi^\mu_{-t}(p)}{x^\nu}u^\nu(\phi_t(p))\right)\right|_{t=0} &= \left.\TD{}{t}u^\nu(\phi_t(p))\right|_{t=0}\delta\ud{\mu}{\nu} + \left.\TD{}{t} \PD{\phi^\mu_{-t}(p)}{x^\nu}\right|_{t=0} u^\nu(p) \notag\\
	&= \PD{u^\mu(p)}{x^\lambda} \left.\TD{\phi^\lambda(p)}{t}\right|_{t=0} - \PD{v^\mu(p)}{x^\nu} u^\nu(p)\notag\\
	&= \PD{u^\mu(p)}{x^\lambda} v^\lambda(p) - \PD{v^\mu(p)}{x^\nu} u^\nu(p) \notag\\
	&= v^\lambda \partial_\lambda u^\mu - u^\nu \partial_\nu v^\mu\,.
\end{align}
In components free notation, we can introduce the \textbf{Lie bracket $\boldsymbol{[v,u]}$} to express the above result more concisely:
\begin{align}
	[v,u] \ce vu - uv \ce v^\mu \partial_\mu u^\nu \partial_\nu - u^\nu \partial_\nu v^\mu \partial_\mu\,.
\end{align}
 If this expression seems cryptic, remember that vector fields act on scalars and they produce their directional derivative. Thus, a more precise way to write the Lie bracket would be
 \begin{align}
 	[v,u](f) \ce v(u(f)) - u(v(f)) = v^\mu \partial_\mu u^\nu \partial_\nu f - u^\nu \partial_\nu v^\mu \partial_\mu f\,.
 \end{align}
This allows us to interpret the Lie bracket in a neat geometric fashion: First of all, notice that $v(f)$ is a scalar. In fact, $v(f) = v^\mu \partial_\mu f$ is just the directional derivative of $f$ along $v$. Because $v(f)$ is a scalar, the operation $u(v(f))$ is well-defined. It simply means taking the directional derivative of the scalar $v(f)$ along the direction $u$. Now recall that if $f$ is dragged along by a flow $\phi_t$ for an infinitesimal amount of ``time''~$t$, it changes to first order by $t\,v(f)$. Thus, $u(v(f))$ tells us by how much $f$ changes when we first flow along $v$ for a little while and then along $u$. Conversely, $v(u(f))$ tells us about the change in $f$ if we first flow it along $u$ and then along $v$ for a small amount of time. The Lie bracket is then simply a measure for the discrepancy between the two procedures. Since $v(u(f))$ and $u(v(f))$ land on different points, we can visualize the situation by a parallelogram which does not close (cf. Figure~\ref{fig:7_LieBracket}). That parallelograms built in this way do not close has nothing to do with curvature or torsion and is true even in Euclidean geometry.
The concept of determining how much a tensor field changes to first order when it is dragged along by a flow generated by a vector field is sufficiently important that it deserves its own name: It is called a \textbf{Lie derivative $\boldsymbol{\mathcal{L}_v}$ along $\boldsymbol{v}$}. Formally, the Lie derivative is defined by pulling-back tensor fields by the flow $\phi_t$ generated by~$v$. Since this is again tantamount to essentially considering an infinitesimal change of coordinates, one can work out that the coordinate expression for the Lie derivative of a $(p, q)$ tensor reads
\begin{align}\label{eq:DefLieD}
	\mathcal{L}_v T\ud{\mu_{1}\cdots \mu_{p}}{\nu_{1} \cdots \nu_{q}} &= v^\lambda \partial_\lambda T\ud{\mu_{1}\cdots \mu_{p}}{\nu_{1} \cdots \nu_{q}} - \partial_\lambda v^{\mu_1} T\ud{\lambda\cdots \mu_{p}}{\nu_{1} \cdots \nu_{q}} - \dots - \partial_\lambda v^{\mu_{p}}T\ud{\mu_{1}\cdots \lambda}{\nu_{1} \cdots \nu_{q}} \notag\\
	&\phantom{=} + \partial_{\nu_{1}} v^{\lambda} T\ud{\mu_{1}\cdots \mu_{p}}{\lambda \cdots \nu_{q}} + \dots + \partial_{\nu_{q}} v^{\lambda} T\ud{\mu_{1}\cdots \mu_{p}}{\nu_{1} \cdots \lambda}\,.
\end{align}
The Lie derivatives of scalar and vector fields are contained as special cases:
\begin{align}
	\mathcal{L}_vf &= v(f) = v^\mu \partial_\mu f\notag\\
	\mathcal{L}_v u &= [v,u] = v^\mu\partial_\mu u^\nu - u^\mu \partial_\mu v^\nu\,.
\end{align}
The Lie derivative can also be generalized to tensor densities $\tilde{T}$. Since the transformation law of tensor densities differs slightly from the one of regular tensors, one finds that the Lie derivative in components language acquires an additional term compared to~\eqref{eq:DefLieD}:
\begin{align}\label{eq:DefLieDTensorDensity}
	\mathcal{L}_v \tilde{T}\ud{\mu_{1}\cdots \mu_{p}}{\nu_{1} \cdots \nu_{q}} &= v^\lambda \partial_\lambda \tilde{T}\ud{\mu_{1}\cdots \mu_{p}}{\nu_{1} \cdots \nu_{q}} - \partial_\lambda v^{\mu_1} \tilde{T}\ud{\lambda\cdots \mu_{p}}{\nu_{1} \cdots \nu_{q}} - \dots - \partial_\lambda v^{\mu_{p}}\tilde{T}\ud{\mu_{1}\cdots \lambda}{\nu_{1} \cdots \nu_{q}} \notag\\
	&\phantom{=} + \partial_{\nu_{1}} v^{\lambda} \tilde{T}\ud{\mu_{1}\cdots \mu_{p}}{\lambda \cdots \nu_{q}} + \dots + \partial_{\nu_{q}} v^{\lambda} \tilde{T}\ud{\mu_{1}\cdots \mu_{p}}{\nu_{1} \cdots \lambda} + \boldsymbol{w\, \left(\partial_\lambda v^\lambda\right) \tilde{T}\ud{\mu_{1}\cdots \mu_{p}}{\nu_{1} \cdots \nu_{q}}}\,,
\end{align}
where, as we recall, $w$ is the weight of the tensor density. Notice that~\eqref{eq:DefLieDTensorDensity} contains~\eqref{eq:DefLieD} as the $w=0$ special case. Later, in subsection~\ref{ssec:LieDRevisited}, we will see that the Lie derivative can also be defined for connections and, importantly, that the coordinate expression is \textit{not} given by either~\eqref{eq:DefLieD} or~\eqref{eq:DefLieDTensorDensity}. Rather, it is given by~\eqref{eq:LieDConnection}.

\subsection{Covariant Derivatives and the Connection}\label{ssec:CovD}
For a scalar field $f$, we defined the directional derivative as $v(f) = v^\mu\partial_\mu f$. Since the field is directly defined on the smooth manifold $\M$, there is no issue in giving a precise meaning to $\partial_\mu f$. It simply amounts to the usual definition from multivariable calculus:
\begin{align}
	\partial_\mu f \ce \TD{f}{x^\mu} = \lim_{\epsilon\to 0} \frac{f(x^{1}, \dots, x^\mu+\epsilon, \dots, x^n) - f(x^{1}, \dots, x^\mu, \dots, x^n)}{\epsilon}\,.
\end{align}
Now that we have introduced vector fields and other tensors we would like to define a similar notion of taking the derivative of a tensor in the direction of a vector field.

A prime candidate for such a derivative is the Lie derivative, which we discussed in the previous subsection. It tells us how a tensor field changes when dragged infinitesimally along a flow generated by a vector field. However, at closer inspection it does not truly behave like a directional derivative. For instance, if $u_p$ and $v_p$ are two vectors at $p$ and $w_p\ce u_p+ v_p$ their vector sum, then the directional derivative of a scalar field satisfies
\begin{align}
	u_p(f) + v_p(f) = \left(u_p + v_p\right)(f) = w_p(f)\,,
\end{align}
whereas the Lie derivative of a tensor $T_p$ at $p$ fails to have this property,
\begin{align}
	\mathcal{L}_{u_p}T_p + \mathcal{L}_{v_p} T_p \neq \mathcal{L}_{u_p + v_p} T_p =  \mathcal{L}_{w_p} T_p\,.
\end{align}
Thus, the Lie derivative is not linear in this sense. In other words, in the case of a scalar field we can take the derivative in the direction $v_p$, add the derivative in the direction $u_p$ to it, and we are guaranteed that this is the same as if we had taken the derivative in the direction $w_p=u_p+v_p$ to begin with. The Lie derivative does not behave in this way. 

Furthermore, the directional derivative $v_p(f)$ only depends on the properties of $v_p$ and $f$ at the point $p$. It is thus local in this sense. The Lie derivative, on the other hand, depends on the properties of $v_p$ and $T_p$ in a \textit{neighbourhood} of $p$ and is, in this sense, slightly ``non-local''. To illustrate this point, we take and slightly adapt the nice example from~\cite{Margalef-Bentabol:2018}: Let us work in a coordinate chart $\{x,y\}$ and consider the scalar field $f(x,y)$ together with the vector fields $u = \partial_x$, $v= (y+1)\partial_x$, and $w=\partial_y$. Clearly, if we evaluate $u$ and $v$ at the point $p$ with coordinates $(x_0, 0)$, they agree:
\begin{align}
	\left.u\right|_{y=0} = \left.\partial_x\right|_{y=0} = \left.(y+1)\partial_x\right|_{y=0} = \left.v\right|_{y=0}\,.
\end{align}
For the directional derivative of $f$ in the directions $u$ and $v$ evaluated at $(x_0,0)$ we thus find
\begin{align}
	\left.u(f)\right|_{y=0} &= \partial_x f(x,0) &\text{and} && \left.v(f)\right|_{y=0} &= \partial_x f(x,0)\,.
\end{align}
That is, even though we take the derivatives in different directions, they agree with each other because the vector fields happen to agree in that particular point. For the Lie derivative we find instead a disagreement:
\begin{align}
	\left.\mathcal{L}_u w\right|_{y=0} &= \left.u^\mu \partial_\mu w^\nu \partial_\nu\right|_{y=0} - \left.w^\nu\partial_\nu u^\mu\partial_\mu \right|_{y=0}\notag\\
	&= \left. \delta\ud{\mu}{x}\partial_\mu \delta\ud{\nu}{y}\partial_\nu\right|_{y=0} - \left.\delta\ud{\nu}{y}\partial_\nu \delta\ud{\mu}{x}\partial_\mu\right|_{y=0}\notag\\
	&= 0
\end{align}
versus
\begin{align}
	\left.\mathcal{L}_v w\right|_{y=0} &= \left.v^\mu \partial_\mu w^\nu\partial_\nu\right|_{y=0} - \left.w^\nu \partial_\nu v^\mu\partial_\mu\right|_{y=0} \notag\\
	&= \left. (y+1)\delta\ud{\mu}{x}\partial_\mu \delta\ud{\nu}{y}\partial_\nu\right|_{y=0} - \left.\delta\ud{\nu}{y}\partial_\nu\left((y+1)\delta\ud{\mu}{x}\right)\partial_\mu \right|_{y=0} \notag\\
	&= -\partial_x\,.
\end{align}
Hence, even though the vector fields $u$ and $v$ coincide at the point $(x_0, 0)$, the Lie derivatives do not agree!

Since the Lie derivative has not the desired linearity and locality properties we look for in a directional derivative, we might be tempted to mimic the definition of derivative for scalar functions instead. Concretely, we might try to define the directional derivative of a vector field $\nabla_u v$ at a point $p$, as follows:
\begin{align}
	\left.\nabla_u v\right|_p \ce \lim_{\epsilon\to 0} \frac{v_{p+ \epsilon\, u} - v_{p}}{\epsilon}\,.
\end{align}
This poses two problems:
\begin{enumerate}
	\item The point $p$ lives in $\M$, while the vector $u_p$ lives in the tangent space $T_p\M$. The manifold is just a topological space, i.e., a space in which addition of points is not even defined, while the tangent space is a vector space. Thus, the expression ``$p+\epsilon\, u$'' has no mathematical meaning. It is as if we are trying to add apples and oranges.
	\item The second problem is that even if we could give meaning to ``$p+\epsilon\, u$'', we would be subtracting vectors which live in two different spaces. Let's say ``$p+\epsilon\, u$'' represents the point $q$. Then we are effectively asking to compute $v_q - v_p$. But $v_q$ lives in $T_q\M$, while $v_p$ is a vector in $T_p\M$. Again, it is like subtracting apples from oranges.
\end{enumerate}
It is important to highlight that the above definition of directional derivative of a vector field \textit{does} make sense in Euclidean geometry. The reason is that points in the \textit{manifold} $\bbR^n$ can be identified with vectors in the \textit{vector space} $\bbR^n$. Thus, the operation $p+\epsilon\, u$, i.e., adding a vector to a point, becomes meaningful. Furthermore, the tangent space $T_p\bbR^n$ is isomorphic to $\bbR^n$ itself. Since this is true for any point in $\bbR^n$, we find that all tangent spaces are isomorphic to each other. This gives rise to the usual notion of Euclidean geometry that we can add and subtract vectors at different points of space. Thus, $v(q) - v(p)$ is a meaningful operation in Euclidean geometry because there is a canonical way of transporting vectors from point to point.

In subsection~\ref{ssec:ParallelTransport} we will see how both issues can be solved by a direct approach. This will lead us to introduce the notion of \textbf{parallel transport}. In the present subsection, we shall follow a different strategy to overcome the obstacles.  We emulate the axiomatic approach we already used to define the directional derivative of scalar fields. To begin with, we change terminology: Instead of referring to $\nabla$ as directional derivative, we shall call it the \textbf{covariant derivative $\boldsymbol{\nabla}$} from now on. We define the covariant derivative as map from the space of all vector fields on $\M$, $\mathcal{V}(\M)$, times the space of all tensor fields on $\M$, $\mathcal{T}{\M}$, into that same space. Symbolically, we want to define a map
\begin{align}
	\nabla:\mathcal{V}(\M)\times \mathcal{T}(\M) &\to \mathcal{T}(\M) \notag\\
	(v,T) &\mapsto \nabla_v T
\end{align}
This map takes a vector field $v\in \mathcal{V}(\M)$ together with a tensor field $T\in \mathcal{T}(\M)$ as input and produces the tensor field $\nabla_v T\in \mathcal{T}(\M)$ as output. It has to do so obeying the following set of axioms:
\begin{itemize}
	\item[A1]  \;\;\;$\nabla_v f = v(f) \quad $ for all $v\in\mathcal{V}(\M)$ and $f\in C^{\infty}(\M)$
	\item[A2] \;\;\;$\nabla_v(c\, T_1 + T_2) = c\, \nabla_v T_1 + \nabla_v T_2\quad $ for all $v\in \mathcal{V}(\M)$, $T_1, T_2\in \mathcal{T}(\M)$, and $c\in \bbR$
	\item[A3] \;\;\;$\nabla_v(T_1 \otimes T_2) = (\nabla_v T_1)\otimes T_2 + T_1\otimes (\nabla_v T_2)\quad$ for all $v\in \mathcal{V}(\M)$ and $T_1, T_2\in \mathcal{T}(\M)$
	\item[A4] \;\;\;$\nabla_{c\, v_{1} + v_{2}}T =c\,\nabla_{v_{1}} T + \nabla_{v_{2}}T\quad$ for all $v_1, v_2\in \mathcal{V}(\M)$, $T\in \mathcal{T}(\M)$, and $c\in \bbR$.
\end{itemize}
The first axiom simply makes sure that the covariant derivative, which is supposed to generalize the notion of directional derivative acting on scalars, agrees with the definition we have given in subsection~\ref{ssec:VFTFD}. Axiom A2 captures the linearity of the directional derivative, while axiom A3 is the general version of the Leibniz rule. This axiom truly captures the essence of $\nabla$ being a \textit{differential} operator. Notice that if in A3 we have $T_1=f$, then because of $f\otimes T_2 \equiv f\, T_2$, we find as special case
\begin{align}
	\nabla_v(f T) = (\nabla_vf)\,T + f\, \nabla_v T = v(f)\, T +f\, \nabla_v T\,,
\end{align}
where we also made use of A1. Finally, axiom A4 captures the idea we have discussed further above. Namely, that the covariant derivative along $c\, v_1 + v_2$ should be the same as when computing it along $c\, v_1$ and $v_2$ separately and then summing the results. Notice that the Lie derivative also satisfies axioms A1 (action on scalars), A2 (linearity in $\mathcal{T}(\M)$ argument), and A3 (Leibniz rule). What sets the covariant derivative apart from the Lie derivative is axiom A4, which is \textit{not} satisfied by the Lie derivative.

Working with axioms might seem overly abstract, but it is actually quite simple to work out in a coordinate chart how the covariant derivative acts on vectors and $1$-forms. Once this is understood, it is straightforward to generalize its action to any tensor (density). 

Let us begin with deriving a coordinate expression for the covariant derivative of vector fields. To do so, we work with coordinates $\{x^\mu\}$ and we introduce a basis $\{e_\mu\} \ce \{\partial/\partial x^\mu\}$ for $T\M$. The covariant derivative of $v=v^\nu e_\nu$ in the direction of $u=u^\mu e_\mu$ can then be written as
\begin{align}\label{eq:CovDFromAxioms}
	\nabla_u v &\overset{\text{A4}}{=} u^\mu \nabla_{e_{\mu}}(v^\nu e_\nu) \overset{\text{A3, A2}}{=} u^\mu \left(\nabla_{e_{\mu}}(v^\nu)e_\nu + v^\nu \nabla_{e_{\mu}}(e_\nu) \right) \notag\\
	&\overset{\text{A1}}{=} u^\mu \left(\partial_\mu v^\nu e_{\nu} + v^\nu \nabla_{e_{\mu}}e_\nu \right)\,.
\end{align}
In the first line we made use of axiom A4 and the Leibniz rule A3. We have also made implicit use of A2 when applying the Leibniz rule, since $v^\nu e_\nu$ really represents a linear combination and the derivative acts on each term in that sum. Finally, we used that the components $v^\nu$ of a vector field are just smooth functions, which allows us to apply A1 in order to write $\nabla_{e_{\mu}}(v^\nu) = e_{\mu}(v^\nu) = \partial_\mu v^\nu$. That is, the covariant derivative just becomes the directional derivative along $e_\mu$ and since $e_\mu = \partial/\partial x^\mu$ this simply gives us the coordinate derivatives of the scalar functions $v^\nu$.
 
Recall that we defined the covariant derivative as a map from $\mathcal{V}(\M)\times \mathcal{T}(\M)$ to $\mathcal{T}(\M)$. However, the last line of~\eqref{eq:CovDFromAxioms} does not look like an element of $\mathcal{T}(\M)$ since the term $v^\nu \nabla_{e_{\mu}}e_\nu$ is not a linear combination of basis elements $e_{\mu}$. We have already used up all axioms to arrive at~\eqref{eq:CovDFromAxioms}. Therefore, to remedy the situation, we have to introduce a new concept: The \textbf{affine connection $\boldsymbol{\Gamma}$}. Concretely, we demand that in a coordinate chart $\{x^\mu\}$ and with respect to a basis $\{e_\mu\}$ of $T\M$ the $n\times n\times n$ components of the affine connection satisfy
\begin{align}\label{eq:DefAffineConnection}
	\boxed{\nabla_{e_{\mu}}e_\nu = \Gamma\ud{\alpha}{\mu\nu}e_\alpha}
\end{align}
We can take this as the defining equation for the affine connection and use it to simplify equation~\eqref{eq:CovDFromAxioms} to
\begin{align}\label{eq:CovDVectorInBasis}
	\nabla_u v = u^\mu\left(\partial_\mu v^\alpha + \Gamma\ud{\alpha}{\mu\nu} v^\nu\right)e_\alpha\,.
\end{align}
This is manifestly an element of $\mathcal{T}(\M)$ and it has a recognizable form. In fact, we can simply read off the \textbf{component expression} for the covariant derivative of a vector field, which is
\begin{align}\label{eq:CovDVectorField}
	\boxed{\nabla_\mu v^\nu = \partial_\mu v^\nu + \Gamma\ud{\nu}{\mu\lambda}v^\lambda}
\end{align}
Let us briefly pause and comment on the role of equation~\eqref{eq:DefAffineConnection}. The salient point to notice is that the axioms A1--A4 do \textit{not} specify a unique covariant derivative operator $\nabla$! Rather, if someone hands us a concrete differential operator, we can check whether it satisfies the axioms and, if it does, we can use equation~\eqref{eq:DefAffineConnection} to determine the coefficients of the affine connection. However, the logic can also be turned around and the whole paradigm of teleparallel gravity hinges on this mathematical fact: If someone hands us an affine connection $\Gamma\ud{\alpha}{\mu\nu}$, we can \textit{define} a covariant derivative operator $\nabla$ which satisfies the axioms. In fact, it suffices to specify $\Gamma\ud{\alpha}{\mu\nu}$ and to declare that equation~\eqref{eq:CovDVectorField} holds. This unambiguously defines the meaning of the operator $\nabla$ and we know how to apply it to \textit{any} tensor field.

Actually, we have not yet shown that the last part is true, i.e., we still need to show that saying how $\nabla$ acts on a vector field is sufficient in order to know how it acts on all tensor fields. To do so, we also need to work out how $\nabla$ acts on $1$-forms. Recall that $1$-forms live in the dual space $T^*\M$ and thus define a linear map which maps vector fields to scalar fields according to $\langle \omega, v\rangle = \omega_\mu v^\mu$. For the directional derivative of this particular scalar field we find
\begin{align}
	u(\langle \omega, v\rangle) &\overset{\text{A1}}{=} \nabla_u(\langle \omega, v\rangle) \overset{\text{A3}}{=} \langle \nabla_u \omega, v\rangle + \langle \omega, \nabla_u v\rangle\,,
\end{align}
where we first made use of axiom A1 and then A3, the Leibniz rule. We can solve for the first term on the right hand side:
\begin{align}
	\langle \nabla_u \omega, v\rangle = u(\langle \omega, v\rangle) - \langle \omega, \nabla_u v\rangle\,.
\end{align}
Notice that we have reduced the task of finding the covariant derivative of $\omega$ to computing the directional derivative of a scalar and the covariant derivative of a vector. Working again in a coordinate chart $\{x^\mu\}$ and using~\eqref{eq:CovDVectorField}, we can complete our task as follows:
\begin{align}
	\left(\nabla_u w\right)_\mu v^\mu &= u^\mu \partial_\mu \left(\omega_\alpha v^\alpha\right) - \omega_\alpha u^\mu\left(\partial_\mu v^\alpha + \Gamma\ud{\alpha}{\mu\nu}\right) \notag\\
	&= u^\mu (\partial_\mu \omega_\alpha) v^\alpha + u^\mu \omega_\alpha \partial_\mu v^\alpha -\omega_\alpha u^\mu\left(\partial_\mu v^\alpha + \Gamma\ud{\alpha}{\mu\nu}\right) \notag\\
	&= u^\mu \left(\partial_\mu \omega_\alpha - \Gamma\ud{\lambda}{\mu\alpha} \omega_\lambda\right) v^\alpha\,.
\end{align}
From this we can finally read off that the covariant derivative of a $1$-form, expressed in the component language, reads
\begin{align}
	\boxed{\nabla_\mu \omega_\alpha = \partial_\mu \omega_\alpha - \Gamma\ud{\lambda}{\mu\alpha} \omega_\lambda}
\end{align}
All we have used to arrive at this result are the axioms A1--A4 and equation~\eqref{eq:CovDVectorField}. Thus, the covariant derivative of a $1$-form is completely determined once we know what the covariant derivative of a vector field is. Once we know these two covariant derivatives, we can work out the coordinate expression for the covariant derivative of any tensor field, simply by application of the Leibniz rule. The general result reads
\begin{equation}
\boxed{
\begin{aligned}
	\nabla_\alpha T\ud{\mu_1 \dots \mu_p}{\nu_1 \dots \nu_q} &= \partial_\alpha T\ud{\mu_1 \dots \mu_p}{\nu_1 \dots \nu_q} + \Gamma\ud{\mu_1}{\alpha\lambda} T\ud{\lambda \dots \mu_p}{\nu_1 \dots \mu_q} + \dots + \Gamma\ud{\mu_p}{\alpha\lambda} T\ud{\mu_1 \dots \lambda}{\nu_1 \dots \nu_q} \\
	&\phantom{=} - \Gamma\ud{\lambda}{\alpha \nu_1} T\ud{\mu_1 \dots \mu_p}{\lambda \dots \nu_q} - \dots - \Gamma\ud{\lambda}{\alpha \nu_q} T\ud{\mu_1 \dots \mu_p}{\nu_1 \dots \lambda}
\end{aligned}
}
\end{equation}
Let us stress again at this point that the axioms do not specify a unique operator $\nabla$. Infinitely many covariant derivative operators exist and it is ultimately our choice, which one we use. From a mathematical point of view, this also means that we added a new structure. So far, everything we did could be defined on the manifold $\M$ (curves, scalar fields) or on spaces derived from the manifold itself (vectors, $1$-forms, general tensors). However, defining a covariant derivative requires us to add something new by hand. Once we have selected an affine connection, we are working in the framework of an \textbf{affine geometry $\boldsymbol{(\M, \Gamma)}$}. In the next subsection, where we introduce the metric tensor, we will finally arrive at metric-affine geometries.

However, before we do so, we clarify a last point which sometimes causes confusion. The affine connection~$\Gamma\ud{\alpha}{\mu\nu}$ carries three indices, but it should not be mistaken for a tensor! A connection is a different type of object than a $(1,2)$ tensor! To see this more explicitly, one should recall that tensors have a very simple transformation behaviour under change of coordinates. For instance, a $(1,2)$ tensor $S\ud{\alpha}{\mu\nu}$ transforms as
\begin{align}
	S\ud{\alpha}{\mu\nu} \quad \mapsto \quad \tilde{S}\ud{\alpha}{\mu\nu} = \PD{\tilde{x}^\alpha}{x^\beta} \PD{x^\rho}{\tilde{x}^\mu} \PD{x^\sigma}{\tilde{x}^\nu} S\ud{\beta}{\rho\sigma}
\end{align}
under the change of coordinates $x^\mu \mapsto \tilde{x}^\mu(x)$. In contrast, an affine connection transforms under the same change of coordinates as
\begin{align}\label{eq:ConnectionTransformationLaw}
	\boxed{\Gamma\ud{\alpha}{\mu\nu}\quad \mapsto\quad \tilde{\Gamma}\ud{\alpha}{\mu\nu} = \PD{\tilde{x}^\alpha}{x^\beta} \PD{x^\rho}{\tilde{x}^\mu} \PD{x^\sigma}{\tilde{x}^\nu} \Gamma\ud{\beta}{\rho\sigma} + \PD{\tilde{x}^\alpha}{x^\lambda}\PD{^2 x^\lambda}{\tilde{x}^\mu \partial\tilde{x}^\nu}}
\end{align}
We can distinguish between two pieces in this transformation law: A term which transforms homogeneously, like a tensor would, and an inhomogeneous term. The necessity for this second, inhomogeneous piece in the transformation behaviour can be seen by noticing that, for instance, $\nabla_\mu v^\nu$ is a tensor by definition. As we have seen in~\eqref{eq:CovDVectorField}, this can be written as a partial derivative plus the connection. However, the partial derivative of vector field components does \textit{not} transform in a tensorial way. It transforms in an inhomogeneous fashion and the connection compensates for this behaviour, rendering $\nabla_\mu v^\nu$ indeed a proper tensor.

As a final comment, we remark that this non-tensorial transformation behaviour implies that (a) adding a $(1,2)$ tensor $S\ud{\alpha}{\mu\nu}$ to a connection $\Gamma\ud{\alpha}{\mu\nu}$ gives us an equally valid but completely new connection $\hat{\Gamma}\ud{\alpha}{\mu\nu} \ce \Gamma\ud{\alpha}{\mu\nu} + S\ud{\alpha}{\mu\nu}$ and (b) a connection which is not zero in one coordinate system can be made to vanish by a clever change of coordinates. We will re-encounter this fact in~\ref{ssec:STEGR} when we discuss the coincident gauge.

\subsection{The Metric Tensor and the Geodesic Equation}\label{ssec:Metric}
Up to this point, we mostly worked with the manifold $\M$. This is sufficient to talk about events, curves (to model observers and test particles), scalar fields, vector fields, general tensor fields of type $(p,q)$, and tensor densities. This structure alone is also sufficient to introduce flows of tensor fields and define the Lie derivative.

Only in the last subsection did we encounter the necessity to introduce a new structure: An affine connection~$\Gamma\ud{\alpha}{\mu\nu}$. This necessity arose in order to define a covariant derivative for vectors and other tensor fields. The connection is an object which we can freely choose and it defines the covariant derivative of any tensor via the equation~\eqref{eq:CovDVectorField}. A manifold together with an affine connection is referred to as \textbf{affine geometry~$\boldsymbol{(\M, \Gamma)}$}. This pair is sufficient to describe all concepts introduced so far.

However, our description is incomplete. For instance, even tough we can define curves, we cannot answer the question which curve is the shortest one between two points. More in general, we do not know how to measure the length of curves or even the magnitude of vectors. To remedy that, we now introduce the \textbf{metric tensor $\boldsymbol{g}$} and we extend the affine geometry $(\M, \Gamma)$ to the \textbf{metric-affine geometry $\boldsymbol{(\M, g, \Gamma)}$}. The idea behind the metric is to generalize the notion of scalar product between vectors from Euclidean geometry to any kind of geometry. We proceed again in an axiomatic fashion and define $g$ as a map
\begin{align}
	g: T\M\times T\M &\to \bbR\notag\\
	(u,v) &\mapsto g(u,v)
\end{align}
which satisfies the following axioms:
\begin{itemize}
	\item[A1] \;\;\;Linearity in both slots: $g(f\, v_1 + v_2, w) = f\, g(v_1, w) + g(v_2, w)$\newline
		$\phantom{\text{Linearity in both slots: }} 	g(v, f w_1 + w_2) = f\, g(v, w_1) + g(v, w_2)$
	\item[A2] \;\;\;Symmetry: $g(v, w) = g(w, v)$
	\item[A3] \;\;\;Non-degeneracy: If $g(v, w) = 0$ for all $w$, then $v=0$.
\end{itemize}
Thus, the metric tensor is a map which takes two vectors as input and produces a real number. Given a coordinate chart $\{x^\mu\}$ and a basis $\{e_\mu\}$ of $T\M$, this allows us to define the components of the metric tensor $g$ with respect to that chart and that basis as
\begin{align}
	g_{\mu\nu} \ce g(e_\mu, e_\nu)\,.
\end{align}
Together with axiom A1 it then follows that
\begin{align}
	g(v, w) &= g(v^\mu e_\mu, w^\nu e_\nu) \overset{\text{A1}}{=} v^\mu g(e_\mu, w^\nu e_\nu) \notag\\
	&\overset{\text{A1}}{=} v^\mu w^\nu g(e_\mu, e_\nu) = g_{\mu\nu} v^\mu w^\nu\,.
\end{align}
From axiom A2 it follows that $g_{\mu\nu} = g_{\nu\mu}$, while axiom A3 implies the existence of an \textbf{inverse metric}, which we denote by $g^{\mu\nu}$. Importantly, the metric and its inverse satisfy the identity $g_{\mu\lambda}g^{\lambda\nu} = \delta\du{\mu}{\nu}$.

This generalizes the familiar scalar product between vectors from Euclidean to more general geometries. Consequently, once we are given a metric tensor $g$, we can define the norm of vectors, angles between vectors, areas, volumes, and so on. For instance, we define the norm of a vector as
\begin{align}
	\|v\|^2 \ce g(v,v) = g_{\mu\nu} v^\mu v^\nu\,.
\end{align}
It should be emphasized that this norm is not always positive! In fact, depending on the \textbf{signature of the metric}, there can be non-zero vectors for which the norm is positive, zero, or even negative. Concretely, the signature $(p, n)$ is defined by the number of positive ($p$) and negative ($n$) eigenvalues of $g$.

A Euclidean metric has signature $(n,0)$ and the norm of non-zero vectors is always positive. A Lorentzian metric on the other hand has signature $(n-1,1)$ and the norm of non-zero vectors can be positive, negative, or zero. Here we are mostly interested in metrics with Lorentzian signature and we can classify vectors as being \textbf{spacelike}, \textbf{timelike}, and \textbf{null}. The definition goes as follows:
\begin{align}
	\text{A vector $v$ is called } 
	\left\{\begin{array}{c}
		\textbf{spacelike} \\
		\textbf{null} \\
		\textbf{timelike}
	\end{array}\right\}
	\text{ if $g(v,v)$ is }
	\left\{\begin{array}{c}
		> 0 \\
		= 0 \\
		< 0
	\end{array}\right\}\,.
\end{align}
We emphasize that this definition relies on the convention that the signature of $g$ is mostly plus. We could also have defined signature of a Lorentzian metric as $(1,n-1)$, which would invert $>$ to $<$ and vice versa for even $n$ in the above definition. The definitions coincide for both conventions if the number of dimensions $n$ is odd.

This classification can also be extended to curves and hypersurfaces:
\begin{align}
	\text{A curve $\gamma$ is }
	\left\{\begin{array}{c}
		\text{spacelike} \\
		\text{null} \\
		\text{timelike}
	\end{array}\right\}
	\text{ if its \textbf{tangent} vector is everywhere }
		\left\{\begin{array}{c}
		\text{spacelike} \\
		\text{null} \\
		\text{timelike}
	\end{array}\right\}\,.
\end{align}
\begin{align}
	\text{A hypersurface is } 	
	\left\{\begin{array}{c}
		\text{spacelike} \\
		\text{null} \\
		\text{timelike}
	\end{array}\right\}
	\text{ if its \textbf{normal} vector is everywhere }
		\left\{\begin{array}{c}
		\text{timelike}\\
		\text{null} \\
		\text{spacelike}
	\end{array}\right\}\,.
\end{align}
Notice the reversed order in the second bracket! Using this terminology and the concept of a metric, we can define the \textbf{length of a spacelike curve } $\gamma$ as
\begin{align}
	L[\gamma] \ce \int_I \sqrt{g_{\mu\nu}(\gamma) \dot{\gamma}^\mu \dot{\gamma}^\nu}\dd s\,
\end{align}
where $s$ is the parameter along the curve and $\dot{\gamma}^\mu$ its (spacelike) tangent vector. Similarly, we can define the \textbf{proper time} along a timelike curve as
\begin{align}
	T[\gamma] \ce \int_I \sqrt{-g_{\mu\nu}(\gamma) \dot{\gamma}^\mu \dot{\gamma}^\nu} \dd s\,.
\end{align}
The minus sign under the square root is necessary since the scalar product is negative for timelike tangent vectors. Also, up to a dimensionful constant, the proper time gives us the \textbf{action of massive point particles}, namely
\begin{align}
	\S[\gamma] \ce m\, T[\gamma] = m \int_I\sqrt{-g_{\mu\nu}(\gamma) \dot{\gamma}^\mu \dot{\gamma}^\nu} \dd s\,,
\end{align}
where $m>0$ is the mass of the particle. By varying this functional with respect to the (timelike) particle trajectory $\gamma$, one obtains the so-called \textbf{geodesic equation}
\begin{align}
	\frac{\delta S[\gamma]}{\delta \gamma^\alpha} &\overset{!}{=} 0 &  \Longrightarrow && \ddot{\gamma}^\alpha + \LC{\alpha}{\mu\nu} \dot{\gamma}^\mu \dot{\gamma}^\nu &= 0\,,
\end{align}
where we have introduced the \textbf{Christoffel symbols} (aka \textbf{Levi-Civita} connection)
\begin{align}\label{eq:DefLeviCivitaConnection}
	\LC{\alpha}{\mu\nu} \ce \frac12 g^{\alpha\lambda}\left(\partial_\mu g_{\nu \lambda} + \partial_\nu g_{\mu\lambda} - \partial_\lambda g_{\mu\nu}\right)\,.
\end{align}
As is well-known, these symbols do \textbf{not} transform as tensors, despite appearances. In fact, this is a first concrete example for a connection and we can use it to define a covariant derivative $\D$ via the equation
\begin{align}
	\D_\mu v^\nu \ce \partial_\mu v^\nu + \LC{\nu}{\mu \alpha} v^\alpha\,.
\end{align}
We will return to this special connections and its properties in subsection~\ref{ssec:Classification}. The crucial point we wish to emphasize here is the following: Manifolds $\M$ are just mere topological spaces which do not come equipped with metrics. We are free to choose one. Once we have made a choice, we can automatically define a covariant derivative, namely the derivative defined by the Levi-Civita connection~\eqref{eq:DefLeviCivitaConnection}, without any further choices. A geometry based on $\M$ and $g$ is called a \textbf{Riemannian geometry $\boldsymbol{(\M, g)}$}.

Before concluding this subsection, we point out that the determinant of the metric, which we denote by~$g$, is a tensor density of weight $w=+2$. This follows easily from simple linear algebra considerations and the transformation behaviour of a $(0,2)$ tensor. First of all, we note that the metric can be thought of as a $n\times n$ square matrix with components $g_{\mu\nu}$. Thus, the tools of linear algebra certainly apply. Moreover, under a change of coordinates $x^\mu \mapsto x'^\mu(x)$ the metric transforms as
\begin{align}
	g'_{\mu\nu} = \PD{x^\alpha}{x'^\mu}\PD{x^\beta}{x'^\nu} g_{\alpha\beta}\,.
\end{align}
The right hand side is simply the product of three matrices: The metric and two copies of the inverse Jacobian matrix $(J^{-1})\ud{\mu}{\nu} = \PD{x^\alpha}{x'^\nu}$. In terms of matrices, we can thus write
\begin{align}\label{eq:MatrixFormTransformationLaw}
	g' = J^{-1}\, g\, J^{-1}\,.
\end{align}
 From linear algebra we further know that
\begin{align}\label{eq:LinAlgId}
	\det(A\, B) = \det(A)\, \det(B)
\end{align}
for any two square matrices $A$ and $B$. Therefore, by solving~\eqref{eq:MatrixFormTransformationLaw} for the untransformed metric and applying the identity~\eqref{eq:LinAlgId} twice, we find for the determinant of the metric
\begin{align}
	\det(g) = \det(J\, g'\, J ) = \det(J)^2\, \det(g')\,.
\end{align}
According to the convention of~\ref{ssec:VFTFD}, this means that the determinant is a scalar density of weight $w=+2$. This result is important because it implies that the square root of the determinant is a scalar density of weight $w=+1$. If we multiply a scalar field $f$ by $\sqrt{|g|}$ and we integrate it, we are guaranteed that the resulting integral is independent of the choice of coordinates. This plays an important role in the construction of action functionals.

\newpage
\asection{3}{Curvature, Torsion, Non-Metricity: The Fundamental Objects of Metric-Affine Geometries}\label{sec:GeometricObjects}
Metric-affine geometries are characterized by having curvature, torsion, non-metricity, or any combination of these three properties. These properties are all defined in terms of the connection $\Gamma$ and, in the case of non-metricity, in terms of the connection and the metric tensor $g$. In this section we will first properly define these terms by using the concept of parallel transport. This will aid us in gaining an intuitive understanding of curvature, torsion, and non-metricity.

Once these concepts have been clarified, we will deepen our understanding of metric-affine geometries and discuss many important results. These results will come in handy when we formulate and analyze teleparallel theories of gravity in sections~\ref{sec:Trinity},~\ref{sec:TrinityGeneralizations}, and~\ref{sec:f(Q)}.

\subsection{Parallel Transport}\label{ssec:ParallelTransport}
As we discussed in subsection~\ref{ssec:CovD}, there is no canonical way to compare vectors (or tensors in general) at different points on a manifold. This posed an obstacle for defining the directional derivative of vectors and other tensors. We resolved the problem by introducing a connection $\Gamma$. However, we could just as well have chosen an alternative route. Namely, we could have introduced the concept of \textbf{parallel transport}. This notion will prove useful in better understanding metric-affine geometries $(\M, g, \Gamma)$ and it will lead us to a sort of classification of these geometries.

Recall that we faced two problems in defining a covariant derivative for tensor fields: The first problem was that expressions of the form ``$p+\epsilon\, u$'' are nonsensical from a mathematical point of view, since $p$ lives in~$\M$, while $u$ lives in $T_p\M$. In general, these are two completely different spaces. The second problem was that in generalizing the difference quotient of ordinary calculus, we would have to compute the difference~$v_q - v_p$. That is, we are asked to compute the difference of a vector living in $T_p\M$ and one living in $T_q\M$. This is again a meaningless operation.

The concept of parallel transport resolves both of these problems: The first problem is resolved by replacing the nonsensical expression ``$p+\epsilon\, u$'' by $\gamma(s)$, where $\gamma$ is a curve passing through $p$ and $q$ with tangent vector~$u$. The second problem is resolved by \textit{choosing a prescription} of how to move a given vector $v$ from one point~$p$ to another point $q$ along the curve $\gamma$. Importantly, we find again that there is no canonical way of providing such a prescription. We simply have to \textit{choose} one. This is reminiscent of the fact that the covariant derivative is not uniquely determined by the axioms we formulated in~\ref{ssec:CovD}. Infinitely many covariant derivative operators can be chosen which satisfy all the axioms.

To implement the two solutions described above, we define a \textbf{parallel transport map}
\begin{align}
	P(\gamma)^t_s: T_{\gamma(s)}\M &\to T_{\gamma(t)}\M\notag\\
	v_{\gamma(s)}&\mapsto P(\gamma)^t_s v_{\gamma(s)}\,,
\end{align}
where $\gamma(s) = p$, $\gamma(t) = q$ and which satisfies the following axioms
\begin{enumerate}
	\item[A1] \;\;\;$P(\gamma)^t_t = \text{id}$;
	\item[A2] \;\;\;$P(\gamma)^t_u \circ P(\gamma)^{u}_s = P(\gamma)^t_s$;
	\item[A3] \;\;\;$P(\gamma)^t_s$ is smooth in $s$, $t$, and $\gamma$.
\end{enumerate}
Given a vector $v_p$ at $p$ (i.e., a vector which lives in $T_p\M$), we can now transport this vector to $q$ along the curve $\gamma$ and we obtain the new vector
\begin{align}\label{eq:TransportVector}
	P(\gamma)^t_s v_p\,,
\end{align}
which lives in the tangent space $T_q\M$. We emphasize that the choice of the map $P(\gamma)_s^t$ is completely arbitrary, as long as it satisfies the above axioms. Its introduction merely serves the purpose to compare a vector at $p$ to a vector at $q$. This is exactly what is needed when talking about the derivative of a vector and we are therefore led to define the covariant derivative as
\begin{align}\label{eq:CovD_ParallelT}
	\nabla_{w} v \ce  \lim_{s\to 0}\frac{P(\gamma)^0_s v_{\gamma(s)} - v_{\gamma(0)} }{s} = \TD{}{s}\left.\left[ P(\gamma)^{0}_s\, v_{\gamma(s)} \right]\right|_{s=0}\,,
\end{align}
where $w\ce \dot{\gamma}$ is the tangent vector to the curve $\gamma$. Recall that the connection $\Gamma$, which we introduced in subsection~\ref{ssec:CovD} in order to define the covariant derivative, can be freely chosen. This suggests that there is a relation between the parallel transport map $P(\gamma)_s^t$ and the connection $\Gamma$. To see this relation, we choose to work in a coordinate chart $\{x^\mu\}$ and we shall compare the components of~\eqref{eq:CovD_ParallelT} to the components of~\eqref{eq:CovDVectorField}. Equation~\eqref{eq:TransportVector}, which describes the transported vector, reads in component language
\begin{align}
	[P(\gamma)_s^0]\ud{\alpha}{\mu} v^\mu_{\gamma(s)}\,.
\end{align}
Taylor expanding up to first order around $s=0$, and using $\gamma(0) = p$, we find
\begin{align}
	[P(\gamma)_s^0]\ud{\alpha}{\mu} v^\mu_{\gamma(s)} = [P(\gamma)^0_0]\ud{\alpha}{\mu} v^\mu_p + s\, [P(\gamma)^0_0]\ud{\alpha}{\mu} \left.\TD{}{s}v^\mu_{\gamma(s)}\right|_{s=0} + s\, v^\mu_p \left.\TD{}{s} [P(\gamma)^0_s]\ud{\alpha}{\mu} \right|_{s=0} +\O(s^2)\,.
\end{align}
Notice that here we made use of the differentiability of $P(\gamma)^0_s$ in $s$, which is guaranteed by axiom A3. Using $[P(\gamma)^0_0]\ud{\alpha}{\mu} = \delta\ud{\alpha}{\mu}$ (axiom A1 in components language) together with $\left.\TD{}{s}v^\mu_{\gamma(s)}\right|_{s=0} = w^\nu_p \partial_\nu v^\mu_p$, this reduces to
\begin{align}\label{eq:FirstOrderExpansionP}
	[P(\gamma)^{0}_s]\ud{\alpha}{\mu} v^\mu_{\gamma(s)} = v^\alpha_p + s\, \left(w^\nu_p \partial_\nu v^\alpha_p + v^\mu_p \left.\TD{}{s}[P(\gamma)^{0}_s]\ud{\alpha}{\mu}\right|_{s=0}\right) + \O(s^2)\,.
\end{align}
Next, we make again use of axiom A3, which assures us of the differentiability of $P(\gamma)^0_s$ with respect to $\gamma$, in order to compute
\begin{align}\label{eq:DerivativeOfP}
	\left.\TD{}{s}[P(\gamma)_s^0]\ud{\alpha}{\mu}\right|_{s=0} &= \TD{[P(\gamma)_s^0]\ud{\alpha}{\mu}}{\gamma^\nu} \left.\TD{}{s}\gamma^\nu\right|_{s=0} = \left.\TD{[P(\gamma)_s^0]\ud{\alpha}{\mu}}{\gamma^\nu}\right|_{s=0} w^\nu_p \,.
\end{align}
By plugging~\eqref{eq:DerivativeOfP} into~\eqref{eq:FirstOrderExpansionP} we finally find that the covariant derivative~\eqref{eq:CovD_ParallelT} is equal to
\begin{align}
	(\nabla_w v)^\alpha &= w^\nu_p \left(\partial_\nu v^\alpha_p + \left.\TD{[P(\gamma)_s^0]\ud{\alpha}{\mu}}{\gamma^\nu}\right|_{s=0} v^\mu_p  \right)\,.
\end{align}
We left the subscript $p$ on the right hand side to emphasize that everything is defined locally at $p$, as one has to expect from the covariant derivative (see the discussion in subsection~\ref{ssec:CovD}). From the same subsection we recall that the covariant derivative of a vector field (in components) is given by
\begin{align}
	w^\nu \nabla_\nu v^\alpha = w^\nu\left(\partial_\nu v^\alpha + \Gamma\ud{\alpha}{\nu\mu} v^\mu\right)\,.
\end{align}
By comparing the last two equations, we find the relation
\begin{align}\label{eq:RelationGammaParallelT}
	\boxed{\left.\TD{[P(\gamma)_s^0]\ud{\alpha}{\mu}}{\gamma^\nu}\right|_{s=0} = \Gamma\ud{\alpha}{\nu\mu}}
\end{align}
This means that connection and parallel transport are equivalent to each other! Or, more precisely:
\begin{itemize}
	\item Given a connection $\Gamma$, it induces a notion of parallel transport  $P(\gamma)_s^t$ which can be deduced from integrating equation~\eqref{eq:RelationGammaParallelT};
	\item Given a parallel transport map $P(\gamma)_s^t$, we can determine a connection $\Gamma$ associated with it by computing its derivative according to~\eqref{eq:RelationGammaParallelT}. 
\end{itemize}
We point out that because of the axioms listed above, it follows that
\begin{align}
	P(\gamma)^0_s \circ P(\gamma)^s_{0} = P(\gamma)^s_{s} = \text{id}\,.
\end{align}
In other words, to every $P(\gamma)^0_s$ there exists an inverse map $P(\gamma)^s_{0}$. We can therefore view $P(\gamma)^0_s$ as an element of $GL(n, \bbR)$, the group of real-valued, non-degenerate $n\times n$ matrices.

Now that the relation between the covariant derivative, the connection, and the map $P(\gamma)_s^t$ is clarified, we introduce the following definition: A vector $v$ is said to be \textbf{parallel transported along $\boldsymbol{\gamma}$} if
\begin{align}
	P(\gamma)^t_s v_{\gamma(s)} \overset{!}{=} v_{\gamma(t)}\,.
\end{align}
Observe what this equation is saying: Given a vector field $v$, we say that it has been parallel transported if the vector $v_{\gamma(s)}$ at the point $\gamma(s)$ is equal to the vector $v_{\gamma(t)}$ at the point $\gamma(t)$ after having been transported by $P(\gamma)^t_s$ along the curve $\gamma$.

Since this \textbf{parallel transport condition} has to hold for any $s$ and $t$, we can also consider an infinitesimal version of it with $t=s+\epsilon$. Expanding in $\epsilon$ and using the definition~\eqref{eq:CovD_ParallelT}, we find that this can equivalently be formulated as
\begin{align}\label{eq:ParallelTransportCondition}
	\boxed{\nabla_{\dot{\gamma}(s)} v_{\gamma(s)} \overset{!}{=} 0}
\end{align}
We call this the \textbf{parallel transport equation}. In components, this equation reads
\begin{align}\label{eq:InfinitesimalTransport}
	\boxed{\dot{\gamma}^\mu\left(\partial_\mu v^\nu + \Gamma\ud{\nu}{\mu\lambda} v^\lambda\right) \overset{!}{=} 0}
\end{align}
This helps us in understanding how a vector changes when it is being infinitesimally parallel transported. Consider vector $v$ at the point $\gamma(s+\epsilon)$. For small $\epsilon$ we can Taylor expand and obtain
\begin{align}
	v^\nu_{\gamma(s+\epsilon)} &= v^\nu_{\gamma(s)} + \epsilon\, \left.\TD{}{\epsilon}v^\nu_{\gamma(s+\epsilon)}\right|_{\epsilon=0} + \O(\epsilon^2) \notag\\
	&= v^\nu_{\gamma(s)} + \epsilon\, \dot{\gamma}^\mu\partial_\mu v^\nu_{\gamma(s)} + \O(\epsilon^2)\,.
\end{align}
Let us now assume that $v_{\gamma(s+\epsilon)}$ has been generated by \textit{parallel transporting} $v_{\gamma(s)}$. Let us also assume that $\gamma(s) = p$ and $\gamma(s+\epsilon) = q$ are infinitesimally close. Then, using the parallel transport condition~\eqref{eq:InfinitesimalTransport}, we find that the last equation can be written as
\begin{align}\label{eq:InfinitesimalParallelTransport}
	\boxed{v^\nu_q = v^\nu_p - \epsilon\,\Gamma\ud{\nu}{\mu\lambda}(p) \dot{\gamma}^\mu v^\lambda_p}
\end{align}
We can read this equation as saying that starting from $v_p$, parallel transport generates a vector $v_q$ at the infinitesimally close point $q$ by subtracting a term which depends on the connection at $p$, the vector $v_p$, and the infinitesimal displacement vector $\epsilon\, \dot{\gamma}^\mu$, also defined at $p$.

 As we will see in what follows, this infinitesimal version of parallel transport and its interpretation allow us to better understand metric-affine geometries~$(\M, g, \Gamma)$. The idea is very simple: Now that we know how to parallel transport vectors, we can ask how the characteristic properties of vectors are affected by the transport. The characteristic properties of vectors are the following.
\begin{itemize}
	\item[(a)] Every vector has a direction.
	\item[(b)] Vectors can be added together ``tip to tail''. In particular, adding together two vectors which point in different directions results in a new vector which points in yet another direction.
	\item[(c)] Provided the manifold is endowed with a metric, we can assign a magnitude to every vector.
\end{itemize}
Our intuition about vectors is largely rooted in Euclidean geometry, which makes use of a very particular notion of parallel transport and a very particular metric. It is therefore deeply ingrained in our minds that vectors can be moved around at will in $\mathbb R^n$ without affecting their direction nor their magnitude. Also, it is irrelevant whether we add the tail of $\vec{B}$ to the tip of $\vec{A}$ or vice versa; both operations result in the same vector $\vec{C} \ce\vec{A} + \vec{B}$ and we can visualize this using a parallelogram.

However, all of this can change when the notion of parallel transport---which is tantamount to a choice of connection $\Gamma$, as we remind the reader---is more general than the one used in Euclidean geometry. In fact, a generic connection $\Gamma$ has an effect on all three properties listed above, when a vector is being parallel transported. In what follows, we look at each property in turn.

\subsection{Curvature}\label{ssec:Curvature}
\begin{figure}[htb!]
	\centering
	\includegraphics[width=0.75\linewidth]{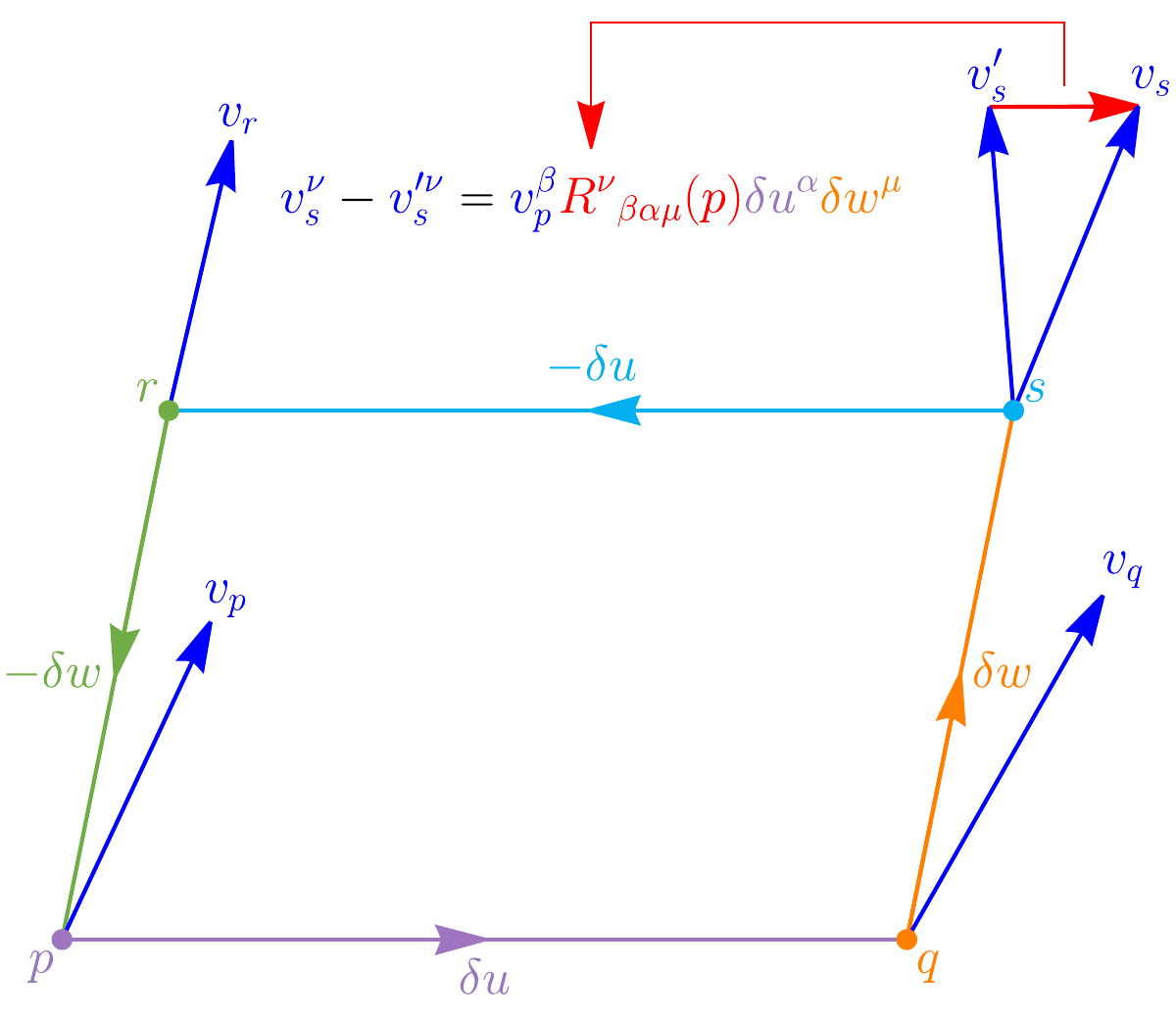}
	\caption{\protect Parallel transporting a vector around a closed loop results in a change of direction for that vector. This change is measured by the curvature tensor $R\ud{\alpha}{\mu\nu\rho}$.  \hspace*{\fill}}
	\label{fig:8_Curvature}
\end{figure}

The first property to be considered is how the direction of a vector changes when it is parallel transported. Clearly, to get a meaningful notion of ``the direction of the vector has changed due to parallel transport'', we have to somehow compare the vector to itself before and after parallel transport. This can only be achieved if we consider a closed curved $\gamma$. In order to make the computation manageable, we shall consider an \textit{infinitesimal} loop consisting of four curve segments, as shown in Figure~\ref{fig:8_Curvature}. There are four points, $p$, $q$, $r$, and $s$, all connected by curve segments. Let $p$ be connected to $q$ via the curve $\gamma_u$, which has a tangent vector $u$. Let $v_p$ be the vector which we shall parallel transport around the loop shown in Figure~\ref{fig:8_Curvature}.

To begin with, we transport $v_p$ from $p$ to $q$ along $\gamma_u$. Since the loop is infinitesimal, equation~\eqref{eq:InfinitesimalParallelTransport} applies and we can think of this process as displacing $v_p$ by $\epsilon\, u$, where $\epsilon\ll 1$. This results in
\begin{align}
	v^\nu_q &= v^\nu_p - \Gamma\ud{\nu}{\mu\lambda}(p)\, \delta u^\mu\, v^\lambda_p\,,
\end{align}
where we defined $\delta u^\mu \ce \epsilon\, u^\mu$ for ease of notation. The subscript $p$ shall remind us that everything on the right hand side is defined at $p$. This becomes important once we parallel transport $v_q$ to $s$ along the curve $\gamma_w$, which has a tangent vector $w$. Again, $q$ and $s$ are infinitesimally close and we are essentially just displacing $v_q$ by the infinitesimal vector $\delta w\ce \lambda\, w$, where $\lambda\ll 1$. By applying again equation~\eqref{eq:InfinitesimalParallelTransport} we obtain
\begin{align}
	v^\nu_s &= v^\nu_q - \Gamma\ud{\nu}{\mu\lambda}(q)\, \delta w^\mu\, v^\lambda_q\notag\\
	&= \underbrace{v^\nu_p - \Gamma\ud{\nu}{\mu\lambda}(p)\, \delta u^\mu\, v^\lambda_p}_{=v^\nu_q} - \underbrace{\left[\Gamma\ud{\nu}{\mu\lambda}(p) + \partial_\rho \Gamma\ud{\nu}{\mu\lambda}(p)\,\delta u^\rho_p\right]}_{= \Gamma\ud{\nu}{\mu\lambda}(q)}\times \underbrace{\left[v^\lambda_p - \Gamma\ud{\lambda}{\alpha\beta}(p)\, \delta u^\alpha \, v^\beta_p\right]}_{= v^\lambda_q}\, \delta w^\mu\,.
\end{align}
In the second line we expressed $v_q$ by quantities defined in $p$, according to the infinitesimal parallel transport equation. Also, we expanded $\Gamma(q)$ up to first order around the point $p$. If we keep only terms which are at most second order in $\delta u$ and $\delta v$, the above equation simplifies to
\begin{align}
	v^\nu_s &= v^\nu_p - \Gamma\ud{\nu}{\mu\lambda}(p)\, \delta u^\mu\, v^\lambda_p - \Gamma\ud{\nu}{\mu\lambda}(p)\, \delta w^\mu\, v^\lambda_p\notag\\
	&\phantom{=} -v^\beta_p \left[\partial_\alpha \Gamma\ud{\nu}{\mu\beta}(p) - \Gamma\ud{\nu}{\mu\lambda}(p) \Gamma\ud{\lambda}{\alpha\beta}(p) \right] \,\delta u^\alpha\, \delta w^\mu\,.
\end{align}
The next step would be to displace $v_s$ to $r$ by the infinitesimal amount $-\delta u$ and then to apply an infinitesimal displacement $-\delta w$ to arrive back at $p$. This would require us to perform many more expansions where we then only keep terms which are at most second order in $\delta u$ and $\delta w$. Since this would make the computations rather cumbersome, we choose a cleverer route. In fact, we can simply displace $v_p$ from $p$ to $r$ along $\delta w$ and then from $r$ to $s$ along $\delta u$. This results in a vector $v'_s$ and the computations are virtually the same as the ones we already did. Thus, we find
\begin{align}
	v'^\nu_s &= v^\nu_p - \Gamma\ud{\nu}{\mu\lambda}(p)\, \delta w^\mu\, v^\lambda_p - \Gamma\ud{\nu}{\mu\lambda}(p)\, \delta u^\mu\, v^\lambda_p \notag\\
	&\phantom{=} - v^\beta_p \left[\partial_\mu \Gamma\ud{\nu}{\alpha\beta}(p) - \Gamma\ud{\nu}{\alpha\lambda}(p)  \Gamma\ud{\lambda}{\mu\beta}(p)\right]\, \delta u^\alpha \, \delta w^\mu\,.
\end{align}
Now we can compare $v_s$ to $v'_s$. First of all, we notice that the zeroth and first order terms are all the same. The two vectors only differ in their second order terms and we find
\begin{align}
	v^\nu_s - v'^\nu_s &= v^\beta_p \left[\partial_\mu \Gamma\ud{\nu}{\alpha\beta}(p) - \partial_\alpha \Gamma\ud{\nu}{\mu\beta}(p) + \Gamma\ud{\nu}{\mu\lambda}(p) \Gamma\ud{\lambda}{\alpha\beta}(p) -  \Gamma\ud{\nu}{\alpha\lambda}(p)  \Gamma\ud{\lambda}{\mu\beta}(p) \right]\, \delta u^\alpha \, \delta w^\mu \notag\\
	&\ec v^\beta_p R\ud{\nu}{\beta\alpha\mu}\, \delta u^\alpha \, \delta w^\mu
\end{align}
where in the last line we introduced the \textbf{curvature tensor}
\begin{equation}\label{eq:DefCurvature}
	\boxed{R\ud{\alpha}{\mu\nu\rho} \ce 2\partial_{[\nu} \Gamma\ud{\alpha}{\rho]\mu} + 2\Gamma\ud{\alpha}{[\nu|\lambda} \Gamma\ud{\lambda}{\rho]\mu} = \partial_\nu \Gamma\ud{\alpha}{\rho\mu} - \partial_\rho \Gamma\ud{\alpha}{\nu\mu} + \Gamma\ud{\alpha}{\nu\lambda}\Gamma\ud{\lambda}{\rho\mu} - \Gamma\ud{\alpha}{\rho\lambda} \Gamma\ud{\lambda}{\nu\mu}}
\end{equation}
From the way we obtained this tensor, it is clear that it measures the change in orientation of $v_p$ when we parallel transport it along a closed loop. Furthermore, we observe that the curvature tensor is anti-symmetric in its last two lower indices:
\begin{align}
	R\ud{\alpha}{\mu(\nu\rho)} = 0\,.
\end{align}  
Notice that the curvature tensor is solely constructed from the connection $\Gamma$. No metric was necessary for its construction. A connection for which the curvature tensor vanishes is called a \textbf{flat connection}. Given that the curvature tensor has four indices and given that it is anti-symmetric in its last two lower indices, there are two traces one can build without invoking a metric. The first one is called the \textbf{Ricci tensor} 
\begin{align}
	\boxed{R_{\mu\nu} \ce R\ud{\lambda}{\mu\lambda\nu}}
\end{align}
The second trace goes by the name of \textbf{homothetic tensor} and is given by
\begin{align}
	\boxed{H_{\mu\nu} \ce R\ud{\lambda}{\lambda\mu\nu}}
\end{align}
As we will see in subsection~\ref{ssec:Identities}, the homothetic tensor can be expressed in terms of the Ricci tensor and the torsion tensor.

If a metric is present, there are two more traces that can be built. First, we can raise the second index of the curvature tensor and then define the \textbf{co-Ricci tensor}
\begin{align}
	\boxed{P\ud{\mu}{\nu} \ce g^{\rho\lambda} R\ud{\mu}{\rho\nu\lambda} = R\ud{\mu\lambda}{\nu\lambda}}
\end{align}
The co-Ricci tensor can also be expressed in terms of the Ricci tensor and the non-metricity tensor. Finally, the last trace that can be built (we ignore the traces of the homothetic and co-Ricci tensor since these tensors are not independent) is the \textbf{Ricci scalar}
\begin{align}
	\boxed{R \ce g^{\mu\nu} R_{\mu\nu} = R\ud{\mu}{\mu} = R\ud{\lambda\mu}{\lambda\mu}}
\end{align}
This completes our discussion of the curvature tensor an its various traces.

\subsection{Torsion}\label{ssec:Torsion}
\begin{figure}[htb!]
	\centering
	\includegraphics[width=1\linewidth]{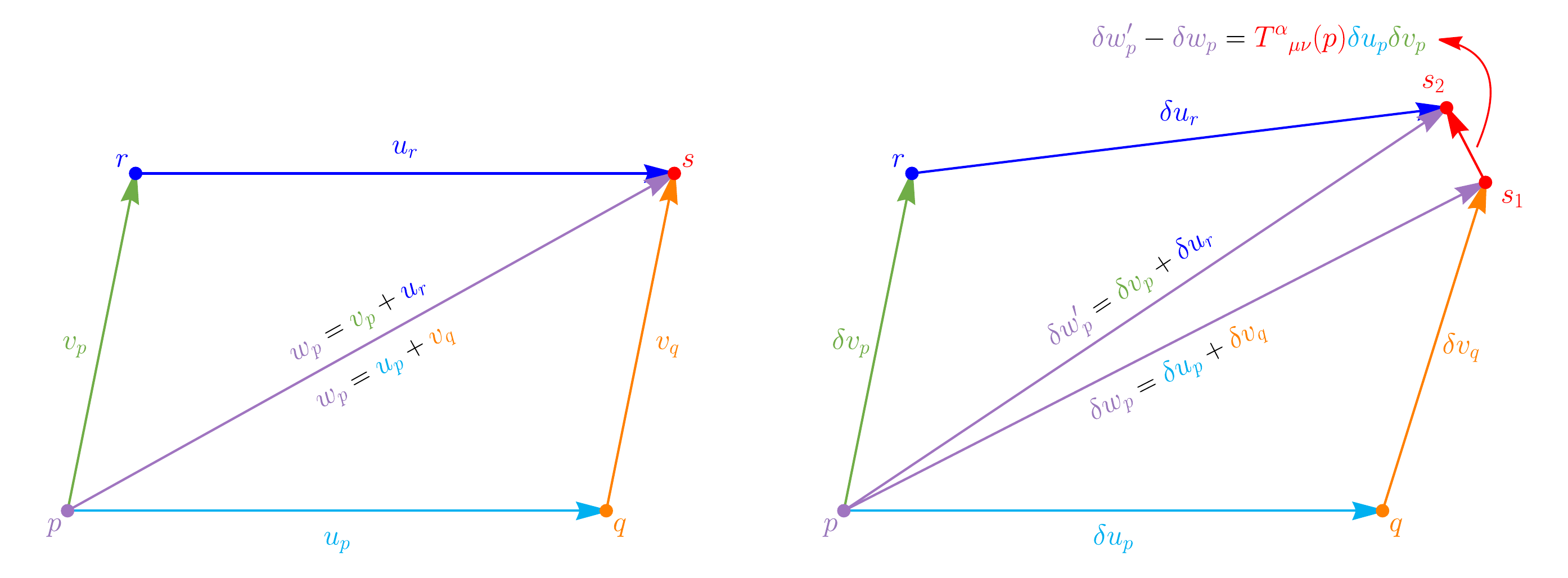}
	\caption{\protect Left panel: The notion of parallel transport used in Euclidean geometry implies that a vector $v_p$ parallel transported along $u_p$ to its tip at $q$ has the same effect as parallel transporting $u_p$ along $v_p$ up to the point $r$. That is, both operations result in a new vector pointing from $p$ to $s$. Right panel: When a generic connection is being used to define parallel transport, the same operation described above (at an infinitesimal level) no longer leads to a closed parallelogram. Rather, the two parallel transports end on different points, $s_1$ and $s_2$. The difference between $\delta w_p$ and $\delta w'_p$ is measured by the torsion tensor $T\ud{\alpha}{\mu\nu}$.   \hspace*{\fill}}
	\label{fig:9_Torsion}
\end{figure}

Let us now turn to property (b) listed above. In Euclidean geometry, the sum of two vectors can be visualized as a parallelogram. Let's say we have two vectors at $p\in\M$, which we call $u_p\in T_p\M$ and $v_p \in T_p\M$, respectively. Then, moving $v_p$ along $u_p$ until the vectors are tip to tail is the same as moving $u_p$ along~$v_p$ until the vectors are tip to tail. More precisely, we can think of both vectors having their tail in $p$. Vector's $u_p$ tip is pointing to $q$, while the tip of $v_p$ is at $r$. Then, moving $v_p$ to $q$ results in a new vector pointing at $s$. The same vector is obtained by moving $u_p$ along $v_p$ to $r$. The total displacement from $p$ to $s$ is given by $w_p \ce u_p + v_p$ (see Figure~\ref{fig:9_Torsion}).
In non-Euclidean geometry, it is conceivable that this will no longer be the case. To analyze the situation, we restrict ourselves to infinitesimal vectors $\delta u_p\ce \epsilon\, u_p$ and $\delta v_p \ce \lambda\, v_p$, with $\epsilon\ll 1$ and $\lambda\ll 1$. An absolutely necessary assumption is that $\delta u_p$ and $\delta v_p$ are linearly independent. For if they are either parallel or anti-parallel, we would never obtain something resembling the parallelogram shown in Figure~\ref{fig:9_Torsion}.

According to equation~\eqref{eq:InfinitesimalParallelTransport}, the infinitesimal parallel transport of $\delta v_p$ to the tip of $\delta u_p$ (i.e., to the point $q$) is given by
\begin{align}
	\delta v^\alpha_q &=  \delta v^\alpha_p - \Gamma\ud{\alpha}{\mu\nu}(p)\,\delta u^\mu_p\, \delta v^\nu_p\,
\end{align}
and the total displacement from $p$ to $s_1$ is
\begin{align}
	\delta w^\alpha_p \ce  \delta u^\alpha_p + \delta v^\alpha_p - \Gamma\ud{\alpha}{\mu\nu}(p)\,\delta u^\mu_p\,\delta v^\nu_p\,.
\end{align}
Conversely, if we first transport $\delta u_p$ to $r$ (i.e., the tip of $\delta v_p$) and then consider the total displacement from $p$ to $s_2$ we obtain
\begin{align}
	\delta w'^\alpha_p \ce \delta v^\alpha_p + \delta u^\alpha_p - \Gamma\ud{\alpha}{\nu\mu}(p)\,\delta u^\mu_p\, \delta v^\nu_p\,.
\end{align}
To measure whether the parallelogram actually closes, i.e., in order to see whether the total displacement results in $s_1 = s_2$, we have to compare $w^\alpha_p$ to $w'^\alpha_p$:
\begin{align}
	w'^\alpha_p - w^\alpha_p &= \left(\Gamma\ud{\alpha}{\mu\nu}(p) - \Gamma\ud{\alpha}{\nu\mu}(p)\right)\, \delta u^\mu_p\, \delta v^\mu_p \notag\\
	&\ec T\ud{\alpha}{\mu\nu}\, \delta u^\mu_p\, \delta v^\mu_p\,,
\end{align}
where in the last line we introduced the \textbf{torsion tensor}
\begin{align}\label{eq:DefTorsion}
	\boxed{T\ud{\alpha}{\mu\nu} \ce 2\Gamma\ud{\alpha}{[\mu\nu]} = \Gamma\ud{\alpha}{\mu\nu} - \Gamma\ud{\alpha}{\nu\mu}}
\end{align}
Notice that the parallelogram closes if and only if $T\ud{\alpha}{\mu\nu} = 0$. That is, it closes precisely when torsion vanishes. If torsion is not zero, it provides us with a measure for the failure of the infinitesimal parallelogram to close. Also, notice that the torsion is anti-symmetric in its lower two indices. This implies that if $\delta u_p$ and $\delta v_p$ are linearly dependent, i.e., if $\delta u_p = f\, \delta v_p$ for some non-zero scalar $f$, then $T\ud{\alpha}{\mu\nu} \,\delta u_p\, \delta v_p = f\, T\ud{\alpha}{\mu\nu} \,\delta v_p\, \delta v_p = 0$, since we are contracting something anti-symmetric with something symmetric. This agrees with our intuition that two linearly dependent vectors do not span a parallelogram, hence there is no ``failure to close'' to be measured. 

On a more technical note, the torsion tensor is simply the anti-symmetric part of the connection. Recall that the connection does not transform in a tensorial manner under coordinate transformations due to the inhomogeneous piece. However, this inhomogeneity cancels when we compute the difference $\Gamma\ud{\alpha}{\mu\nu} - \Gamma\ud{\alpha}{\nu\mu}$, making $T\ud{\alpha}{\mu\nu}$ a genuine tensor.

A connection which is symmetric in the lower indices, i.e., a connection for which $T\ud{\alpha}{\mu\nu}$ is zero, is called \textbf{torsion-free}. Finally, we remark that we can construct the trace of the torsion tensor, even in the absence of a metric, by contracting indices as
\begin{align}\label{eq:DefTraceTorsion}
	\boxed{T_\mu \ce T\ud{\alpha}{\mu\alpha}}
\end{align}
One has to be mindful of the order of the contracted indices, since 
\begin{align}
	T\ud{\alpha}{\alpha\mu} = - T\ud{\alpha}{\mu\alpha} = - T_\mu\,.
\end{align}
Moreover, this is the only trace that can be built. If there is a metric, one could be tempted to construct $g^{\mu\nu} T\ud{\alpha}{\mu\nu}$. However, this contraction is identically zero, since the metric is symmetric in $\mu$ and $\nu$, while the torsion tensor is anti-symmetric.

\subsection{Non-Metricity}\label{ssec:NonMetricity}
\begin{figure}[htb!]
	\centering
	\includegraphics[width=0.75\linewidth]{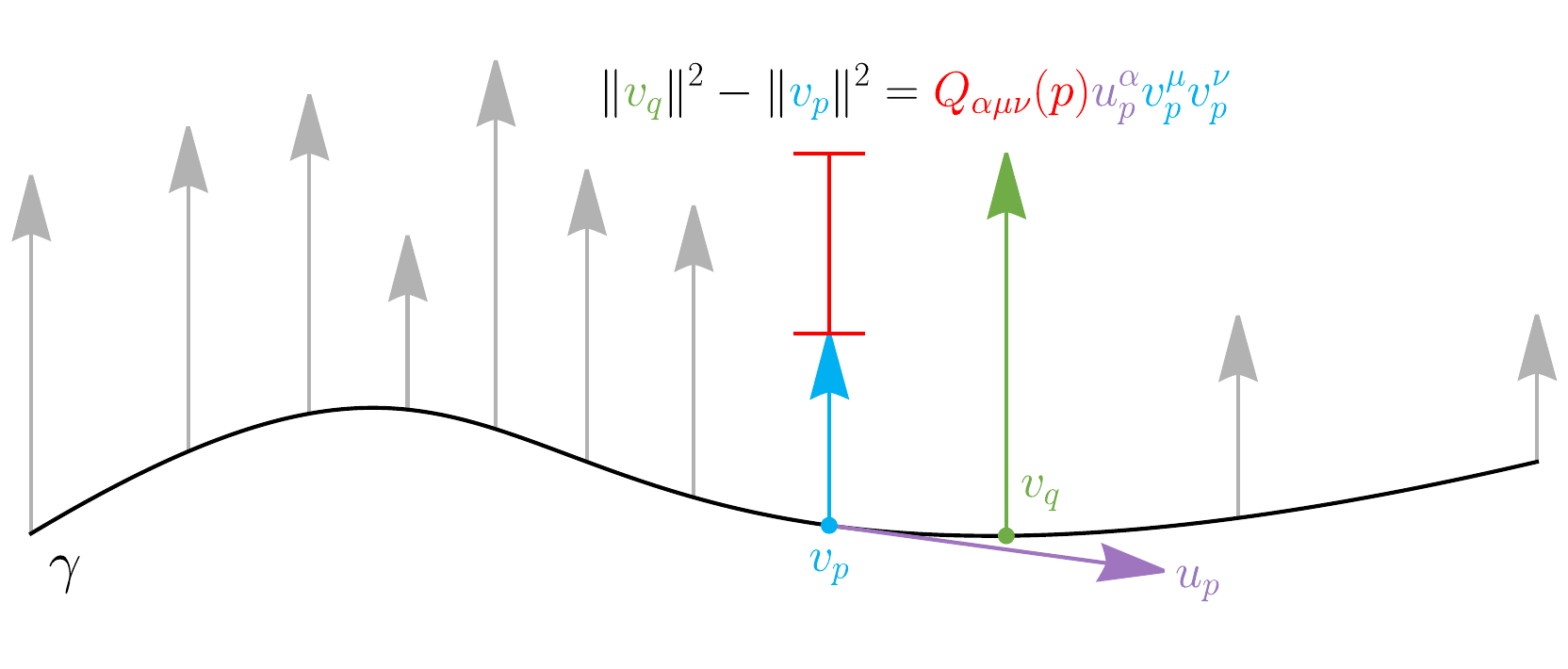}
	\caption{\protect A vector field $v$ is being parallel transported along a curve $\gamma$. The magnitude $\|v\|^2$ changes from point to point along $\gamma$. For two infinitesimally close points $p$ and $q$, the change in magnitude is measured by the non-metricity tensor, $\|v_q\|^2 - \|v_p\|^2 = Q_{\alpha\mu\nu}(p) u^\alpha_p v^\mu_p v^\nu_p$, where $u_p$ is the tangent vector of $\gamma$ at the point $p$.  \hspace*{\fill}}
	\label{fig:10_NonMetricity}
\end{figure}

Finally, we consider the third property associated with vectors: Their magnitude. Let $v\in T\M$ with components $v^\mu$ in a given coordinate chart. In oder to define the magnitude of a vector, we need a metric. Let that metric be $g$ and denote its components in the same coordinate chart as before by $g_{\mu\nu}$. Then, we define the magnitude\footnote{We recall that if the signature of the metric is Lorentzian, the magnitude can be positive, negative, or even zero for $v\neq 0$.} of the vector $v$ as
\begin{align}
	\|v\|^2 \ce g(v, v) = g_{\mu\nu} v^\mu v^\nu\,.
\end{align} 
How does the magnitude of a vector change if we parallel transport it along some curve $\gamma$, as illustrated in Figure~\ref{fig:10_NonMetricity}? To answer this question, we assume that $u=u^\mu \partial_\mu$ is the tangent vector to the curve~$\gamma$. Furthermore, we assume that $v$ is parallel transported along $\gamma$, which means it satisfies the parallel transport equation
\begin{align}
	u^\alpha \nabla_\alpha v^\mu = 0\,
\end{align}
with respect to the covariant derivative induced by $\Gamma\ud{\alpha}{\mu\nu}$.

 This allows us to determine how the magnitude changes under parallel transport. Using Leibniz's rule we find
\begin{align}
	\TD{}{t} P(\gamma)_0^{t}\|v\|^2 &= u^\alpha \nabla_\alpha \left(g_{\mu\nu} v^\mu v^\nu\right) = \left(u^\alpha \nabla_\alpha g_{\mu\nu}\right) v^\mu v^\nu + 2 g_{\mu\nu} \underbrace{\left(u^\alpha \nabla_\alpha v^\mu\right)}_{=0} v^\nu \notag\\
	&\ec Q_{\alpha\mu\nu} u^\alpha v^\mu v^\nu\,,
\end{align}
where the last term on the first line vanishes due to the parallel transport equation and where we have introduced the \textbf{non-metricity tensor}
\begin{align}\label{eq:DefNonMetricity}
	\boxed{Q_{\alpha\mu\nu} \ce \nabla_\alpha g_{\mu\nu}}
\end{align}
If the connection $\Gamma\ud{\alpha}{\mu\nu}$ is such that this tensor is not zero, $Q_{\alpha\mu\nu}$ can be interpreted as a measure for how the magnitude of a vector changes under parallel transport. The condition
\begin{align}\label{eq:MetricityCondition}
	\nabla_\alpha g_{\mu\nu} \overset{!}{=} 0
\end{align}
is called the \textbf{metricity condition} and a connection $\Gamma\ud{\alpha}{\mu\nu}$ which satisfies~\eqref{eq:MetricityCondition} is called \textbf{metric-compatible}. A point which is sometimes overlooked or which causes a bit of confusion is that the non-metricity tensor with its last two indices raised is \textit{not} equal to the covariant derivative of the inverse metric,
\begin{align}
	Q\du{\alpha}{\mu\nu} \neq \nabla_\alpha g^{\mu\nu}\,.
\end{align}
By computing the covariant derivative of the identity $g_{\mu\lambda}g^{\lambda\nu} = \delta\du{\mu}{\nu}$, one can easily show that the correct expression is
\begin{align}
		\boxed{Q\du{\alpha}{\mu\nu} = - \nabla_\alpha g^{\mu\nu}}\,,
\end{align}
i.e., with a minus sign in front of the derivative. For completeness we also remark that non-metricity as measure for the change of magnitude under parallel transport is the simplest geometric interpretation of $Q_{\alpha\mu\nu}$. In more generality we can say that the non-metricity tensor measures how quantities which depend on the metric change when they are parallel transported. For instance, $Q_{\alpha\mu\nu}$ is also a measure for how the angle\footnote{For a definition of angles in Lorentzian geometries of arbitrary dimension see for instance~\cite{DAmbrosioThesis}.} between two vectors change. An other example is the $n$-dimensional volume of a region $\Omega\subset \M$, which is defined as
\begin{align}
	\text{Vol}(\Omega) \ce \int_{\Omega} \sqrt{|g|}\, \dd^n x\,.
\end{align}
If we parallel transport $\text{Vol}(\Omega)$ along $\gamma$ with tangent vector $u$, we obtain
\begin{align}
	\TD{}{t}\text{Vol}(\Omega	) &= \int_{\Omega} u^\alpha \left(\nabla_\alpha \sqrt{|g|}\right)\, \dd^n x = \frac12 \int_{\Omega} \sqrt{|g|}\, u^\alpha g^{\mu\nu} \nabla_\alpha g_{\mu\nu}\, \dd^n x \notag\\
	&= \frac12 \int_{\Omega} \sqrt{|g|} u^\alpha g^{\mu\nu} Q_{\alpha\mu\nu}\, \dd^n x\,.
\end{align}
In the last line, the trace $g^{\mu\nu} Q_{\alpha\mu\nu}$ of the non-metricity tensor appears. It is convenient to properly introduce a symbol for this trace, as it will appear quite frequently. Since the non-metricity tensor has three indices and the last two are symmetric, we can define two independent traces:
\begin{equation}
\boxed{
	\begin{aligned}
	Q_\alpha &\ce g^{\mu\nu} Q_{\alpha\mu\nu} = Q\dud{\alpha}{\mu}{\mu} & \text{ and } && \bar{Q}_\alpha &\ce g^{\mu\nu} Q_{\mu\nu\alpha} = Q\ud{\mu}{\mu\alpha}
\end{aligned}
}
\end{equation}

\subsection{Classification of Metric-Affine Geometries and the Decomposition of the Connection}\label{ssec:Classification}
Given that the connection is not a tensor, it cannot have an intrinsic geometric meaning. That is to say, the connection $\Gamma$ by itself cannot be a measure of some geometric property of a metric-affine geometry $(\M, g, \Gamma)$. However, we have seen that the connection does give rise to true tensorial objects: Curvature, torsion, and non-metricity.

\begin{figure}[htb!]
	\centering
	\includegraphics[width=0.9\linewidth]{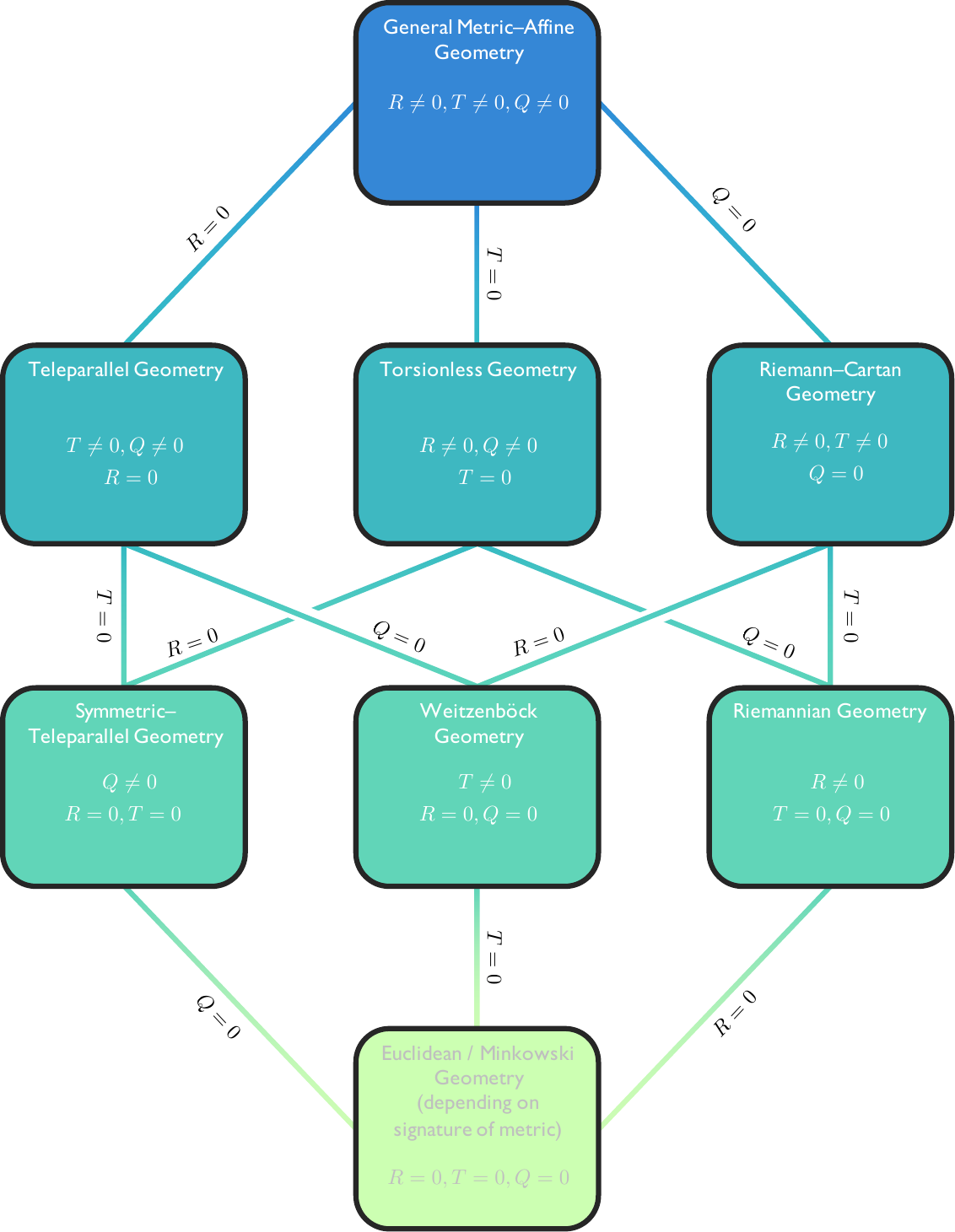}
	\caption{\protect A diagram of all the possible metric-affine geometries $(\M, g, \Gamma)$. The classification is based on whether or not curvature $R\ud{\alpha}{\mu\nu\rho}$, torsion $T\ud{\alpha}{\mu\nu}$, and/or non-metricity $Q_{\alpha\mu\nu}$ vanish.  \hspace*{\fill}}
	\label{fig:11_Diagram}
\end{figure}

Therefore, the connection and the metric \textit{do} carry intrinsic geometric information about a given metric-affine geometry. In fact, we can distinguish between different types of geometries:
\begin{itemize}
	\item[0.] \textbf{Bare manifold:} The simplest type consists simply of a manifold $\M$ without any metric nor connection. This is sufficient to talk about curves, scalar fields, vector fields, and other tensor fields. However, no notion of length or distance or covariant differentiation (except for the scalar field) is defined. From a physics perspective, this is the least useful type of geometry.
	\item[1.] \textbf{Affine geometry:} An affine geometry consists of the couple $(\M, \Gamma)$. One can do all the things one can do with a bare manifold and, on top of that, a covariant derivative can be defined. Given a connection $\Gamma$, one can also compute the curvature and torsion tensors. Thus, affine geometries can have curvature, or torsion, or both. However, because no metric is defined, one lacks a notion of distance or length and, consequently, of geodesics. 
	\item[2.] \textbf{Riemannian geometry:} A Riemannian geometry consists of the pair $(\M, g)$. This type of geometry has the advantage that it comes equipped with a notion of length and distance. Thus, it is possible to talk about geodesics, magnitudes of vectors, as well as areas and volumes and so on. Given a metric, one can compute its Christoffel symbols, aka its Levi-Civita connection. Thus, a Riemannian manifold comes naturally equipped with a covariant derivative. Namely the derivative $\D$ induced by the Levi-Civita connection of the metric. It turns out that this connection is torsion-free and metric-compatible. Therefore, Riemannian geometries are characterized by having curvature, but no torsion and no non-metricity.
	\item[3.] \textbf{Metric-affine geometry:} The most general type is the metric-affine geometry, consisting of the triple $(\M, g, \Gamma)$. All geometric concepts discussed so far are defined for this type of geometry. Furthermore, one can subdivide metric-affine geometries as follows (see also Figure~\ref{fig:11_Diagram}):
		\begin{itemize}
			\item[3.1] $R\ud{\alpha}{\mu\nu\rho} = 0$, $T\ud{\alpha}{\mu\nu} = 0$, $Q_{\alpha\mu\nu} = 0$: When all three geometric tensors vanish, one is left with Euclidean space or Minkowski space (depending on the metric signature).
			\item[3.2]  $R\ud{\alpha}{\mu\nu\rho} \neq 0$, $T\ud{\alpha}{\mu\nu} = 0$, $Q_{\alpha\mu\nu} = 0$: Curvature is the only non-vanishing tensor. This means, unsurprisingly, that Riemannian geometry is a special case of a metric-affine geometry. This is also the mathematical framework within which General Relativity is formulated.
			\item[3.3] $R\ud{\alpha}{\mu\nu\rho} = 0$, $T\ud{\alpha}{\mu\nu} \neq 0$, $Q_{\alpha\mu\nu} = 0$: Torsion is the only non-vanishing tensor. This will be the geometry on which we build TEGR, the Teleparallel Equivalent of General Relativity, and its various extensions.
			\item[3.4] $R\ud{\alpha}{\mu\nu\rho} = 0$, $T\ud{\alpha}{\mu\nu} = 0$, $Q_{\alpha\mu\nu} \neq 0$: Non-metricity is the only non-vanishing tensor. This is the basis on which we construct STEGR, the Symmetric Teleparallel Equivalent of General Relativity, and its extensions.
			\item[3.5] $R\ud{\alpha}{\mu\nu\rho} = 0$, $T\ud{\alpha}{\mu\nu} \neq 0$, $Q_{\alpha\mu\nu} \neq 0$: Torsion and non-metricity are both non-vanishing. This geometry can also be used to construct theories of gravity, namely the General Teleparallel Equivalent of General Relativity, or GTEGR for short.
			\item[3.6] $R\ud{\alpha}{\mu\nu\rho} \neq 0$, $T\ud{\alpha}{\mu\nu} \neq 0$, $Q_{\alpha\mu\nu} = 0$: Curvature and torsion are non-zero. This is a possible geometry, but not one that will be further discussed in this review.
			\item[3.7] $R\ud{\alpha}{\mu\nu\rho} \neq 0$, $T\ud{\alpha}{\mu\nu} = 0$, $Q_{\alpha\mu\nu} \neq 0$: It is also possible to obtain geometries with non-vanishing curvature and non-metricity. This type of geometry will also not be of any interest to us.
			\item[3.8] $R\ud{\alpha}{\mu\nu\rho} \neq 0$, $T\ud{\alpha}{\mu\nu} \neq 0$, $Q_{\alpha\mu\nu} \neq 0$: Clearly, the most general type of geometry is the one where none of the characteristic tensors vanish. 
		\end{itemize}
\end{itemize} 
Even tough the connection does not have a \textit{direct} geometric meaning which is invariant under changes of coordinates (and thus intrinsic to the geometry), geometric information can nevertheless be extracted from it. This begs the question whether the connection can be decomposed in a form which makes the geometric information it carries more evident. That is, can it be brought into a form which shows us whether it gives rise to non-vanishing torsion or non-metricity?

The strategy is as follows: We work in a generic metric-affine geometry $(\M, g, \Gamma)$ and we compute the covariant derivative of the metric, perform cyclic permutations of the indices, and finally isolate the connection coefficients $\Gamma\ud{\alpha}{\mu\nu}$. The final result should definitely know about torsion (since torsion is the anti-symmetric part of the connection), it should know about the Levi-Civita connection, since the Levi-Civita connection is usually obtained in precisely this fashion, and it should also know about non-metricity, since we never imposed the vanishing of $\nabla_\alpha g_{\mu\nu}$. The covariant derivative and its cyclic permutations read
\begin{align}
	\partial_\alpha g_{\mu\nu} - \Gamma\ud{\beta}{\alpha\nu} g_{\mu\beta} - \Gamma\ud{\beta}{\alpha\mu} g_{\beta\nu} &= Q_{\alpha\mu\nu} \notag\\
	\partial_\mu g_{\nu\alpha} - \Gamma\ud{\beta}{\mu\alpha} g_{\nu\beta} - \Gamma\ud{\beta}{\mu\alpha} g_{\beta\alpha} &= Q_{\mu\nu\alpha} \notag\\
	\partial_\nu g_{\alpha\mu} - \Gamma\ud{\beta}{\nu\alpha} g_{\beta\mu} - \Gamma\ud{\beta}{\nu\mu} g_{\alpha\beta} &= Q_{\nu\alpha\mu} \,.
\end{align}
After adding together the first two equations and subtracting the last one, we obtain
\begin{align}
	T\ud{\beta}{\nu\mu}g_{\alpha\beta} + T\ud{\beta}{\nu\alpha}g_{\mu\beta} + T\ud{\beta}{\alpha\mu}g_{\nu\beta}  + \partial_\alpha g_{\mu\nu} + \partial_\mu g_{\nu\alpha} - \partial_\nu g_{\alpha\mu} - 2 \Gamma\ud{\beta}{\alpha\mu}g_{\nu\beta} =  Q_{\alpha\mu\nu} + Q_{\mu\nu\alpha} - Q_{\nu\alpha\mu}\,.
\end{align}
This finally allows us to solve for the connection and we find (after re-labelling some indices)
\begin{align}\label{eq:GammaDecomposition}
	\boxed{\G{\alpha}{\mu\nu} = \LC{\alpha}{\mu\nu} + K\ud{\alpha}{\mu\nu} + L\ud{\alpha}{\mu\nu}}
\end{align}
In this compact form of the decomposed connection we have introduced the \textbf{contorsion tensor $\boldsymbol{K\ud{\alpha}{\mu\nu}}$} and the \textbf{disformation tensor $\boldsymbol{L\ud{\alpha}{\mu\nu}}$}
\begin{equation}
	\boxed{
	\begin{aligned}
		K\ud{\alpha}{\mu\nu} &\ce \frac12 T\ud{\alpha}{\mu\nu} + T\dud{(\mu}{\alpha}{\nu)}\\
		L\ud{\alpha}{\mu\nu} &\ce \frac12 Q\ud{\alpha}{\mu\nu} - Q\dud{(\mu}{\alpha}{\nu)}
	\end{aligned}
	}
\end{equation}
Observe that the contorsion tensor is constructed from the torsion tensor alone, while the disformation tensor only depends on the non-metricity. From this decomposition, one recovers very quickly the well-known fact that a torsion-free (i.e., $T\ud{\alpha}{\mu\nu} = 0$) and metric-compatible (i.e., $Q_{\alpha\mu\nu} = 0$) connection is \textit{uniquely} given by the Levi-Civita connection:
\begin{align}
	\boxed{
	T\ud{\alpha}{\mu\nu} = 0\, \text{ and }\, Q_{\alpha\mu\nu} = 0\qquad\Longrightarrow\qquad \Gamma\ud{\alpha}{\mu\nu} = \LC{\alpha}{\mu\nu}
	}
\end{align}

\subsection{The Lie Derivative Revisited: Symmetries of Metric-Affine Geometries}\label{ssec:LieDRevisited}
The concept of Lie derivative has nothing to do with parallel transport or a connection. It solely relies on flows generated by vector fields. Hence, Lie derivatives can be defined completely intrinsically to $\M$ and no need arises to introduce a connection or a metric. 

Nevertheless, one can establish a relationship between the Lie derivative and the covariant derivative. For instance, it is well-known that in Riemannian geometry one can replace in the computation of the Lie derivative the coordinate derivative $\partial_\mu$ by the covariant derivative $\D_\mu$ without altering the result. The reason for this is that all terms containing a connection cancel out at the end of the computation.

Despite these cancellations, there are advantages to using the covariant derivative $\D_\mu$ when computing the Lie derivative. For instance, in the case of the metric tensor one finds
\begin{align}\label{eq:LieDg}
	\mathcal{L}_v g_{\mu\nu} &= v^\lambda \partial_{\lambda} g_{\mu\nu} + g_{\lambda\nu} \partial_\mu v^\lambda + g_{\mu\lambda} \partial_{\nu} v^\lambda \notag\\
	&= v^\lambda \D_{\lambda} g_{\mu\nu} + g_{\lambda\nu} \D_\mu v^\lambda + g_{\mu\lambda} \D_{\nu} v^\lambda \notag\\
	&= 2 \D_{(\mu}v_{\nu)}\,.
\end{align}
From the first to the second line we replaced $\partial_\mu$ by $\D_\mu$, since this does not alter the result. Then we used the fact that $\D_\mu$ is metric-compatible, $\D_{\lambda}g_{\mu\nu} = 0$, in order to eliminate the first derivative and commute the metric past $\D_{\mu}$ in order to lower the index of the vector field. This gives us the compact result on the last line.

It is sometimes useful, for instance when discussing symmetries of metric-affine geometries, to compute the Lie derivative in terms of a general affine connection. Of particular interest are the Lie derivatives of the metric and the connection, which can be written as~\cite{BeltranJimenez:2020}
\begin{equation}\label{eq:LieDgandG}
	\boxed{
	\begin{aligned}
	\mathcal{L}_v g_{\mu\nu} &= 2 g_{\lambda(\mu}\nabla_{\nu)}v^\lambda + \left(Q_{\lambda\mu\nu} - 2T_{(\mu\nu)\lambda}\right) v^\lambda = 2\D_{(\mu}v_{\nu)} \\
	\mathcal{L}_v \Gamma\ud{\alpha}{\mu\nu} &= \nabla_\mu \nabla_\nu v^\alpha - T\ud{\alpha}{\nu\lambda} \nabla_\mu v^\lambda - \left(R\ud{\alpha}{\nu\mu\lambda} + \nabla_\mu T\ud{\alpha}{\nu \lambda}\right)v^\lambda\,,
\end{aligned}
	}
\end{equation}

where $\nabla$ is a general affine connection. When computing the Lie derivative of the connection, one has to make sure to use the correct formula. Namely,
\begin{align}\label{eq:LieDConnection}
	\mathcal{L}_v \Gamma\ud{\alpha}{\mu\nu} = v^\lambda \partial_\lambda \Gamma\ud{\alpha}{\mu\nu} - \Gamma\ud{\lambda}{\mu\nu} \partial_\lambda v^\alpha + \Gamma\ud{\alpha}{\lambda\nu} \partial_\mu v^\lambda +  \Gamma\ud{\alpha}{\mu\lambda} \partial_\nu v^\lambda + \boldsymbol{\partial_\mu\partial_\nu v^\alpha}\,.
\end{align}
This follows directly from the fact that the connection does \textit{not} transform like a tensor and instead obeys the inhomogeneous transformation law~\eqref{eq:ConnectionTransformationLaw}. Indeed, one recognizes the first four terms to be related to the homogeneous piece in the coordinate transformation of the connection, while the last term is produced by the inhomogeneous piece. Also, one should note that even tough the connection is not a tensor, its Lie derivative \textit{is} a $(1,2)$ tensor! This is also nicely evident from equation~\eqref{eq:LieDgandG}, where the right hand side is completely constructed from tensorial quantities.

The formulas~\eqref{eq:LieDgandG} play a role in characterizing symmetries of metric-affine geometries, as alluded to before. In the works~\cite{Hohmann:2019, DAmbrosio:2021b, Hohmann:2021, DAmbrosio:2021, Heisenberg:2022, DAmbrosio:2023b}, symmetries of metric-affine geometries were defined as follows: Let $\phi_s:\bbR\times\M \to \M$ be a $1$-parameter family of diffeomorphisms which satisfies $\phi_0 = \text{id}$, $\phi_s\circ \phi_t = \phi_{s+t}$, and which is smooth in the parameter $s$. This $1$-parameter family is a \textbf{symmetry of a metric-affine geometry $\boldsymbol{(\M, g, \Gamma)}$} if
\begin{align}
	\begin{cases}
		\phi^*_s g_{\mu\nu} & \overset{!}{=}\quad g_{\mu\nu} \\
		\phi^*_s \Gamma\ud{\alpha}{\mu\nu} &\overset{!}{=}\quad \Gamma\ud{\alpha}{\mu\nu}
	\end{cases}	
\end{align}
This is called the \textbf{symmetry condition}. It demands that neither the metric nor the connection change under the action of the diffeomorphism. It is important that the connection appears in this definition. This ensures that all objects constructed from the connection, such as curvature, torsion, and non-metricity, respect the symmetry generated by $\phi_s$.

Since $\phi_s$ is smooth in $s$, we can also consider the \textbf{infinitesimal symmetry condition}, obtained by expanding the original symmetry condition to first order in $s$ around $s=0$. It reads
\begin{align}
	&\begin{cases}
		\mathcal{L}_\xi g_{\mu\nu} & \overset{!}{=}\quad 0 \\
		\mathcal{L}_\xi \Gamma\ud{\alpha}{\mu\nu} & \overset{!}{=}\quad 0
	\end{cases}
	& \text{with} &&
	\xi \ce \left.\TD{}{s}\phi_s\right|_{s=0}\,,
\end{align}
where $\xi$ is the vector field which generates the flow $\phi_s$. It is often called the \textbf{generating vector field}. These symmetry conditions and the Lie derivatives of metric and connection will reappear when we discuss cosmology and black holes in subsections~\ref{ssec:Cosmology} and~\ref{ssec:BlackHoles}.

\subsection{Integration in the Presence of Torsion and Non-Metricity: The Generlized Gauss Theorem}\label{ssec:Integration}
Integration on manifolds $\M$ is a subject usually covered in courses on calculus of several variables or differential geometry. The theorems of Gauss and Stokes, into which this subject culminates, can be assumed to be familiar to all readers due to their widespread use in physics.

What concerns us here, however, is how non-trivial geometric features, characterized by torsion and non-metricity, affect Gauss' theorem. The importance of this investigation lies in the fact that Gauss' theorem appears in variational principles or discussions of conserved charges~\cite{Delhom:2020}.

Let us begin by recalling that on a Riemannian manifold $(\M, g)$, where torsion and non-metricity both vanish, we are uniquely left with the Levi-Civita connection $\LC{\alpha}{\mu\nu}$ and the covariant derivative operator $\D_\mu$ it induces. Using the easy to verify identity
\begin{equation}\label{eq:LeviCivitaIdentity}
	\LC{\lambda}{\lambda\mu} = \partial_\mu \log\sqrt{|g|}\,,
\end{equation}
one can re-express the divergence of a vector field $v^\mu$ as
\begin{equation}\label{eq:DivergenceIdentity}
	\D_\mu v^\mu = \frac{1}{\sqrt{|g|}}\partial_\mu \left(\sqrt{|g|}v^\mu\right)\,.
\end{equation}
Gauss' theorem, in its familiar form, emerges from this simple identity~\cite{PoissonBook}
\begin{align}\label{eq:GaussTheorem}
	\int_\M \dd^4 x\,\sqrt{|g|}\, \D_\mu v^\mu &= \int_\M \dd^4 x\, \partial_\mu\left(\sqrt{|g|}v^\mu\right) = \oint_{\partial\M}\dd^3 y\, \sqrt{|h|}\,\varepsilon\, n_\mu v^\mu\, .
\end{align}
Here, $\partial\M$ denotes the boundary of $\M$, $n^\mu$ is the outward-pointing normal vector to $\partial\M$, $\varepsilon \ce n_\mu n^\mu = \pm 1$, and $h$ is the determinant of the induced metric on $\partial\M$ (obtained by pulling back $g_{\mu\nu}$ to $\partial\M$).

Clearly, this result crucially depends on the identity~\eqref{eq:LeviCivitaIdentity} for the Levi-Civita connection. Hence, if we change the connection, we can no longer expect to find the same form of Gauss' theorem as the one given in~\eqref{eq:GaussTheorem}.  

However, an analogous theorem holds for a generic metric-affine geometry $(\M, g, \Gamma)$. To see that, we make use of the decomposition~\eqref{eq:GammaDecomposition} of the connection which we encountered in the previous subsection:
\begin{equation}
	\G{\alpha}{\mu\nu} = \LC{\alpha}{\mu\nu} + L\ud{\alpha}{\mu\nu} + K\ud{\alpha}{\mu\nu}\,.
\end{equation}
This allows us to write the divergence of the vector field $v^\mu$ as
\begin{equation}
	\nabla_\mu v^\mu = \D_\mu v^\mu + L\ud{\lambda}{\lambda\mu} v^\mu + K\ud{\lambda}{\lambda\mu} v^\mu\,.
\end{equation}
Next, we use the easy provable relations
\begin{align}
	L\ud{\lambda}{\lambda\mu} &= -\frac12 Q_\mu &\text{and} && K\ud{\lambda}{\lambda\mu} &= -T_\mu\,,
\end{align}
and combine these with the identity~\eqref{eq:DivergenceIdentity} to finally obtain
\begin{equation}\label{eq:DivergenceFinal}
	\nabla_\mu v^\mu = \frac{1}{\sqrt{|g|}}\partial_\mu\left(\sqrt{|g|}v^\mu\right) -\left(\frac12 Q_\mu + T_\mu\right) v^\mu\,.
\end{equation}
This form of the divergence of a vector field lends itself to formulating the analogue of Gauss' theorem for a generic metric-affine geometry. We find that the generalized Gauss' theorem for a generic metric-affine geometry $(\M, g, \Gamma)$ takes the form
\begin{equation}
	\boxed{\int_\M\dd^4 x\, \sqrt{|g|}\,\nabla_\mu v^\mu = \oint_{\partial\M}\dd^3 y\, \sqrt{|h|}\,\varepsilon\, n_\mu v^\mu - \int_\M\dd^4 x\, \sqrt{|g|}\,\left(\frac12 Q_\mu + T_\mu\right)v^\mu}
\end{equation}
This can also be re-cast into a slightly different form by noticing that the covariant derivative of the square root of the metric determinant is given by
\begin{equation}
	\nabla_\mu \sqrt{|g|} = \frac12\sqrt{|g|}\, Q_\mu\,.
\end{equation}
This allows us to phrase Gauss' theorem as a statement about the divergence of a vector \textit{density}, $\nabla_\mu\left(\sqrt{|g|}v^\mu\right)$, rather than about the divergence of a vector field, $\nabla_\mu v^\mu$. Concretely, the generalized Gauss theorem can be equivalently stated as
\begin{equation}
	\boxed{\int_\M\dd^4 x\,\nabla_\mu\left(\sqrt{|g|}\, v^\mu\right) = \oint_{\partial\M}\dd^3 y\, \sqrt{|h|}\,\varepsilon\, n_\mu v^\mu - \int_\M\dd^4 x\, \sqrt{|g|}\, T_\mu v^\mu}
\end{equation}
This concludes our discussion of the generalized Gauss theorem and we proceed with presenting important and useful identities for metric-affine geometries.

\subsection{Collection of Geometric Identities}\label{ssec:Identities}
In concrete computations involving the covariant derivative $\nabla_\mu$ with respect to a generic affine connection $\G{\alpha}{\mu\nu}$, it is often necessary to commute two such operators in order to obtain simpler expressions or expressions with a more transparent geometric meaning.
Recall that the covariant derivative can act on scalar, vector, and tensor fields (or densities). Thus, the simplest case to consider is the action of the commutator $[\nabla_\mu,\nabla_\nu]\ce \nabla_\mu\nabla_\nu - \nabla_\nu \nabla_\mu$ on a scalar field~$f$. Simply by using the basic definitions given in~\eqref{eq:DefTorsion}, one finds
\begin{equation}\label{eq:CommutatorScalarField}
	\boxed{[\nabla_\mu, \nabla_\nu]f = - T\ud{\lambda}{\mu\nu}\partial_\lambda f}
\end{equation}
Thus, the commutator acting on scalar fields vanishes if and only if the torsion tensor vanishes. Next, we consider the commutator $[\nabla_\mu,\nabla_\nu]$ acting on a vector field $v^\alpha$. It is again only necessary to use the basic definitions given in~\eqref{eq:DefCurvature} and~\eqref{eq:DefTorsion}, but the computations become longer. What they boil down to is the identity
\begin{equation}\label{eq:CommutatorVectorField}
	\boxed{[\nabla_\mu, \nabla_\nu] v^\alpha = R\ud{\alpha}{\lambda\mu\nu} v^\lambda - T\ud{\lambda}{\mu\nu} \nabla_\lambda v^\alpha}
\end{equation}
Observe that when torsion vanishes, the above identity reduces to the form familiar from Riemannian geometry:
\begin{align}
	[\nabla_\mu, \nabla_\nu] v^\alpha = R\ud{\alpha}{\lambda\mu\nu} v^\lambda\,.
\end{align}
However, $R\ud{\alpha}{\lambda\mu\nu}$ is \textit{not} the curvature tensor with respect to the Levi-Civita connection, since the connection could be metric-incompatible, i.e., it could have a non-zero non-metricity tensor. 

It is also useful to prove an analogous identity for the commutator acting on a $1$-form $\omega_\alpha$. In this case, one finds
\begin{equation}\label{eq:CommutatorOneForm}
	\boxed{[\nabla_\mu, \nabla_\nu] \omega_\alpha = -R\dud{\alpha}{\lambda}{\mu\nu} \omega_\lambda - T\ud{\lambda}{\mu\nu} \nabla_\lambda \omega_\alpha}
\end{equation}
Note the appearance of a minus sign in front of the curvature tensor! With the identities~\eqref{eq:CommutatorVectorField} and~\eqref{eq:CommutatorOneForm} at our disposal, we can easily prove that the commutator $[\nabla_\mu,\nabla_\nu]$ acting on a tensor field of type $(p,q)$ is given by
\begin{equation}\label{eq:CommutatorGenericTensor}
	\boxed{
	\begin{aligned}
		[\nabla_\mu,\nabla_\nu]S^{\mu_1\dots\mu_p}{}_{\nu_1\dots\nu_q} =& \phantom{..-}R^{\mu_1}{}_{\lambda\mu\nu} S^{\lambda\dots\mu_p}{}_{\lambda\dots\nu_q}	+ \dots + R^{\mu_p}{}_{\lambda\mu\nu} S^{\mu_1\dots\lambda}{}_{\lambda\dots\nu_q} \\
			& - R_{\nu_1}{}^{\lambda}{}_{\mu\nu} S^{\mu_1\dots\mu_p}{}_{\lambda\dots\nu_q} - \dots - R_{\nu_q}{}^{\lambda}{}_{\mu\nu} S^{\mu_1\dots\mu_p}{}_{\nu_1\dots\lambda} \\
			&-T^{\lambda}{}_{\mu\nu}\nabla_\lambda S^{\mu_1\dots\mu_p}{}_{\nu_1\dots\nu_q}
	\end{aligned}
	}
\end{equation}
This identity follows from the previous two mentioned above together with the fact that a $(p,q)$ tensor lives in the tensor product space $T\M^{\otimes p}\otimes T^*\M^{\otimes q}$. 

Identity~\eqref{eq:CommutatorGenericTensor} is very general and covers almost every case of interest. The cases not covered by identity~\eqref{eq:CommutatorGenericTensor} only involve tensor \textit{densities}. To remedy that, we need to understand how the commutator acts on the metric tensor and, more importantly, on its determinant. For the metric tensor, it follows from the definition of the non-metricity tensor and from identity~\eqref{eq:CommutatorGenericTensor} that we can express the commutator as
\begin{equation}
	\boxed{[\nabla_\mu, \nabla_\nu]g_{\alpha\beta} = 2\nabla_{[\mu}Q_{\nu]\alpha\beta} = -2 R_{(\alpha\beta)\mu\nu}-T\ud{\lambda}{\mu\nu}Q_{\lambda\alpha\beta}}
\end{equation}
Using $\nabla_\mu |g|^\frac{w}{2} = \frac{w}{2} |g|^\frac{w}{2} g^{\alpha\beta} \nabla_\mu g_{\alpha\beta}$, where the integer $w\geq 0$ is the density weight introduced in~\ref{ssec:VFTFD}, together with the above identity, one finds that the commutator acting on $|g|^\frac{w}{2}$ is given by
\begin{align}
	\boxed{[\nabla_\mu, \nabla_\nu] |g|^\frac{w}{2} = \frac{w}{2} |g|^{\frac{w}{2}} g^{\alpha\beta} [\nabla_\mu, \nabla_\nu] g_{\alpha\beta} = w\, |g|^{\frac{w}{2}} \nabla_{[\mu}Q_{\nu]\alpha\beta} = w\, |g|^{\frac{w}{2}} \nabla_{[\mu}Q_{\nu]}}
\end{align}
In the last step we used $Q_{\nu} \ce g^{\alpha\beta}Q_{\nu\alpha\beta}$. This finally allows us to determine the action of the commutator on a tensor density of type $(p,q)$ and weight $w$:
\begin{align}
	\boxed{[\nabla_\mu,\nabla_\nu]\left(|g|^{\frac{w}{2}}S^{\mu_1\dots\mu_p}{}_{\nu_1\dots\nu_q}\right) = |g|^{\frac{w}{2}}\left([\nabla_\mu,\nabla_\nu]S^{\mu_1\dots\mu_p}{}_{\nu_1\dots\nu_q} + w\, \nabla_{[\mu}Q_{\nu]}\right)}
\end{align}
where the first commutator on the right hand side is of course given by identity~\eqref{eq:CommutatorGenericTensor}. Observe that all previous identities follow from this one. By setting $w=0$, we recover the identities for tensor fields (as opposed to tensor densities), including the one for scalar fields, which corresponds to $w=0$ together with $p=q=0$. 

Finally, we remark that the covariant derivative $\nabla_\mu$ with respect to a generic affine connection $\G{\alpha}{\mu\nu}$ satisfies the Jacobi identity:
\begin{equation}
	\boxed{[\nabla_\alpha,[\nabla_\beta, \nabla_\gamma]] + [\nabla_\beta,[\nabla_\gamma,\nabla_\alpha]] + [\nabla_\gamma,[\nabla_\alpha,\nabla_\beta]] = 0}
\end{equation}
We now turn to important identities involving the curvature tensor. Some of these identities will play an important role in subsections~\ref{ssec:TEGR} and~\ref{ssec:STEGR}, where they greatly simplify and illuminate the definition of the Teleparallel Equivalent of GR (TEGR) and the Symmetric Teleparallel Equivalent of GR (STEGR), respectively. 

To begin with, we remark that it follows from the definitions of the curvature tensor, the torsion tensor, and the covariant derivative, that the following identities hold
\begin{equation}
\boxed{
\begin{aligned}
	R\ud{\alpha}{\mu(\nu\rho)} &= 0\\
	R\ud{\mu}{[\alpha\beta\gamma]} - \nabla_{[\alpha}T\ud{\mu}{\beta\gamma]} + T\ud{\lambda}{[\alpha\beta}T\ud{\mu}{\gamma]\lambda} &= 0\label{eq:Biacnhi_2}\\
	\nabla_{[\alpha} R\ud{\mu}{|\nu|\beta\gamma]} - T\ud{\lambda}{[\alpha\beta}R\ud{\mu}{|\nu|\gamma]\lambda} &= 0
\end{aligned}
}
\end{equation}
When we discuss teleparallel theories of gravity, it will prove to be useful to know how to relate the curvature tensor $R\ud{\alpha}{\mu\nu\rho}(\Gamma)$ of the affine connection to the curvature tensor $\R\ud{\alpha}{\mu\nu\rho}(g)$ of the Levi-Civita connection. To establish a relationship between the two, we point out that adding a $(1,2)$ tensor to a given connection $\Gamma\ud{\alpha}{\mu\nu}$ results in a new and equally valid connection
\begin{align}
	\hat{\Gamma}\ud{\alpha}{\mu\nu} = \G{\alpha}{\mu\nu} + \Omega\ud{\alpha}{\mu\nu}\,.
\end{align}
This follows directly from the transformation behaviour of a connection under changes of coordinates (cf. equation~\eqref{eq:ConnectionTransformationLaw}). We can then compute the curvature tensor of the connection $\hat{\Gamma}\ud{\alpha}{\mu\nu}$ and express it in terms of the curvature of $\Gamma\ud{\alpha}{\mu\nu}$ as well as contributions coming from the tensor $\Omega\ud{\alpha}{\mu\nu}$. One finds
\begin{align}
	\boxed{\hat{R}\ud{\alpha}{\beta\mu\nu} = R\ud{\alpha}{\beta\mu\nu} + T\ud{\lambda}{\mu\nu} \Omega\ud{\alpha}{\lambda\beta} + 2\D_{[\mu}\Omega\ud{\alpha}{\nu]\beta} + 2 \Omega\ud{\alpha}{[\mu|\lambda|}\Omega\ud{\lambda}{\nu]\beta}}
\end{align}
where $\D$ is the covariant derivative with respect to the Levi-Civita connection, as usual. This identity allows us to easily find a relation between $R\ud{\alpha}{\mu\nu\rho}(\Gamma)$ and $\R\ud{\alpha}{\mu\nu\rho}(g)$. We simply assume that the original connection was the Levi-Civita one, i.e., $\Gamma\ud{\alpha}{\mu\nu} = \LC{\alpha}{\mu\nu}$, while the tensor $\Omega\ud{\alpha}{\mu\nu}$ is the sum of contortion and disformation tensor,  $\Omega\ud{\alpha}{\mu\nu} = K\ud{\alpha}{\mu\nu} + L\ud{\alpha}{\mu\nu}$. Thus, we find that the two curvature tensors are related to each other via
\begin{equation}\label{eq:RelationBetweenCurvatures}
\boxed{
\begin{aligned}
	R\ud{\alpha}{\mu\nu\rho}(\Gamma) &= \R\ud{\alpha}{\mu\nu\rho}(g) + T\ud{\lambda}{\nu\rho}K\ud{\alpha}{\lambda\mu} + 2\D_{[\nu}K\ud{\alpha}{\rho]\mu} + T\ud{\lambda}{\nu\rho}L\ud{\alpha}{\lambda\mu} + 2\D_{[\nu}L\ud{\alpha}{\rho]\mu} \\
	&\phantom{= \R\ud{\alpha}{\mu\nu\rho}} + 2 K\ud{\alpha}{[\nu|\lambda}K\ud{\lambda}{\rho]\mu} + 2 L\ud{\alpha}{[\nu|\lambda}K\ud{\lambda}{\rho]\mu} + 2K\ud{\alpha}{[\nu|\lambda}L\ud{\lambda}{\rho]\mu} + 2 L\ud{\alpha}{\nu[\lambda} L\ud{\lambda}{\rho]\mu}
\end{aligned}
}
\end{equation}
The actual identity which will prove to be useful in teleparallel theories of gravity is the one which relates the Ricci scalars of the two connections. It reads
\begin{align}\label{eq:KeyIdG}
	\boxed{R(\Gamma) = \R(g) + \bbT + \Q + T^{\rho\mu\nu} Q_{\mu\nu\rho} - T^\mu Q_\mu  + T^\mu\bar{Q}_\mu + \D_\alpha \left(Q^\alpha - \bar{Q}^\alpha+2 T^\alpha\right)}
\end{align}
where we have introduced the \textbf{torsion scalar} and \textbf{non-metricity scalar}, respectively defined by
\begin{align}
	\bbT &\ce \frac12 \left(\frac14 T_{\alpha\mu\nu} + \frac12 T_{\mu\alpha\nu} - g_{\alpha\mu} T_\nu \right) T^{\alpha\mu\nu} \notag\\
	\Q &\ce\frac14 Q_{\alpha\mu\nu}Q^{\alpha\mu\nu} - \frac12 Q_{\alpha\mu\nu}Q^{\mu\alpha\nu} - \frac14 Q_\alpha Q^\alpha + \frac12 Q_\alpha \bar{Q}^\alpha\,.
\end{align}
Two special cases of this identity which play a role in TEGR and STEGR, respectively, are
\begin{align}\label{eq:KeyIdT}
	\boxed{R(\Gamma) = \R(g) + \bbT + 2 \D_\alpha T^\alpha}
\end{align}
where we set non-metricity to zero, and
\begin{align}\label{eq:KeyIdQ}
	\boxed{R(\Gamma) = \R(g) + \Q + \D_\alpha\left(Q^\alpha - \bar{Q}^\alpha\right)}
\end{align}
where torsion vanishes. 

Finally, we recall that for a general connections the curvature tensor is not antisymmetric in the first two indices, so one can form the non-zero homothetic tensor $H_{\mu\nu}=R^\lambda{}_{\lambda\mu\nu}$. However, by taking traces of the second  Bianchi identity above, one can show
\begin{equation*}
    H_{\mu\nu}=2R_{[\mu\nu]}+2\nabla_{[\mu}T_{\nu]}+\nabla_\lambda T^\lambda{}_{\mu\nu} +T_\lambda T^\lambda{}_{\mu\nu}\,.  
\end{equation*}
It thus follows that the homothetic tensor is \textit{not} an independent trace of the curvature tensor. It can be expressed with the help of other, already defined tensors. Another trace of the curvature tensor is the co-Ricci tensor $P\ud{\mu}{\nu} = R\ud{\mu\lambda}{\nu\lambda}$. However, using the straightforward identity $\nabla_{[\mu}Q_{\nu]\rho\sigma}=-R_{(\rho\sigma)\mu\nu}$ one can show
\begin{equation*}
    P\ud{\mu}{\nu} = R\ud{\mu}{\nu} - 2\nabla_{[\nu}Q\du{\lambda]}{\mu\lambda}\,.
\end{equation*}
So this trace is also not independent. For the Levi-Civita connection one has by metric-compatibility that $P_{\mu\nu} = R_{\mu\nu}$ as well as $H_{\mu\nu}=0$; from the latter it follows then that the Ricci tensor is symmetric, as one is used to from Riemannian geometry.

\newpage
\asection{4}{The Geometrical Trinity of General Relativity}\label{sec:Trinity}
In 1915, Einstein completed his General Theory of Relativity and he based it on Riemannian geometry. He found this at the time relatively new branch of mathematics to be an adequate language to \textit{(a)} develop a field theoretic description of gravity which cures the action-at-a-distance problem of Newtonian gravity, \textit{(b)} fully explore the consequences of the equivalence principle, and \textit{(c)} implement the idea that the laws of Nature do not depend on our arbitrary choice of coordinate systems. The latter one was an idea which, at the time, was unheard of and revolutionary. Today, we call this the principle of general covariance.

Even though it was never Einstein's intention to ``geometrize gravity''~\cite{Lehmkuhl:2014}, as it is sometimes phrased, the theory he developed lends itself to an interpretation of the phenomena of gravity as the manifestation of the curvature of spacetime. This has been the prevalent interpretation of gravity for the past 100 years.

However, as we saw in sections~\ref{sec:Fundamentals} and~\ref{sec:GeometricObjects}, Riemannian geometry is a special case of the much more general theory of metric-affine geometry. There is no physical principle that we know of which unequivocally selects Riemannian geometry as the only viable description of gravity. In fact, there are three distinct and yet physically equivalent descriptions of gravity, which are rooted in the mathematical framework of metric-affine geometry. These formulations ascribe gravitational phenomena either to non-vanishing curvature, torsion, or non-metricity. These descriptions form the \textbf{geometric trinity of General Relativity} \cite{Heisenberg:2018, BeltranJimenez:2019c}.
\begin{figure}[htb!]
	\centering
	\includegraphics[width=0.8\linewidth]{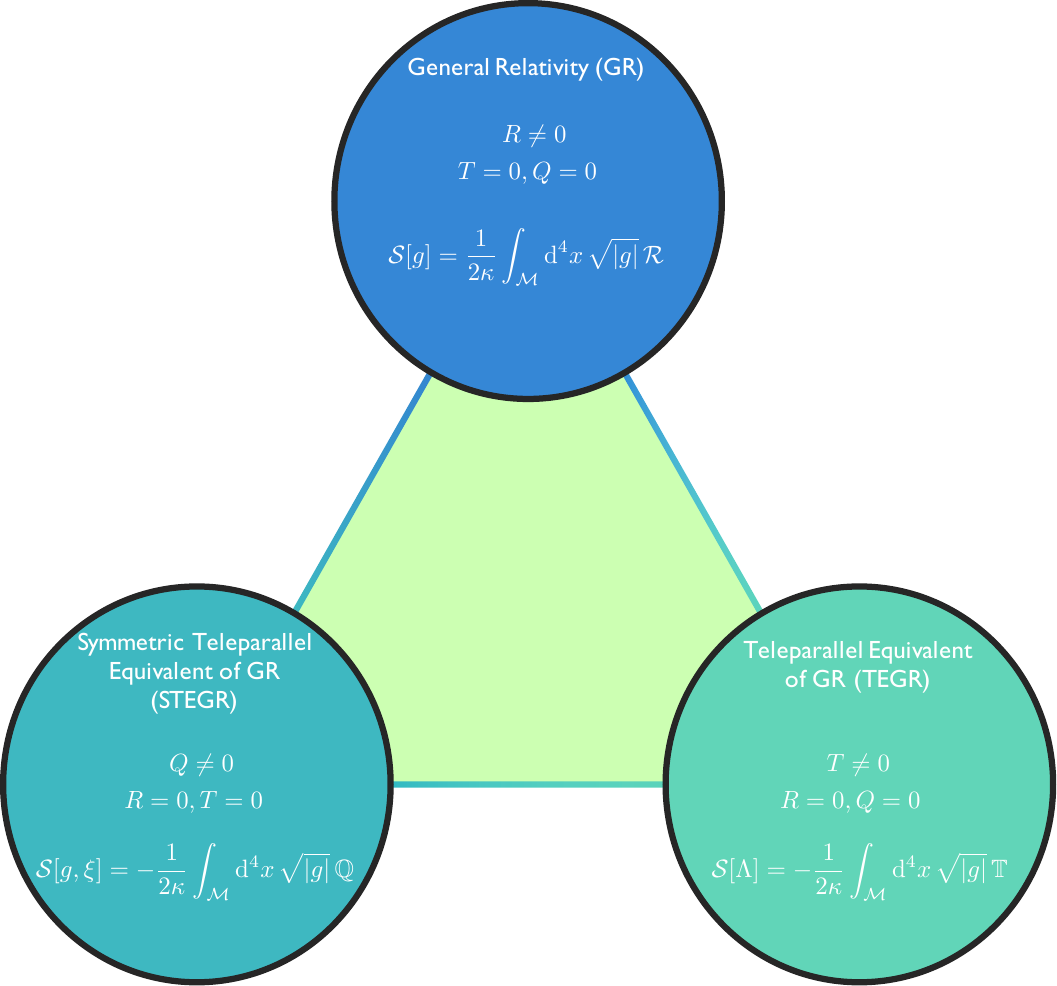}
	\caption{\protect Three different but equivalent representations of gravity: General Relativity (based on curvature), the Symmetric Teleparallel Equivalent of GR (based on non-metricity), and the Teleparallel Equivalent of GR (based on torsion).  \hspace*{\fill}}
	\label{fig:12_TrinityGR}
\end{figure}
The next three subsections are dedicated to the corners of the triangle shown in Figure~\ref{fig:12_TrinityGR}: GR, TEGR, and STEGR. We also discuss CGR as a gauge fixed version of STEGR, as well as GTEGR from which TEGR and STEGR emerge and which can be thought of as the lower edge of the triangle in~Figure~\ref{fig:12_TrinityGR}.

First, we review Einstein's original formulation in order to establish some basic facts and notations. This will also facilitate the comparison with the other two formulations, which we present in subsections~\ref{ssec:TEGR} (TEGR) and~\ref{ssec:STEGR} (STEGR). In all three formulations, the starting point is a generic metric-affine geometry and we follow a strict structure in order to construct the theory and work out its main features: \textit{a.} The Geometric Postulates; \textit{b. Form of the Connection}; \textit{c. Construction of the Action Functional}; \textit{d. The Metric and Connection Field Equations}; \textit{e. The Palatini Formulation of the Action Principle}; \textit{f. The Bianchi Identities}; and finally \textit{g. Counting Degrees of Freedom}.
We deviate from this basic structure when we discuss CGR as the gauge fixed version of STEGR in subsection~\ref{ssec:CGR}. Similarly, a different approach is used in subsection~\ref{ssec:GTEGR}, where the focus is on GTEGR, its special properties, and how TEGR and STEGR emerge from this more general theory.\newline

\subsection{Einstein's Original Formulation of General Relativity}
\paragraph{\textbf{The Geometric Postulates}}
We start with a metric-affine geometry $(\M, g, \Gamma)$ and stress that at this stage, neither $\M$ nor $g$ are fixed. This means that we do not choose a particular manifold nor do we choose a particular metric on that manifold. Both entities, $\M$ and $g$, will be determined later as solutions of Einstein's field equations. However, we need to select a connection $\Gamma$ (or, as we will see later, select at least a class of connections) in order to formulate the theory. Thus, we postulate that $\Gamma$ satisfies
\begin{align}
	T\ud{\alpha}{\mu\nu} &\overset{!}{=} 0 && \text{and} & Q_{\alpha\mu\nu} &\overset{!}{=} 0\, .
\end{align}
These two postulates leave the curvature tensor $\R\ud{\alpha}{\mu\nu\rho}$ as the only non-zero tensor which characterizes the spacetime geometry. It will be the main building block for GR.\newline

\paragraph{\textbf{Form of a Torsionless, Metric-Compatible Connection}}
It is well-known, and it can also be checked using~\eqref{eq:GammaDecomposition}, that these two geometric postulates are satisfied if and only if the connection is given by the Levi-Civita connection,
\begin{equation}\label{eq:LC}
	\Gamma\ud{\alpha}{\mu\nu} \equiv \LC{\alpha}{\mu\nu} = \frac12 g^{\alpha\lambda}\left(\partial_\mu g_{\nu\lambda} + \partial_\nu g_{\mu\lambda} - \partial_\lambda g_{\mu\nu}\right)\, .
\end{equation} 
Because the connection is completely determined by the metric, we will omit $\Gamma$ from the triple $(\M, g, \Gamma)$ and simply say that spacetime is modelled by the pair $(\M, g)$, where it is silently understood that $\Gamma$ is given by~\eqref{eq:LC}.\newline

\paragraph{\textbf{Construction of the Action Functional}} 
The action functional which defines GR is the famous Einstein-Hilbert (EH) action plus the equally famous Gibbons-Hawking-York (GHY) boundary term~\cite{York:1972, Gibbons:1977}. Including a cosmological term and a matter action for completeness, we can define GR by the functional
\begin{align}\label{GR_pureMetric}
	\S_{\rm GR}[g, \Psi] &\ce \S_{\rm EH}[g] + \S_{\rm GHY}[h] + \S_{\rm matter}[g, \Psi] \notag\\
		&\ce \frac{1}{2\kappa}\int_\M \dd^4 x\, \sqrt{|g|} \left(\R - 2\Lambda\right) + \frac{1}{\kappa}\oint_{\partial\M}\dd^3 y\, \sqrt{|h|}\,\varepsilon\, \mathcal K + \S_{\rm matter}[g, \Psi]\,,
\end{align}
with $\kappa \ce 8\pi G$. The fist integral on the second line is the aforementioned EH action (including a cosmological constant~$\Lambda$), the second integral is the GHY boundary term, and~$\S_{\rm matter}[g, \Psi]$ is the action of (tensorial) matter fields~$\Psi$ which are minimally coupled to the gravitational field $g_{\mu\nu}$. Spinorial fields are not described by the action~\eqref{GR_pureMetric}. To describe Fermions, we would have to describe gravity in terms of a tetrad field, rather than a metric tensor.

As we will discuss shortly, the GHY term is necessary whenever the manifold $\M$ has a boundary $\partial\M$, otherwise the variational principle is ill-defined and would not yield any field equations. In the above boundary  integral, $\varepsilon$ is defined as $\varepsilon\ce n_\mu n^\mu = \pm 1$, where $n^\mu$ is the normal vector to $\partial\M$. This vector is normalized to either $\varepsilon=+1$ (when $\partial\M$ is timelike) or $\varepsilon=-1$ (when $\partial\M$ is spacelike). Furthermore, $h$ denotes the determinant of the metric intrinsic to $\partial\M$, while $\mathcal{K}$ is the trace of the extrinsic curvature of $\partial\M$ viewed as hypersurface embedded into $\M$. For a didactical discussion of hypersurfaces, embeddings, and the concept of extrinsic curvature, we refer the reader to Poisson's book~\cite{PoissonBook}.\newline

\paragraph{\textbf{The Field Equations}}
Our next task is to find field equations for the metric which contain at most second order derivatives of $g$ and which are of the form
\begin{align}
	\mathcal{E}(g)_{\mu\nu} &= 8\pi G\, \T_{\mu\nu} &\text{with} && \D_\mu \mathcal E\ud{\mu}{\nu} &= 0\, ,
\end{align}
and where $\mathcal T_{\mu\nu}$ stands for the energy-momentum tensor of matter fields. This form of the field equations can be motivated by considering the Newtonian limit. The requirement of second order field equations is necessary for having a well-posed initial value problem and the divergence-freeness of the tensor $\mathcal E_{\mu\nu}$ ensures that the covariant conservation of the matter energy-momentum tensor is a consequence of the field equations.

Given only these requirements, Lovelock showed~\cite{Lovelock:1969} that the left hand side of the field equations has to be
\begin{equation}
	\mathcal E(g)_{\mu\nu} = a\, G_{\mu\nu} + \Lambda\, g_{\mu\nu}\, ,
\end{equation}
where $a$ and $\Lambda$ are real constants and $G_{\mu\nu}$ is the so-called Einstein tensor, which is explicitly given by
\begin{equation}
	G_{\mu\nu} \ce \R_{\mu\nu} - \frac12 \R\, g_{\mu\nu}\, .
\end{equation}
This tensor indeed only contains first and second order derivatives of the metric. Taking again into consideration the Newtonian limit, we find that the constant $a$ is equal to $1$ and $\Lambda$, the so-called cosmological constant, has to be very small. Measurements performed by the Planck collaboration~\cite{Planck:2015, Planck:2018} revealed that the cosmological constant is positive and of the order of $\Lambda\sim 10^{-52}\unit{m^{-2}}$, in SI units. Thus, the Einstein field equations take on the form
\begin{equation}\label{eq:EinsteinFieldEq}
	\boxed{\R_{\mu\nu} -\frac12 \R \, g_{\mu\nu} + \Lambda\, g_{\mu\nu} = \kappa\, \mathcal{T}_{\mu\nu}}
\end{equation}
Given a set of initial data on a three-dimensional Cauchy surface, these equations determine the metric and the manifold, $(\M, g)$, up to diffeomorphisms~\cite{HawkingEllisBook, WaldBook}. This is analogous to Maxwell's equations, which determine the vector potential $A^\mu$ up to gauge transformations. For more details and a mathematically robust formulation of the initial value problem of GR, see for instance~\cite{HawkingEllisBook, WaldBook}.

The field equations~\eqref{eq:EinsteinFieldEq} follow from the action functional~\eqref{GR_pureMetric} by taking a variation with respect to the inverse metric $g^{\mu\nu}$ and demanding that this variation vanishes. As mentioned above, if $\M$ has a boundary the variational principle is ill-defined unless we add the GHY boundary term. The necessity of this term was first realized by York~\cite{York:1972} and shortly afterwards also by Gibbons and Hawking~\cite{Gibbons:1977}. Its origin is easy to understand if we consider the variation of the Einstein-Hilbert action with respect to $g^{\mu\nu}$, which results in
\begin{align}\label{eq:VariatioEH}
	\delta_g\S_{\rm EH}[g] =& \phantom{-}\frac{1}{2\kappa}\int_\M\dd^4 x\, \sqrt{|g|}\, \left(\R_{\mu\nu} - \frac12 \R\, g_{\mu\nu} + \Lambda\, g_{\mu\nu}\right)\delta g^{\mu\nu}  \notag\\
		 & - \frac{1}{2\kappa}\oint_{\partial\M}\dd^3 y\, \sqrt{|h|}\,\varepsilon\, \delta h^{\mu\nu} n^\alpha \partial_\alpha\left(\delta g_{\mu\nu}\right)\, .
\end{align}
Recall that in the calculus of variation it is assumed that the variation $\delta g^{\mu\nu}$ is fixed at the boundary, $\left.\delta g^{\mu\nu}\right|_{\partial\M} = 0$, but otherwise arbitrary. The condition $\left.\delta g^{\mu\nu}\right|_{\partial\M} = 0$ implies that derivatives of   $\delta g^{\mu\nu}$ in directions tangential to $\partial\M$ vanish, but it does \textit{not} imply that derivatives in the direction normal to $\partial\M$ vanish. In particular, one can conclude that
\begin{equation}
	\left.n^\alpha\partial_\alpha\left(\delta g^{\mu\nu}\right)\right|_{\partial\M} \ne 0\,,
\end{equation}
which in turn implies that the boundary integral in~\eqref{eq:VariatioEH} does not vanish in general. Hence, the variation of the EH action with respect to $g^{\mu\nu}$ under the boundary condition $\left.\delta g^{\mu\nu}\right|_{\partial\M} = 0$ does \textit{not} imply the Einstein field equations, unless one also imposes the \textit{additional} boundary condition $\left.n^\alpha\partial_\alpha\left(\delta g^{\mu\nu}\right)\right|_{\partial\M} = 0$.

Gibbons, Hawking, and York realized that this problem can be circumvented by the introduction of a boundary integral, whose variation precisely cancels the boundary integral in~\eqref{eq:VariatioEH}. Indeed, the variation of the GHY functional reads
\begin{align}
	\delta_g \S_{\rm GHY}[h] = \frac{1}{2\kappa} \oint_{\partial\M}\dd^3 y\, \sqrt{|h|}\, \varepsilon\, \delta h^{\mu\nu} n^\alpha \partial_\alpha\left(\delta g_{\mu\nu}\right)\,,
\end{align}
which then implies that the total variation of the GR action is given by
\begin{align}
	\delta_g \S_{\rm GR}[g, \Psi] = \frac{1}{2\kappa} \int_\M\dd^4 x\, \sqrt{|g|}\, \left(\R_{\mu\nu} - \frac12 \R\, g_{\mu\nu} + \Lambda\, g_{\mu\nu}\right)\delta g^{\mu\nu} - \frac12 \int_{\M}\dd^4 x\, \sqrt{|g|}\, \T_{\mu\nu} \delta g^{\mu\nu}\,,
\end{align}
where we have defined the energy-momentum tensor of matter fields as
\begin{align}\label{eq:DefEMT}
	\boxed{\T_{\mu\nu} \ce -\frac{2}{\sqrt{|g|}}\frac{\delta \S_{\rm matter}}{\delta g^{\mu\nu}}}
\end{align}
Thus, only once one has supplemented the action by an appropriate boundary term does one reproduce the celebrated Einstein field equations. As we will see, neither in TEGR nor STEGR are boundary terms needed.\newline	

\paragraph{\textbf{The Palatini Formulation of the Action Principle}}
Einstein's General Relativity can also be formulated in the framework of a \textit{general} metric-affine geometry $(\M, g, \Gamma)$, where $\Gamma$ a priori possesses non-trivial curvature, torsion, and non-metricity. What is needed is a slight adaptation of the action principle. In the so-called \textbf{Palatini formalism}, metric and connection are regarded as two independent fields and the action is varied with respect to both. As we will see below, even if we start with a completely general $\Gamma$, the connection field equations turn out to be purely algebraic equations which fix $\Gamma$ to be the Levi-Civita connection up to a projective symmetry. This gives us back Einstein's original connection and its original field equations for the metric.

The Palatini action functional for GR in absence of a cosmological constant is defined as
\begin{equation}\label{eq:GR_alaPalatini}
	\S_{\rm GR}[g,\Gamma] \ce \frac{1}{2\kappa}\int_\M \dd^4 x\,\sqrt{|g|} g^{\mu\nu}R_{\mu\nu}(\Gamma)+\S_{\rm matter}\,,
\end{equation}
where we recognize the first integral as being the EH action but written in terms of the Ricci scalar of the general affine connection $\Gamma$, rather than the Ricci scalar $\R$ of the Levi-Civita connection. Notice also that it is not necessary to include a boundary term \`{a} la Gibbons-Hawking-York, since there are no second order derivatives. The variational principle is thus well-defined.

In fact, the metric field equations are determined along the same lines as in the standard GR case, but without the complication of boundary terms. By performing the variation of $\S_{\rm GR}[g, \Gamma]$ with respect to the inverse metric---while keeping the connection fixed---we find
\begin{align}
	\left.\delta_g \S_{\rm GR}[g, \Gamma]\right|_{\Gamma} = \frac{1}{2\kappa} \int_{\M} \dd^4 x\, \left[\delta_g\left(\sqrt{|g|}\right)g^{\mu\nu} R_{(\mu\nu)}(\Gamma) + \sqrt{|g|}\, \delta g^{\mu\nu} R_{\mu\nu}(\Gamma)\right]  + \left. \delta_g \S_{\rm matter}\right|_{\Gamma}\,.
\end{align}
The Ricci tensor is not varied with respect to the metric, since it is constructed exclusively from the affine connection $\Gamma$. Also, note that the Ricci tensor is in general \textit{not} symmetric but that, due to the contraction with $g^{\mu\nu}$, only its symmetric part contributes. Using the well-known identity
\begin{align}
	\delta_g\left(\sqrt{|g|}\right) = -\frac12 \sqrt{|g|}\, g_{\mu\nu}\delta g^{\mu\nu}
\end{align}
together with definition~\eqref{eq:DefEMT} we find that the metric field equations can be written as
\begin{align}\label{eq:MetricFEGR}
	R_{(\mu\nu)}(\Gamma) - \frac12 R(\Gamma)\, g_{\mu\nu}  = \kappa\, \T_{\mu\nu}\,.
\end{align}
These equations have the same form as Einstein's field equations, but the Ricci tensor and Ricci scalar depend on the affine connection $\Gamma\ud{\alpha}{\mu\nu}$, rather than on the Levi-Civita connection $\LC{\alpha}{\mu\nu}$.

Next, we turn our attention to the variation with respect to $\Gamma$, while keeping the metric fixed:
\begin{align}
	\left.\delta_\Gamma \S_{\rm GR}[g,\Gamma]\right|_{g} = \frac{1}{2\kappa} \int_\M \dd^4 x\, \sqrt{|g|}\, g^{\mu\nu}\delta_{\Gamma} R_{(\mu\nu)}(\Gamma) + \left. \delta_\Gamma \S_{\rm matter}\right|_{g}\,.
\end{align}
The variation of the Ricci tensor is given by the Palatini identity,
\begin{align}
	\delta_\Gamma R_{\mu\nu}(\Gamma) = \nabla_\alpha \delta \Gamma\ud{\alpha}{\nu\mu} - \nabla_\nu \delta \Gamma\ud{\alpha}{\alpha\mu} - T\ud{\alpha}{\beta\nu} \delta\Gamma\ud{\beta}{\alpha\mu}.
\end{align}
With the help of the Palatini identity and an integration by parts in order to move the covariant derivative off $\nabla\delta\Gamma$, we find that the variation with respect to the connection can be written as
\begin{align}
	\left.\S_{\rm GR}[g, \Gamma]\right|_{g} &= - \frac{1}{2\kappa} \int_{\M} \dd^4 x\, \left[\nabla_\alpha \left(\sqrt{|g|}\, g^{\mu\nu}\right)\delta\Gamma\ud{\alpha}{(\mu\nu)} - \nabla_{(\mu}\left(\sqrt{|g|}\, g^{\mu\nu}\right)\delta \Gamma\ud{\alpha}{\alpha|\nu)} - \sqrt{|g|}\, g^{\mu\nu} T\ud{\alpha}{\beta(\mu} \delta \Gamma\ud{\beta}{\alpha|\nu)}\right] \notag\\
	&\phantom{=} + \frac{1}{2\kappa} \int_{\M} \dd^4 x\, \left[\nabla_\alpha\left(\sqrt{|g|}\, g^{\mu\nu} \delta\Gamma\ud{\alpha}{(\mu\nu)}\right) - \nabla_{(\mu}\left(\sqrt{|g|}\, g^{\mu\nu}\, \delta\Gamma\ud{\alpha}{\alpha|\nu)}\right)\right] + \left.\delta_\Gamma\S_{\rm matter}\right|_{g}\,,
\end{align}
where we have kept the total divergences. We cannot simply drop these, as we would usually do, since in a general metric-affine geometry they do \textit{not} give rise to pure boundary terms. In fact, the generalized Gauss theorem of subsection~\ref{ssec:Integration} tells us that
\begin{align}
	\int_{\M}\dd^4 x \nabla_\alpha\left(\sqrt{|g|}\, g^{\mu\nu} \delta\Gamma\ud{\alpha}{(\mu\nu)}\right) &= \oint_{\partial\M} \dd^3 y\, \sqrt{|h|}\, \epsilon\, n_\alpha g^{\mu\nu}\delta\Gamma\ud{\alpha}{(\mu\nu)} - \int_\M \dd^4 x\, \sqrt{|g|}\, T_\alpha g^{\mu\nu}\delta\Gamma\ud{\alpha}{(\mu\nu)}\notag\\
	\int_{\M}\dd^4 x \nabla_{(\mu}\left(\sqrt{|g|}\, g^{\mu\nu}\, \delta\Gamma\ud{\alpha}{\alpha|\nu)}\right) &= \oint_{\partial\M} \dd^3 y\, \sqrt{|h|}\, \epsilon\, n_{(\mu} g^{\mu\nu}\, \delta\Gamma\ud{\alpha}{\alpha|\nu)} - \int_{\M}\dd^4 x\sqrt{|g|}\, T_{(\mu} g^{\mu\nu}\, \delta\Gamma\ud{\alpha}{\alpha|\nu)}\,.
\end{align}
Both boundary integrals vanish because of the standard boundary condition $\left.\delta\Gamma\right|_{\partial\M} = 0$. That is, the boundary integrals vanish because the variations are being kept fixed at the boundary of the integration region. However, the bulk integrals on the right side contribute to the variation of the action and we therefore find
\begin{align}
	\left.\S_{\rm GR}[g, \Gamma]\right|_{g} &= - \frac{1}{2\kappa} \int_{\M} \dd^4 x\, \left[\nabla_\alpha \left(\sqrt{|g|}\, g^{\mu\nu}\right)\delta\Gamma\ud{\alpha}{(\mu\nu)} - \nabla_{(\mu}\left(\sqrt{|g|}\, g^{\mu\nu}\right)\delta \Gamma\ud{\alpha}{\alpha|\nu)} - \sqrt{|g|}\, g^{\mu\nu} T\ud{\alpha}{\beta(\mu} \delta \Gamma\ud{\beta}{\alpha|\nu)}\right] \notag\\
	&\phantom{=} + \frac{1}{2\kappa}\int_{\M}\dd^4 x\sqrt{|g|}\, T_{(\mu} g^{\mu\nu}\, \delta\Gamma\ud{\alpha}{\alpha|\nu)} - \frac{1}{2\kappa} \int_\M \dd^4 x\, \sqrt{|g|}\, T_\alpha g^{\mu\nu}\delta\Gamma\ud{\alpha}{(\mu\nu)} + \left.\delta_\Gamma\S_{\rm matter}\right|_{g}
\end{align}
After some index reshuffling, we can factor out the common factor $\delta\Gamma$ and read off the connection field equations
\begin{align}\label{eq:ConnectionFEGR}
	\nabla_\alpha \left(\sqrt{|g|}\, g^{\mu\nu}\right) - \delta\ud{\mu}{\alpha}\nabla_\beta\left(\sqrt{|g|}\, g^{\beta\nu}\right) = \sqrt{|g|}\left[g^{\mu\nu} T_\alpha + g^{\beta\nu}T\ud{\mu}{\alpha\beta} - \delta\ud{\mu}{\alpha}g^{\beta\nu}T_\beta\right] + \tilde{\mathcal{H}}\du{\alpha}{\mu\nu}\,,
\end{align}
where we have also introduced the \textbf{hypermomentum of matter}\footnote{Notice that it follows from the definition that $\tilde{\mathcal{H}}\ud{\alpha}{\mu\nu}$ is a tensor density of weight $w=+1$ and that equation~\eqref{eq:ConnectionFEGR} is thus self-consistent.},
\begin{align}\label{eq:DefHyperMomentum}
	\tilde{\mathcal{H}}\du{\alpha}{\mu\nu} \ce 2\kappa \frac{\delta\S_{\rm matter}}{\delta \Gamma\ud{\alpha}{\mu\nu}}\,.
\end{align}
These are the connection field equations of GR. Fermionic fields naturally couple to torsion and do therefore contribute to the hypermomentum. Against expectation, if torsion is present, bosonic fields also do contribute to the hypermomentum. A detailed discussion of matter coupling in metric-affine geometries can be found in~\cite{BeltranJimenez:2020}.

For simplicity, we shall first assume that torsion and hypermomentum both vanish. Then, the connection field equations~\eqref{eq:ConnectionFEGR} reduce to
\begin{align}\label{eq:SimplifiedCFE}
	\nabla_\alpha \left(\sqrt{|g|}\, g^{\mu\nu}\right) - \delta\ud{\mu}{\alpha}\nabla_\beta\left(\sqrt{|g|}\, g^{\beta\nu}\right) = 0\,.
\end{align}
By taking the $\alpha=\mu$ trace of this equation, we obtain
\begin{align}\label{eq:FirstTraceFEQ}
	\nabla_\beta\left(\sqrt{|g|}\, g^{\beta\nu}\right) = 0\qquad\Longleftrightarrow\qquad Q^\nu - 2\bar{Q}^\nu = 0\,,
\end{align}
where we have written the equation also in terms of the non-metricity tensor and its two traces, $Q_\nu = g^{\alpha\beta}Q_{\nu\alpha\beta}$ and $\bar{Q}_\nu = g^{\alpha\beta}Q_{\alpha\beta\nu}$. Plugging this result back into equation~\eqref{eq:SimplifiedCFE} yields
\begin{align}\label{eq:CFEQwithNM}
	\nabla_{\alpha}\left(\sqrt{|g|}\, g^{\mu\nu}\right) = 0\qquad \Longleftrightarrow\qquad g^{\mu\nu}\,Q_\alpha - 2Q\du{\alpha}{\mu\nu} = 0 \,.
\end{align}
After contracting this equation with $g_{\mu\nu}$, it follows that
\begin{align}
	4 Q_\alpha - 2g_{\mu\nu}Q\du{\alpha}{\mu\nu} = 2 Q_\alpha = 0\,.
\end{align}
Hence, equation~\eqref{eq:CFEQwithNM} finally tells us that
\begin{align}
	Q\du{\alpha}{\mu\nu} = 0\,.
\end{align}
Recall that we assumed $T\ud{\alpha}{\mu\nu} = 0$ and this simplifying assumption led us to uncover that the connection field equation reduces to $Q_{\alpha\mu\nu} = 0$. We can therefore conclude that the connection is torsionless and metric-compatible, which uniquely fixes it to be the Levi-Civita connection. This means that the metric field equations~\eqref{eq:MetricFEGR} become the standard Einstein equations.

The simplifying assumptions that torsion and hypermomentum vanish can be lifted and even in full generality it is found that the connection field equations are purely algebraic equations for the connection which, at the end of the day, can be completely solved. It is found that the connection is given by the Levi-Civita connection, up to a projective transformation $\Gamma\ud{\alpha}{\mu\nu} \mapsto \Gamma\ud{\alpha}{\mu\nu} + \delta\ud{\alpha}{\nu}\xi_\mu$. For the general case, we refer the reader to~\cite{BeltranJimenez:2017}.

The key lesson here is the following: As long as the action has the same form as the EH action, changing the geometric framework will \textit{not} lead to a new formulation of GR. General Relativity arises naturally if the dynamics is described by an action of the EH form, even if the variational principle is formulated \`{a} la Palatini. Hence, if we wish to develop teleparallel theories of gravity, we do not only have to change the geometric framework, we also have to change the action such that it has a genuinely \textit{different} form, but is nevertheless equivalent to the EH action. Equivalent means that the field equations for the metric are the same and that the theories all propagate the same number of physical degrees of freedom.\newline

\paragraph{\textbf{The Bianchi Identities}}
The integrand of the Einstein-Hilbert action is $\sqrt{|g|}\, \R$, which is a scalar density of weight $w=+1$. Recall that under a change of coordinates $x^\mu \mapsto x'^\mu(x)$ a scalar density of weight one transforms as (see equation~\eqref{eq:TensorDensity})
\begin{align}
	\sqrt{|g(x)|}\, \R(x) = \det\left(J\right)\, \sqrt{|g(x')|}\, \R(x')\,,
\end{align}
where $J$ is the Jacobian matrix with components $J\ud{\mu}{\nu} = \PD{x'^\mu}{x^\nu}$. It therefore follows from the change of integration variables formula of calculus that
\begin{align}
	\int_{\M}\dd^4 x'\, \sqrt{|g(x')|}\, \R(x') \det\left(J\right) = \int_{\M} \dd^4 x \sqrt{|g(x)|}\, \R(x)\,.
\end{align}
In other words, the Einstein-Hilbert action is invariant under diffeomorphisms. Since this is true for \textit{any} diffeomorphism, we can just as well consider a $1$-parameter family of diffeomorphisms: Let $\phi_s:\bbR\times\M \to\M$ be such a $1$-parameter family of diffeomorphisms with $\phi_{s=0} = \text{id}$ and with generating vector field $v\ce \left.\TD{\phi_s}{s}\right|_{s=0}$. This family of diffeomorphisms can be read as a family of changes of coordinates, i.e, $x^\mu \mapsto \phi^\mu_s(x)$ for every value of $s$. The EH action is invariant under all these diffeomorphisms. Recall from subsection~\ref{ssec:LieDerivative} that a $1$-parameter family of diffeomorphisms generates a flow and that the infinitesimal change of a tensor under a flow is measured by the Lie derivative. In the case of the metric we have
\begin{align}
	\phi^*_s g_{\mu\nu} = g_{\mu\nu} + s\,\mathcal{L}_v g_{\mu\nu}\,,
\end{align}
where the parameter $|s|\ll 1$ is infinitesimally small and $\phi^*_s g_{\mu\nu}$ shall be understood as saying ``we applied the diffeomorphism to the metric''. Applying this $1$-parameter family of diffeomorphisms to the EH action is tantamount to considering
\begin{align}
	\S_{\rm EH}[\phi^*_s g]\,.
\end{align}
Due to the invariance we have
\begin{align}
	\S_{\rm EH}[\phi^*_s g] = \S_{\rm EH}[g]\,
\end{align}
and if we expand this equation in $s$ we find
\begin{align}
	\S_{\rm EH}[g] + s\, \delta_v \S_{\rm EH}[g] = \S_{\rm EH}[g]\qquad\Longrightarrow\qquad \delta_v \S_{\rm EH}[g] = 0\,,
\end{align}
where the variation $\delta_v \S_{\rm EH}[g]$ is defined as
\begin{align}
	2\kappa\, \delta_v \S_{\rm EH}[g] &= \int_{\M} \frac{\delta}{\delta g_{\mu\nu}}\left(\sqrt{|g|}\,\R\right)\, \delta_v g_{\mu\nu} \dd^4 x\notag\\
	&= \int_{\M} \frac{\delta}{\delta g_{\mu\nu}}\left(\sqrt{|g|}\,\R\right)\, 	\mathcal{L}_v g_{\mu\nu} \dd^4 x\,. 
\end{align} 
Of course we already know the variation of $\sqrt{|g|}\, \R$ with respect to $g_{\mu\nu}$. Up to boundary terms, this is simply the Einstein tensor with raised indices multiplied by the square root of $|g|$, i.e., $\sqrt{|g|}\,G^{\mu\nu}$. By recalling from equation~\eqref{eq:LieDg} that
\begin{align}
	\mathcal{L}_v g_{\mu\nu} = 2\D_{(\mu}v_{\nu)}\,,
\end{align}
we can further simplify the form of the variation $\delta_v \S_{\rm EH}[g]$ and we find
 \begin{align}
 	 2\int_{\M}\dd^4 x\, \sqrt{|g|}\,G^{\mu\nu}\D_{(\mu}v_{\nu)} = 2\int_{\M}\dd^4 x\, \sqrt{|g|}\,\D_{\mu} G^{\mu\nu} v_\nu  = 0\,,
 \end{align}
where we integrated by parts and dropped the boundary term. Since this has to hold for \textit{any} diffeomorphism (i.e., for any generating vector field $v^\mu$), we finally find
\begin{equation}
	\boxed{
	\begin{aligned}
	\S_{\rm EH}[g] \text{ invariant under diffeomorphisms } & &\Longrightarrow & & \D_\mu G^{\mu\nu} = 0
\end{aligned}
	}
\end{equation}
These are the \textbf{Bianchi identities} and they are a consequence of the diffeomorphism invariance of the theory. These equations imply that not all of Einstein's field equations are dynamical. This affects the counting of degrees of freedom, as we will now see.\newline

\paragraph{\textbf{Counting Degrees of Freedom}}
The basic variable considered in GR is the metric. It has a total of ten independent components and thus we have an upper bound of ten physical degrees of freedom for the gravitational field. There is an equal number of second order partial differential equations, namely $G_{\mu\nu} = \kappa \, \T_{\mu\nu}$.

However, since Einstein's equations are generally covariant, we are free to perform diffeomorphisms. Each diffeomorphism provides us with four choices, which in turn grants us the freedom to fix four components of the metric. Moreover, the Bianchi identities tell us that four of Einstein's field equations are actually constraints, rather than dynamical equations. This follows from expanding the Bianchi identities as
\begin{align}\label{eq:BianchiExpanded}
\D_\mu G^{\mu\nu} &= \D_0 G^{0\nu} + \D_i G^{i\nu} \notag\\
	&=  \partial_0 G^{0\nu} + \LC{\mu}{\mu \lambda} G^{\lambda\nu} + \LC{\nu}{\mu\lambda}G^{\mu\lambda} + \partial_i G^{i\nu} = 0\,,	
\end{align}
where the index $i$ stands for spatial derivatives and is summed over the numbers $1,2,3$. Let us now determine the order of spatial and temporal derivatives of the metric in each term. To that end, we recall that $G_{\mu\nu}$ is second order in all derivatives, while the Levi-Civita connection is only first order in all derivatives. Hence, it follows that
\begin{align}
	\LC{\mu}{\mu \lambda} G^{\lambda\nu} + \LC{\nu}{\mu\lambda}G^{\mu\lambda}  \qquad\leadsto\qquad \text{Highest derivatives contained: $\partial^2_0 g_{\mu\nu}$, $\partial_0\partial_i g_{\mu\nu}$, and $\partial_i \partial_j g_{\mu\nu}$}\,.
\end{align}
Furthermore, since $\partial_i$ increases the order of spatial derivatives, we find that
\begin{align}
	\partial_i G^{i\nu} \qquad\leadsto\qquad \text{Highest derivatives contained: $\partial^2_0 \partial_i g_{\mu\nu}$, $\partial_0 \partial_i \partial_j g_{\mu\nu}$, $\partial_i \partial_j \partial_k g_{\mu\nu}$, etc. }
\end{align}
In other words, these two terms contain third order derivatives. However, the temporal derivatives are at most second order! This is an important realization because $\partial_0 G^{0\nu}$ contains third order time derivatives, \textit{provided that $G^{0\nu}$ contains second order time derivatives}. However, the Bianchi identities tell us that this cannot be the case. None of the other terms we analyzed contains third order time derivatives. The highest order is two. Hence, there is nothing which could cancel the presumed third order time derivatives in $\partial_0 G^{0\nu}$, which is necessary for the Bianchi identities to hold. 

It follows that the assumption that $G^{0\nu}$ contains second order time derivatives is wrong! At most, it can contain first order time derivatives (and indeed it does). The important conclusion is that the four equations $G^{0\nu} = \kappa \, \T^{0\nu}$ constitute constraints on the initial data, rather than dynamical equations for the metric. So we are finally left with
\begin{align*}
	10 \text{ metric components } - 4 \text{ diffeomorphisms } - 4 \text{ constraints } = 2 \text{ physical degrees of freedom}.
\end{align*}\newline

\subsection{The Teleparallel Equivalent of General Relativity (TEGR)}\label{ssec:TEGR}
In the previous subsection we saw how GR emerges from an action principle in conjunction with the geometric postulates of vanishing torsion and vanishing non-metricity. We also saw that the postulates can be dropped and that GR emerges from a Palatini variational principle, where the connection $\Gamma$ is assumed to be an independent field, but which is ultimately fixed by the field equations to be precisely the Levi-Civita connection. This fact hinges on the form of the action: As long as the action has the EH form, GR emerges naturally. It follows that in order to obtain an equivalent but \textit{different} geometric formulation of GR, we need to change the geometric framework as well as the action principle. In the following, we show how the so-called Teleparallel Equivalent of GR, or TEGR for short, achieves this.\newline

\paragraph{\textbf{The Geometric Postulates}}
TEGR attributes the effects of gravity to a non-vanishing torsion tensor. The starting point is a metric-affine geometry $(\M, g, \Gamma)$, where the connection is postulated to satisfy
\begin{align}
	R\ud{\alpha}{\mu\nu\rho} &\overset{!}{=} 0 && \textsf{and} & Q_{\alpha\mu\nu\phantom{\rho}} &\overset{!}{=} 0\, .
\end{align}
This may raise the question, in what sense GR and TEGR could even be equivalent to each other, given that curvature is postulated to vanish in the latter one. The key observation to resolve this apparent tension is the following: What is postulated to vanish in TEGR is the curvature tensor $R\ud{\alpha}{\mu\nu\rho}$ with respect to the affine connection~$\Gamma$, not the curvature tensor $\R\ud{\alpha}{\mu\nu\rho}$ with respect to the Levi-Civita connection. Moreover, recall that the two curvature tensors are related to each other by the identity~\eqref{eq:RelationBetweenCurvatures}. Starting from this identity, we have seen that an important special case emerges when non-metricity is set to zero. Namely equation~\eqref{eq:KeyIdT}, which relates the two Ricci scalars and which we repeat here for convenience:
\begin{equation}\label{eq:RicciScalarsMT}
	R(\Gamma) = \R(g) + \bbT + 2\D_\alpha T^\alpha\,.
\end{equation}
The torsion scalar $\bbT$ is explicitly given by
\begin{equation}\label{eq:TorsionScalarRing}
	\bbT \ce \frac12 \left(\frac14 T_{\alpha\mu\nu} + \frac12 T_{\mu\alpha\nu} - g_{\alpha\mu} T_\nu \right) T^{\alpha\mu\nu}\, ,
\end{equation}
and $\D_\mu$ still denotes the covariant derivative with respect to the Levi-Civita connection, \textit{not} the covariant derivative $\nabla_\mu$ with respect to the more general connection $\Gamma$. This equation will prove to be the key to formulate GR in terms of torsion, rather than curvature. But before that, we study the form of the connection more closely.\newline

\paragraph{\textbf{Form of a Flat, Metric-Compatible Connection}}
The geoemtric postulates demand that the connection be flat and metric-compatible. These requirements do not completely fix the connection, unlike the postulates used in GR. Rather, we end up with a whole class of connections which satisfy the postulates of flatness and metric-compatibility. To see how this comes about, we start with the observation that the trivial connection, i.e., the connection $\hat{\Gamma}\ud{\alpha}{\mu\nu} = 0$ is obviously flat. Now recall that under a change of coordinates a connection transforms inhomogeneously (cf. equation~\eqref{eq:ConnectionTransformationLaw}). For our trivial connection, we find that a change of coordinates from $\hat{x}^\mu$ to $x^\mu(\hat{x})$ leads to
\begin{align}
	\hat{\Gamma}\ud{\alpha}{\mu\nu} \qquad\mapsto\qquad \Gamma\ud{\alpha}{\mu\nu} &=  \PD{x^\alpha}{\hat{x}^\beta} \PD{\hat{x}^\rho}{x^\mu} \PD{\hat{x}^\sigma}{x^\nu} \underbrace{\hat{\Gamma}\ud{\beta}{\rho\sigma}}_{= 0} + \PD{x^\alpha}{\hat{x}^\lambda}\PD{^2 \hat{x}^\lambda}{x^\mu \partial x^\nu} = \PD{x^\alpha}{\hat{x}^\lambda} \partial_\mu \PD{\hat{x}^\lambda}{x^\nu}\,.
\end{align}
If we read $\PD{\hat{x}^\mu}{x^\nu}$ as the components of a matrix $\Lambda$, we can write the last equation as
\begin{align}\label{eq:FlatConnection}
	\boxed{\Gamma\ud{\alpha}{\mu\nu} = \left(\Lambda^{-1}\right)\ud{\alpha}{\lambda}\partial_\mu \Lambda\ud{\lambda}{\nu}}
\end{align}
This is a key equation for all teleparallel theories of gravity for the following two facts:
\begin{enumerate}
	\item If the curvature tensor is zero in one coordinate system, it is zero in any other coordinate system. Since it is zero for the trivial connection $\hat{\Gamma}\ud{\alpha}{\mu\nu} = 0$, it is also zero for the connection $\Gamma\ud{\alpha}{\mu\nu}$ in equation~\eqref{eq:FlatConnection}, since this connection has been obtained by a change of coordinates. 
	\item The change of coordinates is completely arbitrary and the vanishing of $R\ud{\alpha}{\mu\nu\rho}$ for the connection in~\eqref{eq:FlatConnection} does \textit{not} depend on the details of the transformation. Thus, we may as well ``forget'' the origin of~\eqref{eq:FlatConnection}. That is to say, we can conclude that any connection of the form~\eqref{eq:FlatConnection}, where $\Lambda\ud{\mu}{\nu}$ is a matrix belonging to the general linear group $GL(4, \bbR)$ is flat.
\end{enumerate}
Now we turn to the second postulate, which demands metric-compatibility. By plugging the flat connection~\eqref{eq:FlatConnection} into $Q\du{\alpha}{\mu\nu} = 0$, we find
\begin{equation}\label{eq:MetricityMT}
	g^{\lambda(\mu}\partial_\alpha\Lambda\ud{\nu)}{\rho}(\Lambda^{-1})\ud{\rho}{\lambda} =\frac12 \partial_\alpha g^{\mu\nu}\,.
\end{equation} 
This equation allows us to eliminate the metric and express it in terms of the $\Lambda\ud{\mu}{\nu}$. Observe that the metric has ten components while the matrix $\Lambda\ud{\mu}{\nu}$ has $4\times 4 = 16$ components. This redundancy in the description is well-understood~\cite{Golovnev:2017, Krssak:2018, BeltranJimenez:2019c}: Six of the components of $\Lambda\ud{\mu}{\nu}$ reflect the freedom to perform local Lorentz transformations. This is a symmetry of the theory.

We reach the following conclusions: Choose a matrix $\Lambda\in GL(4,\mathbb R)$ and write the connection as in~\eqref{eq:FlatConnection}. The so defined connection is guaranteed to be flat. Furthermore, impose~\eqref{eq:MetricityMT} in order to obtain a metric-compatible connection. This turns the metric into an auxiliary field. For completeness, we remark that the torsion tensor is given by
\begin{equation}
	T\ud{\alpha}{\mu\nu} = 2 \left(\Lambda^{-1}\right)\ud{\alpha}{\lambda}\,\partial_{[\mu}\Lambda\ud{\lambda}{\nu]}\,.
\end{equation}
for any flat connection parametrized by $\Lambda\in GL(4, \bbR)$.\newline

\paragraph{\textbf{Construction of the Action Functional}} 
The construction of an action functional which is equivalent but not equal to the one of GR is fairly straightforward. As alluded to above, the key observation is that the Ricci scalar of a metric-compatible connection is related to the Ricci scalar of the Levi-Civita connection via equation~\eqref{eq:RicciScalarsMT}. If we also impose flatness, which amounts to $R(\Gamma) = 0$, we find
\begin{align}\label{eq:RicciT}
	\R(g) = - \bbT(\Lambda) - 2\D_\alpha T^\alpha(\Lambda)\,.
\end{align}
The notation $\bbT(\Lambda)$ emphasizes that $\bbT$ depends on $\Lambda$ and that the metric has been integrated out from the metricity condition. This equation now allows us to simply replace the Ricci scalar in the EH action by the right hand side of equation~\eqref{eq:RicciT}. However, such an action would be strictly \textit{equal} to the original EH action, because the connection carries a Levi-Civita piece and a torsion piece. The scalars $\bbT$ and $\D_\alpha T^\alpha$ conspire in such a way, that the torsion piece drops out and only the Levi-Civita part contributes to the action.

However, by dropping $\D_\alpha T^\alpha$, which amounts to a mere boundary term, we obtain an action which is genuinely \textit{different} from the EH action, but which leads to the same field equations. Thus, we define the action of TEGR as
\begin{align}\label{eq:BoxedActionTEGR}
	\boxed{\S_{\rm TEGR}[\Lambda] \ce -\frac{1}{2\kappa}\int_{\M} \dd^4 x\, \sqrt{|g|}\, \bbT(\Lambda) + \S_{\rm matter}}
\end{align}
It is silently understood that the metric can be expressed in terms of $\Lambda$. However, observe that $\Lambda$ has $16$ components, while the metric has only ten. Consequently, we find \textit{more} field equations than in GR. As we will see later, ten of these equations are precisely the Einstein field equations. The remaining six are Bianchi identities related to the local Lorentz symmetry expressed through six of the components of $\Lambda$.  

A further consequence of dropping $\D_\alpha \bbT^\alpha$ is that the action functional only depends on first order derivatives. Thus, the variational principle is well-defined without having to add boundary terms \`{a} la Gibbons-Hawking-York. This is one of the features we had anticipated in the previous subsection.\newline

The action for TEGR could also have been constructed using a different strategy, which does not rely on the geometric identity~\eqref{eq:RicciT}. Rather, the strategy which we shall briefly sketch relies on counting degrees of freedom: Given the postulates of vanishing curvature and vanishing non-metricity, the only remaining tensor which can play a fundamental role is the torsion tensor. Thus, our task is to construct a scalar from the torsion tensor, which is then used to define the action. Clearly, this scalar cannot be linear in the torsion tensor. It has to be at least quadratic. As it turns out (we will discuss this point in more detail in subsection~\ref{ssec:QuadraticTorsion}), there are precisely three independent scalars one can build from contractions of the torsion tensor with itself and with the help of the metric. Thus, the most general scalar assumes the form
\begin{align}
    \hat{\bbT} \ce c_1\, T_{\alpha\mu\nu}T^{\alpha\mu\nu} + c_2\, T_{\mu\alpha\nu} T^{\alpha\mu\nu} + c_3\, T_\mu T^\mu\,,
\end{align}
where $c_1$, $c_2$, and $c_3$ are arbitrary, real constants. Using this scalar, it is easy to derive field equations and perform a counting of degrees of freedom around a Minkowski background. In order to obtain precisely two degrees of freedom, as in GR, one finds that the parameters have to be chosen as
\begin{align}
    c_2 &= 2 c_1 & & \text{and} & c_3 = -4 c_1\,.
\end{align}
Up to an over all normalization, this reproduces the torsion scalar~\eqref{eq:TorsionScalarRing} and thus the action~\eqref{eq:BoxedActionTEGR}.\\

\paragraph{\textbf{The Palatini Formulation of the Action Principle}}
This action can also be written in a manifestly covariant form, which highlights which type of metric-affine geometry is being considered, i.e., which geometric postulates are being implemented. This action functional reads
\begin{equation}\label{eq:ActionTeleparallel}
	\S_\textsf{TEGR}[g, \Gamma; \tilde{\Pi}, \tilde{\chi}] \ce -\int_\M \dd^4 x\, \left(\frac{1}{2\kappa} \sqrt{|g|}\, \bbT + \tilde\Pi\du{\alpha}{\mu\nu\rho}\, R\ud{\alpha}{\mu\nu\rho} + \tilde{\chi}\ud{\alpha}{\mu\nu}\, Q\du{\alpha}{\mu\nu}\right) + \S_{\rm matter}\, ,
\end{equation}
where $\tilde{\Pi}\du{\alpha}{\mu\nu\rho}$ and $\tilde{\chi}\ud{\alpha}{\mu\nu}$ are tensor densities of weight $w=+1$ which act as Lagrange multipliers. These multipliers enforce the postulates of vanishing curvature and vanishing non-metricity. It should also be noted that the Lagrange multipliers possess the symmetries $\tilde{\Pi}\du{\alpha}{\mu\nu\rho} = \tilde{\Pi}\du{\alpha}{\mu[\nu\rho]}$ and $\tilde{\chi}\ud{\alpha}{\mu\nu} = \tilde{\chi}\ud{\alpha}{(\mu\nu)}$, which they inherit from the curvature tensor and the non-metricity tensor, respectively. Just as in the Palatini formulation of GR, the connection $\Gamma$ refers to a generic affine connection. A priori, it has nothing to do with the previous connection which is parametrized by the matrix $\Lambda$. When it comes to working out the field equations, the Palatini formalism offers some advantages.\newline

\paragraph{\textbf{The Metric and Connection Field Equations}}
Based on the Palatini action given in~\eqref{eq:ActionTeleparallel}, we can perform four independent variations. These are
\begin{align}
	\frac{\delta \S_{\rm TEGR}}{\delta g^{\mu\nu}} &\overset{!}{=} 0, & && \frac{\delta \S_{\rm TEGR}}{\delta \tilde{\Pi}\du{\alpha}{\mu\nu\rho}} &\overset{!}{=} 0\notag\\
	\frac{\delta \S_{\rm TEGR}}{\delta \Gamma\ud{\alpha}{\mu\nu}} &\overset{!}{=} 0 & &&  \frac{\delta \S_{\rm TEGR}}{\delta \tilde{\chi}\ud{\alpha}{\mu\nu}} &\overset{!}{=} 0,.
\end{align}
The variations with respect to the Lagrange multipliers are the most straightforward ones. The variations with respect to the metric and the connection require more work. Also, the field equations that follow from the variational principle are highly coupled in the sense that the Lagrange multipliers appear with covariant derivatives acting on them. Untangling the field equations, cleanly implementing the conditions of vanishing curvature and vanishing non-metricity, and bringing the equations into a simple form requires some effort. For a detailed derivation we refer the reader to~\cite{BeltranJimenez:2018}. The end result is
\begin{equation}
\boxed{
\begin{aligned}
	&\textsf{Metric field equations:} & (\nabla_\alpha + T_\alpha)S\du{(\mu\nu)}{\alpha} + t_{\mu\nu} -\frac12 \bbT\,g_{\mu\nu} &= \kappa\, \T_{\mu\nu}\notag\\
	&\textsf{Connection field equations:} & (\nabla_\alpha + T_\alpha)\left[\sqrt{|g|} {S}\dud{[\mu}{\alpha}{\nu]}\right] &= 0
\end{aligned}}
\end{equation}
where we have introduced the \textbf{torsion conjugate $\boldsymbol{S}\du{\alpha}{\mu\nu}$} and the symmetric tensor $t_{\mu\nu}$,
\begin{align}
	S\du{\alpha}{\mu\nu} &\ce \PD{\bbT}{T\ud{\alpha}{\mu\nu}} = \frac14\, T\du{\alpha}{\mu\nu} + \frac12\, T\udu{[\mu}{\alpha}{\nu]} -   \delta\du{\alpha}{[\mu}T^{\nu]}   \notag\\
	t_{\mu\nu} &\ce \PD{\bbT}{g^{\mu\nu}} = \frac12 S\du{(\mu|}{\lambda\kappa}\, T_{\nu)\lambda\kappa} - T\ud{\lambda\kappa}{(\mu}\, S_{\lambda\kappa|\nu)}\,.
\end{align}
It is also assumed that the hypermomentum density, which enters as $(\nabla_\alpha + T_\alpha)\tilde{\mathcal{H}}\dud{[\mu}{\alpha}{\nu]}$ in the field equations is either identically zero (i.e., the matter content and the matter couplings have been chosen such that there is no contribution to this tensor density) or that it is conserved in the sense that
\begin{align}
	\left(\nabla_\alpha + T_\alpha\right)\tilde{\mathcal{H}}\dud{\mu}{\alpha}{\nu} = 0\,.
\end{align}
This conservation law holds by virtue of the gauge invariance of the matter sector and is confirmed in all standard cases where a non-trivial hypermomentum arises from coupling matter fields to the connection. For more details, see~\cite{BeltranJimenez:2018, BeltranJimenez:2020}.

As one can show, the metric field equations are simply the Einstein field equations in disguise, while the connection field equations arise as Bianchi identities. As shown in~\cite{BeltranJimenez:2018}, this is a consequence of the curvature tensor being the curvature of a $GL(4, \bbR)$ connection. This has important consequences: The connection field equations carry no dynamical information. Put differently, these equations do not determine the metric nor the connection. In fact, they are just trivially satisfied. The dynamics of the theory is solely determined by the metric field equations, which are the Einstein equations.\newline

\paragraph{\textbf{The Bianchi Identities}}
Bianchi identities arise quite naturally whenever an action is invariant under a certain local symmetry. Since the actions~\eqref{eq:ActionTeleparallel} and~\eqref{eq:ActionTeleparallel} are generally covariant, it is no surprise that Bianchi identities can be found. Starting from an action of the form $\S[g, \Gamma]$, where $\Gamma$ is a generic affine connection, and assuming that the action is diffeomorphism invariant, it was shown in~\cite{BeltranJimenez:2020} that the Bianchi identities in a metric-affine geometry take the form
\begin{align}\label{eq:GeneralBianchiId}
	\boxed{2\D_\mu \tilde{\M}\ud{\mu}{\lambda} - \hat{\nabla}_\nu \hat{\nabla}_\mu \tilde{\C}\du{\lambda}{\mu\nu} + T\ud{\alpha}{\lambda\nu} \hat{\nabla}_\mu \tilde{\C}\du{\alpha}{\mu\nu} + \left(R\ud{\alpha}{\nu\mu\lambda} - T_\mu T\ud{\alpha}{\nu\lambda}\right) \tilde{\C}\du{\alpha}{\mu\nu} \equiv 0}
\end{align}
where
\begin{align}
	\tilde{\M}^{\mu\nu} &\ce \frac{\delta \S[g, \Gamma]}{\delta g_{\mu\nu}} & \text{and} && \tilde{\C}\du{\alpha}{\mu\nu} &\ce \frac{\delta \S[g, \Gamma]}{\delta \Gamma\ud{\alpha}{\mu\nu}}\,.
\end{align}
are placeholders for the the metric and connection field equations (these are tensor densities of weight one) and where
\begin{align}
	\hat{\nabla}_\mu \ce \nabla_\mu + T_\mu\,.
\end{align}
The Bianchi identities of GR follow from this general identity as a special case. Moreover, if we fix the connection to be flat and metric-compatible, we find 
\begin{align}
	2\D_\mu \tilde{\M}\ud{\mu}{\lambda} - \hat{\nabla}_\nu \hat{\nabla}_\mu \tilde{\C}\du{\lambda}{\mu\nu} + T\ud{\alpha}{\lambda\nu} \hat{\nabla}_\mu \tilde{\C}\du{\alpha}{\mu\nu} - T_\mu T\ud{\alpha}{\nu\lambda} \tilde{\C}\du{\alpha}{\mu\nu} \equiv 0\,.
\end{align}
Since in TEGR the metric field equations are the same as Einstein's, i.e., since $\tilde{\M}_{\mu\nu} = \sqrt{|g|}\, G_{\mu\nu}$, this simplifies further to
\begin{align}
	 \boxed{\hat{\nabla}_\nu \hat{\nabla}_\mu \tilde{\C}\du{\lambda}{\mu\nu} - T\ud{\alpha}{\lambda\nu} \hat{\nabla}_\mu \tilde{\C}\du{\alpha}{\mu\nu} + T_\mu T\ud{\alpha}{\nu\lambda} \tilde{\C}\du{\alpha}{\mu\nu} \equiv 0}
\end{align}
due to $\D_{\mu}(\sqrt{|g|}\, G^{\mu\nu}) = 0$. Thus, only the connection field equations remain and they have to satisfy the above Bianchi identity.\newline

\paragraph{\textbf{Counting Degrees of Freedom}}
If we start with a metric $g_{\mu\nu}$ and a general affine connection $\Gamma\ud{\alpha}{\mu\nu}$ we have a total of $10+64$ fields. The flatness condition $R\ud{\alpha}{\mu\nu\rho} = 0$ drastically reduces this number. Since any flat connection can be written as $\Gamma\ud{\alpha}{\mu\nu} = (\Lambda^{-1})\ud{\alpha}{\lambda}\partial_\mu \Lambda\ud{\lambda}{\nu}$, where $\Lambda\ud{\mu}{\nu}$ is a $GL(4, \bbR)$ matrix, we find that the connection carries at most $4\times 4 = 16$ degrees of freedom, rather than $64$.  Finally, we also have to take into account the postulate of vanishing non-metricity,
\begin{align}
    2(\Lambda^{-1})\ud{\lambda}{\kappa}\partial_\alpha\Lambda\ud{\kappa}{(\mu}g_{\nu)\lambda} = \partial_\alpha g_{\mu\nu}\,,
\end{align}
which relates the metric and the matrix $\Lambda$. This equation is solved by
\begin{align}
    g_{\mu\nu} = \Lambda\ud{\alpha}{\mu}\Lambda\ud{\beta}{\nu} c_{\alpha\beta}\,,
\end{align}
where $c_{\alpha\beta}$ is a symmetric, constant tensor. It is only natural to choose $c_{\alpha\beta} = \eta_{\mu\nu}$, where the latter denotes the Minkowski metric, since we are interested in metrics with Lorentzian signature. Notice that this also means that instead of potentially $10+16$ degrees of freedom, we now have at most $16$, since the connection as well as the metric can be parametrized with the $16$ components of $\Lambda\ud{\mu}{\nu}$.

At this point it should be noted that flat connections possess a gauge symmetry. Namely, transformations of the form $\Lambda \mapsto \Lambda\mathcal{U}$, where $\mathcal{U}\in GL(4, \bbR)$, leave the curvature tensor and thus the flatness condition invariant. If it also has to respect the postulate of vanishing non-metricity, then it has to leave the metric invariant, which means
\begin{align}
    \mathcal{U}\ud{\alpha}{\kappa}\mathcal{U}\ud{\beta}{\lambda}\Lambda\ud{\kappa}{\mu}\Lambda\ud{\lambda}{\nu}\eta_{\alpha\beta}\overset{!}{=} \Lambda\ud{\alpha}{\mu}\Lambda\ud{\beta}{\nu}\eta_{\alpha\beta}\,.
\end{align}
This implies that $\mathcal{U}$ belong to the proper orthochronous Lorentz group, since this guarantees that the Minkowski metric is left invariant and it is the part of the Lorentz group which is connected to the identity. 

We therefore learn that six components of $\Lambda\ud{\mu}{\nu}$ simply represent Lorentz transformations and that these transformations are pure gauge. That is, they do not change the form of the metric, nor do they affect the flatness postulate. We have therefore a maximal number of $16-6 = 10$ degrees of freedom. However, TEGR is a generally covariant theory and diffeomorphisms remove $2\times 4$ degrees of freedom. Hence, we are finally left with only two degrees of freedom, as we expected.\\

\subsection{The Symmetric Teleparallel Equivalent of General Relativity (STEGR)}\label{ssec:STEGR}

\paragraph{\textbf{The Geometric Postulates}}
We now turn to the third geometric formulation of GR, which ascribes gravitational phenomena to non-metricity~\cite{Nester:1998}: The so-called Symmetric Teleparallel Equivalent of GR (STEGR)~\cite{BeltranJimenez:2017b}. The starting point is again a metric-affine geometry $(\M, g, \Gamma)$, but this time restricted by the geometric postulates
\begin{align}
	R\ud{\alpha}{\mu\nu\rho} &\overset{!}{=} 0 &&\text{and} & T\ud{\alpha}{\mu\nu\phantom{\rho}} &\overset{!}{=} 0\, .
\end{align}
The postulate of vanishing curvature may, just as in TEGR, raise the question of how the theory we seek to construct can possibly be equivalent to standard GR, where curvature plays an essential role. However, the resolution to this apparent tension is the same as in TEGR: What is postulated to vanish is the curvature of the affine connection $\Gamma$, not the curvature of the Levi-Civita connection on which GR is based.\newline

\paragraph{\textbf{Form of a Flat, Torsionless Connection and the Coincident Gauge}}
Before constructing an action functional for STEGR, let us work out what a flat and torsionless connection looks like. From the previous subsection, we recall that a flat connection can always be written as
\begin{equation}
	\G{\alpha}{\mu\nu} = \left(\Lambda^{-1}\right)\ud{\alpha}{\lambda}\,\partial_\mu\Lambda\ud{\lambda}{\nu}\,,
\end{equation}
where $\Lambda\ud{\mu}{\nu}$ are the components of a matrix belonging to the general linear group $GL(4, \mathbb R)$. The postulate of vanishing torsion can then be rephrased as
\begin{align}
	T\ud{\alpha}{\mu\nu} &\overset{!}{=} 0 &&\Longrightarrow & \partial_{[\mu}\Lambda\ud{\alpha}{\nu]} &\overset{!}{=} 0\,.
\end{align}
This last condition implies that the matrix $\Lambda\ud{\mu}{\nu}$ can be written as $\Lambda\ud{\mu}{\nu} = \partial_\nu \xi^\mu$, where $\xi^\mu$ denotes a collection of four arbitrary functions of the coordinates $x^\mu$, \textit{not a vector field}! We conclude that a flat, torsionless connection can be written as
\begin{equation}\label{eq:FlatTorsionfreeConnection}
	\G{\alpha}{\mu\nu} = \frac{\partial x^\alpha}{\partial \xi^\lambda}\partial_\mu\partial_\nu\xi^\lambda\,,
\end{equation}
where $\partial x^\alpha/\partial\xi^\lambda$ should be understood as the inverse of the Jacobian matrix $\partial\xi^\lambda/\partial x^\alpha$. This result means that in any given coordinate system $\{x^0, x^1, x^2, x^3\}$ we can choose four independent functions $\{\xi^0, \xi^1, \xi^2, \xi^3\}$, such that the Jacobian matrix $\partial \xi^\mu / \partial x^\nu$ is invertible (i.e., has a non-zero determinant) and this allows us then to construct a flat and torsionless connection via equation~\eqref{eq:FlatTorsionfreeConnection}.

Moreover, equation~\eqref{eq:FlatTorsionfreeConnection} reveals that flat and torsionless connections have a remarkable property: They can be set to zero globally by an appropriate choice of coordinates. In fact, given any flat and torsionless connection, it necessarily has the form~\eqref{eq:FlatTorsionfreeConnection} with some functions $\xi^\mu$. Therefore, if we choose our coordinates such that $x^\mu = \xi^\mu$, the connection is exactly equal to zero because $\partial_\mu\partial_\nu \xi^\lambda = 0$. This is known as the \textbf{coincident gauge}\footnote{More generally, one could also choose the functions $\xi^\mu$ to be of the form $\xi^\mu = M\ud{\mu}{\nu} x^\nu + \xi^\mu_0$, where $M\ud{\mu}{\nu}$ is a non-degenerate matrix with constant entries and $\xi^\mu_0$ are constants~\cite{BeltranJimenez:2018}. This is also known as \textbf{coincident gauge}.}.

We emphasize that the coincident gauge can always be chosen and that it has nothing to do with an action principle. It is available as long as the postulates of vanishing curvature and vanishing torsion are in place. However, we also stress that there are caveats one has to be aware of when it comes to working in a fixed coordinate system and wanting to use the coincident gauge. We will discuss these caveats in subsections~\ref{ssec:Cosmology} and~\ref{ssec:BlackHoles}.\newline

\paragraph{\textbf{Construction of the Action Functional}} 
To construct an action functional for STEGR, we follow the same strategy as in the case of TEGR. The key observation is that the curvature tensor of the affine connection is related to the curvature tensor of the Levi-Civita connection by the identity~\eqref{eq:KeyIdQ} we discussed in~\ref{ssec:Identities}. We rewrite this identity here for convenience:
\begin{align}\label{eq:IdRicciQ}
	R(\Gamma) = \R(g) + \Q + \D_\alpha\left(Q^\alpha - \bar{Q}^\alpha\right)\,,
\end{align}
where $\D$ is the covariant derivative with respect to the Levi-Civita connection, the two traces of the non-metricity tensor are given by
\begin{align}
	Q_\alpha &\ce Q\du{\alpha\lambda}{\lambda} & & \text{and}& \bar{Q}_\alpha &\ce Q\ud{\lambda}{\lambda\alpha}\, ,
\end{align}
and the non-metricity scalar $\Q$ is defined as
\begin{align}
	\Q &\ce\frac14 Q_{\alpha\mu\nu}Q^{\alpha\mu\nu} - \frac12 Q_{\alpha\mu\nu}Q^{\mu\alpha\nu} - \frac14 Q_\alpha Q^\alpha + \frac12 Q_\alpha \bar{Q}^\alpha\,.
\end{align}
The latter can also be expressed in terms of the disformation tensor $L\ud{\alpha}{\mu\nu} \ce \frac12 Q\ud{\alpha}{\mu\nu} - Q\dud{(\mu}{\alpha}{\nu)}$ as
\begin{align}\label{eq:QinLsquared}
	\Q = g^{\mu\nu}\left(L\ud{\alpha}{\alpha\beta} L\ud{\beta}{\mu\nu} - L\ud{\alpha}{\beta\mu} L\ud{\beta}{\nu\alpha}\right)\,.
\end{align}
Recall that the identity~\eqref{eq:IdRicciQ} is valid only when torsion vanishes. Thus, one of the geometric postulates is already implemented. The second postulate of STEGR, which demands that the curvature of the affine connection vanishes, then implies that
\begin{equation}\label{eq:IdentityRicciST}
	\R(g) = -\Q - \D_\alpha\left(Q^\alpha - \bar{Q}^\alpha\right)\,.
\end{equation}
In other words, the Ricci scalar of the Levi-Civita connection can be expressed in terms of the non-metricity scalar and a divergence term. This allows us to replace $\R(g)$ in the Einstein-Hilbert action by the right hand side of the identity~\eqref{eq:IdentityRicciST}. Thus, in the Symmetric Teleparallel Equivalent of GR gravity is described by the action functional
\begin{equation}\label{eq:STEGRAction}
	\S_{\rm STEGR}[g, \xi] = -\frac{1}{2\kappa}\int_\M\dd^4 x\, \sqrt{|g|}\,\Q(g, \xi) + \mathcal{S}_\textsf{matter}\, .
\end{equation}
We have dropped the divergence $\D_\alpha\left(Q^\alpha - \bar{Q}^\alpha\right)$ since, by the generalized Gauss theorem which we discussed in subsection~\ref{ssec:Integration}, this term amounts to a mere boundary term which can thus have no influence on the field equations. Moreover, as we have discussed in the GR and TEGR sections, changing the action is a necessary step in order to arrive at a genuinely new formulation.

Notice that the action is a functional of the metric $g_{\mu\nu}$ \textit{and} the four functions $\xi^\alpha$ which parametrize the flat, torsionless connection~\eqref{eq:FlatTorsionfreeConnection}. \newline

The candidate for the STEGR action could also have been constructed without knowing the geometric identity~\eqref{eq:IdentityRicciST}. Just as in TEGR, one can start with the most general Lagrangian which is quadratic in the non-metricity tensor. Due to the symmetry of the non-metricity tensor, one finds that there are precisely five independent scalars one can construct from contractions of the non-metricity tensor (this will be discussed in more details in subsection~\ref{ssec:QuadraticNonMetricity}). The most general Lagrangian then reads
\begin{align}
    \hatQ = c_1\, Q_{\alpha\mu\nu}Q^{\alpha\mu\nu} + c_2\, Q_{\mu\alpha\nu}Q^{\alpha\mu\nu} + c_3\, Q_\mu Q^\mu + c_4\, \bar{Q}_\mu \bar{Q}^\mu + c_5\, Q_\mu \bar{Q}^\mu\,,
\end{align}
where $c_1$, $c_2$, $c_3$, $c_4$, and $c_5$ are arbitrary, real constants. By expanding the theory around a Minkowski background and demanding that it propagates two degrees of freedom, which is tantamount to demanding that the linearized theory is invariant under linearized diffeomorphisms, one finds that the parameters $c_i$ have to satisfy
\begin{align}
	c_3 &= -c_1, &  c_4 &= -2c_1 -c_2, &  c_5 = 2c_1\,.
\end{align}
These relations are satisfied by the parameter values which reproduce the STEGR action and they leave $c_1$ and $c_2$ free. The linearized theory cannot fix these parameters, but other considerations can reproduce the STEGR action up to an overall normalization. For instance, demanding that the full, non-linear theory satisfies the contracted Bianchi identity $\D_\mu \M\ud{\mu}{\nu} =0$, where $\M\ud{\mu}{\nu}$ stands for the metric field equations, requires $c_4$ to vanish. Thus, we find
\begin{align}
    c_2 &= -2c_1, & c_3 &= -c_1, & c_4 &= 0, & c_5 &= 2c_1\,,
\end{align}
which indeed reproduces the action~\eqref{eq:STEGRAction} up to an overall normalization constant.\\

\paragraph{\textbf{The Palatini Formulation of the Action Principle}}
Just as in TEGR, we can employ the Palatini formalism in order to express the action principle in a manifestly covariant way which also highlights which type of metric-affine geometry is being considered. This action is defined as
\begin{equation}\label{eq:STEGRActionCovariant}
	\S_{\rm STEGR}[g, \Gamma] \ce -\int_\M\dd^4 x\, \left(\frac{1}{2\kappa} \sqrt{|g|}\,\Q(g, \Gamma) + \tilde{\Pi}\du{\alpha}{\mu\nu\rho}\, R\ud{\alpha}{\mu\nu\rho} + \tilde{\chi}\du{\alpha}{\mu\nu}\, T\ud{\alpha}{\mu\nu}\right) + \mathcal{S}_\textsf{matter}\,,
\end{equation}
where the Lagrange multipliers $\tilde{\Pi}\du{\alpha}{\mu\nu\rho}$ and $\tilde{\chi}\du{\alpha}{\mu\nu}$ are tensor densities of weight $w=+1$. These Lagrange multipliers inherit the symmetries $\tilde{\Pi}\du{\alpha}{\mu\nu\rho} = \tilde{\Pi}\du{\alpha}{\mu[\nu\rho]}$ and $\tilde{\chi}\du{\alpha}{\mu\nu} = \tilde{\chi}\du{\alpha}{[\mu\nu]}$ from the curvature and torsion tensor, respectively. 

Notice that the action is a functional of the metric $g_{\mu\nu}$ and a generic affine connection $\Gamma\ud{\alpha}{\mu\nu}$. By varying the above action with respect to the Lagrange multiplier densities, one obtains two constraints on the connection. Namely, the connection is restricted to be flat and torsionless. Since these constraints do not completely fix the connection, we still have the freedom to choose four arbitrary functions $\xi^\alpha$ in order to parametrize the connection in agreement with equation~\eqref{eq:FlatTorsionfreeConnection}.

As a final comment we add that the curvature tensor measures the change in direction of a vector which is being parallel transported around a closed loop (cf. subsection~\ref{ssec:Curvature}). Hence, when curvature vanishes, there is no change in direction and the vector remains, in this sense, parallel to itself. This justifies the use of the term \textbf{teleparallel}. Moreover, the vanishing of torsion implies that the connection $\Gamma\ud{\alpha}{\mu\nu}$ is symmetric in its lower two indices. Hence the use of the word \textbf{symmetric} in \textit{Symmetric Teleparallel Equivalent of GR}.\newline

\paragraph{\textbf{The Metric and Connection Field Equations}}
To obtain the field equations of STEGR we can either take the action~\eqref{eq:STEGRAction} as starting point or the action~\eqref{eq:STEGRActionCovariant}. In either case we make the observation that the non-metricity tensor is linear in first order derivatives of the metric and that the non-metricity scalar $\Q$ is consequently quadratic in first order derivatives. Due to the absence of second order derivatives in either action principle, there is no need to add boundary terms \`{a} la Gibbons-Hawking-York. Both variational principles are well-defined as they stand.

If we choose to work with the Palatini formalism, we have to vary the action~\eqref{eq:STEGRActionCovariant} with respect to the metric $g_{\mu\nu}$, the general affine connection $\Gamma\ud{\alpha}{\mu\nu}$, as well as the Lagrange multiplier densities $\tilde{\Pi}\du{\alpha}{\mu\nu\rho}$ and $\tilde{\chi}\du{\alpha}{\mu\nu}$.

If instead we work with the action~\eqref{eq:STEGRAction}, we only need to perform variations with respect to $g_{\mu\nu}$ and $\xi^\alpha$. The first approach turns out to be simpler, despite the additional variations one has to perform. The computations have been carried out in great detail in~~\cite{BeltranJimenez:2018}. For the variation with respect to the inverse metric, one obtains the metric field equations
\begin{equation}
	\boxed{\frac{2}{\sqrt{|g|}}\nabla_\alpha\left[\sqrt{|g|}\, P\ud{\alpha}{\mu\nu}\right] + q_{\mu\nu} -\frac12 \Q\, g_{\mu\nu} = \kappa\, \T_{\mu\nu}}
\end{equation}
As always, $\T_{\mu\nu}$ denotes the energy-momentum tensor of matter fields and we have introduced the \textbf{non-metricity conjugate $\boldsymbol{P\ud{\alpha}{\mu\nu}}$}  and the symmetric tensor $q_{\mu\nu}$, respectively defined by
\begin{align}
	P\ud{\alpha}{\mu\nu} &\ce \frac12 \frac{\partial\Q}{\partial Q\du{\alpha}{\mu\nu}} = -\frac14\, Q\ud{\alpha}{\mu\nu} + \frac12 \,Q\dud{(\mu}{\alpha}{\nu)} + \frac14\, g_{\mu\nu} Q^\alpha   - \frac14\,\left(g_{\mu\nu}\bar{Q}^\alpha + \delta^\alpha{}_{(\mu} Q_{\nu)}\right) \notag\\
	q_{\mu\nu} &\ce \frac{\partial \Q}{\partial g^{\mu\nu}} = P_{(\mu|\alpha\beta}Q\du{\nu)}{\alpha\beta} - 2 P\ud{\alpha\beta}{(\nu}Q_{\alpha\beta|\mu)}\,.
\end{align}
It should be noted that the non-metricity scalar $\Q$ can be expressed with the help of the non-metricity tensor and its conjugate as
\begin{equation}
	\Q = P_{\alpha\mu\nu}Q^{\alpha\mu\nu}\,.
\end{equation}
The variations with respect to the Lagrange multipliers and the general affine connection boil down to a connection field equation of the form
\begin{equation}
	\boxed{\nabla_\mu\nabla_\nu \left(\sqrt{|g|}\,P\ud{\mu\nu}{\alpha}\right) = 0}
\end{equation}
Here, just as in TEGR, we have assumed that the hypermomentum density~\eqref{eq:DefHyperMomentum} vanishes. Alternatively, we could have demanded that it is conserved, $\nabla_\mu H\du{\alpha}{\mu\nu} = 0$.

In both field equations the connection is flat and torsionless, as required by the geometric postulates. It is thus parametrized by the four functions $\xi^\alpha$. Moreover, observe that we have ten metric field equations and four connection field equations. These numbers match the number of fields in the theory, namely ten metric components and four $\xi$'s. However, just as in the case of GR and TEGR, not all equations are independent due to the diffeomorphism invariance of the theory.\newline

\paragraph{\textbf{The Bianchi Identities}}
The action~\eqref{eq:STEGRActionCovariant} is manifestly invariant under diffeomorphisms. This follows from the fact that $\sqrt{|g|}\, \Q$ is a scalar density of weight $w=+1$. Also, since the Lagrange multipliers have density weight $w=+1$ and they are fully contracted, the curvature and torsion constraints are also scalar densities with the correct weight. Correct means that the integrand transforms in such a way under diffeomorphisms that the integral remains invariant. Following the same considerations as in GR, but now also taking the transformation behaviour of the connection into account, we find the following identities for STEGR:
\begin{align}
	\D_\mu \M\ud{\mu}{\nu} + \mathcal{C}_\nu \equiv 0\,,
\end{align}
where we have defined
\begin{align}
	\M_{\mu\nu} &\ce\frac{2}{\sqrt{|g|}}\nabla_\alpha\left[\sqrt{|g|}\, P\ud{\alpha}{\mu\nu}\right] + q_{\mu\nu} -\frac12 \Q\, g_{\mu\nu}\notag\\
	\C_\alpha &\ce\nabla_\mu\nabla_\nu \left(\sqrt{|g|}\,P\ud{\mu\nu}{\alpha}\right)
\end{align}
The tensor $\M_{\mu\nu}$ is simply the expression that appears on the left side of the metric field equations, while~$\C_\alpha$ represents the left  side of the connection field equations. As a consequence of these Bianchi identities, it follows that if the metric field equations are satisfied, i.e., if $\M_{\mu\nu} = \kappa\, \T_{\mu\nu}$, then
\begin{align}
	\kappa\,\D_\mu \T\ud{\mu}{\nu} + \C_\nu \equiv 0\,.
\end{align}
By invoking the covariant conservation of energy-momentum of matter fields, i.e., $\D_\mu \T\ud{\mu}{\nu}$, we find
\begin{align}
	\text{If } \M_{\mu\nu} = \kappa\, \T_{\mu\nu} \text{ is satisfied, then } \C_\nu \equiv 0\,.
\end{align}
In other words, if the metric field equations are satisfied, then the connection field equations become mere identities. That is, the connection field equations are trivially satisfied and carry no dynamical information.

Since one can show that $\M_{\mu\nu} = G_{\mu\nu}$, where the right hand side is the Einstein tensor without cosmological constant, one can reach an even stronger conclusion: The Einstein tensor satisfies the Bianchi identity $\nabla_\mu G\ud{\mu}{\nu}$ also off-shell, i.e., when the Einstein equations are not satisfied. By combining this fact with the Bianchi identity of STEGR, one reaches the conclusion that
\begin{align}
	\boxed{\C_\nu \equiv 0}
\end{align}
is \textit{always} true! Since $\M_{\mu\nu} = G_{\mu\nu}$ implies that $\M_{\mu\nu}$ is \textit{independent} of $\xi^\alpha$ (it only knows about the Levi-Civita part of the connection and nothing else) it follows that $\M_{\mu\nu} = \kappa\, \T_{\mu\nu}$ are purely equations for the metric. Furthermore, since $\C_\nu = 0$ is always identically satisfied, there are no equations which determine the four functions $\xi^\alpha$! They remain completely arbitrary. What these considerations show is the following:
\begin{itemize}
	\item STEGR is equivalent to GR in the sense that both theories possess the same field equations and consequently the same solution space. They are nevertheless rooted in different mathematical frameworks, they used different fields in their formulation, and this opens the door to conceptual and philosophical differences between the two theories.
	\item There is a sense in which STEGR is invariant under two copies of the diffeomorphism group. First its action is manifestly diffeomorphism invariant and its field equations are manifestly general covariant. Thus, performing a diffeomorphism which changes the metric and the connection simultaneously does not affect the theory. Secondly, we have the freedom to choose the four functions $\xi^\alpha$ at will. The metric field equations do not depend upon this choice and there are no dynamical equations which determine the $\xi$'s. Thus, this constitutes a second freedom.
\end{itemize}
As we will see later, the independence of $\M_{\mu\nu}$ from the $\xi$'s hinges on a carefully balanced cancellation. If one considers the most general non-metricity scalar $\hat{\Q}$, as we do in~\ref{ssec:QuadraticNonMetricity}, this independence is lost unless one chooses certain parameters in the theory in a careful way. Also, we will see that in $f(\Q)$ gravity the $\xi$'s are no longer arbitrary. They come with their own dynamical field equations.\newline

\paragraph{\textbf{Counting Degrees of Freedom}}
After the discussion of the Bianchi identities it comes as no surprise that STEGR propagates two physical degrees of freedom. Concretely, the counting goes as follows: The theory is formulated in terms of a metric $g_{\mu\nu}$ with ten components and a general affine connection $\Gamma\ud{\alpha}{\mu\nu}$ with $4\times 4\times 4 = 64$ components. By either postulating the vanishing of curvature and torsion or by solving the constraints that arise from the Palatini formulation of the theory, one finds that the connection carries four potential degrees of freedom (the $\xi$'s).

This leaves us with $10+4 = 14$ variables and an equal number of field equations. However, the metric field equations only contain the metric and no $\xi$'s. Also, there are no dynamical equation for the $\xi$'s. They remain completely arbitrary and do not constitute anything physical. In fact, as we will see in the next subsection, they play the role of St\"{u}ckelberg fields, which ensure that the theory is generally covariant. 

This leaves us with at most ten dynamical variables, namely the metric components. However, since the metric field equations are simply the Einstein field equations, the same counting as in GR assures us that only two of these components represent physical degrees of freedom.

Alternatively, one could also argue as follows: STEGR is a diffeomorphism invariant theory and it is also invariant under the replacement $\xi^\alpha \mapsto \hat{\xi}^\alpha$, where $\hat{\xi}^\alpha$ is a new set of four functions which parametrize the flat, torsionless connection. Thus, the $\xi$'s play no dynamical role and since diffeomorphisms remove $2\times 4$ degrees of freedom one finds again $14 - 4 - 2\times 4 = 2$ physical degrees of freedom.

Thus, in either case, we conclude that STEGR propagates the same two degrees of freedom as GR, as had to be expected. This will no longer be true when we consider generalizations of STEGR in subsections~\ref{ssec:QuadraticNonMetricity} and~\ref{ssec:NonLinearExtensions} and in particular in section~\ref{sec:f(Q)}.

\subsection{Coincident General Relativity (CGR)}\label{ssec:CGR}
Coincident General Relativity, or CGR for short, refers to a special case of STEGR. In fact, CGR is simply STEGR in coincident gauge. Even tough this might seem trivial, CGR has played an important role in applications such as $f(\Q)$ cosmology~\cite{BeltranJimenez:2019, DAmbrosio:2020b, Bajardi:2020, Ayuso:2020, Frusciante:2021, Atayde:2021, Anagnostopoulos:2021, Capozziello:2022}. Furthermore, by comparing and contrasting CGR with full STEGR and what is nowadays called the Einstein action, one is lead to a deeper understanding of the role played by the flat and torsionless connection (or, equivalently, by the $\xi$'s).

Let us begin by introducing the action of CGR. As mentioned above, CGR makes use of the coincident gauge, which means the flat and torsionless connection $\Gamma\ud{\alpha}{\mu\nu}$ vanishes globally. Upon using the decomposition~\eqref{eq:GammaDecomposition}, we find that this implies
\begin{align}
	\Gamma\ud{\alpha}{\mu\nu} = \LC{\alpha}{\mu\nu} + L\ud{\alpha}{\mu\nu} &\overset{\text{CG}}{=} 0 & \Longrightarrow && L\ud{\alpha}{\mu\nu}  \overset{*}{=} - \LC{\alpha}{\mu\nu}\,.
\end{align}
The star on top of the equal sign shall remind us that this relation only holds in the coincident gauge. This last equality is particularly useful if we recall that the STEGR action can be written in terms of the disformation tensor $L\ud{\alpha}{\mu\nu}$ alone (cf. equation~\ref{eq:QinLsquared}). Hence, the CGR action takes the form
\begin{align}\label{eq:CGRAction}
	\S_\text{CGR}[g] \equiv \S_\text{STGR}[g, \Gamma=0] &= \frac{1}{2\kappa}\int_\M\dd^4 x\, \sqrt{|g|}\,g^{\mu\nu}\left(L\ud{\alpha}{\alpha\beta} L\ud{\beta}{\mu\nu} - L\ud{\alpha}{\beta\mu} L\ud{\beta}{\nu\alpha}\right) \notag\\
	&\overset{*}{=} \frac{1}{2\kappa}\int_\M\dd^4 x\, \sqrt{|g|}\,g^{\mu\nu}\left(\LC{\alpha}{\alpha\beta} \LC{\beta}{\mu\nu} - \LC{\alpha}{\beta\mu} \LC{\beta}{\nu\alpha}\right)\,.
\end{align}
The first integral is valid in complete generality, while the second one holds only in the coincident gauge. Observe further that the action is a functional of the metric alone. This is one of the features which make CGR attractive for studying applications: The connection has been dealt with and even globally trivialized, such that one only has to work with the metric. However, when using CGR as the starting point for defining non-linear modifications such as in $f(\Q)$ gravity, one is prone to encounter subtleties related to having fixed the coincident gauge. These subtleties have to do with assuming some background symmetries (spherical symmetry, homogeneity and isotropy, ...) and we will discuss them in more detail in subsections~\ref{ssec:Cosmology} and~\ref{ssec:BlackHoles}. 

Here, we wish to highlight an other feature of CGR. Namely that its action is exactly equal to the so-called Einstein action, which in turn is simply the Einstein-Hilbert action \textit{without} the second order derivatives. This can easily be seen by recalling that the Ricci tensor is given by
\begin{align}\label{eq:RicciTensorLC}
	\R_{\mu\nu} = \partial_\alpha \LC{\alpha}{\nu\mu} - \partial_\nu \LC{\alpha}{\alpha\mu} + \LC{\alpha}{\alpha\beta} \LC{\beta}{\mu\nu} - \LC{\alpha}{\beta\mu} \LC{\beta}{\nu\alpha}\,.
\end{align}
By comparing~\eqref{eq:RicciTensorLC} to the action~\eqref{eq:CGRAction} it follows that the Einstein action is given by the Ricci scalar $\R = g^{\mu\nu}\R_{\mu\nu}$ minus the term
\begin{align}
	g^{\mu\nu}\left(\partial_\alpha \LC{\alpha}{\nu\mu} - \partial_\nu \LC{\alpha}{\alpha\mu}\right)\,
\end{align}
which contains the second order derivatives of the metric. Hence, neither the CGR nor the Einstein action require the GHY boundary term. Both of them give rise to a well-defined variational principle.

However, despite looking the same, there is a crucial difference between the Einstein action and the CGR action. The former is \textit{not} diffeomorphism invariant. This follows from the fact that the connection does not transform like a tensor under coordinate transformations (see equation~\eqref{eq:ConnectionTransformationLaw}) and one can check that the Einstein action picks up boundary terms under such transformations. 

The CGR action, on the other hand, is the gauge-fixed version of a perfectly covariant functional, namely the STEGR action. Hence, we can interpret the connection as a St\"{u}ckelberg field which restores the general covariance of the Einstein action.

\subsection{The General Teleparallel Equivalent of General Relativity (GTEGR)}\label{ssec:GTEGR}

So far we have seen that gravity can be described from three different perspectives: Following Einstein's original path, we can encode gravity in the curvature of spacetime while setting torsion and non-metricity to zero. Or we can describe gravity using torsion in a flat and metric-compatible spacetime. The third option is to work in a flat and torsionless spacetime, but with non-zero non-metricity. 

We can think of these three descriptions as the three corners of a triangle, as illustrated in Figure~\ref{fig:12_TrinityGR}. We can also give meaning to the edges of the triangle. Of particular interest to us is the lower edge which connects TEGR and STEGR. In fact, there exists yet another teleparallel theory of gravity, called the General Teleparallel Equivalent of GR (GTEGR)~\cite{BeltranJimenez:2019d, Hohmann:2022, Heisenberg:2022}, which subsumes TEGR and STEGR in the sense that these two theories are the gauge-fixed offsprings of a more general parent theory.

To construct the theory, we start again with a general metric-affine geometry $(\M, g, \Gamma)$ and a single geometric postulate,
\begin{align}
    R\ud{\alpha}{\mu\nu\rho} \overset{!}{=} 0\,.
\end{align}
The geometric identity~\eqref{eq:KeyIdG} we encountered in subsection~\ref{ssec:Identities} then allows us to write the Ricci scalar of the Levi-Civita connection as
\begin{align}
    -\R(g) &= \bbT + \Q + T^{\rho\mu\nu}Q_{\mu\nu\rho} - T^\mu Q_\mu + T^\mu \bar{Q}_\mu + \D_\alpha\left(Q^\alpha - \bar{Q}^\alpha + 2 T^\alpha\right)\notag\\
    &= \bbG + \D_\alpha\left(Q^\alpha - \bar{Q}^\alpha + 2 T^\alpha\right)\,,
\end{align}
where in the last equation we have introduced the scalar
\begin{align}
    \bbG \ce \bbT + \Q + T^{\rho\mu\nu}Q_{\mu\nu\rho} - T^\mu Q_\mu + T^\mu \bar{Q}_\mu\,.
\end{align}
We then define the following action
\begin{align}\label{eq:GTEGRAction}
    \S_{\rm GTEGR}[g, \Lambda] \ce -\frac{1}{2\kappa}\int_{\M}\dd^4 x\, \sqrt{|g|}\, \bbG(g, \Lambda) + \S_{\rm matter}\,,
\end{align}
where $\Lambda\in GL(4, \bbR)$ is the matrix used to parametrize the flat connection. Since the EH action and the action of GTEGR only differ by a total derivative, it comes as no surprise that both actions describe the same theory. However, notice that while GR only makes use of a metric, GTEGR also involves the matrix $\Lambda\in GL(4, \bbR)$ in its definition. This mismatch in the number of fields is no reason for concern, since GTEGR enjoys an additional symmetry. In fact, $\delta_\Lambda \S_{\rm GTEGR} = 0$ is satisfied off-shell, which means that the connection is not dynamical~\cite{BeltranJimenez:2019d}. Put differently, this means that only the metric carries physical degrees of freedom while the connection is pure gauge. Furthermore, since the metric field equations obtained from~\eqref{eq:GTEGRAction} have to be the Einstein field equations, the metric propagates exactly the same two degrees of freedom as GR.

Observe further that in the absence of non-metricity the scalar $\bbG$ reduces to $\bbT$ and TEGR is recovered. Similarly, when torsion is absent, $\bbG$ reduces to the non-metricity scalar $\Q$ and STEGR emerges. Demanding either the vanishing of torsion or the vanishing of non-metricity amounts to imposing additional conditions on the connection, as we have seen in previous subsections. It is in this sense that we can think of TEGR and STEGR as partially gauge-fixed versions of the more general theory GTEGR: The pure gauge connection of GTEGR can be partially fixed by either imposing $Q_{\alpha\mu\nu} = 0$ or $T\ud{\alpha}{\mu\nu} = 0$, which simply amounts to working with TEGR or STEGR, respectively.

\subsection{Non-flat combinations in the edges and in the dot}\label{ssec:Edges}
\begin{figure}[htb!]
	\centering
	\includegraphics[width=0.75\linewidth]{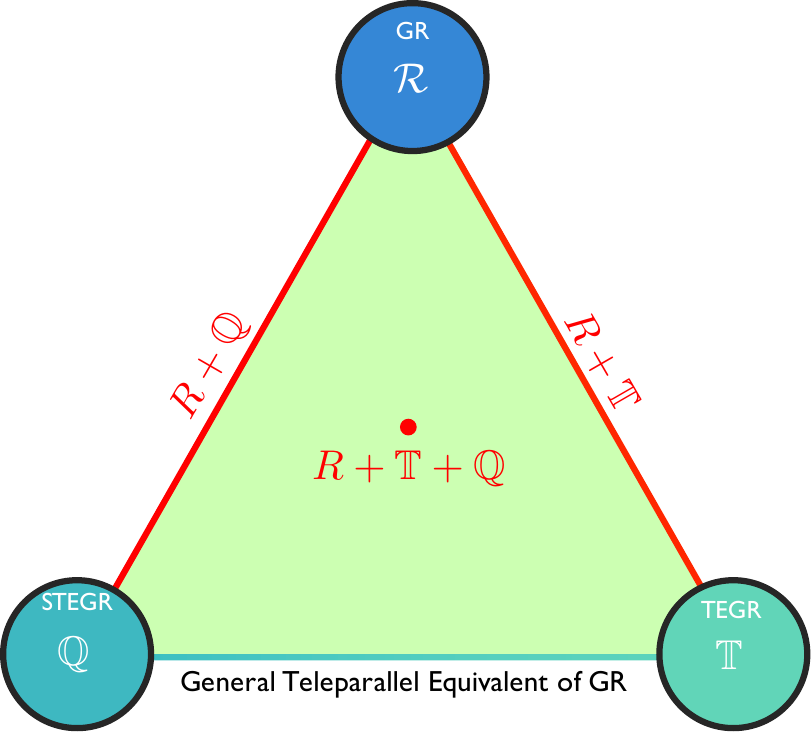}
	\caption{\protect Of the metric-affine theories of gravity which live on the edges and in the center of the triangle, only the General Teleparallel theory is equivalent to GR. \hspace*{\fill}}
	\label{fig:13_EdgesAndCenter}
\end{figure}

So far we have discussed the three corners of Figure~\ref{fig:12_TrinityGR} as well as one edge. This corresponds to four different formulations of General Relativity: Standard GR based on curvature and the teleparallel theories TEGR, STEGR, and GTEGR which are all based on the postulate of vanishing curvature. It is only natural to ask whether other equivalent formulations are possible. In particular, there are two more edges present in Figure~\ref{fig:12_TrinityGR}. These would correspond to non-flat geometries with either torsion or non-metricity (but not both at the same time). Finally, we can also imagine a dot in the center of the triangle, which represents a theory based on non-vanishing curvature, torsion, and non-metricity. A modified version of Figure~\ref{fig:12_TrinityGR} could look like~\ref{fig:13_EdgesAndCenter}.
Notice that in all three new cases we wish to discuss, the postulate of teleparallelism (i.e., the condition $R\ud{\alpha}{\mu\nu\rho} = 0$) is \textit{not} imposed. This has far-reaching consequences.

Recall that in TEGR, STEGR, and GTEGR the crucial step was to impose $R\ud{\alpha}{\mu\nu\rho} = 0$, which immediately implies that the connection has the form $\Gamma = (\Lambda)^{-1}\partial\Lambda$. Since the Lagrangians which define these three theories are all quadratic in $T$ and $Q$, we find that they all possess something akin to a ``kinetic term'', $T^2 \sim (\partial\Lambda)^2$ and $Q^2 \sim (\partial\Lambda)^2$. However, if the flatness condition is \textit{not} imposed, we lose this ``kinetic term''. In particular, the actions
\begin{align}
    \S_{\rm Einstein-Cartan}[g, \Gamma] &= \frac{1}{2\kappa}\int_{\M} \dd^4x\, \sqrt{|g|}\left(R+\bbT\right) + \S_{\rm matter}[g, \Gamma, \Psi]\notag\\
    \S[g,\Gamma] &= \frac{1}{2\kappa} \int_\M\dd^4 x\, \sqrt{|g|}\left(R + \Q\right) + \S_{\rm matter}[g, \Gamma, \Psi] \notag\\
    \S[g, \Gamma] &= \frac{1}{2\kappa} \int_\M\dd^4 x\, \sqrt{|g|}\left(R + \bbT + \Q \right) + \S_{\rm matter}[g, \Gamma, \Psi]
\end{align}
are all deprived of this ``kinetic term''. The first one corresponds to the left edge in Figure~\ref{fig:13_EdgesAndCenter} and is also known as \textbf{the Einstein-Cartan action}. The action in the middle represents the theory living on the right edge, while the action on the bottom corresponds to the dot in Figure~\ref{fig:13_EdgesAndCenter}. As it turns out, in all three cases the connection is a mere auxiliary field which can be integrated out. After integrating out the connection, the resulting actions are \textit{not} equivalent to GR! Rather, one obtains three modified gravity theories. 

Furthermore, one can also show that integrating out the connection changes the way matter fields couple in these theories, leading again to non-GR behaviour. This is in the same spirit as what was shown in~\cite{BeltranJimenez:2017} for more general Lagrangians based on the Palatini formalism. 

\subsection{Matter Coupling}\label{ssec:MatterCoupling}
Our discussion of the geometric trinity and the equivalence between teleparallel theories of gravity and GR was so far limited to the pure gravity sector. Does the equivalence between teleparallel theories and GR also hold in the presence of matter fields? 

In GR, the coupling of the gravitational field to matter fields follows the so-called \textbf{minimal coupling principle}. It states that a matter theory formulated in Minkowski space is promoted to a matter theory coupled to the gravitational field $g_{\mu\nu}$ by replacing $\eta_{\mu\nu}\mapsto g_{\mu\nu}$ and $\partial_\mu \mapsto \D_\mu$, provided that the matter fields only couple to $g_{\mu\nu}$, $g^{\mu\nu}$, and $\sqrt{|g|}$, but not derivatives of the metric. Is the minimal coupling principle preserved in TEGR and STEGR?

Let us first consider TEGR and naively apply the minimal coupling principle in the form $\eta_{\mu\nu}\mapsto g_{\mu\nu}$ and $\partial_\mu \mapsto \nabla_\mu$, where $\nabla_\mu$ is the covariant derivative operator with respect to the connection $\Gamma\ud{\alpha}{\mu\nu}$. As a specific example, we consider the electromagnetic potential $A_\mu$ and its associated Maxwell $2$-form $F_{\mu\nu} \ce \partial_\mu A_\nu - \partial_\nu A_\mu$. According to the minimal coupling principle, the Maxwell $2$-form becomes
\begin{align}
    F_{\mu\nu} = \nabla_\mu A_\nu - \nabla_\nu A_\mu = \partial_\mu A_\nu - \partial_\nu A_\mu - T\ud{\alpha}{\mu\nu} A_\alpha\,.
\end{align}
We immediately conclude that the minimal coupling principle fails, since the Maxwell action picks up terms proportional to the torsion tensor, thus spoiling the equivalence between TEGR and GR. For fermionic fields, one obtains a similar failure of the minimal coupling principle. The Dirac Lagrangian is directly affected by the connection in the presence of an axial torsion. 

In STEGR the situation is quite different: The minimal coupling principle is preserved even in the presence of non-metricity. In the case of the electromagnetic field $A_\mu$ it is straightforward to verify that non-metricity does not contribute to $F_{\mu\nu}$ due to its symmetry and thus one finds
\begin{align}
    F_{\mu\nu} = \nabla_\mu A_\nu - \nabla_\nu A_\mu = \partial_\mu A_\nu - \partial_\nu A_\mu\,,
\end{align}
just as in GR. For fermions, this property remains unchanged. The non-metricity drops out completely from the Dirac Lagrangian due to the symmetry of the non-metricity tensor. For a more detailed analysis of matter couplings in TEGR and STEGR we refer the reader to~\cite{BeltranJimenez:2020}.

The key message here is that in the presence of matter fields, the equivalence is only maintained between STEGR and GR. TEGR coupled to matter fields is no longer equivalent to GR.

\newpage
\asection{5}{The Geometrical Trinity of Modified Gravity Theories}\label{sec:TrinityGeneralizations}
In section~\ref{sec:Trinity} we introduced three different geometric approaches to formulate the theory of General Relativity. This so-called geometric trinity of GR has conceptual advantages. For instance, the teleparallel theories TEGR and STEGR possess well-defined variational principles~\cite{BeltranJimenez:2018, BeltranJimenez:2019c} without the need of adding a GHY boundary term. Furthermore, STEGR and CGR have inspired new approaches to define the elusive gravitational energy-momentum~\cite{BeltranJimenez:2017b, BeltranJimenez:2019b, BeltranJimenez:2021, Koivisto:2022}, it is possible to compute black hole entropy without adding counter terms to the action~\cite{BeltranJimenez:2018, BeltranJimenez:2019b, Heisenberg:2022b}, and the coincident gauge might open a new avenue toward the quantization of the gravitational field. 

However, since the field equations of TEGR and STEGR are identical to the Einstein field equations, these theories cannot address any phenomenological questions which elude GR---such as the accelerated expansion of the Universe or the shape of galactic rotation curves.

Such questions are typically addressed by theories of \textbf{modified gravity} and the geometric trinity of GR presented in the previous section can be used as starting point for developing such modifications. There are two approaches which are commonly considered in the literature:
\begin{enumerate}
	\item The actions of TEGR and STEGR are quadratic in the torsion and the non-metricity tensors, respectively. One can thus try to construct the most general scalar which is quadratic in the torsion or non-metricity tensor and take this scalar to define an action functional. In the case of torsion, one finds a three-parameter family of theories described by an action which is quadratic in the torsion tensor. In the case of non-metricity, one finds a five-parameter family of quadratic Lagrangians. These generalizations are discussed in subsections~\ref{ssec:QuadraticTorsion} and~\ref{ssec:QuadraticNonMetricity}, respectively.
	\item An other popular direction is to consider non-linear extensions of the form $f(\mathbb{T})$ and $f(\Q)$, where~$f$ is some function which is only subjected to the condition that its first derivative does not vanish. Non-linear extensions of this type are the subject of subsection~\ref{ssec:NonLinearExtensions}. In section~\ref{sec:f(Q)} we will have a closer look at $f(\Q)$, its application to cosmology, black hole physics, and the question of how many degrees of freedom the theory propagates.
\end{enumerate}
Since these modifications of GR are based on the framework of metric-affine geometry, we will sometimes refer to them as the \textbf{geometrical trinity of modified gravity theories}.\\

\subsection{Quadratic Actions for Torsion Theories}\label{ssec:QuadraticTorsion}
Recall from subsection~\ref{ssec:TEGR} that the action of TEGR is constructed solely from quadratic contractions of the torsion tensor. Concretely, we defined the so-called torsion scalar as
\begin{align}
	\bbT \ce \frac12 \left(\frac14 T_{\alpha\mu\nu} + \frac12 T_{\mu\alpha\nu} - g_{\alpha\mu} T_\nu \right) T^{\alpha\mu\nu}\, .
\end{align}
Now we are interested in constructing the \textit{most general scalar} which is quadratic in the torsion tensor. To that end, we need to consider the symmetries of $T\ud{\alpha}{\mu\nu}$. A priori, a tensor with three indices can be contracted in six different ways with itself; one just has to perform all possible permutations of indices. However, because~$T\ud{\alpha}{\mu\nu}$ is antisymmetric in its lower indices, only two of these contractions are independent:
\begin{align}
	&T_{\alpha\mu\nu}T^{\alpha\mu\nu} && \text{and} & &T_{\mu\alpha\nu} T^{\alpha\mu\nu}\,.
\end{align}
The next thing to consider is the trace of the torsion tensor. Due to its antisymmetry, the torsion tensor possesses only one trace; $T_\mu \ce T\ud{\alpha}{\mu\alpha}$. Thus, the only other quadratic contraction we can build out of the torsion tensor is
\begin{equation}
	T_\mu T^\mu\,.
\end{equation}
With this we have exhausted all options and we conclude that the most general scalar which is quadratic in the torsion tensor is a linear combination of the three terms discussed above:
\begin{align}\label{eq:TScalar}
	\hatT \ce c_1\, T_{\alpha\mu\nu}T^{\alpha\mu\nu} + c_2\, T_{\mu\alpha\nu} T^{\alpha\mu\nu} + c_3\, T_\mu T^\mu \, ,
\end{align}
where $c_1$, $c_2$, and $c_3$ are \textit{arbitrary}, real constants. It is easy to see that the scalar $\hatT$ reduces to $\bbT$ for the parameter choice $c_1 = \frac14$, $c_2 = \frac12$, $c_3 = -1$. Using the general torsion scalar defined in~\eqref{eq:TScalar}, we can now write the action functional of \textbf{Teleparallel Gravity (TG)}\footnote{The theory defined by this action is sometimes referred to as New General Relativity in the literature (for instance in~\cite{Hayashi:1979, BeltranJimenez:2017b, Blixt:2018}).} as
\begin{equation}\label{eq:ActionTG}
	\S_\textsf{TG}[g, \Gamma, \Psi] \ce -\int_\M \dd^4 x\, \left(\frac{1}{\kappa}\sqrt{|g|}\, \hatT + \tilde\Pi\du{\alpha}{\mu\nu\rho}\, R\ud{\alpha}{\mu\nu\rho} + \tilde{\chi}\ud{\alpha}{\mu\nu}\, Q\du{\alpha}{\mu\nu}\right) + \S_\textsf{matter}[g, \Psi]\,,
\end{equation}
where the matter fields $\Psi$ are assumed to be minimally coupled and having a vanishing hypermomentum. This action looks deceptively similar to the action of TEGR, since we have only replaced $\bbT$ by the more general~$\hatT$. Indeed, even the field equations look very similar:
\begin{align}
	&\textsf{Metric field equations:} & (\nabla_\alpha + T_\alpha)\hat{S}\du{(\mu\nu)}{\alpha} + \hat{t}_{\mu\nu} -\frac12 \hatT\,g_{\mu\nu} &= \kappa\, \T_{\mu\nu}\notag\\
	&\textsf{Connection field equations:} & (\nabla_\alpha + T_\alpha)\left[\sqrt{|g|} \hat{S}\dud{[\mu}{\alpha}{\nu]}\right] &= 0\,,
\end{align}
where the hatted torsion conjugate $\hat{S}\du{\alpha}{\mu\nu}$ and the symmetric tensor $\hat{t}_{\mu\nu}$ are defined as
\begin{align}
	\hat{S}\du{\alpha}{\mu\nu} &\ce \PD{\hatT}{T\ud{\alpha}{\mu\nu}} = c_1\, T\du{\alpha}{\mu\nu} + c_2\, T\udu{[\mu}{\alpha}{\nu]} + c_3\,  \delta\du{\alpha}{[\mu}T^{\nu]}   \notag\\
	\hat{t}_{\mu\nu} &\ce \PD{\hatT}{g^{\mu\nu}} = \frac12 \hat{S}\du{(\mu|}{\lambda\kappa}\, T_{\nu)\lambda\kappa} - T\ud{\lambda\kappa}{(\mu}\, \hat{S}_{\lambda\kappa|\nu)}\,.
\end{align}
Unsurprisingly, also the Bianchi identities posses the same form we encountered before and it is still possible to parametrize the connection as described in subsection~\ref{ssec:TEGR}, since the latter one has nothing to do with an action principle. Thus, TEGR and TG are theories which look very similar. However, there are important differences when it comes to the number of physical degrees of freedom. As we have seen in subsection~\ref{ssec:TEGR}, TEGR possesses, as expected, two degrees of freedom.

For TG, the situation can look quite different in this regard. Even though the field equations have a similar form and are still second order partial differential equations, the number of degrees of freedom depends on how one chooses the parameters $c_1$, $c_2$, and $c_3$. That is because these parameters appear in certain combinations in the field equations and there are choices which can make some second order time derivatives of the metric disappear. Thus, some equations can be turned into constraints, rather than dynamical equations, which has an impact on the number of degrees of freedom. Similarly, equations which are constraints in TEGR can be turned into dynamical equations by detuning the parameters $c_1$, $c_2$, and $c_3$. Moreover, it is possible that for certain parameter combinations further constraint equations appear as integrability conditions, thus affecting the number of degrees of freedom even more.

These considerations become slightly more transparent through the lens of a Hamiltonian analysis. In~\cite{Blixt:2018, Blixt:2019}, the first step of such an analysis was carried out (see also~\cite{Blixt:2020} for a detailed review of the results obtained so far). More precisely, it has been investigated how many independent so-called primary constraints appear through the vanishing of certain combinations of the parameters $c_1$, $c_2$, $c_3$. The analysis of these primary constraints revealed that the three-parameter family of theories described by the action~\eqref{eq:ActionTG} compartmentalizes into nine different sectors (which we dub the \textbf{primary sectors}, following the nomenclature of~\cite{DAmbrosio:2020}). Each sector is characterized by a different number of primary constraints (cf. Table~\ref{tab:PrimarySectorsTG}).

\begin{table}[hbt]
\begin{tabular}{|p{0.15\textwidth}|cccp{0.16\textwidth}<{\centering}|p{0.15\textwidth}<{\centering}|}
\hline
\textbf{Primary}    & \multicolumn{4}{c|}{\textbf{Parameter combinations}}                                                                                                                      & \multicolumn{1}{l|}{\textbf{\# of primary}} \\
\textbf{sector}     & \multicolumn{1}{p{0.16\textwidth}|}{\centering$\mathbf{2c_1 + c_2 + c_3}$} & \multicolumn{1}{p{0.16\textwidth}|}{\centering$\mathbf{2c_1 - c_2}$} & \multicolumn{1}{p{0.16\textwidth}|}{\centering$\mathbf{2c_1 + c_2}$} & $\mathbf{2c_1 + c_2 + 3c_3}$ & \multicolumn{1}{l|}{\textbf{constraints}}   \\ \hline\hline
$\mathbf{0}$        & \multicolumn{1}{c|}{$\neq 0$}                    & \multicolumn{1}{c|}{$\neq 0$}              & \multicolumn{1}{c|}{$\neq 0$}              & $\neq 0$                     & $0$                                         \\ \hline
\textbf{I}          & \multicolumn{1}{c|}{$\red{=0}$}                   & \multicolumn{1}{c|}{$\neq 0$}              & \multicolumn{1}{c|}{$\neq 0$}              & $\neq 0$                     & $3$                                         \\ \hline
\textbf{II}         & \multicolumn{1}{c|}{$\neq 0$}                    & \multicolumn{1}{c|}{$\red{=0}$}             & \multicolumn{1}{c|}{$\neq 0$}              & $\neq 0$                     & $3$                                         \\ \hline
\textbf{III}        & \multicolumn{1}{c|}{$\neq 0$}                    & \multicolumn{1}{c|}{$\neq 0$}              & \multicolumn{1}{c|}{$\red{=0}$}             & $\neq 0$                     & $5$                                         \\ \hline
\textbf{IV}         & \multicolumn{1}{c|}{$\neq 0$}                    & \multicolumn{1}{c|}{$\neq 0$}              & \multicolumn{1}{c|}{$\neq 0$}              & $\red{=0}$                    & $1$                                         \\ \hline
\textbf{V (TEGR)}          & \multicolumn{1}{c|}{$\red{=0}$}                   & \multicolumn{1}{c|}{$\red{=0}$}             & \multicolumn{1}{c|}{$\neq 0$}              & $\neq 0$                     & $6$                                         \\ \hline
\textbf{VI}         & \multicolumn{1}{c|}{$\neq 0$}                    & \multicolumn{1}{c|}{$\red{=0}$}             & \multicolumn{1}{c|}{$\red{=0}$}             & $\neq 0$                     & $8$                                         \\ \hline
\textbf{VII} & \multicolumn{1}{c|}{$\neq 0$}                    & \multicolumn{1}{c|}{$\red{=0}$}             & \multicolumn{1}{c|}{$\neq 0$}              & $\red{=0}$                    & $4$                                         \\ \hline
\textbf{VIII}       & \multicolumn{1}{c|}{$\red{=0}$}                   & \multicolumn{1}{c|}{$\neq 0$}              & \multicolumn{1}{c|}{$\red{=0}$}             & $\red{=0}$                    & $9$                                         \\ \hline
\end{tabular}
\caption{\protect Each primary sector is defined by the vanishing of certain parameter combinations. The vanishing of one or more of these parameter combinations leads to the appearance of primary constraints. The number of independent primary constraints is listed in the last column. TEGR is contained in sector V. This table was inspired by~\cite{Blixt:2020}, but we adopt a slightly different nomenclature in order to be consistent with~\cite{DAmbrosio:2020}.\hspace*{\fill}}
\label{tab:PrimarySectorsTG}
\end{table}
Primary constraints reduce the number of degrees of freedom. Hence, the more primary constraints, the fewer degrees of freedom there are. However, the exact number of physical degrees of freedom within each sector has not yet been determined. In fact, the Hamiltonian analysis could not be carried out to completion and it is not even known whether there are secondary constraints. In~\cite{DAmbrosio:2023} it was argued that the standard Hamiltonian method for constrained systems (the so-called Dirac-Bergmann algorithm) is in general not applicable to teleparallel theories of gravity.  We briefly touch upon this subject in subsection~\ref{ssec:Hamiltonian} and refer the reader to~\cite{DAmbrosio:2023} for more details on this important open question.

Before concluding this subsection, we emphasize that TEGR has a special place among the theories described by the action~\eqref{eq:ActionTG}. In fact, one has to ask what distinguishes  the particular choice of parameters which turns TG into TEGR from all other possible choices. The answer: Enhanced symmetries. Perturbation theory around Minkowski space shows~\cite{Koivisto:2018, BeltranJimenez:2019c} that a self-consistent theory requires $2 c_1 + c_2 + c_3 = 0$. If this condition is satisfied, one is left with a $1$-parameter family of theories (up to an overall normalization) which propagate one additional degrees of freedom besides the graviton. Among this $1$-parameter family of theories, the one which satisfies $2c_1 - c_2$ enjoys an additional symmetry and it looses the additional degree of freedom. One is then left with TEGR. Removing either one of these parameter conditions leads to a loss of symmetry accompanied by an increase in degrees of freedom, not all of which are healthy.\newline

\subsection{Quadratic Actions for Non-Metricity Theories}\label{ssec:QuadraticNonMetricity}
In subsection~\ref{ssec:STEGR} we constructed STEGR's action functional from the non-metricity scalar
\begin{equation}
	\Q = \frac14 Q_{\alpha\mu\nu}Q^{\alpha\mu\nu} - \frac12 Q_{\alpha\mu\nu}Q^{\mu\alpha\nu} - \frac14 Q_\alpha Q^\alpha + \frac12 Q_\alpha \bar{Q}^\alpha\,.
\end{equation} 
Now we want to define a new class of theories, which we subsume under the umbrella term \textbf{Symmetric Teleparallel Gravity (STG)}, by using the most general scalar which is quadratic in the non-metricity tensor. To that end, we need to consider all possible independent contractions of $Q_{\alpha\mu\nu}$ with itself. There are six contractions one can build this way, since we have six possible index permutations. However, because $Q_{\alpha\mu\nu}$ is symmetric in its last two indices, this cuts down the number to three. Using again the symmetry of $Q_{\alpha\mu\nu}$, one can then show that only the contractions
\begin{align}
	&Q_{\alpha\mu\nu}Q^{\alpha\mu\nu} && \text{and} & &Q_{\mu\alpha\nu}Q^{\alpha\mu\nu}
\end{align}
are independent. Next, we consider the traces of the non-metricity tensor. Because of its symmetry, there are two such traces:
\begin{align}
	Q_\alpha &\ce Q\du{\alpha\lambda}{\lambda} & & \text{and}& \bar{Q}_\alpha &\ce Q\ud{\lambda}{\lambda\alpha}\, .
\end{align}
Using these traces, we can build three more contractions which are quadratic in the non-metricity tensor:
\begin{align}
	& Q_\mu Q^\mu\, , && & \bar{Q}_\mu \bar{Q}^\mu\, , && \text{and} && & Q_\mu \bar{Q}^\mu\,.
\end{align}
With this we have exhausted all possibilities and we finally conclude that the most general scalar which is quadratic in the non-metricity tensor is
\begin{align}
	\hatQ \ce c_1\, Q_{\alpha\mu\nu}Q^{\alpha\mu\nu} + c_2\, Q_{\mu\alpha\nu}Q^{\alpha\mu\nu} + c_3\, Q_\mu Q^\mu + c_4\, \bar{Q}_\mu \bar{Q}^\mu + c_5\, Q_\mu \bar{Q}^\mu\, ,
\end{align}
where $c_1$, $c_2$, $c_3$, $c_4$, and $c_5$ are arbitrary, real constants. Just as in the case of TG, the action of STG is obtained by replacing $\Q$ with $\hatQ$ in the action of STEGR. This results in the functional\footnote{The theory described by this action is sometimes referred to as Newer General Relativity. See for instance~\cite{BeltranJimenez:2017b}.}
\begin{equation}\label{eq:ActionSTG}
	\S_\textsf{STG}[g, \Gamma, \Psi] \ce -\int_\M\dd^4 x\, \left(\frac{1}{\kappa}\sqrt{|g|}\,\hatQ + \tilde{\Pi}\du{\alpha}{\mu\nu\rho}\, R\ud{\alpha}{\mu\nu\rho} + \tilde{\chi}\du{\alpha}{\mu\nu}\, T\ud{\alpha}{\mu\nu}\right) + \mathcal{S}_\textsf{matter}[g, \Psi]\,,
\end{equation}
where eventual matter fields $\Psi$ are assumed to be minimally coupled and having a vanishing hypermomentum~$\mathcal{H}\du{\alpha}{\mu\nu}$. Not only the action looks deceptively similar to the one of STEGR, also the field equations have virtually the same form:
\begin{align}\label{eq:STGFEQ}
	&\textsf{Metric field equations:} & \frac{2}{\sqrt{|g|}}\nabla_\alpha\left[\sqrt{|g|}\hat{P}\ud{\alpha}{\mu\nu}\right] + \hat{q}_{\mu\nu} -\frac12 \hatQ\, g_{\mu\nu} &= \kappa\, \T_{\mu\nu} \notag\\
	&\textsf{Connection field equations:} & \nabla_\mu\nabla_\nu \left(\sqrt{|g|} \hat{P}\ud{\mu\nu}{\alpha}\right) &= 0\,.
\end{align} 
The hatted non-metricity conjugate $\hat{P}\ud{\alpha}{\mu\nu}$ and the symmetric tensor $\hat{q}_{\mu\nu}$ are defined as
\begin{align}
	\hat{P}\ud{\alpha}{\mu\nu} &\ce \frac12 \PD{\hatQ}{Q\du{\alpha}{\mu\nu}} = c_1\, Q\ud{\alpha}{\mu\nu} + c_2\, Q\dud{(\mu}{\alpha}{\nu)} + c_3\, g_{\mu\nu} Q^\alpha + c_4\, \delta\ud{\alpha}{(\mu}\bar{Q}_{\nu)} + \frac12 c_5\,\left(g_{\mu\nu}\bar{Q}^\alpha + \delta\ud{\alpha}{(\mu} Q_{\nu)} \right)    \notag\\
	\hat{q}_{\mu\nu} &\ce \PD{\hatQ}{g^{\mu\nu}} = P_{(\mu|\lambda\kappa}Q\du{\nu)}{\lambda\kappa} - 2P\ud{\lambda\kappa}{(\mu}Q_{\lambda\kappa|\nu)}
\end{align}
and it is still true that
\begin{align}
	\hatQ = \hat{P}_{\alpha\mu\nu} Q^{\alpha\mu\nu}\,.
\end{align}
Moreover, the Bianchi identities, which derive from the diffeomorphism invariance of the action~\eqref{eq:ActionSTG}, read
\begin{align}
	\D_\nu \hatM\ud{\nu}{\mu} + \hat{\C}_\mu \equiv 0\,,
\end{align}
where $\hatM_{\mu\nu}$ and $\hat{\C}_\alpha$ represent the left hand side of the field equations~\eqref{eq:STGFEQ}, i.e.,
\begin{align}
	\hatM_{\mu\nu} &\ce \frac{2}{\sqrt{|g|}}\nabla_\alpha\left[\sqrt{|g|}\hat{P}\ud{\alpha}{\mu\nu}\right] + \hat{q}_{\mu\nu} -\frac12 \hatQ\, g_{\mu\nu}\notag\\
	\hat{\C}_\alpha &\ce \nabla_\mu\nabla_\nu \left(\sqrt{|g|} \hat{P}\ud{\mu\nu}{\alpha}\right)\,.
\end{align} 
Thus, it follows that when the metric field equations are satisfied, the connection field equations are identically satisfied as a consequence of the Bianchi identities:
\begin{align}
	\hatM_{\mu\nu} = \kappa\,\T_{\mu\nu}\, \text{ satisfied}\quad\Longrightarrow\quad \D_\nu\hatM\ud{\nu}{\mu} = 0\quad\Longrightarrow\quad \hat{\C}_\mu \equiv 0\,.
\end{align}
What distinguishes STG from STEGR is the number of physical degrees of freedom. As we know, STEGR propagates the same two degrees of freedom as GR. When it comes to STG, the number of degrees of freedom depends on how one chooses the parameters $c_1$, $c_2$, $c_3$, $c_4$, and $c_5$. The reason is the same as in the case of TG: It is possible to tune the parameters such that certain second order time-derivatives of the metric drop out from the field equations, thus turning some of the equations into constraints. The more independent constraints there are, the lower the number of degrees of freedom. Conversely, it is also possible that equations which appear as constraints in STEGR are turned into dynamical equations because the parameters are no longer finely tuned to lead to certain cancellations. This has the effect of increasing the number of degrees of freedom and it can also lead to pathologies.

In~\cite{DAmbrosio:2020}, the first steps of a Hamiltonian analysis were carried out. After performing an ADM decomposition of the metric and after applying the coincident gauge, the momenta conjugate to lapse, shift, and intrinsic metric were studied. The full expressions, which can be found in~\cite{DAmbrosio:2020} are quite long. However, if we only consider the kinetic part of the Lagrangian of STG, which reads
\begin{align}
    \mathcal{L}_{\rm kinetic} &= -\sqrt{h}\left(\frac{2\tilde{c}}{N^3}\dot{N}^2 + \frac{c_{35}}{N^2}\, \dot{N}\, h^{ab} \dot{h}_{ab} - \frac{\hat{c}}{2N^3}h_{ab} \dot{N}^{a} \dot{N}^{b} + \frac{1}{2N}\left\{c_1 h^{ac} h^{db} \dot{h}_{ad}\dot{h}_{cb} + c_3 h^{ac} h^{bd}\dot{h}_{ac}\dot{h}_{bd}\right\}\right)\notag\\
    \text{with } & \tilde{c} \ce c_1 + c_2 + c_3 + c_4 + c_5,\qquad  \hat{c} \ce 2c_1 + c_2 + c_4, \qquad\text{and}\qquad c_{35}\ce 2c_3 + c_5\,,
\end{align}
we can already gain important insights. In fact, the momenta conjugate to lapse, shift, and intrinsic metric have the form
\begin{align}
    \tilde{\pi} &= -\frac{\sqrt{h}}{N^2}\left(4\,\tilde{c}\, \dot{N} + c_{35}\,N\, h^{ab}\dot{h}_{ab}\right) + \text{terms without time derivatives}\notag\\
    \tilde{\pi}_a &= \frac{\sqrt{h}}{N^3}\,\hat{c}\, h_{ab} \dot{N}^b + \text{terms without time derivatives}\notag\\
    \tilde{\pi}^{ab} &= -\frac{\sqrt{h}}{N^2}\left(c_1 h^{ac} h^{bd}\dot{h}_{cd} N + h^{ab}\left\{c_2h^{cd}\dot{h}_{cd} N + c_{35}\dot{N}\right\}\,\right) + \text{terms without time derivatives}\,.
\end{align}
Evidently, the momentum conjugate to lapse is turned into a so-called primary constraint if the parameters are chosen such that $\tilde{c} = $ and $c_{35} = 0$. Similarly, the momentum conjugate to shift becomes a constraint when $\hat{c} = 0$. These choices correspond precisely to the so-called primary sectors I and II shown in Table~\ref{tab:PrimarySectors}. More constraints can be identified through a systematic analysis based on the kinetic matrix, which is composed of the following submatrices:
\begin{align}
    \frac{\delta^2 \mathcal{L}}{\delta \dot{N}\delta\dot{N}} &= - 4\frac{\sqrt{h}}{N^3}\tilde{c} &  \frac{\delta^2 \mathcal{L}}{\delta \dot{N}^{a}\delta\dot{N}^{b}} &= \frac{\sqrt{h}}{N^3} \hat{c}\, h_{ab} \notag\\
    \frac{\delta^2 \mathcal{L}}{\delta \dot{N}^{a}\delta\dot{N}} &= 0 & \frac{\delta^2 \mathcal{L}}{\delta \dot{h}_{bc}\delta\dot{N}^{a}} &= 0\notag\\
    \frac{\delta^2 \mathcal{L}}{\delta \dot{h}_{ab}\delta\dot{N}} &= -\frac{\sqrt{h}}{N^2}c_{35} h^{ab} & \frac{\delta^2 \mathcal{L}}{\delta \dot{h}_{cd}\delta\dot{h}_{ab}} &= -\frac{\sqrt{h}}{2N}\left(c_1\,h^{ad}h^{bc} + c_1 h^{ac}h^{bd} + 2 c_3 h^{ab}h^{cd} \right)\,.
\end{align}
It is found that the determinant of the kinetic matrix $\mathcal{K}$ is given by
\begin{align}
    \det{\mathcal{K}} = 8\frac{h^2}{N^{18}}\,c^5_1\, \hat{c}^3\left(3c^2_{35} - 4\left(c_1+3c_3\right)\tilde{c}\right)\,.
\end{align}
By demanding that the determinant vanishes, i.e., demanding that the matrix is degenerate, one finds additional primary sectors. In fact, one finds that there are four independent solutions to the above equations. These solutions are
\begin{align}
    &\text{Sector I:} &  \tilde{c} &= 0 \text{ and } c_{35} = 0\notag\\
    &\text{Sector II:} & \hat{c} &= 0\notag\\
    &\text{Sector III:} & c_1 &= 0\notag\\
    &\text{Sector IV:} & c_3 &= -\frac{c_1}{3} + \frac{c^2_{35}}{4\tilde{c}}
\end{align}
To determine the number of constraints in each sector, we only need to compute $10-\text{rank}(\mathcal{K})$ in each sector. For the first four, we find $1$, $3$, $5$, and again $1$ primary constraints, respectively.

Even more sectors can be identified by combining the different parameter conditions in the different sectors, so as to create new and independent sectors with more constraints. This process is described in detail in~\cite{DAmbrosio:2020} and ultimately leads to Table~\ref{tab:PrimarySectors}.
\begin{table}[hbt]
\begin{tabular}{|p{0.15\textwidth}|ccccp{0.125\textwidth}<{\centering}|p{0.15\textwidth}<{\centering}|}
\hline
\textbf{Primary}   & \multicolumn{5}{c|}{\textbf{Parameter combinations}}                                                                                                                                                       & \multicolumn{1}{l|}{\textbf{\# of primary}} \\
\textbf{sector}    & \multicolumn{1}{p{0.125\textwidth}|}{\centering$\mathbf{\tilde{c}}$} & \multicolumn{1}{p{0.125\textwidth}|}{\centering$\mathbf{\hat{c}}$} & \multicolumn{1}{p{0.125\textwidth}|}{\centering$\mathbf{c_{35}}$} & \multicolumn{1}{p{0.125\textwidth}|}{\centering$\mathbf{c_1}$} & $\mathbf{\frac{c_1}{3} + c_3 -\frac{c^2_{35}}{\tilde{c}}}$ & \multicolumn{1}{l|}{\textbf{constraints}}   \\ \hline\hline
$\mathbf{0}$       & \multicolumn{1}{c|}{$\neq 0$}             & \multicolumn{1}{c|}{$\neq 0$}           & \multicolumn{1}{c|}{$\neq 0$}          & \multicolumn{1}{c|}{$\neq 0$}       & $\neq 0$                              & $0$                                         \\ \hline
\textbf{I}         & \multicolumn{1}{c|}{$\red{=0}$}            & \multicolumn{1}{c|}{$\neq 0$}           & \multicolumn{1}{c|}{$\red{=0}$}         & \multicolumn{1}{c|}{$\neq 0$}       & N/A                         & $1$                                         \\ \hline
\textbf{II}        & \multicolumn{1}{c|}{$\neq 0$}             & \multicolumn{1}{c|}{$\red{=0}$}          & \multicolumn{1}{c|}{$\neq 0$}          & \multicolumn{1}{c|}{$\neq 0$}       & $\neq 0$                              & $3$                                         \\ \hline
\textbf{III}       & \multicolumn{1}{c|}{$\neq 0$}             & \multicolumn{1}{c|}{$\neq 0$}           & \multicolumn{1}{c|}{$\neq 0$}          & \multicolumn{1}{c|}{$\red{=0}$}      & $\neq 0$                              & $5$                                         \\ \hline
\textbf{IV}        & \multicolumn{1}{c|}{$\neq 0$}             & \multicolumn{1}{c|}{$\neq 0$}           & \multicolumn{1}{c|}{$\neq 0$}          & \multicolumn{1}{c|}{$\neq 0$}      & $\red{=0}$                             & $1$                                         \\ \hline
\textbf{V (STEGR)} & \multicolumn{1}{c|}{$\red{=0}$}            & \multicolumn{1}{c|}{$\red{=0}$}          & \multicolumn{1}{c|}{$\red{=0}$}         & \multicolumn{1}{c|}{$\neq 0$}       & N/A                         & $4$                                         \\ \hline
\textbf{VI}        & \multicolumn{1}{c|}{$\red{=0}$}             & \multicolumn{1}{c|}{$\neq 0$}          & \multicolumn{1}{c|}{$\red{=0}$}          & \multicolumn{1}{c|}{$\red{=0}$}      &  N/A                             & $6$                                         \\ \hline
\textbf{VII}       & \multicolumn{1}{c|}{$\neq 0$}             & \multicolumn{1}{c|}{$\red{=0}$}          & \multicolumn{1}{c|}{$\neq 0$}          & \multicolumn{1}{c|}{$\red{=0}$}      & $\neq 0$                             & $8$                                         \\ \hline
\textbf{VIII}      & \multicolumn{1}{c|}{$\neq  0$}            & \multicolumn{1}{c|}{$\red{=0}$}          & \multicolumn{1}{c|}{$\neq 0$}         & \multicolumn{1}{c|}{$\neq 0$}      & $\red{=0}$                         & $4$                                        \\ \hline
\textbf{IX}      & \multicolumn{1}{c|}{$\neq  0$}            & \multicolumn{1}{c|}{$\neq 0$}          & \multicolumn{1}{c|}{$\neq 0$}         & \multicolumn{1}{c|}{$\red{= 0}$}      & $\red{=0}$                         & $6$                                        \\ \hline
\textbf{X}      & \multicolumn{1}{c|}{$\red{=0}$}            & \multicolumn{1}{c|}{$\red{=0}$}          & \multicolumn{1}{c|}{$\red{=0}$}         & \multicolumn{1}{c|}{$\red{=0}$}      & N/A                         & $10$                                        \\ \hline
\textbf{XI}      & \multicolumn{1}{c|}{$\neq 0$}            & \multicolumn{1}{c|}{$\red{=0}$}          & \multicolumn{1}{c|}{$\neq 0$}         & \multicolumn{1}{c|}{$\red{=0}$}      & $\red{=0}$                         & $9$                                        \\ \hline
\end{tabular}
\caption{\protect Each primary sector (first column) is defined by the vanishing of certain combinations of the parameters $c_1$, $c_2$, $c_3$, $c_4$, and $c_5$ (columns two through six). For brevity, and following~\cite{DAmbrosio:2020}, we have defined $\tilde{c}\ce c_1+c_2+c_3+c_4+c_5$, $\hat{c} \ce 2c_1 + c_2 + c_4$, and $c_{35} \ce 2c_3 + c_5$. The vanishing of these parameter combinations (or the vanishing of combinations thereof) correspond to the appearance of one or more primary constraints (the number of independent primary constraints is shown in the last column). STEGR is contained in sector V.\hspace*{\fill}}
\label{tab:PrimarySectors}
\end{table}
Notice that in sector V, which harbours STEGR as a special case, the number of primary constraints matches the one of GR. However, just as in TG, it is currently unknown in which sector secondary constraints occur and how many there are. Hence, the exact number of degrees of freedom is not known for most sectors. At most, we can currently say that sector $0$ propagates ten degrees of freedom, but it is also a highly pathological theory. Sector X propagates no degrees of freedom, while sector XI has less degrees of freedom than GR. Both sectors are therefore uninteresting. Finally, sector V contains STEGR, which has two degrees of freedom, but it is unclear whether other theories with a different number of degrees of freedom can co-habitate that sector.

The reason these questions have remained unanswered thus far is because of challenges posed by the Dirac-Bergmann algorithm, as mentioned in the previous subsection. These challenges seem to afflict all teleparallel theories of gravity, as has recently been argued in~\cite{DAmbrosio:2023}, and the development of new methods---or at least the exploration of other known methods but applied to teleparallel theories---seems to be necessary.

Finally, we remark that STEGR distinguishes itself from the other possible theories within the five-parameter family by having an enhanced set of symmetries. In~\cite{BeltranJimenez:2019c}, perturbations around Minkowski space were studied. The perturbative ansatz $g_{\mu\nu} = \eta_{\mu\nu} + h_{\mu\nu}$, with $|h_{\mu\nu}|\ll 1$, leads to the quadratic Lagrangian
\begin{align}
	L = c_1 \partial_\alpha h_{\mu\nu} \partial^\alpha h^{\mu\nu} + \left(c_2+c_4\right)\partial_\alpha h_{\mu\nu}\partial^\mu h^{\alpha\nu} + c_3 \partial_\alpha h \partial^\alpha h + c_5 \partial_\mu h\ud{\mu}{\nu}\partial^\nu h\,, 
\end{align}
where $h\ce \eta^{\mu\nu} h_{\mu\nu} = h\ud{\mu}{\mu}$ is the trace of the perturbations. This is nothing but the most general Lagrangian for a spin-$2$ field. As is well known from the Fierz-Pauli analysis of this Lagrangian, symmetries have to be imposed in order to remove ghostly degrees of freedom. Demanding that the theory is invariant under
\begin{align}\label{eq:TDiff}
	h_{\mu\nu}\qquad \mapsto\qquad h_{\mu\nu} + 2\partial_{(\mu}\xi_{\nu)}\qquad \text{ for some vector $\xi^\mu$ which satisfies } \partial_\mu \xi^\mu = 0\,,
\end{align} 
so-called \textbf{transversal diffeomorphisms}, leads to the condition
\begin{align}
	2c_1 + c_2 + c_4 = 0\,,
\end{align}
which is indeed satisfied by the STEGR parameters\footnote{This is simply the condition $\hat{c} = 0$, which defines Sector II and which is also a part of Sector V.}. In order to recover the two propagating degrees of freedom of a massless spin-$2$ field, one can further impose \textbf{linearized diffeomorphisms},
\begin{align}
	h_{\mu\nu}\qquad \mapsto\qquad h_{\mu\nu} + 2\partial_{(\mu}\xi_{\nu)}\,,
\end{align}
where the vector field $\xi^\mu$ is now unrestricted. This leads to
\begin{align}
	2 c_1 = -2c_3 = c_5\,.
\end{align}
Both conditions taken together then imply
\begin{align}
	c_3 &= -c_1, &  c_4 &= -2c_1 -c_2, &  c_5 = 2 c_1\,,
\end{align}
which is equivalent to
\begin{align}
	\tilde{c} &= 0, & \hat{c} &= 0, & c_{35} &= 0\,.
\end{align}
These are precisely the defining equations of sector V in Table~\ref{tab:PrimarySectors}. STEGR, which inhabits sector V, is therefore distinguish through its symmetries and the healthy degrees of freedom it propagates. Instead of imposing linearized diffeomorphisms, one could have also imposed the \textbf{linearized Weyl symmetry}
\begin{align}
	h_{\mu\nu} \qquad\mapsto\qquad h_{\mu\nu} + \phi\, \eta_{\mu\nu}\,,
\end{align}
for some arbitrary scalar field and in addition to the transverse diffeomorphisms~\eqref{eq:TDiff}. Demanding this symmetry implies
\begin{align}
	c_3 &= -\frac{3}{8}c_1, & c_5 &= 2c_1\,.
\end{align}
This describes a linearized version of unimodular gravity, which is essentially GR plus the constraint $\sqrt{|g|}=1$. As a consequence, in unimodular gravity the cosmological constant emerges from an integration constant~\cite{Alvarez:2023}. Notice that Sector V does not respect the linearized Weyl symmetry. This symmetry only seems to be respected by Sector IX, which has nothing to do with STEGR or GR. However, it should be pointed out that the classification was obtained \textit{without} restricting the metric through the condition~$\sqrt{|g|} = 1$. \newline

\subsection{Non-Linear Extensions: \texorpdfstring{$f(\R)$, $f(\bbT)$, $f(\Q)$}, and \texorpdfstring{$f(\bbG)$ T} Theories}\label{ssec:NonLinearExtensions}
As we discussed in section~\ref{sec:Trinity}, one can set up a geometric trinity to describe gravity. Einstein's original formulation based on non-vanishing curvature is equivalent to TEGR, which is based on non-vanishing torsion, and both theories are in turn equivalent to STEGR, which is built on a non-vanishing non-metricity tensor. The General Teleparallel Equivalent of GR unifies the torsion and non-metricity description of gravity and is also equivalent to GR.

These four formulations are equivalent in the sense that they posses the same field equations, propagate the same degrees of freedom, and therefore ultimately possess the same solution space. Each formulation of the trinity can be derived from an action principle. We recall that these actions are given by
\begin{align}
	\S_{\rm EH}[g] &= \phantom{-}\frac{1}{2\kappa}\int_{\M}\sqrt{|g|}\, \R \, \dd^4 x \notag\\
	\S_{\rm TEGR}[\Lambda] &= - \frac{1}{2\kappa}\int_{\M}\sqrt{|g|}\,\bbT\, \dd^4 x\notag\\
	\S_{\rm STEGR}[g, \xi] &= - \frac{1}{2\kappa}\int_{\M}\sqrt{|g|}\,\Q\, \dd^4 x\,.\notag\\
	\S_{\rm GTEGR}[g, \Lambda] &= -\frac{1}{2\kappa} \int_{\M}\sqrt{|g|}\, \bbG\, \dd^4 x\,.
\end{align}
The actions are \textit{equivalent}, in the sense spelled out above, but they are \textit{not equal}. In fact, they depend on different fields and they all differ by boundary terms. This opens the door for yet another generalization of the geometrical trinity of gravity. Namely, we can replace the scalars $\R$, $\bbT$, $\Q$, and $\bbG$ by arbitrary functions and obtain the following action functionals:
\begin{align}
	\S_{f(\R)}[g] &\ce \phantom{-}\frac{1}{2\kappa}\int_{\M}\sqrt{|g|}\, f(\R) \, \dd^4 x \notag\\
	\S_{f(\bbT)}[\Lambda] &\ce - \frac{1}{2\kappa}\int_{\M}\sqrt{|g|}\,f(\bbT)\, \dd^4 x\notag\\
	\S_{f(\Q)}[g, \xi] &\ce - \frac{1}{2\kappa}\int_{\M}\sqrt{|g|}\,f(\Q)\, \dd^4 x\,\notag\\
	\S_{f(\bbG)}[g, \Lambda] &\ce -\frac{1}{2\kappa} \int_{\M}\sqrt{|g|}\, f(\bbG)\, \dd^4 x
\end{align}
The motivation for this non-linear extension is that the added freedom in choosing a function $f$ may help in explaining the accelerated expansion of the universe, structure formation, and other phenomena which in the trinity of GR require the introduction of dark energy and dark matter.

Indeed, given that the original functionals differed by boundary terms, one has to conclude that the resulting non-linear extensions are \textit{no longer equivalent to each other}! In particular this means that each one of the above functionals gives rise to its own peculiar field equations with its own number of propagating degrees of freedom.

Probably the most studied and best understood theory among these three is $f(\R)$ gravity, since it was first proposed by Buchdahl in 1970~\cite{Buchdahl:1970}. Given the extensive literature and the fact that our focus is on~$f(\Q)$ gravity, we shall just discuss some basic aspects of $f(\R)$ gravity and refer the reader to the extensive review articles~\cite{Sotiriou:2008, Capozziello:2011} and references therein.\newline

\paragraph{$\boldsymbol{f(\R)}$ \textbf{Gravity}} 
Following the same route that led to Einstein's field equations, it is straightforward to deduce the equations of $f(\R)$ gravity. They are
\begin{align}
	f'(\R)\, \R_{\mu\nu} - \frac12 f(\R) \, g_{\mu\nu} + \left(g_{\mu\nu}\Box - \D_\mu \D_\nu\right)f'(\R) = \kappa\, \T_{\mu\nu}\,,
\end{align}
where we have defined $f'(\R) \ce \TD{f(\R)}{\R}$ and $\Box \ce g^{\mu\nu}\D_\mu \D_\nu$. If we choose $f(\R) = \R$, the equations reproduce to Einstein's field equations, as they should. In order to avoid this trivial case, we shall now assume $f''(\R) \neq 0$. Then one sees that the above field equations are actually fourth order non-linear equations for the metric, due to the second order differential operator $g_{\mu\nu}\Box - \D_\mu \D_\nu$ acting on $f'(\R)$ (which itself already contains second order derivatives of the metric). 

What may seem alarming at first sight is actually not that troublesome. One can show~\cite{Sotiriou:2008, Capozziello:2011} that the theory propagates three healthy degrees of freedom: Two degrees of freedom corresponding to a massless graviton and one scalar degree of freedom.\newline

\paragraph{$\boldsymbol{f(\bbT)}$ \textbf{Gravity}} 
Starting from the $f(\bbT)$ action coupled to matter fields $\Psi$,
\begin{align}\label{eq:f(T)Action}
	\S_{f(\bbT)}[g, \Gamma] \ce -\frac{1}{2\kappa}\int_{\M} \dd^4 x\, \sqrt{|g|}\, f(\bbT) + \S_{\rm matter}[g, \Psi]\,,
\end{align}
one finds a set of metric and connection field equations
\begin{align}
	\left(\nabla_\alpha + T_\alpha\right)\left[f'(\bbT)\, S\du{(\mu\nu)}{\alpha}\right] + f'(\bbT)\, t_{\mu\nu} - \frac12 f(\bbT) g_{\mu\nu} &= \kappa\, \T_{\mu\nu}\notag\\
	\left(\nabla_\mu + T_\mu\right) \left[f'(\bbT) \, S\dud{[\alpha}{\mu}{\beta]}\right] &= 0\,.
\end{align}
It should be noted that in contrast to the $f(\R)$ field equations, the metric field equations of $f(\bbT)$ gravity are second order. Furthermore, in the case $f(\bbT) = \bbT$ the field equations reduce to the equations of TEGR, as had to be expected. In practice it is often helpful to re-write the metric field equations in the form
\begin{align}\label{eq:f(T)FEQ}
	f'(\bbT)\, G_{\mu\nu} - \frac12 \left(f(\bbT) -f'(\bbT)\, \bbT\right) + f''(\bbT)S\du{(\mu\nu)}{\alpha}\partial_\alpha \bbT &= \kappa \,\T_{\mu\nu}\,.
\end{align}
In this form it is evident that the case $f''(\bbT) = 0$ with $f'(\bbT) = 1$, which is equivalent to $f(\bbT) = \bbT + $const, simply reproduces Einstein's equations with a cosmological constant $\Lambda = - \frac{\text{const.}}{2}$. This form of the equations also highlights that the dynamics will be modified whenever $f''(\bbT) \neq 0$. However, despite some efforts, it has so far not been possible to determine the precise number of degrees of freedom propagated by the theory. The number ranges between three~\cite{Ferraro:2018} and five~\cite{Li:2011, Blagojevic:2020}.\newline

\paragraph{$\boldsymbol{f(\Q)}$ \textbf{Gravity}}
The $f(\Q)$ action, which includes minimally coupled matter fields $\Psi$, reads
\begin{align}\label{eq:f(Q)Action}
	\S_{f(\Q)}[g, \Gamma] \ce -\frac{1}{2\kappa}\int_{\M} \dd^4 x\, \sqrt{|g|}\, f(\Q) + \S_{\rm matter}[g, \Psi]
\end{align}
and it gives rise to the following metric and connection field equations:
\begin{align}
	\frac{2}{\sqrt{|g|}}\nabla_\alpha\left[\sqrt{|g|}\,f'(\Q)\, P\ud{\alpha}{\mu\nu}\right] + f'(\Q)\, q_{\mu\nu} - \frac12 f(\Q)\, g_{\mu\nu} &= \kappa\, \T_{\mu\nu} \notag\\
	\nabla_\mu\nabla_\nu\left(\sqrt{|g|}\,f'(\Q)\,P\ud{\mu\nu}{\alpha}\right) &= 0
\end{align}
These field equations are structurally very similar to the field equations of STEGR. However, it is possible to re-write the metric field equations as~\cite{DAmbrosio:2021, DAmbrosio:2021b}
\begin{align}\label{eq:f(Q)FEQ}
	f'(\Q)\, G_{\mu\nu} - \frac12\left(f(\Q) - f'(\Q)\, \Q\right)g_{\mu\nu} + 2 f''(\Q) P\ud{\alpha}{\mu\nu}\partial_\alpha \Q &= \kappa\, \T_{\mu\nu}\,.
\end{align}
In this form, it is evident that $f''(\Q) = 0$ with $f'(\Q) = \text{const}.$ reproduces the Einstein field equations with a cosmological constant. It is also clear that the dynamics will be considerably modified by the last term on the left hand side. In fact, we will see in subsection~\ref{ssec:Hamiltonian} that this term has an effect on the counting of primary constraints and thus also impacts the number of physical degrees of freedom. It is however important to emphasize that the question how many degrees of freedom $f(\Q)$ propagates has not yet been answered satisfactorily. The current state will be discussed in more details in subsection~\ref{ssec:Hamiltonian}.

Luckily, not knowing the number of physical degrees of freedom does not constitute an obstacle when it comes to applying the theory to cosmology or black holes physics. In this context, or more generally whenever we want to study specific spacetimes, it can be useful to notice that $f(\Q)$ gravity can contain the GR solutions as special cases. In fact, if we impose the condition
\begin{align}
    \Q = \Q_0 = \text{const.}
\end{align}
and if we assume that we can actually satisfy this condition, then it follows that the metric field equations take the form
\begin{align}
    G_{\mu\nu} - \frac12 \frac{f(\Q_0)-f'(\Q_0) \Q_0}{f'(\Q_0)}\, g_{\mu\nu} = \frac{\kappa}{f'(\Q_0)}\, \T_{\mu\nu}\,.
\end{align}
Formally, this can be read as Einstein's field equations with an effective cosmological constant and a re-scaled energy-momentum tensor
\begin{align}
    \Lambda_{\rm eff} &\ce -\frac12 \frac{f(\Q_0)-f'(\Q_0) \Q_0}{f'(\Q_0)}\notag\\
    \hat{\T}_{\mu\nu} &\ce \frac{1}{f'(\Q_0)} \T_{\mu\nu}\,.
\end{align}
Thus, it is possible to recover certain GR solutions in $f(\Q)$ gravity, even when $f''(\Q)\neq 0$, i.e., even when we are \textit{not} in the GR sector of the theory.

For applications and formal considerations it can also be useful to know the Bianchi identities of $f(\Q)$ gravity. Given that the theory is generally covariant, it is possible to find such Bianchi identities by following the same reasoning as in GR (or TEGR and STEGR). One finds the identity
\begin{align}\label{eq:f(Q)Bianchi}
	\D_\mu \M\ud{\mu}{\nu} + \C_\nu &\equiv 0\,,
\end{align}
where we have defined
\begin{align}\label{eq:f(Q)MC}
	\M_{\mu\nu} &\ce \frac{2}{\sqrt{|g|}}\nabla_\alpha\left[\sqrt{|g|}\,f'(\Q)\, P\ud{\alpha}{\mu\nu}\right] + f'(\Q)\, q_{\mu\nu} - \frac12 f(\Q)\, g_{\mu\nu}\notag\\
	\C_\alpha &\ce \nabla_\mu\nabla_\nu\left(\sqrt{|g|}\,f'(\Q)\,P\ud{\mu\nu}{\alpha}\right)\,.
\end{align}
We emphasize that in contrast to STEGR, $\M_{\mu\nu}$ does \textit{not} satisfy the identity $\D_\mu \M\ud{\mu}{\nu} = 0$ and thus the connection field equations are \textit{not} just trivial identities. Quite on the contrary, the connection field equations are now dynamical equations for the connection, which can have physical degrees of freedom.

What one can conclude, however, is that when the metric field equations are satisfied, i.e., when $\M_{\mu\nu} = \kappa\,\T_{\mu\nu}$ holds, then the connection field equations are also satisfied, due to the Bianchi identities. In fact, we easily find
\begin{align}
    \D_\mu \M\ud{\mu}{\nu} + \C_\nu = \kappa\, \underbrace{\D_\mu \T\ud{\mu}{\nu}}_{=0} + \C_\nu = \C_\nu \equiv 0\,.
\end{align}
This fact can, for instance, be used to simplify the Hamiltonian analysis of the theory~\cite{DAmbrosio:2023}. \newline

\paragraph{\textbf{$\boldsymbol{f(\bbG)}$ Gravity}}
As discussed in subsection~\ref{ssec:GTEGR}, the General Teleparallel Equivalent of GR encompasses TEGR and STEGR at the same time. That is, TEGR and STEGR emerge from this more general theory as partially gauge-fixed theories. It is therefore no surprise that one can also consider the non-linear extension $\bbG \mapsto f(\bbG)$ and that this modification has some relations to $f(\bbT)$ and $f(\Q)$ gravity. Following~\cite{Hohmann:2022, Heisenberg:2022}, it is convenient to first introduce the auxiliary tensors
\begin{align}
    M\ud{\alpha}{\mu\nu} &\ce \Gamma\ud{\alpha}{\mu\nu} - \LC{\alpha}{\mu\nu} =  K\ud{\alpha}{\mu\nu} + L\ud{\alpha}{\mu\nu}\notag\\
    Z\du{\alpha}{\mu\nu} &= - M\du{\alpha}{\mu\nu} - M\udu{\nu}{\alpha}{\mu} + M\ud{\rho}{\alpha\rho}g^{\mu\nu} + M\ud{\mu\rho}{\rho} \delta\ud{\nu}{\alpha} \,.
\end{align}
With their help, one can express the field equations of $f(\bbG)$ in the relatively compact form
\begin{align}
	&f'(\bbG)\,G_{\mu\nu} - \frac12\left(f(\bbG)-f'(\bbG)\,\bbG\right)g_{\mu\nu} + \D_{(\mu}f'(\bbG)\, M\ud{\sigma}{\nu)\sigma} + f''(\bbG)\left(\,M\ud{[\rho\sigma]}{\sigma}\, g_{\mu\nu}-\,M\ud{\rho}{(\mu\nu)}\right)\partial_\rho \bbG  = \kappa\, \T_{\mu\nu}\notag\\
	&\nabla_\rho\left(f'(\bbG)\, Z\du{\mu}{\nu\rho}\right) - f'(\bbG)\, M\ud{\lambda}{\rho\lambda}Z\du{\mu}{\nu\rho} = 0\,.
\end{align}
It can be verified that by imposing either $Q_{\alpha\mu\nu} = 0$ or $T\ud{\alpha}{\mu\nu} = 0$, which we should read as partial gauge-fixing conditions for the connection, one recovers the $f(\bbT)$ and $f(\Q)$ field equations, respectively. Moreover, just as before, the metric field equations reveal that choosing $f''(\bbG) = 0$ with $f'(\bbG) \neq 0$ simply yields Einstein's field equations. Finally, if we impose the condition $\bbG = \bbG_0 = \text{const}$, we find that $f(\bbG)$ can contain some of the GR solutions, since then the field equations reduce to
\begin{align}
    G_{\mu\nu} -\frac12 \frac{f(\bbG_0) - f'(\bbG_0)\bbG_0}{f'(\bbG_0)} g_{\mu\nu} &= \frac{\kappa}{f'(\bbG_0)}\T_{\mu\nu}\,.
\end{align}
That is, we obtain Einstein's field equations with an effective cosmological constant and a rescaled energy-momentum tensor. All of this is unsurprising, since all these properties hold in $f(\bbT)$ and $f(\Q)$ gravity. However, since $f(\bbG)$ does not make use of a gauge-fixing condition such as $Q_{\alpha\mu\nu} = 0$ or $T\ud{\alpha}{\mu\nu} = 0$, it is possible that it leaves more freedom to find interesting beyond-GR solutions. A first attempt at finding cosmological solutions has been carried out in~\cite{Heisenberg:2022}. The fact that removing gauge-fixing conditions can have advantages has also been shown in~\cite{Gomes:2023}, where the so-called canonical frame has been scrutinized in the context of a general teleparallel cosmology.

\newpage
\asection{6}{\texorpdfstring{$f(\Q)$}{fQ} Gravity}\label{sec:f(Q)}
The non-metricity formulation of gravity, and in particular the non-linear extension $f(\Q)$, have witnessed a flurry of research activities over the past few years. Most of these activities concern applications to cosmology and black hole physics. This is natural, considering that one of the motivations for studying non-linear extensions is the possibility to explain phenomena which in standard GR require the introduction of dark energy, the inflaton field, and dark matter. In this section we give an overview over the most important results which have been obtained in applications of $f(\Q)$ gravity to cosmology and black holes. We also briefly touch upon the question of how many degrees of freedom the theory propagates.

\subsection{Cosmology in \texorpdfstring{$f(\Q)$}{fQ}}\label{ssec:Cosmology}
Given that the coincident gauge can always be used in symmetric teleparallel theories of gravity, the simplest thing to do when working on applications of the theory is to use this particular gauge plus a fixed background metric. This was precisely the ansatz in~\cite{BeltranJimenez:2019}, where $f(\Q)$ cosmology was studied for the first time. Specifically, the authors used the ansatz
\begin{align}
	\Gamma\ud{\alpha}{\mu\nu} &= 0 & \text{and} && g_{\mu\nu} &= \begin{pmatrix}
		-N(t)^2 & 0 & 0 & 0\\
		0 & a(t)^2 & 0 & 0\\
		0 & 0 & a(t)^2 & 0\\
		0 & 0 & 0 & a(t)^2
	\end{pmatrix}\,,
\end{align}
where $N(t)$ and $a(t)$ are the usual lapse function and scale factor of the FLRW spacetime. As it turns out, the non-metricity scalar for this ansatz is  simply given by
\begin{align}\label{eq:QinSimpleCosmology}
	\Q = 6\frac{H^2}{N^2}\,,
\end{align}
where $H\ce \frac{\dot{a}}{a}$ is the usual Hubble function. It is evident that the symmetry-reduced action
\begin{align}
	\S[N, a] \ce -\frac{1}{2\kappa} \int_{\M}\dd t\, \dd^3\vec{x}\, a(t)^3\, N(t)\, f(\Q)\,,
\end{align}
which is the $f(\Q)$ action evaluated on the FLRW metric and in coincident gauge, has a residual time-reparametrization invariance~\cite{BeltranJimenez:2017b,BeltranJimenez:2019}. By exploiting this reparametrization freedom, we can fix the lapse function to unity, $N(t) = 1$. The idea now is to study the resulting cosmological equations
\begin{align}
	6 f'\, H^2 -\frac12 f &= \rho\notag\\
	\left(12 H^2\, f'' + f'\right)\dot{H}  &= -\frac12 (\rho + p)\,,
\end{align}
where $\rho$ and $p$ denote the density and pressure, respectively, and where we defined $f' \ce \TD{f}{\Q}$ and $f''\ce \TD{^2 f}{\Q^2}$. As always, standard matter fields also satisfy the continuity equation
\begin{align}
	\dot{\rho} = -3 H\left(\rho+p\right)\,.
\end{align}
A particularly interesting class of theories emerges if we demand that $f$ satisfies the equation
\begin{align}\label{eq:ConditionSimpleCosmology}
	6 f'\, H^2 - \frac12 f = \frac{1}{2\kappa}\Q\,,
\end{align}
with $\kappa = 8\pi G$, since this gives the same background evolution as GR, but the evolution of perturbations is subjected to modifications. Using equation~\eqref{eq:QinSimpleCosmology}, we can rewrite the condition~\eqref{eq:ConditionSimpleCosmology} equivalently as
\begin{align}
	\Q\, f'(\Q) - \frac12 f(\Q) &= \frac{1}{2\kappa}\Q\,,
\end{align}
which is a simple first order differential equation for $f$, which is solved by
\begin{align}
	f(\Q) = \frac{1}{\kappa} \left(\Q + M\, \sqrt{\Q}\right)\,.
\end{align}
Here, $M$ is an integration constant and clearly the special case $M=0$ corresponds to STEGR, while $M\neq 0$ leads to a $1$-parameter family of modified theories. As mentioned before, the background evolution for this family of theories is the same as in GR. In order to discriminate between different values of $M$, it is necessary to study perturbations, which exhibit \textit{different} behaviour than in GR. 

Another interesting ansatz for studying $f(\Q)$ cosmology is a power-law modification of STEGR:
\begin{align}
	f(\Q) = \frac{1}{\kappa} \left[\Q - 6\lambda\, M^2 \left(\frac{\Q}{6M^2}\right)^\alpha\right]\,,
\end{align}
where $\lambda$ and $\alpha$ are dimensionless parameters. The modified Friedmann equation for this ansatz reads
\begin{align}
	H^2 \left[1 + (1 - 2\alpha)\lambda\left(\frac{H^2}{M^2}\right)^{\alpha-1} \right] &= \frac{\kappa}{3}\rho
\end{align}
The previous $f$ is contained as a special case for the choice $\alpha = \frac12$, while STEGR emerges from $\alpha = 1$. By inspecting the form of the modified Friedmann equation one can infer that for $\alpha < 1$ the corrections to the GR evolution become important at low curvature, while for $\alpha > 1$ corrections become relevant in the high curvature regime. In other words, theories with $\alpha > 1$ play a role in the early Universe and theories with $\alpha < 1$ provide us with corrections to late-time cosmology. This opens the possibility for modified inflationary scenarios or a description of the late-time Universe without dark energy. In fact, various $f(\Q)$ cosmology models have been studied and applied to questions pertaining to the late-time Universe~\cite{Lymperis:2022, Paul:2022, Narawade:2023, Narawade:2023b}, large scale structures~\cite{Sokoliuk:2023}, relativistic versions of MOND~\cite{Milgrom:2019, DAmbrosio:2020b}, bouncing cosmologies~\cite{Bajardi:2020, Agrawal:2021, Gadbail:2023}, and quantum cosmology~\cite{Dimakis:2021, Bajardi:2023}. A lot of effort has also gone into constraining or testing $f(\Q)$ models~\cite{Dialektopoulos:2019, Ayuso:2020, Barros:2020, Frusciante:2021, Aggarwal:2022, De:2022, Albuquerque:2022, Ferreira:2022, Koussour:2023, Najera:2023, Bouali:2023, Ferreira:2023, Subramaniam:2023b}.

The majority of the literature on $f(\Q)$ cosmology makes use of the coincident gauge. However, at this point we would like to recall the discussion on the $f(\Q)$ field equations from subsection~\ref{ssec:NonLinearExtensions}, which showed that the connection field equations are no longer trivial identities (this was the case in STEGR). From this it can be expected that $f(\Q)$ propagates more than the two degrees of freedom of GR. When working in the coincident gauge and using the FLRW metric as an ansatz, interesting cosmological models can emerge, but we become largely oblivious to the additional degrees of freedom because of these overly restrictive choices we made. There are two ways out of this. The first is perturbation theory around FLRW, while still using the coincident gauge. This avenue was explored in~\cite{BeltranJimenez:2019} and it led to the insight that $f(\Q)$ gravity propagates \textit{at least} two additional degrees of freedom.

The second option is to abandon the coincident gauge and instead work with a metric and connection which are both compatible with the cosmological principles of homogeneity and isotropy. The advantage of this method is that the connection is not completely trivial and it can enrich the phenomenology to be studied. A systematic study of this approach was undertaken in~\cite{Hohmann:2021, DAmbrosio:2021} and we shall briefly review the main steps and results.\newline

\paragraph{\textbf{Symmetries and symmetry-reduction of the metric}}
Following~\cite{Hohmann:2019, DAmbrosio:2021b, Hohmann:2021, DAmbrosio:2021, Heisenberg:2022, DAmbrosio:2023b}, we define a (continuous) symmetry of a metric-affine geometry as follows: Let $\phi_s: \bbR \times \M \to \M$ be a $1$-parameter family of diffeomorphisms with $\phi_{s=0} = \text{id}$, which is smooth in $s$ and which has a generating vector field $v \ce \left.\TD{\phi_s}{s}\right|_{s=0}$. We say that $\phi_s$ is a continuous symmetry of the metric-affine geometry if and only if
\begin{align}
	\begin{cases}
		\phi^*_s g_{\mu\nu} &\overset{!}{=}\quad g_{\mu\nu}\\
		\phi^*_s \Gamma\ud{\alpha}{\mu\nu} &\overset{!}{=}\quad \Gamma\ud{\alpha}{\mu\nu}
	\end{cases}\,.
\end{align}
These are the \textbf{symmetry conditions}. In case there are also tensorial matter fields $\Psi$ present, we have to impose the additional condition
\begin{align}
	\phi^*_s \Psi \quad\overset{!}{=}\quad \Psi
\end{align}
because otherwise the field equations would be inconsistent. Heuristically, this can also be understood as follows: The right hand side of the $f(\Q)$ field equations contain the energy-momentum tensor of the matter fields. It sources the gravitational field described by $(g_{\mu\nu}, \Gamma\ud{\alpha}{\mu\nu})$. If the matter sources do \textit{not} respect certain symmetries, it is hard to see how they could give rise to a gravitational field which \textit{does} respect these symmetries.

Given that the family of diffeomorphisms $\phi_s$ is smooth in $s$, we can re-write the symmetry conditions equivalently as
\begin{align}\label{eq:SymmCond}
	\begin{cases}
		\mathcal{L}_v g_{\mu\nu} &\overset{!}{=}\quad 0 \\
		\mathcal{L}_v\Gamma\ud{\alpha}{\mu\nu} &\overset{!}{=}\quad 0\\
		\mathcal{L}_v \Psi &\overset{!}{=}\quad 0
	\end{cases}\,,
\end{align}
where $\mathcal{L}_v$ denotes the Lie derivative along the vector field $v$ which generates the symmetry $\phi_s$. For a spacetime which is homogeneous and isotropic, the symmetry generators (written in spherical coordinates) are
\begin{align}
	\R_x &\ce \sin\theta \,\partial_\theta + \frac{\cos\phi}{\tan\theta}\,\partial_\phi & \T_x &\ce \rchi\,\sin\theta\,\cos\phi\,\partial_r + \frac{\rchi}{r}\,\cos\theta\,\cos\phi\,\partial_\theta - \frac{\rchi}{r}\,\frac{\sin\phi}{\sin\theta}\,\partial_\phi \notag\\
	\R_y &\ce -\cos\phi\,\partial_\theta + \frac{\sin\phi}{\tan\theta}\,\partial_\phi & \T_y &\ce \rchi\,\sin\theta\,\sin\phi\,\partial_r + \frac{\rchi}{r}\,\cos\theta\,\sin\phi\,\partial_\theta + \frac{\rchi}{r}\,\frac{\cos\phi}{\sin\theta}\, \partial_\phi \notag\\
	\R_z &\ce -\partial_\phi & \T_z &\ce \rchi\,\cos\theta\,\partial_r - \frac{\rchi}{r}\,\sin\theta\,\partial_\theta\,,
\end{align} 
where $\R_i$ are the \textbf{generators of spatial rotations}, $\T_i$ are the \textbf{generators of spatial translations}, and where we have introduced
\begin{align}
	\rchi \ce \sqrt{1-k\, r^2}\,.
\end{align}
As explained in~\cite{Hohmann:2019, DAmbrosio:2021}, it actually suffices to only use $\R_x$, $\R_y$, $\R_z$, and $\T_x$, since the remaining two generators can be obtained by taking Lie brackets of these four. Moreover, imposing the conditions
\begin{align}
	\mathcal{L}_{\R_i} g_{\mu\nu} &\overset{!}{=} 0 & \text{and} && \mathcal{L}_{\T_i} g_{\mu\nu} \overset{!}{=} 0
\end{align} 
leads to the well-known result
\begin{align}
	g_{\mu\nu} =
	\begin{pmatrix}
		g_{tt}(t) & 0 & 0 & 0 \\
		0 & \frac{g_{rr}(t)}{\rchi} & 0 & 0 \\
		0 & 0 & g_{rr}(t)\, r^2 \\
		0 & 0 & 0 & g_{rr}(t)\,r^2\,\sin^2\theta
	\end{pmatrix}\,.
\end{align}
Thus, the initially ten independent components of the metric are reduced to merely two independent components, namely $g_{tt}$ and $g_{rr}$, which can only depend on time. Also, the metric has a simple diagonal form and the parameter $k\in\bbR$ famously determines the spatial curvature: If $k=0$, then the spatial sections are all flat. For $k>0$ one obtains spherical sections, while $k<0$ describes hyperbolic spatial sections.\newline

\paragraph{\textbf{Symmetry-reduction of the connection}}
According to our definition of symmetries for metric-affine geometries, we have to impose the conditions
\begin{align}
	\mathcal{L}_{\R_i} \Gamma\ud{\alpha}{\mu\nu} &\overset{!}{=} 0 & \text{and} &&  \mathcal{L}_{\T_i} \Gamma\ud{\alpha}{\mu\nu} &\overset{!}{=} 0	
\end{align}
on the connection. The resulting equations are numerous and long, but straightforward to solve. One finds~\cite{Hohmann:2021, DAmbrosio:2021}
\begin{align}\label{eq:SymRedConnectionCosmology}
	\Gamma\ud{t}{\mu\nu} &= 
	\begin{pmatrix}
		C_1 & 0 & 0 & 0 \\
		0 & \frac{C_2}{\rchi^2} & 0 & 0 \\
		0 & 0 & C_2\, r^2 & 0 \\
		0 & 0 & 0 & C_2\, r^2\, \sin^2\theta
	\end{pmatrix} &
	\Gamma\ud{r}{\mu\nu} &=
	\begin{pmatrix}
		0 & C_3 & 0 & 0 \\
		C_3 & \frac{k\,r}{\rchi^2} & 0 & 0 \\
		0 & 0 & -r\,\rchi^2 & -C_5 \, r^2 \,\rchi^2\,\sin\theta \\
		0 & 0 & -C_5 \, r^2 \,\rchi^2\,\sin\theta & - r\,\rchi^2\, \sin^2\theta
	\end{pmatrix} \notag\\
	\Gamma\ud{\theta}{\mu\nu} &= 
	\begin{pmatrix}
		0 & 0 & C_3 & 0 \\
		0 & 0 & \frac{1}{r} & \frac{C_5\,\sin\theta}{\rchi} \\
		C_4 & \frac{1}{r} & 0 & 0 \\
		0 & -\frac{C_5\,\sin\theta}{\rchi} & 0 & -\sin\theta\,\cos\theta
	\end{pmatrix} &
	\Gamma\ud{\phi}{\mu\nu} &=
	\begin{pmatrix}
		0 & 0 & 0 & C_3 \\
		0 & 0 & -\frac{C_5\,\csc\theta}{\rchi} & \frac{1}{r} \\
		0 & \frac{C_5\,\csc\theta}{\rchi} & 0 & \cot\theta \\
		C_4 & \frac{1}{r} & \cot\theta & 0
	\end{pmatrix}\,,
\end{align}
where $C_1$, $C_2$, $C_3$, $C_4$, and $C_5$ are arbitrary functions of time. It should be noted that the initially $4\times 4\times 4 = 64$ independent components of the connection have been reduced to these five functions and a few trigonometric functions. However, it should also be noted that the connection is \textit{not} symmetric and thus not torsionless. In fact, we have not yet implemented the postulates of vanishing torsion and vanishing curvature.\newline

\paragraph{\textbf{Implementing the postulates of vanishing torsion and curvature}}
The vanishing of torsion is straightforward to implement. We simply have to demand that the symmetry-reduced connection~\eqref{eq:SymRedConnectionCosmology} is symmetric, which leads to the two conditions
\begin{align}
	C_3 - C_4 &= 0 & \text{and} && C_5 = 0\,.
\end{align}
This leaves us with $C_1$, $C_2$, and $C_3$ as free functions. Given that so many connection components are zero and that the free functions only depend on time, it is not surprising that the condition of vanishing curvature leaves us algebraic equations and first order differential equations. Specifically, $R\ud{\alpha}{\mu\nu\rho} \overset{!}{=} 0$ is equivalent to the set of equations
\begin{align}\label{eq:CurvatureConditions}
	C_1\, C_3 - C^2_3 - \dot{C}_3 &= 0 \notag\\
	C_1\,C_2 - C_2\, C_3 + \dot{C}_2 & = 0\notag\\
	k + C_2\, C_3 &= 0\,.
\end{align}
Notice that the spatially flat case is special, since then we have $C_2\, C_3 = 0$, which has three possible solutions:
\begin{align*}
	&\text{Case I:} & &C_2 = 0,\ C_3\neq 0\,. \\
	&\text{Case II:} & &C_2 \neq 0,\ C_3 = 0\,. \\
	&\text{Case III:} & &C_2 = 0,\ C_3 = 0\,.
\end{align*}
If $k\neq 0$, the situation is considerably simpler. Since neither $C_2$ nor $C_3$ can be zero, we obtain
\begin{align}
	C_3 = - \frac{k}{C_2}\,.
\end{align}
Using this result, the two differential equations~\eqref{eq:CurvatureConditions} reduce to a single equation:
\begin{align}
	k + C_1\, C_2 + \dot{C}_2 &= 0\,.
\end{align}
Given that $C_2\neq 0$, we can solve this last equation for $C_1$, obtaining
\begin{align}
	C_1 = - \frac{k+\dot{C}_2}{C_2}\,.
\end{align}
We finally arrive at the conclusion that a connection which respects homogeneity and isotropy, and which is also torsionless and flat under the assumption that $k\neq 0$ has the form
\begin{align}
	\Gamma\ud{t}{\mu\nu} &=
	\begin{pmatrix}
		-\frac{k + \dot{C}_2}{C_2} & 0 & 0 & 0 \\
		0 & \frac{C_2}{\rchi^2} & 0 & 0 \\
		0 & 0 & r^2\,C_2 & 0 \\
		0 & 0 & 0 & r^2\,C_2\,\sin^2\theta
	\end{pmatrix} & 
	\Gamma\ud{r}{\mu\nu} &=
	\begin{pmatrix}
		0 & -\frac{k}{C_2} & 0 & 0 \\
		-\frac{k}{C_2} & \frac{k\,r}{\rchi^2} & 0 & 0 \\
		0 & 0 & -r\,\rchi^2 & 0 \\
		0 & 0 & 0 & -r\,\rchi^2\,\sin^2\theta
	\end{pmatrix} \notag\\
	\Gamma\ud{\theta}{\mu\nu} &=
	\begin{pmatrix}
		0 & 0 & -\frac{k}{C_2} & 0 \\
		0 & 0 & \frac{1}{r} & 0 \\
		-\frac{k}{C_2} & \frac{1}{r} & 0 & 0 \\
		0 & 0 & 0 & -\sin\theta\,\cos\theta
	\end{pmatrix} &
	\Gamma\ud{\phi}{\mu\nu} &= 
	\begin{pmatrix}
		0 & 0 & 0 & -\frac{k}{c} \\
		0 & 0 & 0 & \frac{1}{r} \\
		0 & 0 & 0 & \cot\theta \\
		-\frac{k}{C_2} & \frac{1}{r} & \cot\theta & 0
	\end{pmatrix}\,.
\end{align}
We dub this connection $\Gamma^{(k)}$. Now we consider to spatially flat sections, $k=0$, case by case. For Case I, defined by $C_2=0$ under the assumption that $C_3\neq 0$, we obtain the connection $\Gamma^{(\rm I)}$, which is of the form
\begin{align}
	\Gamma\ud{t}{\mu\nu} &=
	\begin{pmatrix}
		C_3 + \frac{\dot{C}_3}{C_3} & 0 & 0 & 0 \\
		0 & 0 & 0 & 0 \\
		0 & 0 & 0 & 0 \\
		0 & 0 & 0 & 0
	\end{pmatrix} & 
	\Gamma\ud{r}{\mu\nu} &=
	\begin{pmatrix}
		0 & C_3 & 0 & 0 \\
		C_3 & 0 & 0 & 0 \\
		0 & 0 & -r & 0 \\
		0 & 0 & 0 & -r\,\sin^2\theta
	\end{pmatrix} \notag\\
	\Gamma\ud{\theta}{\mu\nu} &=
	\begin{pmatrix}
		0 & 0 & C_3 & 0 \\
		0 & 0 & \frac{1}{r} & 0 \\
		C_3 & \frac{1}{r} & 0 & 0 \\
		0 & 0 & 0 & -\sin\theta\,\cos\theta
	\end{pmatrix} &
	\Gamma\ud{\phi}{\mu\nu} &= 
	\begin{pmatrix}
		0 & 0 & 0 & C_3 \\
		0 & 0 & 0 & \frac{1}{r} \\
		0 & 0 & 0 & \cot\theta \\
		C_3 & \frac{1}{r} & \cot\theta & 0
	\end{pmatrix}\,,
\end{align}
This connection depends on the free function $C_3(t)$. In the second case, which is based on the assumption $C_2 \neq 0$, we obtain the connection $\Gamma^{(\rm{II})}$, parametrized as
\begin{align}
	\Gamma\ud{t}{\mu\nu} &=
	\begin{pmatrix}
		-\frac{\dot{C}_2}{C_2} & 0 & 0 & 0 \\
		0 & C_2 & 0 & 0 \\
		0 & 0 & r^2\, C_2 & 0 \\
		0 & 0 & 0 & r^2\,C_2\,\sin^2\theta
	\end{pmatrix} & 
	\Gamma\ud{r}{\mu\nu} &=
	\begin{pmatrix}
		0 & 0 & 0 & 0 \\
		0 & 0 & 0 & 0 \\
		0 & 0 & -r & 0 \\
		0 & 0 & 0 & -r\,\sin^2\theta
	\end{pmatrix} \notag\\
	\Gamma\ud{\theta}{\mu\nu} &=
	\begin{pmatrix}
		0 & 0 & 0 & 0 \\
		0 & 0 & \frac{1}{r} & 0 \\
		0 & \frac{1}{r} & 0 & 0 \\
		0 & 0 & 0 & -\sin\theta\,\cos\theta
	\end{pmatrix} &
	\Gamma\ud{\phi}{\mu\nu} &= 
	\begin{pmatrix}
		0 & 0 & 0 & 0 \\
		0 & 0 & 0 & \frac{1}{r} \\
		0 & 0 & 0 & \cot\theta \\
		0 & \frac{1}{r} & \cot\theta & 0
	\end{pmatrix}\,.	
\end{align}
Finally, the third case, which is clearly the simplest, gives us the connection $\Gamma^{(\rm{III})}$, which can explicitly be written as
\begin{align}
	\Gamma\ud{t}{\mu\nu} &=
	\begin{pmatrix}
		-C_1 & 0 & 0 & 0 \\
		0 & 0 & 0 & 0 \\
		0 & 0 & 0 & 0 \\
		0 & 0 & 0 & 0
	\end{pmatrix} & 
	\Gamma\ud{r}{\mu\nu} &=
	\begin{pmatrix}
		0 & 0 & 0 & 0 \\
		0 & 0 & 0 & 0 \\
		0 & 0 & -r & 0 \\
		0 & 0 & 0 & -r\,\sin^2\theta
	\end{pmatrix} \notag\\
	\Gamma\ud{\theta}{\mu\nu} &=
	\begin{pmatrix}
		0 & 0 & 0 & 0 \\
		0 & 0 & \frac{1}{r} & 0 \\
		0 & \frac{1}{r} & 0 & 0 \\
		0 & 0 & 0 & -\sin\theta\,\cos\theta
	\end{pmatrix} &
	\Gamma\ud{\phi}{\mu\nu} &= 
	\begin{pmatrix}
		0 & 0 & 0 & 0 \\
		0 & 0 & 0 & \frac{1}{r} \\
		0 & 0 & 0 & \cot\theta \\
		0 & \frac{1}{r} & \cot\theta & 0
	\end{pmatrix}\,.	
\end{align}
In conclusion, we find that a connection which is homogeneous, isotropic, torsionless, and flat can be parametrized in four distinct ways. The connections $\Gamma^{(k)}$, $\Gamma^{(\rm{I})}$, $\Gamma^{\rm{II}}$, and $\Gamma^{(\rm III)}$ could be the source of interesting and rich cosmological models. Indeed, for the connection $\Gamma^{(\rm{II})}$ with the choice $f(\Q) = \Q^\kappa$ (assuming $\kappa\geq 2$), exact vacuum solutions were obtained~\cite{DAmbrosio:2021} which can reproduce the scale factor of a fluid with equation of state $p = w\,\rho$, for some constant $w$. The same exact vacuum solution can also mimic de Sitter space. This could be of interest for investigations concerning the early Universe, since this solution can naturally drive inflation.

The effects of using different connections in $f(\Q)$ cosmology have been studied in~\cite{DAmbrosio:2021, Dimakis:2022, Heisenberg:2022, Shabani:2023, Subramaniam:2023}, but a large untapped potential to discover new interesting solutions remains.

\subsection{Black Holes in \texorpdfstring{$f(\Q)$}{fQ}}\label{ssec:BlackHoles}
It is tempting to start an investigation into black hole solutions in $f(\Q)$ gravity by following the same strategy initially used in $f(\Q)$ cosmology. In fact, the simplest possible strategy one can think of is to fix the coincident gauge and choose a metric ansatz which is stationary and spherically symmetric. In formulas:
\begin{align}
	\Gamma\ud{\alpha}{\mu\nu} &= 0 &\text{and} && g_{\mu\nu} =
	\begin{pmatrix}
		-A(r) & 0 & 0 & 0\\
		0 & B(r) & 0 & 0\\
		0 & 0 & r^2 & 0\\
		0 & 0 & 0 & r^2 \sin^2\theta
	\end{pmatrix}\,,
\end{align}
where $A$ and $B$ are arbitrary functions of the radial coordinate $r>0$. This metric ansatz is written in the coordinate system $(t, r, \theta, \phi)$, where $(r, \theta, \phi)$ are the standard spherical coordinates. The idea is then to plug the ansatz metric into the $f(\Q)$ field equations~\eqref{eq:f(Q)FEQ} and work out solutions. As long as $f''\neq 0$, one expects to obtain black hole solutions which deviate from the standard GR solution.

However, as it turns out, this expectation is not only wrong, the entire strategy is not viable! In fact, one finds without much trouble that the field equations of $f(\Q)$ become inconsistent for the above ansatz metric in coincident gauge, except in the special case where $f''=0$. In other words, the field equations themselves tell us that the inconsistency disappears precisely when we are in the regime of the theory which is equivalent to GR. But in that regime we can only recover GR solutions and nothing else! After all, the equivalence of GR and STEGR is based on having the same field equations with the same solution space.

In order to overcome this difficulty, a systematic analysis of stationary and spherically symmetric spacetimes was carried out in~\cite{DAmbrosio:2021b}. This means that no special metric nor any special gauge for the connection is assumed. Rather, the strategy of~\cite{DAmbrosio:2021b} was to follow three simple steps:
\begin{enumerate}
	\item Symmetries: Find the most general metric \textit{and} connection which are stationary (i.e., time-translation invariant) and spherically symmetric (i.e., invariant under spatial rotations around the origin);
	\item Geometric postulates: Use the metric and connection found above and implement the postulate of vanishing curvature and vanishing torsion, $R\ud{\alpha}{\mu\nu\rho} = 0$ and $T\ud{\alpha}{\mu\nu} = 0$;
	\item Field equations: Take the metric and connection which satisfy all symmetries and geometric postulates and plug them into the $f(\Q)$ field equations.
\end{enumerate}
We provide a brief sketch of the individual steps. The ultimate goal is to find the \textit{simplest} representation of a stationary, spherically symmetric metric-affine geometry $(\M, g, \Gamma)$, before studying the field equations of $f(\Q)$ gravity. For details we refer the reader to~\cite{DAmbrosio:2021b}. \newline

\paragraph{\textbf{Symmetries}}
In the present context we are interested in finding solutions which are stationary and spherically symmetric. Given the notion of spacetime symmetry discussed in the previous subsection, this means that we have to impose
\begin{align}\label{eq:SymmCondBH}
	\begin{cases}
		\mathcal{L}_v g_{\mu\nu} &\overset{!}{=}\quad 0 \\
		\mathcal{L}_v\Gamma\ud{\alpha}{\mu\nu} &\overset{!}{=}\quad 0\\
		\mathcal{L}_v \Psi &\overset{!}{=}\quad 0
	\end{cases}\,,
\end{align}
for the vector fields which generate temporal translations and spatial rotations around the origin:
\begin{align}
	\T &\ce \partial_t & \text{(generator of time-translations)} \notag\\
	\R_x &\ce \sin\phi\, \partial_\theta + \frac{\cos\phi}{\tan\theta}\partial_\phi & \notag\\
	\R_y &\ce -\cos\phi \partial_\theta + \frac{\sin\phi}{\tan\theta}\partial_\phi & \text{(generators of rotations)} \notag\\
	\R_z &\ce -\partial_\phi &
\end{align}
Obtaining time-translation invariance and invariance with respect to rotations in the $\phi$-direction is easy: All metric and connection components have to be independent of the coordinates $t$ and $\phi$. Invariance with respect to the $\R_x$ and $\R_y$ generators requires a little more work. However, the result for the metric is well-known (see for instance~\cite{WaldBook, CarrollBook}): A metric which is time-translation and rotationally invariant necessarily has the form
\begin{align}\label{eq:SymRedMet}
	g_{\mu\nu} = \begin{pmatrix}
		g_{tt} & g_{tr} & 0 & 0 \\
		g_{tr} & g_{rr} & 0 & 0 \\
		0 & 0 & g_{\theta\theta} & 0 \\
		0 & 0 & 0 & g_{\theta\theta} \sin^2\theta
	\end{pmatrix}
\end{align}
with respect to the coordinates $(t, r, \theta, \phi)$. In particular, we point out that $g_{tt}, g_{tr}, g_{rr}$, and $g_{\theta\theta}$ are only functions of $r$. 

In the case of the connection it is easier to first impose the postulate of vanishing torsion and then to work out the remaining two symmetry conditions. Since torsion is the anti-symmetric part of the connection, a torsionless connection is simply one that is symmetric in its lower two indices. Imposing this condition also has the effect of reducing the number of independent components of the connection from $4\times 4\times 4 = 64$ to $4\times \frac{4\times (4+1)}{2} = 40$.

Furthermore, it is more convenient to consider the following linear combinations when imposing the remaining symmetry conditions:
\begin{align}
	\cos\phi\, \mathcal{L}_{\R_x} \Gamma\ud{\alpha}{\mu\nu} + \sin\phi\, \mathcal{L}_{\R_y} \Gamma\ud{\alpha}{\mu\nu} &\overset{!}{=} 0\notag\\
	\sin\phi\, \mathcal{L}_{\R_x} \Gamma\ud{\alpha}{\mu\nu} - \cos\phi\, \mathcal{L}_{\R_y} \Gamma\ud{\alpha}{\mu\nu} &\overset{!}{=} 0\,.
\end{align}
We emphasize that imposing these conditions is strictly equivalent to imposing $\mathcal{L}_{\R_x}\Gamma\ud{\alpha}{\mu\nu} \overset{!}{=} 0$ and $\mathcal{L}_{\R_y}\Gamma\ud{\alpha}{\mu\nu} \overset{!}{=} 0$. By imposing the first linear combination we learn that
\begin{itemize}
	\item[a)] Twenty of the $40$ components of $\Gamma\ud{\alpha}{\mu\nu}$ are exactly zero;
	\item[b)] Two components are given in terms of trigonometric functions;
	\item[c)] Six components are determined through algebraic relations to other components of the connection. 
\end{itemize}
Hence, out of the initially $64$ independent components of the connection, three of the symmetry conditions and the postulate of vanishing torsion bring this number down to only $40-20-2-6 = 12$ independent components which are functions of $r$ and $\theta$.

Finally, the second linear combination implements the last symmetry condition. It leads to a set of twelve first order partial differential equations for precisely the twelve independent connection components we are left with after imposing the first three symmetry conditions. These equations can be solved, but because these are partial differential equations with respect to $\theta$, the solutions all depend on $r$. Hence, we find that the symmetry conditions together with the postulate of vanishing torsion leave us with twelve independent connection components, all of which are purely functions of $r$ and nothing else.\newline

\paragraph{\textbf{Geometric postulates}}
Since we have already implemented the postulate of vanishing torsion, we are left with imposing the postulate of vanishing curvature. As can be expected from the form of the curvature tensor and the fact that $20$ connection components vanish, this will lead to a set of algebraic equations and a set of first order partial differential equations. The detailed process of how to consistently solve all algebraic and differential equations is explained in~\cite{DAmbrosio:2021b}, where it is found that this ultimately leads to two different sets of solutions. The first solution set is defined as follows: All connection components can be expressed in terms of the three arbitrary functions $\Gamma\ud{t}{rr}(r)$, $\Gamma\ud{r}{rr}(r)$, $\Gamma\ud{\phi}{r\phi}(r)$, the real constant $c\neq 0$, and trigonometric functions. Concretely, the connection takes the form
\begin{align}
	\Gamma\ud{t}{\mu\nu} &=
	\begin{pmatrix}
		c & \Gamma\ud{\phi}{r\phi} & 0 & 0 \\
		\Gamma\ud{\phi}{r\phi} & \Gamma\ud{t}{rr} & 0 & 0 \\
		0 & 0 & -\frac{1}{c} & 0 \\
		0 & 0 & 0 & -\frac{\sin^2\theta}{c}
	\end{pmatrix} &
	\Gamma\ud{r}{\mu\nu} &=
	\begin{pmatrix}
		0 & 0 & 0 & 0 \\
		0 & \Gamma\ud{r}{rr} & 0 & 0 \\
		0 & 0 & 0 & 0 \\
		0 & 0 & 0 & 0
	\end{pmatrix}\notag\\
	\Gamma\ud{\theta}{\mu\nu} &= 
	\begin{pmatrix}
		0 & 0 & c & 0 \\
		0 & 0 & \Gamma\ud{\phi}{r\phi} & 0 \\
		c & \Gamma\ud{\phi}{r\phi} & 0 & 0 \\
		0 & 0 & 0 & -\sin\theta\, \cos\theta
	\end{pmatrix} &
	\Gamma\ud{\phi}{\mu\nu} &= 
	\begin{pmatrix}
		0 & 0 & 0 & c \\
		0 & 0 & 0 & \Gamma\ud{\phi}{r\phi} \\
		0 & 0 & 0 & \cot\theta \\
		c& \Gamma\ud{\phi}{r\phi} & \cot\theta & 0
	\end{pmatrix}\,.
\end{align}
Furthermore, the derivative of $\Gamma\ud{\phi}{r\phi}$ can be written as
\begin{align}
	\TD{}{r}\Gamma\ud{\phi}{r\phi} = c\, \Gamma\ud{t}{rr} - \Gamma\ud{\phi}{r\phi}\left(\Gamma\ud{\phi}{r\phi} + \Gamma\ud{r}{rr}\right)\,.
\end{align}
These are all the defining properties of solution set 1. For solution set 2 one finds instead that all connection components can be expressed in terms of the four arbitrary functions $\Gamma\ud{t}{rr}(r)$, $\Gamma\ud{t}{\theta\theta}(r)$, $\Gamma\ud{r}{rr}(r)$, $\Gamma\ud{r}{\theta\theta}(r)\neq 0$, the two real constants $c$ and $k$, and trigonometric functions. The connection is explicitly given by
\begin{align}
	\Gamma\ud{t}{\mu\nu} &=
	\begin{pmatrix}
		k - c - c\tilde{c}\,\Gamma\ud{t}{\theta\theta} & \frac{\tilde{c}\, \hat{\Gamma}\ud{t}{\theta\theta}\,\Gamma\ud{t}{\theta\theta}}{\Gamma\ud{r}{\theta\theta}} & 0 & 0 \\
		\frac{\tilde{c}\,\hat{\Gamma}\ud{t}{\theta\theta}\,\Gamma\ud{t}{\theta\theta}}{\Gamma\ud{r}{\theta\theta}} & \Gamma\ud{t}{rr} & 0 & 0 \\
		0 & 0 & \Gamma\ud{t}{\theta\theta} & 0 \\
		0 & 0 & 0 & \Gamma\ud{t}{\theta\theta}\,\sin^2\theta
	\end{pmatrix} &
	\Gamma\ud{r}{\mu\nu} &=
	\begin{pmatrix}
		-c\tilde{c}\,\Gamma\ud{r}{\theta\theta} & c + c \,\tilde{c}\, \Gamma\ud{t}{\theta\theta} & 0 & 0 \\
		c + c \,\tilde{c}\, \Gamma\ud{t}{\theta\theta} & \Gamma\ud{r}{rr} & 0 & 0 \\
		0 & 0 & \Gamma\ud{r}{\theta\theta} & 0 \\
		0 & 0 & 0 & \Gamma\ud{r}{\theta\theta}\,\sin^2\theta
	\end{pmatrix} \notag\\
	\Gamma\ud{\theta}{\mu\nu} &= 
	\begin{pmatrix}
		0 & 0 & c & 0\\
		0 & 0 & -\frac{\hat{\Gamma}\ud{t}{\theta\theta}}{\Gamma\ud{r}{\theta\theta}} & 0 \\
		c & -\frac{\hat{\Gamma}\ud{t}{\theta\theta}}{\Gamma\ud{r}{\theta\theta}} & 0 & 0\\
		0 & 0 & 0 & -\sin\theta\,\cos\theta
	\end{pmatrix} &
	\Gamma\ud{\phi}{\mu\nu} &= 
	\begin{pmatrix}
		0 & 0 & 0 & c \\
		0 & 0 & 0 & -\frac{\hat{\Gamma}\ud{t}{\theta\theta}}{\Gamma\ud{r}{\theta\theta}} \\
		0 & 0 & 0 & \cot\theta \\
		c & -\frac{\hat{\Gamma}\ud{t}{\theta\theta}}{\Gamma\ud{r}{\theta\theta}} & \cot\theta & 0
	\end{pmatrix}\,,
\end{align}
where we have defined  $\tilde{c} \ce 2c-k$ and $\hat{\Gamma}\ud{t}{\theta\theta} \ce 1 + c\, \Gamma\ud{t}{\theta\theta}$ in order to compactify the notation. Moreover, the derivatives of $\Gamma\ud{t}{\theta\theta}$ and $\Gamma\ud{r}{\theta\theta}$ can be expressed in terms of the other free functions. Concretely, one finds
\begin{align}
	\TD{}{r}\Gamma\ud{t}{\theta\theta} &= - \frac{\left\{\left[c\,(2c-k)\,\Gamma\ud{t}{\theta\theta} + 3c -k\right]\Gamma\ud{t}{\theta\theta} + 1\right\} \Gamma\ud{t}{\theta\theta}}{\Gamma\ud{r}{\theta\theta}} - \Gamma\ud{r}{\theta\theta} \Gamma\ud{t}{rr}\notag\\
	\TD{}{r}\Gamma\ud{r}{\theta\theta} &= - c\left((2c-k)\Gamma\ud{t}{\theta\theta} + 2\right)\Gamma\ud{t}{\theta\theta} - \Gamma\ud{r}{\theta\theta} \Gamma\ud{r}{rr} - 1\,.
\end{align}
Observe that in both solution sets the derivatives of $\Gamma\ud{t}{rr}$ and $\Gamma\ud{r}{rr}$ cannot be expressed in terms of other connection components. Thus, in both cases only these two components should be regarded as the unknowns to be solved for in the connection field equations. 

It was also shown in~\cite{DAmbrosio:2021b} that the two solution sets are related to each other by a double scaling limit. However, it should be emphasized that outside of this particular limit, the two solution sets are genuinely different and they describe different physics. We elaborate more on this point further below.\newline

\paragraph{\textbf{Simplest possible form of a stationary, spherically symmetric geometry $\boldsymbol{(\M, g, \Gamma)}$}}
Recall that our task is not only to find expressions for the metric and the connection which satisfy the various symmetries and the geometric postulates. We also wish to find the \textit{simplest possible form}, as that will hopefully help in analyzing and solving the field equations. To simplify the form of the metric, we make use of the diffeomorphism invariance of the theory. This is possible, since we did not yet fix any particular gauge. As is well-known, it is possible to find a diffeomorphism which brings the symmetry-reduced metric~\eqref{eq:SymRedMet} into the simple diagonal form
\begin{align}\label{eq:SymRedMetSimple}
	g_{\mu\nu} = 
	\begin{pmatrix}
		g_{tt}(r) & 0 & 0 & 0\\
		0 & g_{rr}(r) & 0 & 0\\
		0 & 0 & r^2 & 0 \\
		0 & 0 & 0 & r^2 \sin^2\theta 
	\end{pmatrix}\,.
\end{align}
This is of course nothing but the standard form of a metric which is stationary and spherically symmetric, which can be found in textbooks on GR~\cite{HawkingEllisBook, WaldBook, CarrollBook}. However, in the context of metric-affine geometries, the diffeomorphism which achieves this transformation has of course also to be applied to the connection. What is remarkable, is that even tough this diffeomorphism in general changes the connection, it maps solution set 1 onto itself and it also maps solution set 2 onto itself!

This means that when we study the field equations of $f(\Q)$ gravity, we can use the metric in its simple symmetry-reduced form~\eqref{eq:SymRedMetSimple} together with a connection which either belongs to solution set 1 or solution set 2. This is the simplest possible form of a stationary and spherically symmetric metric-affine geometry!\newline

\paragraph{\textbf{A cautionary remark on the coincident gauge}}
It is worth pausing at this point and discussing why the first approach, namely the approach based on a metric of the form~\eqref{eq:SymRedMetSimple} and the coincident gauge, $\Gamma\ud{\alpha}{\mu\nu} = 0$, fails. This comes simply from the fact that \textit{if} the metric has the form~\eqref{eq:SymRedMetSimple}, \textit{then} the connection cannot be identically zero if it also has to satisfy the symmetry conditions. This follows immediately from the two solution sets. Recall that these two solution sets tell us the possible forms a symmetry-reduced connection can have. Both sets exclude the possibility $\Gamma\ud{\alpha}{\mu\nu} = 0$, because in both sets there are components which are purely expressed in terms of trigonometric function and in both sets there are certain components which are not allowed to vanish. 

Does this mean we cannot use the coincident gauge? No, the coincident gauge can always be used. But one has to be careful in \textit{how} one uses it. Our systematic implementation of symmetries and geometric postulates has shown what form the metric and the connection are allowed to have \textit{in the coordinate system ${t, r, \theta, \phi}$}. What the coincident gauge tells us, is that there exists a \textit{different coordinate system} where $\Gamma\ud{\alpha}{\mu\nu}=0$, but where the metric will no longer have its simple diagonal form! A diffeomorphism which trivializes the connection will necessarily complicate the metric. In a sense, all the information which resided in the symmetry-reduced connection is ``moved'' onto the metric by the diffeomorphism. Hence, nothing is gained by using the coincident gauge, which is why we prefer to stick to the two solution sets described above. In the context of stationary and spherically symmetric spacetimes, the transformations which produced the coincident gauge for both solution sets have been worked out~\cite{Bahamonde:2022}.\newline

\paragraph{\textbf{Symmetry-reduced form of the field equations}}
The symmetry-reduced form of the field equations are obtained by plugging the metric ansatz~\eqref{eq:SymRedMetSimple} and either the connection from solution set 1 or the connection from solution set 2 into the $f(\Q)$ field equations~\eqref{eq:f(Q)FEQ}. In both cases we find that the structure of the field equations is
\begin{align}
	&\text{Structure of metric field equations:} &   &\begin{pmatrix}
		\M_{tt} & \M_{tr} & 0 & 0 \\
		\M_{tr} & \M_{rr} & 0 & 0 \\
		0 & 0 & \M_{\theta\theta} & 0 \\
		0 & 0 & 0 & \M_{\theta\theta} \sin^2\theta
	\end{pmatrix}\notag\\
	& \text{Structure of connection field equations:} & & \begin{pmatrix}
		\C_t \\
		\C_r \\
		0 \\
		0
	\end{pmatrix}
\end{align}
Of course, the components of these tensors are different for the two different solution sets of the connection. However, in both cases it turns out to be highly advantageous to first study the off-diagonal component of the metric field equations, i.e., $\M_{tr} = 0$. This leads to two very similar and yet still different equations:
\begin{itemize}
	\item For solution set 1: $\M_{tr} = 0 \quad\longrightarrow\quad c\, \partial_r \Q \, f''(\Q) = 0$.
	\item For solution set 2:   $\M_{tr} = 0 \quad\longrightarrow\quad \left(k-2c(2c-k)\Gamma\ud{t}{\theta\theta}\right)\partial_r \Q \, f''(\Q) = 0$.
\end{itemize}
We observe that both equations admit $\partial_r \Q$ and $f''(\Q) = 0$ as solutions. The first option amounts to saying that the non-metricity scalar is constant. In fact, the metric and the connection for both solution sets only depend on $r$ and $\theta$, but, as was shown in~\cite{DAmbrosio:2021b}, the non-metricity scalar does \textit{not} inherit the $\theta$-dependence. Thus, $\partial_r\Q = 0$ is really saying that the non-metricity scalar is a constant. It is then easy to see that this does \textit{not} yield any solutions which go beyond GR. In fact, the $f(\Q)$ field equations for $\Q = $const. simply become
\begin{align}
	f'(\Q_0) G_{\mu\nu} + \frac12 \left(f(\Q_0)-f'(\Q_0)\, \Q_0\right)g_{\mu\nu} = \kappa\, \T_{\mu\nu}\,,
\end{align}
where $\Q_0$ is a constant number. These equations can be re-written in the more suggestive form
\begin{align}
	G_{\mu\nu} + \Lambda_{\rm eff}g_{\mu\nu} = \kappa\, \bar{\T}_{\mu\nu}\,,
\end{align}
where we have introduced
\begin{align}
	\Lambda_{\rm eff} &\ce \frac12 \frac{f(\Q_0) - f'(\Q_0) \Q_0}{f'(\Q_0)} & \text{and} && \bar{\T}_{\mu\nu} &\ce \frac{1}{f'(\Q_0)} \T_{\mu\nu}\,.
\end{align}
Thus, we obtain the Einstein field equations with an effective cosmological constant and a re-scaled energy-momentum tensor! Notice that the re-scaling and the effective cosmological constant are well-defined since we always assume $f'\neq 0$. Otherwise, one would end up with a trivial, non-dynamical theory. Thus, we conclude that solving the off-diagonal metric field equation with $\Q = $const. does not yield beyond GR solutions.

The second option is to solve $\M_{tr} = 0$ by $f''(\Q) = 0$. However, we already know that this means that $f(\Q) = a\, \Q + b$, where $a$ and $b$ are two real constants. In other words, this option just produces STEGR plus a cosmological constant. Give that STEGR is equivalent to GR, with this option we just recover GR solutions and nothing else. Hence, also in this case we learn that we can only obtain GR solutions for both solution sets of the connection.

This leads us to the third option, which is to impose the constraint equations
\begin{align}\label{eq:ConstraintEqsSolSets}
	\M_{tr} &= 0 & \quad \longrightarrow\quad && c &= 0 & & \text{(for solution set 1)} \notag\\
	\M_{tr} &= 0 &\quad \longrightarrow\quad && \left(k-2c(2c-k)\Gamma\ud{t}{\theta\theta}\right) &= 0 && \text{(for solution set 2)}\,.
\end{align}
A quick glace at the defining properties of solution set 1 reveals that $c=0$ is not possible. In fact, solution set 1 is only valid if $c\neq 0$. Hence, we reach the important conclusion that \textbf{solution set 1 only contains the GR solutions}! If we wish to find beyond GR solution, our only hope is solution set 2. Indeed, the constraint equation~\eqref{eq:ConstraintEqsSolSets} for solution set 2 does have interesting solutions. As it turns out~\cite{DAmbrosio:2021b}, there are two branches.
\begin{align}
	&\text{Branch 1:} & && \Gamma\ud{t}{\theta\theta} &= \frac{k}{2c(2c-k)}& && &\text{for $c\neq 0$ and $k\neq 2c$}\notag\\
	& & && \Gamma\ud{t}{rr} &= \frac{k (8c^2 + 2 c k -k^2)}{8 c^2(2c-k)^2 \left(\Gamma\ud{r}{\theta\theta}\right)^2}\notag\\
	\notag\\
	&\text{Branch 2:} & && \Gamma\ud{t}{rr} &= -\frac{\Gamma\ud{t}{\theta\theta}}{\left(\Gamma\ud{r}{\theta\theta}\right)^2} \notag\\
	& & && c &= k = 0\,.
\end{align}
Both branches are viable in the sense that they lead to self-consistent field equations, as has been shown in~\cite{DAmbrosio:2021b}. Moreover, it has also been shown that both branches lead to beyond-GR solutions. Some solutions have been derived explicitly.\newline

\paragraph{\textbf{Overview of different developments and outlook}}
Let us summarize the situation thus far: We began with a systematic implementation of stationarity and spherical symmetry. This drastically restricted the form of the metric and of the connection. Then, we proceeded with imposing the geometric postulates. In particular, the postulate of vanishing curvature led to further restrictions on the connection and we found that there are two possible parametrizations for a symmetry-reduced connection which also satisfies the geometric postulates. We dubbed these parametrizations solution set 1 and solution set 2.

Remarkably, it is possible to diagonalize the metric and bring it into the standard form of a stationary and spherically symmetric metric \textit{without} spoiling the solution sets. That is, the diffeomorphism which brings the metric into its simplest form maps solution set 1 onto itself and solution set 2 onto itself. Thus, the metric~\eqref{eq:SymRedMetSimple} together with solution sets 1 and 2 for the connection provide us with the simplest representation of a stationary and spherically symmetric metric-affine geometry $(\M, g, \Gamma)$. The solution sets also allow us to understand why the coincident gauge leads to inconsistent field equations, if we simultaneously insist that the metric ansatz has the form~\eqref{eq:SymRedMetSimple}.

By studying the symmetry-reduced metric field equations, we finally learned that solution set 1 only contains the standard GR solutions. If one wishes to find beyond-GR solutions, one has to work with solution set 2. Within this solution set, one finds that the field equations allow for two branches. That is, the off-diagonal equation $\M_{tr} = 0$ imposes a constraint on the connection which admits two genuinely different solutions. Both solutions are fully consistent and can be used to further study the field equations.

This leads us to the question of what can be achieved with these different branches and modified gravity equations. In~\cite{DAmbrosio:2021b}, different methods were used to find beyond-GR black hole solutions. Some exact, but rather unphysical solutions were found. Perturbative techniques led to approximate solutions of the field equations which are asymptotically flat, but which lead to multiple horizons and black hole masses which depend on the connection. Regular black holes, black bounces, and quasi normal modes within the context of $f(\Q)$ gravity were studied in~\cite{Junior:2023, Gogoi:2023}.

Besides black holes, the stationary and spherically symmetric spacetimes considered here have inspired a flurry of investigations into wormholes in $f(\Q)$ gravity~\cite{Banerjee:2021, Mustafa:2021, Parsaei:2022, Hassan:2022b, Hassan:2022, Hassan:2022c, Venkatesha:2023, Jan:2023, Godani:2023, Mustafa:2023, Mishra:2023} as well as modified stellar solutions~\cite{Wang:2021, Maurya:2022b, Maurya:2022, Errehymy:2022, Sokoliuk:2022, Calza:2022, Chanda:2022, Bhar:2023, Ditta:2023, Maurya:2023}. Some thought has also been given to the question, how observational data could be used to constrain $f(\Q)$ gravity~\cite{DAgostino:2022}. The beyond-GR black hole and stellar solutions could play an important role in this regard.

\subsection{Hamiltonian Analysis and Degrees of Freedom of \texorpdfstring{$f(\Q)$}{fQ} Gravity}\label{ssec:Hamiltonian}
The question how many degrees of freedom are being propagated in $f(\Q)$ is currently under debate. Findings from cosmological perturbation theory performed in~\cite{BeltranJimenez:2019} revealed that $f(\Q)$ possesses \textit{at least} two additional degrees of freedom compared to GR. Based on this insight, and the expectation that the primary constraints of $f(\Q)$ gravity are all second class due to its general covariance, led to the educated guess that the theory propagates six degrees of freedom~\cite{DAmbrosio:2020c}.

A more systematic approach based on the Hamiltonian analysis, performed in coincident gauge, was attempted in~\cite{Hu:2022}. The authors concluded that there are eight degrees of freedom. However, this conclusion was challenges by~\cite{DAmbrosio:2023}, who put an upper bound of seven degrees of freedom using a kinetic matrix approach. In the same paper, mistakes in the analysis of~\cite{Hu:2022} were brought to light and general issues with the Hamiltonian analysis were discussed. In particular, it was pointed out that the standard approach due to Dirac~\cite{DiracBook, Dirac:1950} and Bergmann~\cite{Bergmann:1951} encounters severe obstacles and new methods, such as the kinetic matrix approach, have to be employed. Finally, yet another Hamiltonian analysis was attempted by~\cite{Tomonari:2023}, who concluded that there are six degrees of freedom. This is in agreement with the upper bound of~\cite{DAmbrosio:2023} and the authors claim to have overcome the obstacles of the Dirac-Bergmann algorithm which were pointed out in~\cite{DAmbrosio:2023}. However, as we will discuss further below, the resolution is not beyond doubt.

At the moment, only three things seem clear: (a) The theory propagates at least four degrees of freedom, (b) there are at most seven degrees of freedom, and (c) there is confusion about what the precise number might be.

To better understand this unsatisfying state of affairs we shall briefly review the main results on which everyone agrees. Then we discuss the points where mistakes were made or where opinions drift apart. \newline

\paragraph{\textbf{ADM formulation and primary constraints}}
In order to perform the Hamiltonian analysis, it is advantageous to employ the ADM formalism. Under the (weak) assumption that $\M$ has the topology $\M\simeq \bbR\times \Sigma$, where $\Sigma$ is a three-dimensional spacelike hypersurface, we can split the coordinates $\{x^\mu\}$ into one temporal and three spatial coordinates, $\{t, x^{a}\}$. The spatial index takes values in $\{1,2,3\}$. Moreover, the metric can be written as
\begin{align}
	g_{\mu\nu} =
	\begin{pmatrix}
		-N^2 + h_{ab}N^{a} N^{b} & h_{ab} N^{b} \\
		h_{ab} N^{b} & h_{ab}
	\end{pmatrix}\,,
\end{align}
where $N>0$ is the lapse function, $N^{a}$ is called the shift vector field, and $h_{ab}$ is the three-dimensional metric intrinsic to $\Sigma$. Spatial indices  are raised and lowered with $h_{ab}$. Also, we refer to $\{N, N^{a}, h_{ab}\}$ collectively as ADM variables. From now on, we work exclusively in coincident gauge. Hence, $\Gamma\ud{\alpha}{\mu\nu} = 0$ globally and consequently covariant derivatives are turned into partial derivatives, $\nabla_\mu = \partial_\mu$. 

The first step in the Hamiltonian analysis then consists in determining the momentum densities $\tilde{\pi}_0$, $\tilde{\pi}_a$, and $\tilde{\pi}^{ab}$ conjugate to lapse, shift, and intrinsic metric, respectively. The second step is to determine which of the momentum densities can be solved for the velocities $\dot{N}$, $\dot{N}^{a}$, and $\dot{h}_{ab}$. Momenta which are independent of any velocities, i.e., which are of the form $\tilde{\pi} = \tilde{f}(N, N^{a}, h_{ab})$, give rise to \textbf{primary constraints $\boldsymbol{\tilde{C}}$} of the form $\tilde{C} \ce \tilde{\pi} - \tilde{f}$. They put constraints on the physical field configurations and have thus the effect of lowering the number of degrees of freedom.

 In $f(\Q)$, however, one encounters an obstacle in determining primary constraints if the action functional~\eqref{eq:f(Q)Action} is used: Since the momenta are defined by taking variations of $\S_{f(\Q)}$ with respect to $\dot{N}$, $\dot{N}^{a}$ and $\dot{h}_{ab}$, one finds that they are all proportional to $f'(\Q)$. Thus, it is impossible to solve for the velocities without specifying a concrete function $f$.
 
 This obstacle is overcome by introducing an auxiliary scalar field $\phi$ and instead considering the equivalent action functional
 \begin{align}\label{eq:ADMAction}
 	\S[N, N^{a}, h_{ab}, \phi] \ce \int_{\M} \dd^4 x\, \sqrt{|h|}\, N\, \left[f(\phi) - f'(\phi)\left(\phi- \Q\right) \right]\,.
 \end{align}
The field equations derived from this functional are
\begin{align}
	\frac{2}{\sqrt{|g|}} \partial_\alpha \left[\sqrt{|g|}P\ud{\alpha}{\mu\nu} f'(\phi)\right] + f'(\phi)\, q_{\mu\nu} - \frac12 \left[f(\phi) - f'(\phi) \left(\phi - \Q \right)\right] &= 0 \notag\\
	f''(\phi)\left(\phi - \Q\right) &= 0\,.
\end{align}
The first equation, which is obtained from varying the action with respect to the metric variables, has almost the form~\eqref{eq:f(Q)Action}, while the second equation is purely algebraic and admits two solutions:
\begin{align}
	f''(\phi) &= 0 &\text{or} && \phi - \Q = 0\,.
\end{align}
In the first case, we can conclude that $f(\phi) = a\,\phi + b$, for some real constants $a$ and $b$. We can always rescale the action such that $a=1$ and then we find that the first equation reduces precisely to the metric field equation of STEGR plus a cosmological constant $\Lambda\propto b$. The second case is even simpler, since it straightforwardly reproduces the metric field equations of $f(\Q)$ gravity. Thus, we conclude that the field equations are equivalent to the field equations of $f(\Q)$ for any $f$, after we have solved the equations for~$\phi$. The action~\eqref{eq:ADMAction} can thus be regarded as equivalent to the action~\eqref{eq:f(Q)Action}.

The benefit of working with this action is that $\Q$ is ``pulled out'' of $f$, which allows us to study the momenta more easily. The momentum densities computed from the action~\eqref{eq:ADMAction} are given by~\cite{Hu:2022, DAmbrosio:2023,Tomonari:2023}
\begin{align}
	\tilde{\pi}_0 &\ce \frac{\delta \S}{\delta\dot{N}} = 0, & && \tilde{\pi}^{ab} &\ce \frac{\delta \S}{\delta \dot{h}_{ab}} = \sqrt{h}\, f'\left(K_{ab} - K\, h_{ab}\right) \notag\\
	\tilde{\pi}_a &\ce \frac{\delta \S}{\delta \dot{N}^{a}} = - \frac{\sqrt{h}}{N} f'' \partial_a \phi & && \tilde{\pi}_\phi &\ce \frac{\delta \S}{\delta \dot{\phi}} = \frac{\sqrt{h}}{N} f'' \partial_a N^{a}\,,
\end{align}
where $K_{ab}$ and $K$ are the extrinsic curvature and its trace, with the former defined as
\begin{align}
	K_{ab} \ce \frac{1}{2N}\left(\D_{(a}N_{b)} - \dot{h}_{ab}\right)\,.
\end{align}
It is important to note that these momenta have been obtained \textit{after} having performed a series of partial integrations in order to bring the action~\eqref{eq:ADMAction}	into a nicer form, which gives rise to simpler momenta. Performing integrations by parts and dropping boundary terms is allowed, since this does not alter the field equations and, consequently, does not alter the number of degrees of freedom.

Notice that in the special case $f''=0$, which corresponds to STEGR, these momenta reduce precisely to the momenta found in the Hamiltonian analysis of STEGR in~\cite{DAmbrosio:2020c} in the coincident gauge. From now on, we shall always assume $f''\neq 0$, since we are only interested in the degrees of freedom of the modified theory.

From the form of the momenta we can immediately infer that there are five primary constraints. These are
\begin{align}
	\tilde{C} &\ce \tilde{\pi}_0 \approx 0, & && \tilde{C}_a &\ce \tilde{\pi}_a + \frac{\sqrt{h}}{N}f'' \partial_a \phi \approx 0, & && \tilde{C}_\phi &\ce \tilde{\pi}_\phi - \frac{\sqrt{h}}{N}f''\partial_a N^{a} \approx 0\,,
\end{align}
 where $\approx$ stands for ``weak equality'' in the sense of Dirac and Bergmann~\cite{Dirac:1950, DiracBook, Bergmann:1951} (see also~\cite{SundermeyerBook, HenneauxBook, Wipf:1993}). Up to this point, there is complete agreement between~\cite{Hu:2022, DAmbrosio:2023, Tomonari:2023}. \newline
 
 \paragraph{\textbf{Primary Hamiltonian and consistency conditions}}
 The authors of~\cite{Hu:2022, DAmbrosio:2023, Tomonari:2023} also agree on the form of the primary Hamiltonian, which is
 \begin{align}
 	H_{\rm P}(\Sigma_t) = H_0(\Sigma_t) + \int_{\Sigma_t} \dd^3 x\,\left(\lambda^0 \tilde{C}_0 + \lambda^{a} \tilde{C}_a + \lambda^\phi \tilde{C}_\phi\right)\,,
 \end{align}
where $\lambda^0$, $\lambda^{a}$, and $\lambda^\phi$ are Lagrange multipliers which enforce the primary constraints and where $H_0(\Sigma_t)$ is defined as
\begin{align}
	H_0(\Sigma_t) &\ce \int_{\Sigma_t}\dd^3 x\, \left(\dot{N}\tilde{\pi}_0 + \dot{N}^{a} \tilde{\pi}_a + \dot{h}_{ab} \tilde{\pi}^{ab} - \mathcal{L}\right)\,.
\end{align}
Here, $\Sigma_t$ refers to a Cauchy surface, which is simply a leaf in the foliation of $\M$, i.e., a section of $\bbR\times\Sigma$. In yet other words, $\Sigma_t$ corresponds to a $t=$const. spacelike hypersurface. 

The Dirac-Bergmann algorithm demands that the primary constraints be preserved under the time evolution generated by the primary Hamiltonian. This means that the following Poisson brackets have to vanish when the constraints are satisfied:
\begin{align}\label{eq:ConsistencyCondition}
	\{H_{\rm P}, C_I\} = \{H_0, C_I\} + \int_{\Sigma_t}\dd^3 x\,\{C_J, C_I\} \lambda^{J} \overset{!}{\approx} 0\,,
\end{align}
where the Poisson brackets are defined as
\begin{align}
	\{F(\Psi^{a}, \tilde{\pi}_a), G(\Psi^{A}, \tilde{\Pi}_A)\} \ce \int_{\Sigma_t}\dd^3 x\, \left(\frac{\delta F}{\delta \Psi^{A}}\frac{\delta G}{\delta \tilde{\Pi}_A} - \frac{\delta F}{\delta \tilde{\Pi}_A}\frac{\delta G}{\delta \Psi^{A}}\right)\,,
\end{align}
for some fields $\Psi^{A}$ and their conjugate momentum densities $\tilde{\Pi}_A$. Equation~\eqref{eq:ConsistencyCondition}, also called \textbf{consistency condition}, can give rise to \textbf{secondary constraints}. That is, it can put additional constraints on the physical field configurations and thus reduce the number of degrees of freedom even further. It is also possible that it determines the Lagrange multipliers. This is precisely the point where differences in the works of~\cite{Hu:2022,DAmbrosio:2023, Tomonari:2023} start to emerge. 

In~\cite{Hu:2022} it was argued that~\eqref{eq:ConsistencyCondition} leads to one secondary constraint and a system of linear equations for the Lagrange multipliers. It was further argued that these equations possess unique solutions, hence preventing the appearance of further constraints. It thus follows that there are $22-6 = 16$ phase space degrees of freedom or, equivalently, eight configuration space degrees of freedom.

This conclusion was challenged by~\cite{DAmbrosio:2023, Tomonari:2023}. It was first realised in~\cite{DAmbrosio:2023}	that the analysis of~\cite{Hu:2022}	contains an error. Namely, the equations for the Lagrange multipliers are first order partial differential equations (PDEs), rather than linear algebraic equations. This fact was overlooked in~\cite{Hu:2022} and it drastically changes the situation. First of all, the original Dirac-Bergmann algorithm for counting degrees of freedom does \textit{not} foresee the possibility that the Lagrange multipliers are constrained by PDEs. It is silently assumed that the equations are always linear algebraic equations.

That PDEs can arise has been observed also by other authors (see in particular~\cite{SundermeyerBook}) and it is understood that this problem is due to the presence of spatial derivatives of field variables in the primary constraints. The partial integrations necessary for computing the Poisson brackets in the consistency conditions~\eqref{eq:ConsistencyCondition} can  move partial derivatives from the field variables onto the Lagrange multipliers. Unfortunately, the issue has received relatively little attention and no general procedure is known for how to deal with this scenario. In certain simple cases it is possible to solve the PDEs and to reach sensible conclusions from a modified version of the Dirac-Bergmann algorithm. But the general case is far from under control.

Moreover, it was shown in~\cite{DAmbrosio:2023} that the PDEs for the Lagrange multipliers are not all independent, thus leading potentially to further complications. Several other issues were pointed out in the same work, which is why a different route was ultimately selected to give at least an upper bound on the degrees of freedom. Before discussing these issues and the upper bound in more detail, we turn our attention to~\cite{Tomonari:2023}. The authors of~\cite{Tomonari:2023} propose a method to avoid having to deal with PDEs for the Lagrange multipliers. We quote directly from their text:
\begin{quotation}
	\textit{``For some field $A(x)$ on a $(n+1)$-dimensional spacetime, the term $\sqrt{h}A(x)\partial^{(x)}_I\delta^{(n)}(\vec{x} - \vec{y})$, where $I$ runs from $1$ to the dimension of the hypersurface $n$, in PB-algebras can be neglected by setting properly spatial boundary conditions of $A(x)$ in the variational principle, where $h$ is the determinant of the metric of the $n$-dimensional hypersurface.''}
\end{quotation}
The hypersurface the authors refer to is $\Sigma_t$ and the term $\sqrt{h}A(x)\partial^{(x)_I}\delta^{(n)}(\vec{x} - \vec{y})$ has the generic form of the terms which lead to the aforementioned issue. That is, terms of this form lead to PDEs for the Lagrange multipliers. By dropping all terms of this form from the constraint algebra, the authors find indeed a linear system of equations for the Lagrange multipliers. Their analysis leads them to uncover three secondary and two tertiary constraints. They also conclude that all constraints are second class, eventually leading to $\frac12(22-5-3-2) = 6$ degrees of freedom for $f(\Q)$. 

However, as we mentioned above, this procedure is not beyond doubt. Shortly before the quoted passage, the authors of~\cite{Tomonari:2023} assert that
\begin{quotation}
	\textit{``[...] when taking into account that the spatially boundary terms can always be neglected by imposing appropriate spatial boundary conditions in the variational principle and it never affects the dynamics (time evolution).''}
\end{quotation}
It is correct that, given an action functional, one is allowed to drop or neglect boundary terms because such terms do not change the field equations. In this sense, boundary terms do indeed not affect the dynamics. However, it is \textit{not} true that spatial boundary conditions in the variational principle do not affect the dynamics. In fact, boundary conditions constrain the solution space of a theory! This can readily be seen from the following example: Take one of the actions of the trinity and derive the field equations without any further assumptions. One obtains Einstein's field equations which, in particular and among many others, admit the Schwarzschild and FLRW spacetimes as solutions. Now, take the same action but demand that the fields are asymptotically flat. This is a boundary condition and it has the effect of eliminating certain solutions. The equations one obtains are still Einstein's field equations, but the FLRW spacetime is no longer in the solution space because it does not satisfy the boundary condition (i.e., it is not asymptotically flat). Thus, the solution space has been changed by the imposition of boundary conditions.

Moreover, the term $\sqrt{h} A(x) \partial^{(x)}_I\delta^{(n)}(\vec{x} - \vec{y})$ is being dropped from the  \textit{Poisson bracket algebra}, rather than from the action. It is not clear that such a modification does not affect the dynamics. In particular, since the integrals in questions are integrals over Cauchy surfaces $\Sigma_t$, rather than actual boundary integrals. There is nothing which prevents a Cauchy surface to cross through the bulk of a spacetime through regions of intense field strength. In other words, Cauchy surfaces have nothing to do with the boundary surfaces of spacetimes, where fields are generically assumed to be weak and thus negligible.

In conclusion, the approach of~\cite{Tomonari:2023} does indeed allow one to carry out the Dirac-Bergmann analysis of $f(\Q)$ gravity to completion and count degrees of freedom. However, the method used to achieve this feat is not beyond all doubts.\newline

\paragraph{\textbf{Issues of the Dirac-Bergmann algorithm}}
We have mentioned issues with the Dirac-Bergmann algorithm already several times. Specifically, what was point out in~\cite{DAmbrosio:2023} is that the standard algorithm does not foresee consistency conditions involving PDEs for the Lagrange multipliers. Rather, it only foresees systems of linear equations of the form
\begin{align}
	M\, \vec{\lambda} + \vec{v} \overset{!}{\approx} 0\,,
\end{align}
where $\vec{\lambda}$ contains all $r$ Lagrange multipliers coming from $r$ primary constraints, $\vec{v}$ is a vector built from the fields, their conjugate momenta, and their derivatives, and $M$ is a $r\times r$ matrix. The symbol $\overset{!}{\approx}$ means that this equations has to be imposed and that it only has to hold if the primary constraints hold. Three scenarios can now emerge\footnote{For more details on the Hamiltonian analysis of constrained systems and the Dirac-Bergmann algorithm see, for instance, \cite{SundermeyerBook, HenneauxBook, Wipf:1993}. See also the more recent~\cite{Blixt:2020, DAmbrosio:2023}.}:
\begin{enumerate}
	\item If $\det{M}\not\approx 0$, the matrix $M$ is invertible and we can solve for all Lagrange multipliers, $\vec{\lambda} = -M^{-1}\vec{v}$.
	\item If $\det{M} \approx 0$, it is not possible to solve for all Lagrange multipliers. If $\text{rank}(M) = m < r$, there are $r-m$ vectors $\vec{u}_D$, with $D\in\{1,\dots, r-m\}$ which are null vectors of $M$. That is, these vectors satisfy $M\vec{u}_D = 0$. One can show that one can consistently solve for some of the Lagrange multipliers if and only if $\vec{u}_D^\transpose \vec{v} \approx 0$. If this last equation does not hold, one has to impose it. This leads to additional, so-called secondary constraints. 
	\item If $\det{M} \approx 0$ and $\vec{u}_D^\transpose \vec{v} \approx 0$, it is possible that the consistency condition is trivially satisfied or that it leads to secondary constraints.
\end{enumerate}
It should be noted that in the cases 2. and 3., some of the Lagrange multipliers inevitably remain \textit{undetermined}. Since these multipliers appear in the primary Hamiltonian, which generates the dynamics, it means that there is some indeterminacy in the time evolution of the system. This indeterminacy is well-understood to be related to gauge symmetries. Thus, because of this connection to gauge symmetry, it is not alarming when the primary Hamiltonian depends on some arbitrary fields.

This brings us now to the case of PDEs for Lagrange multipliers. These PDEs can arise when the constraints contain spatial derivatives of field variables. Because one has to perform an integration by parts in order to compute the second Poisson bracket in~\eqref{eq:ConsistencyCondition}, one ends up with terms of the form $\partial \lambda$. We emphasize that the presence of partial derivatives in the constraints is only a necessary but not a sufficient condition. After all, also the constraints of electromagnetism and GR possess spatial derivatives, but they do not cause any problems. This has also been discussed in~\cite{DAmbrosio:2023}	.

However, if it happens that the partial derivative has been moved onto the Lagrange multiplier, the system of PDEs has generically the form
\begin{align}
	\sum_{i=1}^{d}M^{(i)}\partial_i\vec{\lambda} + N\vec{\lambda} + \vec{v} \overset{!}{\approx} 0\,.
\end{align}
We have assumed that there are $d$ spatial dimensions and consequently there are $d$ matrices $M^{(i)}$ of dimensions $r\times r$ which multiply the $d$ different first order spatial derivatives $\partial_i \vec{\lambda}$. We have also introduced a $r\times r$ matrix $N$ and a $r$-dimensional vector $\vec{v}$. The $r$ Lagrange multipliers $\vec{\lambda}$ all depend on the $d$ spatial coordinates and time.

As is well-known, in order to obtain a unique solution to a PDE one has to impose boundary conditions or initial value conditions. But this raises the question: How do these initial value or boundary conditions affect the primary Hamiltonian? To be more explicit: We are completely free in choosing these conditions. But no matter what we choose, this choice will affect the primary Hamiltonian and it will depend on the field values we arbitrarily chose for $\vec{\lambda}$. In turn, these field values will show up in the time evolution of the system. Is there a relation to gauge transformation, as there is one in the standard case discussed above? 

If there is, it is not completely clear how it will manifest. Observe that there is a difference between $\vec{\lambda}$ not being completely determined by the linear equations and $\vec{\lambda}$ depending on arbitrary choices for its initial values or boundary values: In the first case, we are forced to introduce arbitrary fields which depend on space and time. In the second case, we arbitrarily fix for instance the $x^{1}$ axis as ``initial surface'' and specify initial values on that surface. This amounts to specifying functions of time and $d-1$ spatial coordinates, since one coordinate is fixed. Albeit, the fixation of $x^{1}$ was arbitrary.

Nevertheless, we are confronted with open questions and the answers are not clear. This means that we have no \textit{reliable} way of dealing with these PDEs such that we can count the degrees of freedom in a way we can trust.

There is a further problem, also brought to light through the analysis of~\cite{DAmbrosio:2023}. Namely, the PDEs for the Lagrange multipliers that emerge in $f(\Q)$ do \textit{not} give rise to a well-posed initial value formulation. This means that the PDEs are under-determined. Or, in yet other words, even if we prescribe initial values for $\vec{\lambda}$, it is not possible to find a unique solution. We can only determine some of the components of $\vec{\lambda}$ and they will depend on the un-determined components. 

It is known that this happens in gauge theories and that this issue is related to the freedom of performing gauge transformations. For instance, in electromagnetism formulated in terms of a vector potential $A^\mu$, the field equations are under-determined. This is tantamount to saying that the initial value problem is not well-posed. The resolution is to realize that the field equations determine all components of $A^\mu$, except one. Thus, by imposing a gauge fixing, this issue is resolved and one obtains a unique solution. Furthermore, as is well-known, this arbitrary gauge fixing does not affect physical observables.

However, what does this under-determination of the PDEs mean in the context of Lagrange multipliers? Is there a connection to gauge symmetries of the theory? All these questions deserve more attention and a detailed analysis, so that we can trust the results obtained from a modified version of the Dirac-Bergmann algorithm.\newline

\paragraph{\textbf{Upper and lower bound on the degrees of freedom}}
Given the obstacles mentioned above, which emerge from applying the Dirac-Bergmann algorithm to $f(\Q)$ gravity, the authors of~\cite{DAmbrosio:2023} opted for a different approach. Using the so-called \textbf{kinetic matrix}, it was shown that $f(\Q)$ propagates \textit{at most} seven degrees of freedom. Together with the four degrees of freedom found through  cosmological perturbation theory in~\cite{BeltranJimenez:2019}, we have a clear lower and upper bound.\newline
The kinetic matrix approach sidesteps the issues discussed so far since it is independent of the Hamiltonian analysis and it is directly concerned with the field equations. The basic idea can easily be explained with a simple example: Consider a field theory in $1+1$ dimensions with second order field equations. Let's say that the field $\Psi$ in question has two components, $\Psi = (\Psi_1, \Psi_2)$, which are functions of the coordinates $x^\mu = (x^0, x^1)$. The coordinate $x^{0}$ plays the role of a time coordinate, while $x^{1}$ is the spatial coordinate. Then, the second order field equations can be written as
\begin{align}
	&\underbrace{\begin{pmatrix}
		\mathcal{K}_{11} & \mathcal{K}_{12} \\
		\mathcal{K}_{21} & \mathcal{K}_{22}
	\end{pmatrix}}_{\ec \mathcal{K}}
	\partial^2_0
	\begin{pmatrix}
		\Psi_1 \\
		\Psi_2
	\end{pmatrix}
	+
	\underbrace{\begin{pmatrix}
		\M^{(1)}_{11} & \M^{(1)}_{12} \\
		\M^{(1)}_{21} & \M^{(1)}_{22}
	\end{pmatrix}}_{\ec \M^{(1)}}
	\partial_0 \partial_1
	\begin{pmatrix}
		\Psi_1 \\
		\Psi_2
	\end{pmatrix}
	+
	\underbrace{\begin{pmatrix}
		\mathcal{P}^{(11)}_{11} & \mathcal{P}^{(11)}_{12} \\
		\mathcal{P}^{(11)}_{21} & \mathcal{P}^{(11)}_{22}
	\end{pmatrix}}_{\ec \mathcal{P}^{(11)}}
	\partial^2_1
	\begin{pmatrix}
		\Psi_1 \\
		\Psi_2
	\end{pmatrix}\notag\\
	& + \text{ lower order derivatives }
	= 0\,.
\end{align}
We have introduced three different matrices which multiply the three different second order derivatives, $\mathcal{K}$, $\M^{(1)}$, and $\mathcal{P}^{(11)}$. The notation will become clear later on.

Now, what does it mean to solve this system of PDEs? First of all, since the system is second order, we have to prescribe two initial value conditions, if we hope to find a unique solution. These conditions are
\begin{align}
	\left.\Psi\right|_{t=t_0} &= F(x^{1}) & \text{and} && \left.\partial_0 \Psi\right|_{t=t_0} &= G(x^{1})\,.
\end{align}
In other words, we prescribe $\Psi$ on a $t=t_0$ hypersurface and we prescribe what its time derivative is on that surface. Observe that if we evaluate the above equation on that particular surface, we know every term, except the first one. In fact, we find
\begin{align}
	&\left.\mathcal{K}\, \partial_0^2\Psi\right|_{t=t_0} + \M^{(1)}\partial_1 G(x^{1}) + \mathcal{P}^{(11)}\partial^2_1 F(x^{1}) + \text{ other terms which we know on $t=t_0$ } = 0\,.
\end{align}
Notice that since $F(x^{1})$ and $G(x^{1})$ are \textit{known} functions of $x^{1}$, we also know what their derivatives with respect to $x^{1}$ are. What we do not know, is what $\partial^2_2\Psi$ equals to on the $t=t_0$ surface. That is where the field equations come into play. We can find out what $\partial^2_2\Psi$ is, if we can solve the above equations for the second order time derivatives. That is, if we can write
\begin{align}
	\left.\partial^2_0 \Psi\right|_{t=t_0} = -\mathcal{K}^{-1}\left(\M^{(1)}\partial_1 G(x^{1}) + \mathcal{P}^{(11)}\partial^2_1 F(x^{1}) + \text{ the lower order terms}\right)\,,
\end{align}
where $\mathcal{K}^{-1}$ is the inverse of $\mathcal{K}$. Hence, if we can invert $\mathcal{K}$, we can formally integrate the PDE and find out what $\Psi$ is away from the $t=t_0$ surface (see also~\cite{DAmbrosio:2023} for a more technical and detailed explanation of this point). What happens if we can \textit{not} invert the matrix $\mathcal{K}$? To answer the question, consider the following case:
\begin{align}
	&\begin{pmatrix}
		\mathcal{K}_{11} & \mathcal{K}_{12} \\
		0 & 0
	\end{pmatrix} \partial^2_0 \Psi
	+
	\M^{(1)}\partial_0 \partial_1\Psi
	+
	\mathcal{P}^{(11)}\partial^2_1\Psi + \text{ lower order derivatives } = 0\,.
\end{align}
Clearly, $\mathcal{K}$ has only rank one and is therefore not invertible. Observe that this has two implications:
\begin{enumerate}
	\item If we explicitly write out the vector-matrix product, we see that the second equation has no second order time derivatives. It is thus just a constraint equation, rather than a dynamical equation.
	\item The first equation can still be solved for, say, $\partial^2_0 \Psi_1$, but $\partial^2_0 \Psi_2$ then appears on the right hand side. Since there is no equation which determines $\partial^2_0 \Psi_2$, we have to prescribe it by hand. Otherwise we cannot integrate the equation for $\partial^2_0 \Psi_1$. This is what generically happens in gauge theories.
\end{enumerate}
We learn an important lesson from this simple example: Whether a given second order PDE can be solved or not is determined by the matrix $\mathcal{K}$ which multiplies the second order time derivatives. We can generalize this insight in the following way.

Let spacetime be $d+1$ dimensional and let $\Psi$ be a vector which contains the $n$ components of a tensor field (that could be a vector, or a metric, or any other tensor). Then we can write the second order PDE for the field in question as
\begin{align}
	\mathcal{K}\, \partial^2_0 \Psi + \sum_{i=1}^{d} \M^{(i)}\partial_0\partial_i \Psi + \sum_{i = 1, i\leq j}^{d}\sum_{j=1}^{d}\mathcal{P}^{(ij)}\partial_i \partial_j \Psi + \text{ lower order terms } = 0 \,,
\end{align}
where we have introduced the $n\times n$ \textbf{kinetic matrix $\boldsymbol{\mathcal{K}}$}, $d$ so-called \textbf{mixing matrices $\M^{(i)}$}, each of dimension $n\times n$, and $\frac{d(d+1)}{2}$ \textbf{potential matrices $\boldsymbol{\mathcal{P}^{(ij)}}$}, also each of dimension $n\times n$. If the kinetic matrix is invertible, we obtain a unique solution for the PDE. However, if $\mathcal{K}$ is not invertible, we find constraint equations. From our simple example it is clear that the number of constraint equation is the same as the number of rows of $\mathcal{K}$ which are zero, in some sense. Of course, a matrix $\mathcal{K}$ which is degenerate does not always have rows filled with zero. Rather, it has rows which are linear combinations of other rows. Thus, the mathematically precise statement is this\footnote{Further mathematical details can be found in~\cite{MiersemannBook, DAmbrosio:2023} and in the Appendix of~\cite{DAmbrosio:2022}, which also provides ample illustrations and examples.}:
\begin{align}
	 \text{If rank}(\mathcal{K}) = r \leq n\quad\Longrightarrow\quad \text{There are $n-r$ constraint equations.}
\end{align}
It thus follows that by determining the rank of the kinetic matrix, we can infer how many constraints there are \textit{at least}. There can be more constraints than just the $n-r$ which follow from the rank of $\mathcal{K}$ since integrability conditions can occur. Given that each constraint reduces the number of degrees of freedom by one, we finally reach the conclusion
\begin{align}
	\text{If rank}(\mathcal{K}) = r \leq n\quad\Longrightarrow\quad \text{There are at most $r$ degrees of freedom.}
\end{align}
In~\cite{DAmbrosio:2023}, this insight was used to show that the number of degrees of freedom in $f(\Q)$ is at most seven. The argument is as follows: The basic variables to consider are the ten metric components $g_{\mu\nu}$ and the four functions $\xi^\alpha$, which parametrize the flat and torsionless connection. Furthermore, if the metric field equations are satisfied, then the Bianchi identities imply that the connection field equations are automatically satisfied as well. Thus, we can completely focus on the metric field equations. If we work in coincident gauge, we remove the four functions $\xi^\alpha$ from our considerations without accidentally killing degrees of freedom. In fact, all degrees of freedom are now encoded in the ten metric components, whose dynamics is described by the metric field equations. Thus, from the originally $10+4$ potential degrees of freedom, we are left with only ten.

It was then shown that the rank of the kinetic matrix of the metric field equations is seven, provided that $f''\neq 0$. If $f''=0$, the rank is six, which had to be expected since the Einstein equations contain $10-6 = 4$ constraints. This is a nice consistency check.

As a final remark, we point out that the kinetic matrix approach can in principle also be used to figure out the precise number of degrees of freedom. This involves also considerations regarding the mixing matrices and the potential matrices with highly involved computations. For more details on this outlook we refer the reader to~\cite{DAmbrosio:2023}.

\newpage
\asection{7}{Summary}\label{sec:Summary}
Gravitational phenomena arise from curved spacetime, a concept made possible by the equivalence principle. This implies that gravity is independent of matter type. Within the framework of geometry, curvature is just one aspect of a manifold's affine properties. In addition to curvature, there are two other fundamental objects associated with the connection of a metric space: torsion and non-metricity. In standard General Relativity following Einstein, both non-metricity and torsion are absent. Embracing the geometric nature of gravity as advocated by the equivalence principle prompts us to explore different ways to represent gravity. In one equivalent description of General Relativity, we envision a flat spacetime with a metric but an asymmetric connection, where gravity is solely attributed to torsion. Alternatively, we can construct a third equivalent representation of GR on a flat spacetime without torsion, attributing gravity to non-metricity. Thus, the same fundamental physical theory, GR, can be articulated through the Einstein-Hilbert action, the Teleparallel Equivalent of GR action, or the Symmetric Teleparallel Equivalent of GR action \cite{Heisenberg:2018,BeltranJimenez:2019c}.

The fundamental foundation of these geometric interpretations paves the way for innovative approaches to modified gravity. These equivalent descriptions of General Relativity involving curvature, torsion, and non-metricity provide diverse starting points for modified gravity theories when scalar quantities are transformed into arbitrary functions. It's worth noting that quadratic non-metricity and torsion Lagrangian with detuned arbitrary 5 and 3 parameters, respectively, can also be considered, albeit with anticipated complexities. In this review, our primary focus lay on $f(\Q)$ theories \cite{BeltranJimenez:2017b}.

We began by establishing the foundational elements of geometry. Starting with the basic manifold, we incorporated coordinates, points, and curves. Tensor fields, including scalars and vectors, were introduced on this manifold. To facilitate the comparison of vector fields at different points, we introduced the affine connection, delving into its general properties and the associated tensor quantities: curvature and torsion tensors. To incorporate the concept of distance, we introduced the metric, which in turn allowed us to define the non-metricity tensor. With these components in place, we were well-prepared to delve into the core principles of General Relativity. We've clearly demonstrated that the theory of General Relativity can be formulated in three distinct ways: as a curvature theory, a torsion theory, or a non-metricity theory. We've examined the key distinctions, addressed subtle nuances, and explored the consistent coupling of matter fields within these frameworks. In doing so, we've identified cases where the minimal coupling principle proves inadequate.

Next, we examined strategies for departing from the principles of General Relativity in a consistent manner. We explored two complementary approaches: one involving generic quadratic Lagrangians with arbitrary parameters, and the other by transforming GR scalars into nonlinear functions. These approaches led us to derive various theories of modified gravity. Given our primary focus on $f(Q)$ theories, we provided an overview of the fundamental characteristics of various modifications before returning our attention to $f(\Q)$ theories. Specifically, we introduced the defining Lagrangian, derived the corresponding field equations, and delved into discussions regarding its symmetries and Bianchi identities. Having gained a solid grasp of the overarching principles of the covariant theory, our focus shifted towards practical applications in cosmology and astrophysics. We specifically examined both cosmological and spherically symmetric backgrounds, utilizing symmetry reduction principles to establish the necessary conditions for the metric and the connection that align with the background symmetries. This systematic approach enabled us to explore the potential derivation of novel cosmological and black hole solutions within the framework of $f(\Q)$ theories.\\

Our motto is Qravity: Gravity with $\Q$. We firmly believe that the intricate structure inherent in the geometric framework of gravity can unlock fresh and captivating perspectives, leading us into uncharted realms and confronting the challenges of conventional formulations. Let us wholeheartedly embrace this captivating new geometry.


\section*{Acknowledgements}
LH is supported by funding from the European Research Council (ERC) under the European Unions Horizon 2020 research and innovation programme grant agreement No 801781. LH further acknowledges support from the Deutsche Forschungsgemeinschaft (DFG, German Research Foundation) under Germany's Excellence Strategy EXC 2181/1 - 390900948 (the Heidelberg STRUCTURES Excellence Cluster).

\newpage
\bibliographystyle{utcaps}
\bibliography{Bibliography}
\end{document}